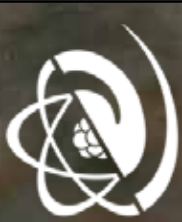

# ORIGINS
Space Telescope

From first stars to life

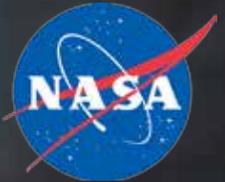

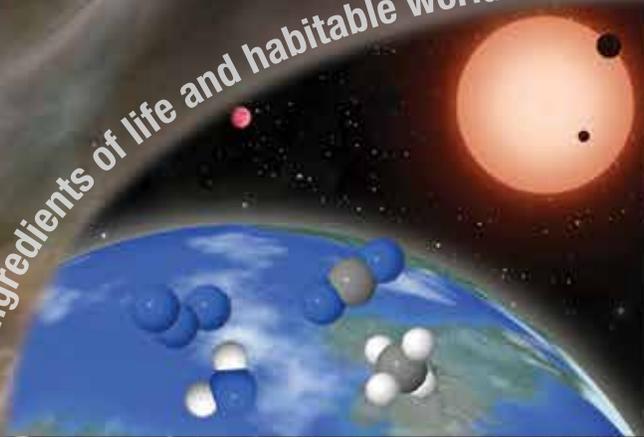

Ingredients of life and habitable worlds

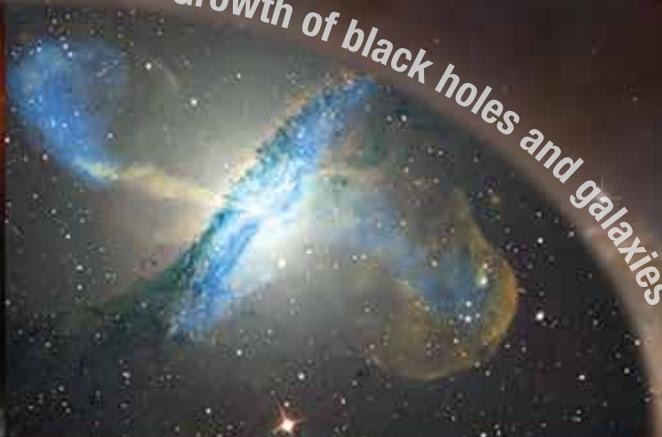

Growth of black holes and galaxies

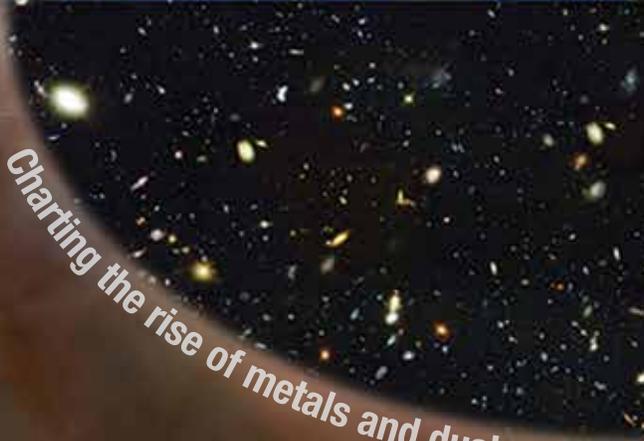

Charting the rise of metals and dust

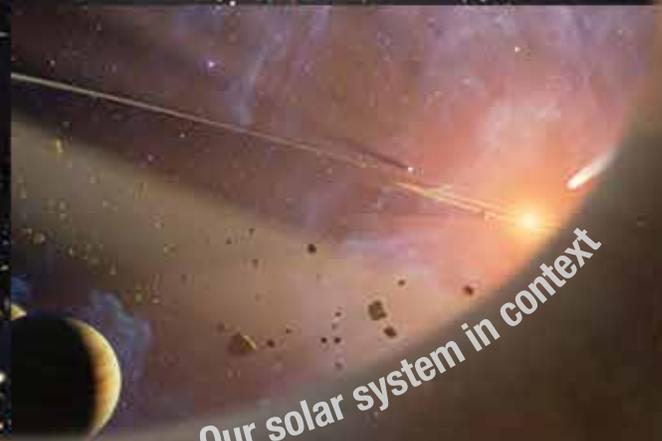

Our solar system in context

*2018 Interim Report*

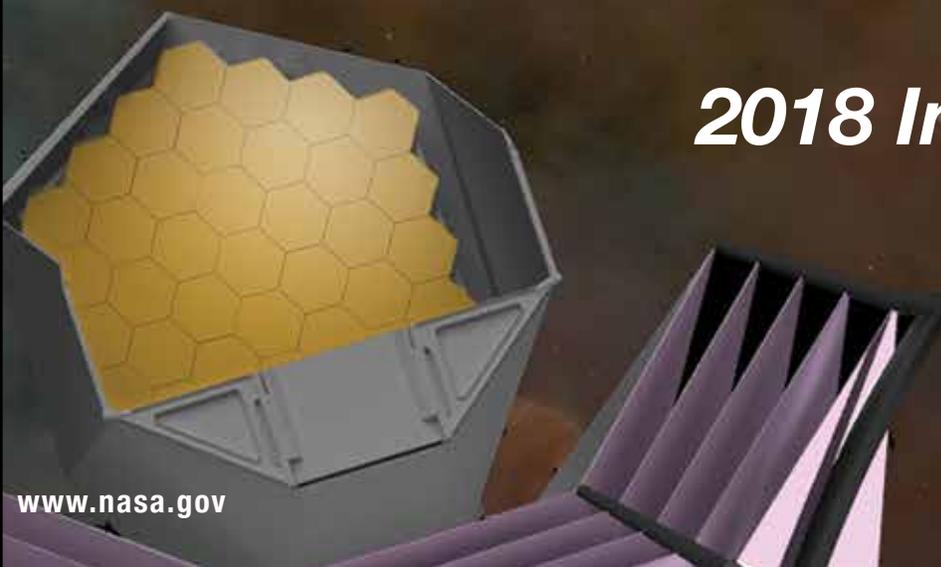

# OST Mission Concept 1*

## Observatory

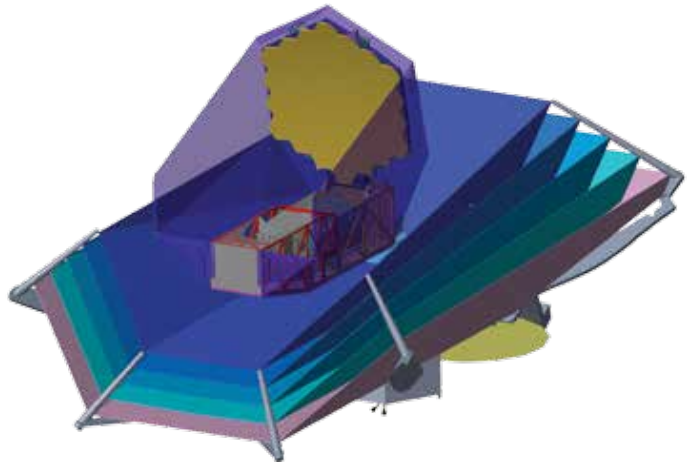

- 9.1 m off-axis primary mirror
- Cold (4K) telescope
- Wavelengths 5 – 660 µm
- 5 science instruments
- Launch 2030s
- Mission operations at Sun-Earth L2
- Data rate: 348 Mb/s
- 5 year lifetime, 10 year goal

\* OST is an evolving concept for the Far-IR Surveyor mission in NASA's visionary astrophysics roadmap. Stay tuned for Concept 2, coming fall of 2018.

## Instruments

| | | Wavelength (µm) | Observing Modes |
|---|---|---|---|
| **MISC** | Mid-Infrared Imager, Spectrometer, Coronagraph | 5-38 | • Imaging, spectroscopy<br>• Coronagraphy (10⁻⁶ contrast)<br>• Transit Spectrometer < 10 ppm stability) |
| **MRSS** | Medium Resolution Survey Spectrometer - IFU | 30-660 | • Multi-band Spectroscopy |
| **FIP** | Far-Infrared Imager and Polarimeter | 40, 80, 120, 240 | • Broadband imaging<br>• Field of view: 2.5'x5', 7.5'x15'<br>• Differential polarimetric imaging |
| **HERO** | Heterodyne Receiver for OST | 63-66 , 111-610 | • Multi-beam spectroscopy |
| **HRS** | High Resolution Spectrometer | 25-200 | • Spectroscopy |

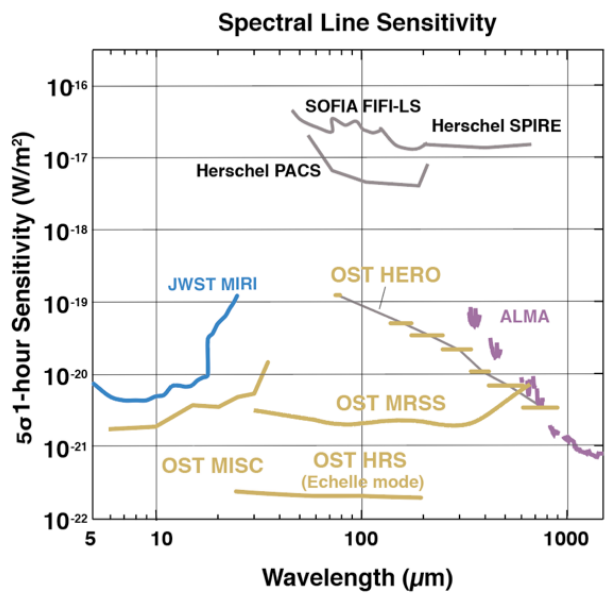

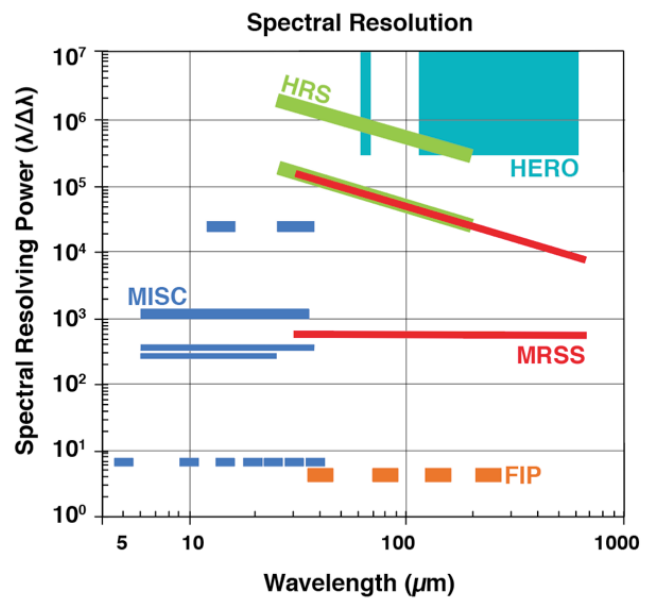

Leveraging improvements in detector technology and a large, cold primary mirror, OST will offer two to four order of magnitude improvement in sensitivity over Herschel. OST will bring arc-second imaging and unprecedented spectroscopic capabilities to the infrared universe.

# Contents

Appendices

## Executive Summary

The Origins Space Telescope (OST) will transform our understanding of the Universe, from formation of the earliest galaxies to the creation of habitable worlds. OST's Mission Concept 1 features a cold (4 K) large-aperture telescope (9.1 m) and five powerful instruments providing imaging and spectroscopy over a wavelength range of 5 to 660 µm. OST's dramatic increase in sensitivity over previous missions ensures breakthrough discoveries in a broad range of astrophysics disciplines.

OST is a mission concept study for the large Far-Infrared Surveyor mission described in NASA's *Astrophysics Roadmap, Enduring Quests, Daring Visions*. The mission's short wavelength coverage starts at 5 µm to support exoplanet science, and hence the mission study was renamed with 'Origins' to reflect this science focus. The OST team's goal was to design a scientifically-compelling mission that is executable in the 2030s. OST builds on a rich tradition of infrared missions, including IRAS, Spitzer, WISE, Herschel, and, JWST, which also employed improved detector technologies and increased aperture sizes to open scientific discovery space. This report has two intertwined components: a scientific case that motivates the observatory design, and an engineering study with enough detail to demonstrate feasibility. OST, operating in a wavelength range of 5 to 660 µm, is conceived as a general-purpose observatory available to all astronomers via peer reviewed proposals, with its data being stored in a public archive.

Enabled by 4 K optics, OST will far exceed previous missions' sensitivity, by up to 4 orders of magnitude in the far-infrared, and provide groundbreaking new capabilities. This interim report describes the science and engineering for OST Mission Concept 1. The final report will also present a Mission Concept 2 that further optimizes the engineering design to support the science in an efficient and powerful observatory. The OST study is a work in progress and this interim report is but a snapshot. The OST team invites all reviewers to participate in OST's ongoing development by attending the team's telecons and visiting the OST websites at: http://origins.ipac.caltech.edu/ and https://asd.gsfc.nasa.gov/firs/events.

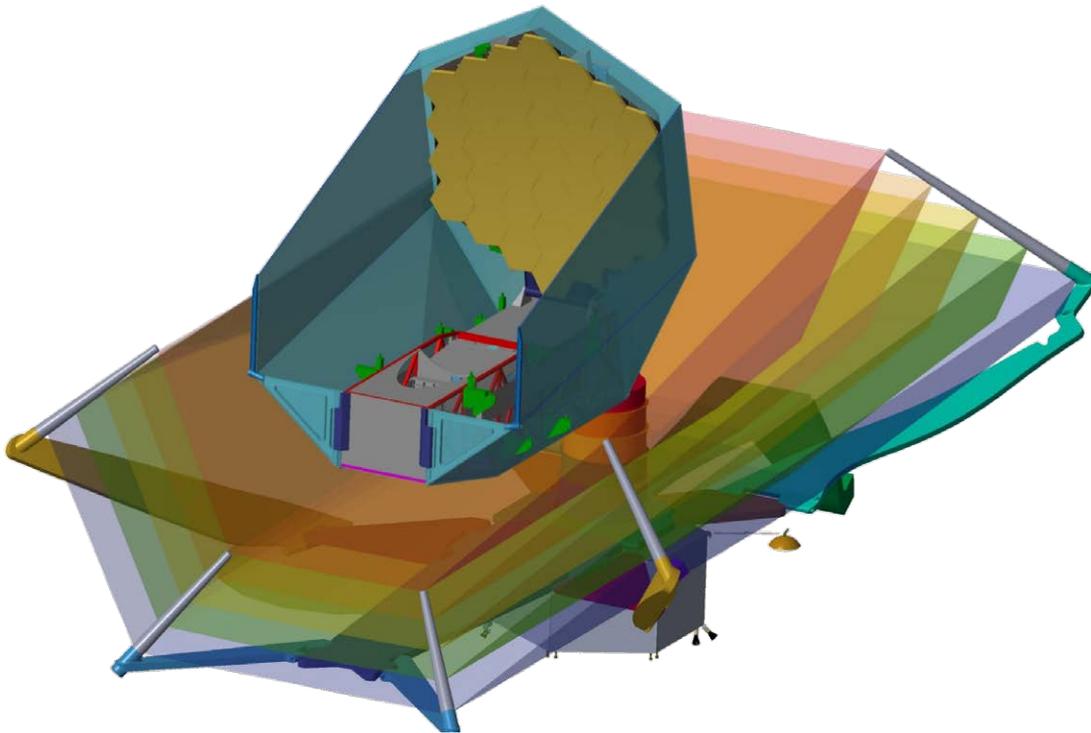

**Figure ES-1:** The OST observatory design is based on proven technology, as well as advances that will be readily available in the mission's 2030 time frame.

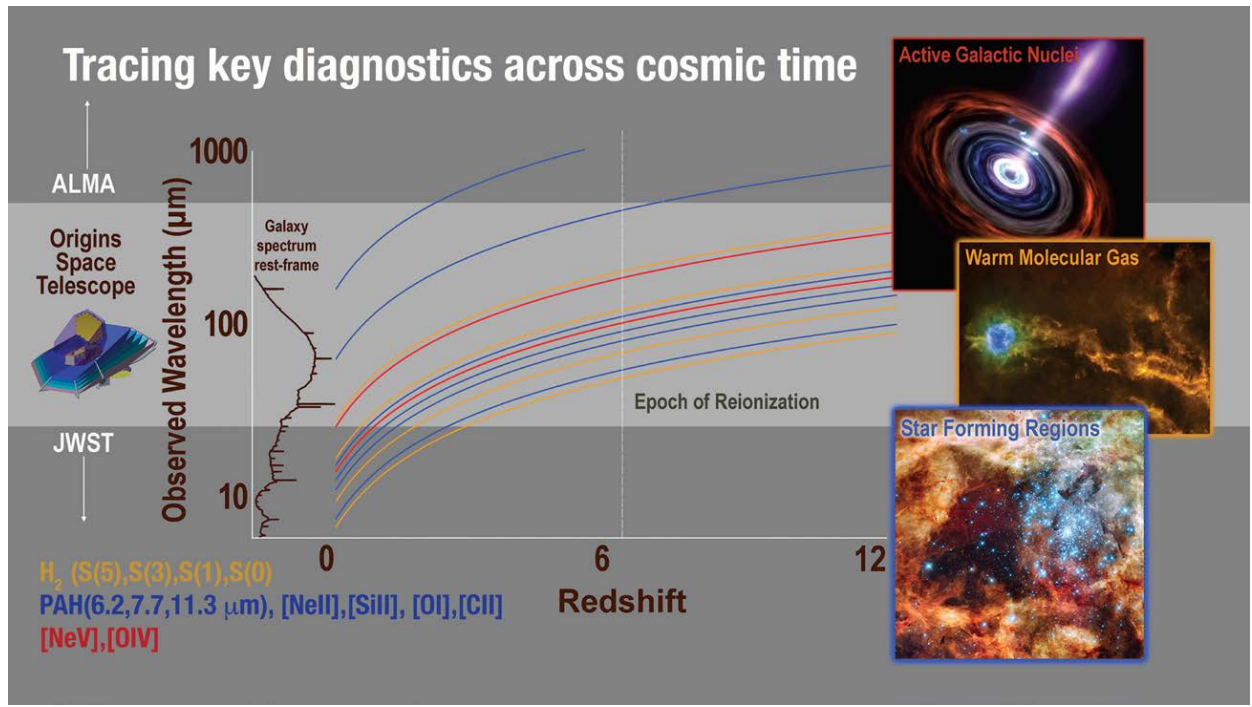

**Figure ES-2:** The OST observatory enables the tracing of key diagnostics across cosmic time.

In Mission Concept 1, OST's 9.1-meter diameter telescope is cooled to 4 K and equipped with next generation detectors, and thus limited only by the natural background across its 5 to 660 μm operating range. It launches in an SLS 8.4 m faring, and will orbit at Sun-Earth L2. OST has a 5-year lifetime, 10 year goal, and like many othe NASA missions, could be serviceable to enable an even longer mission. OST Concept 1 has five powerful instruments: Medium Resolution Survey Spectrometer (MRSS) at 30-660 μm, High Resolution Spectrometer (HRS) at 25-200 μm, Heterodyne Receiver for OST (HERO) at 63-66, 111-641 μm, Far Infrared Imager/Polarimeter (FIP) at 40, 80, 120, 240 μm, and Mid-infrared Imager, Spectrometer, Coronagraph (MISC) at 5-38 μm. The observatory has the agility to conduct large surveys and the pointing stability to support pointed observations. OST Concept 1's tremendous improvement in sensitivity for imaging, spectroscopy, coronagraphy, and transit spectroscopy for exoplanets enables observations no other methodology or any other planned mission can match. OST's wide discovery space will spur new science yet-to-be-conceived.

OST addresses the three key questions from NASA's Roadmap: How does our Universe work?, How did we get here?, and, Are we alone? The OST mission will investigate the formation and evolution of galaxies over cosmic time by observing simple molecules, dust, polycyclic aromatic hydrocarbons (PAHs), and atoms in the interstellar medium that fuels star formation. Using molecular and atomic lines that cannot be seen from the ground with the Atacama Large Millimeter/submillimeter Array (ALMA), and an aperture diameter and angular resolution 2.6 times better than the European Space Agency's (ESA) Herschel Space Observatory (Herschel), OST is designed to map specific species, such as $H_2O$, HD, O, Ne, and C+, that trace cooling gas clouds that collapse into stars and the feedback from those stars on the interstellar medium. Penetrating the extreme dust extinctions in the densest regions, OST observes the formation of galactic nuclei and black holes through tracers, molecules, and fine structure lines that distinguish star formation from accretion disks. OST will measure water vapor and ice that traces the trail of water from the interstellar medium, its location in developing planetary systems, and its delivery to habitable worlds. OST observes the small bodies of our Solar System, the comets, asteroids, and satellites left over from the formation of the Solar System, measuring the chemical abundances and isotope ratios that reveal fractionation and enrichment processes. With a transit spectroscopy channel that measures thermal emission, as well as transmission spectra, OST can reach temperate, terrestrial planets orbiting M dwarfs, as well as any larger transiting planet. OST has the capability to reveal the molecular constituents of exoplanet atmospheres, with excellent sensitivity to biosignature molecules ($O_3$, $CH_4$) and bioindicator molecules ($H_2O$, $CO_2$) – important signs of life on potentially-habitable planets.

The five OST instruments could be built today with prototype detectors. Current detectors match the required sensitivity for imaging, but without the desired number of pixels. Far-infrared spectroscopy requires an increase in sensitivity. Mid-infrared detectors need stability improvements for transit spectroscopy. Detector development programs, in progress and proposed, will improve the detectors' sensitivity and stability and expand them to the desired large formats while ensuring resource requirements for mass, volume, power, cooling, cables, and data are met. The technology required to cool the telescope is achievable and based on heritage work for the JWST/Mid-Infrared Instrument (MIRI) cryocooler. The OST Technology Roadmap shows the plan for advancing all technologies to TRL 5 by 2025, with the detectors as the long lead items. The OST project office could manage these developments using JWST's model of sponsoring competitions for specific objectives. The wide range of proposed instrument capabilities is ideal for a mission planned for operation 20 to 30 years in the future, when critical scientific questions will likely be quite different. The necessary development also offers a way to include international contributions that can augment the resources of a NASA-led mission.

OST is designed to achieve astrophysics breakthroughs ranging from details about the formation and growth of the first galaxies and black holes, to the trail of water during the birth of stars and planetary systems, and potentially life-supporting properties of solar system objects and exoplanets. Although it requires forethought and planning, OST has tractable technology hurdles that make it a feasible large mission for the astronomical community in the 2030s.

# 1 - SCIENCE INVESTIGATION

> OST is designed to revolutionize astrophysics by studying the universe from a time when dust and heavy elements permanently altered the cosmic landscape to the formation of habitable planets and the search for life on exoplanets.

## 1.1 Science Goals and Objectives

### 1.1.1 How does the Universe work? The formation and evolution of galaxies and black holes

> OST probes the birth and evolution of galaxies and black holes over all cosmic time with imaging and spectroscopic surveys of molecular gas, dust, PAHs, and atomic fine structure lines.

A core OST science goal is to study the cosmological history of star, galaxy, and structure formation into the Epoch of Reionization (EoR). Utilizing the unique power of infrared fine-structure emission lines, OST traces the rise of metals from the first galaxies until today. The mission will quantify the dust enrichment history of the Universe, uncover its composition and physical conditions, reveal the first cosmic sources of dust, and probe properties of the earliest star formation. OST will provide a detailed astrophysical probe into the condition of the intergalactic medium at z>6 and the galaxies that dominate the EoR (**Figure 1-1**).

A persistent question in galaxy evolution is why and how galaxies and their central supermassive black holes co-evolve. The infrared uniquely probes these components of galaxy growth and interaction between them. Due to limiting sensitivity and lack of instrumentation providing three-dimensional (3D) mapping capabilities, previous infrared facilities are unable to fully exploit these diagnostics of coeval star formation and black hole growth. OST, equipped with FIP and MRSS, quantifies the rate of star formation and black hole accretion, measures the feedback, and probes the physics in the interstellar medium in galaxies across the cosmic web, from the Milky Way back to the first billion years of cosmic time.

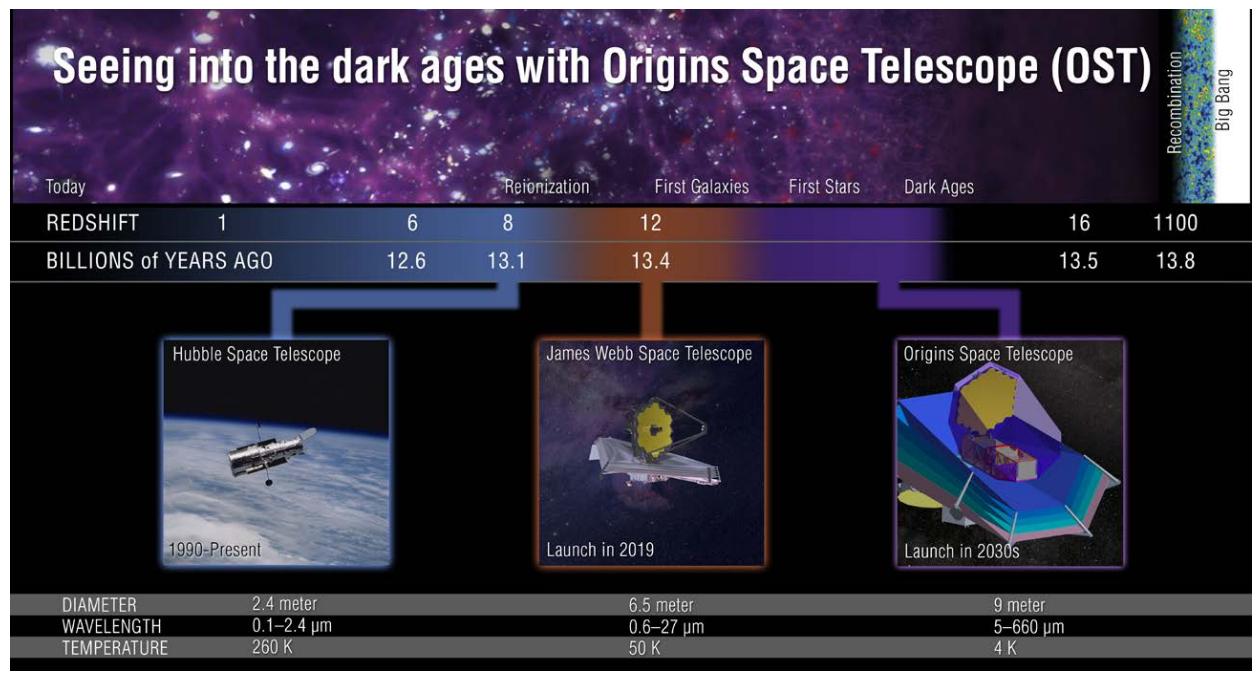

**Figure 1-1:** OST probes over 99% of cosmic time, from the cosmic Dark Ages, through the epoch of peak galaxy growth, to the present day.

With a large, cold aperture and a suite of extremely sensitive instruments that reach into the far-infrared part of the spectrum, OST will be able to probe unique signatures of collapsing structures during the Universe's Dark Ages, as well as the build up of stars and the growth of super-massive black holes over more than 99% of cosmic time.

## Star Formation, Black Hole Accretion, and ISM Fuel Over Cosmic Time

Large area, deep surveys in the infrared [e.g., Berta et al., 2010; Magnelli et al., 2013], have revealed that the bulk of star formation in galaxies occurred between redshifts of z=1-3, and that the vast majority of the UV light from young stars was reprocessed by dust into the far infrared, forming a large component of the cosmic infrared background [Puget et al., 1996; Hauser et al., 1998]. At these epochs, a consensus is emerging that galaxies at a fixed stellar mass have higher star formation rates than those in the local Universe. Why is early star formation apparently more efficient than present-day star formation? What physics in the ISM determines star formation efficiency across the history of star formation? Perhaps radiative feedback from massive stars regulates star formation, mechanical energy from supernovae, or turbulence generated by accretion of intergalactic streams. Each of these processes leave imprints on the ISM accessible to OST.

Another evolutionary trend of critical importance to understanding the evolution of galaxies is the growth of supermassive black holes (BH). Star formation and black hole accretion rate densities (SFRD, BHARD) show similar trends out to z~3 [Madau and Dickinson, 2014], suggesting they grow together. This, along with the local BH-stellar mass correlation [Magorrian et al., 1998; Marconi and Hunt, 2003] imply co-evolution of galaxies and black holes over most of cosmic time. It is not clear how this arises, or what the shapes of the SFRD or BHARD are for z>3. Recent results show obscured star formation dominates, even down to relatively low mass (logM ~9.4) galaxies for z<3 [Whitaker et al., 2017, Dunlop et al., 2017]. The deepest Herschel results suggest the SFRD, as traced in the FIR, may only slowly decline from z=3-5 [Rowan-Robinson et al., 2016].

With its unique ability to perform extremely deep and wide spectroscopic surveys in the infrared, OST uniquely unravels the true history of obscured star formation in galaxies back to the EoR using multiple diagnostics of ionized gas and dust, which has been proven effective in even the most obscured galaxies, and thereby generating samples over large areas that reveal the full population at each epoch. OST is the only planned observatory that can measure the star-forming properties of millions of main-sequence galaxies, starbursts, and low- and high-luminosity AGN, providing a complete picture of the co-evolution of black holes and galaxies.

OST uniquely measures key extinction-free diagnostics of star formation and AGNs in galaxies out to z > 6. Only the rest-frame mid- and far-IR OST provides access to diagnostics of the star formation rate (SFR), black hole accretion rate (BHAR), neutral and ionized atomic gas, and warm molecular gas and dust, in all sources, including those that are highly-obscured and even Compton thick (**Figure 1-2**). Key spectral diagnostics required to measure SFR (SFRD) and BHAR densities (BHARD) are fluxes of bright FIR fine structure lines ([NeII] 12.8, [NeIII] 15.5, [OIV] 25.9, [SIII] 18.7, 33.5) and PAH features at 3.3, 6.2, 7.7, 8.6, 11.3, and 12.7 μm, which OST provides.

The [NeII], [NeIII], and [SIII] lines have relatively-low ionization potentials (~20-40 eV) and provide a direct measure of the SFR [e.g., Ho and Keto, 2007; Inami et al., 2013]. The bright [OIV] line and fainter [NeV] 14.3, 24.3 lines have ionization potentials of 55-97 eV, and are greatly enhanced in AGN (along with the fainter [MgV] 5.6 and [NeVI] 7.6 lines), which have ionization potentials above 100 eV. The IR line ratios effectively provide an extinction-free measure of the hardness of the ambient radiation field and are well-calibrated tools for finding and quantifying the presence of an AGN in dusty galaxies.

Stochastically-heated PAHs are efficient tracers of the ultraviolet (UV) flux, and because the mid-in-frared continuum emission traces the thermal emission from the hottest dust, the equivalent width of individual features can be used to measure the relative power of young stars and an AGN in dusty galaxies. Although at low metallicities, PAHs are expected to be weak [Engelbracht et al., 2006], the team expects OST will detect ~1000 per square degree with stellar masses above 109 M¤ in the EoR. Because PAH features can have fluxes of a few percent of the FIR continuum, OST easily sees them out to z>10, providing an efficient means to determine the redshifts of starburst galaxies and importance of AGN in deeply obscured systems.

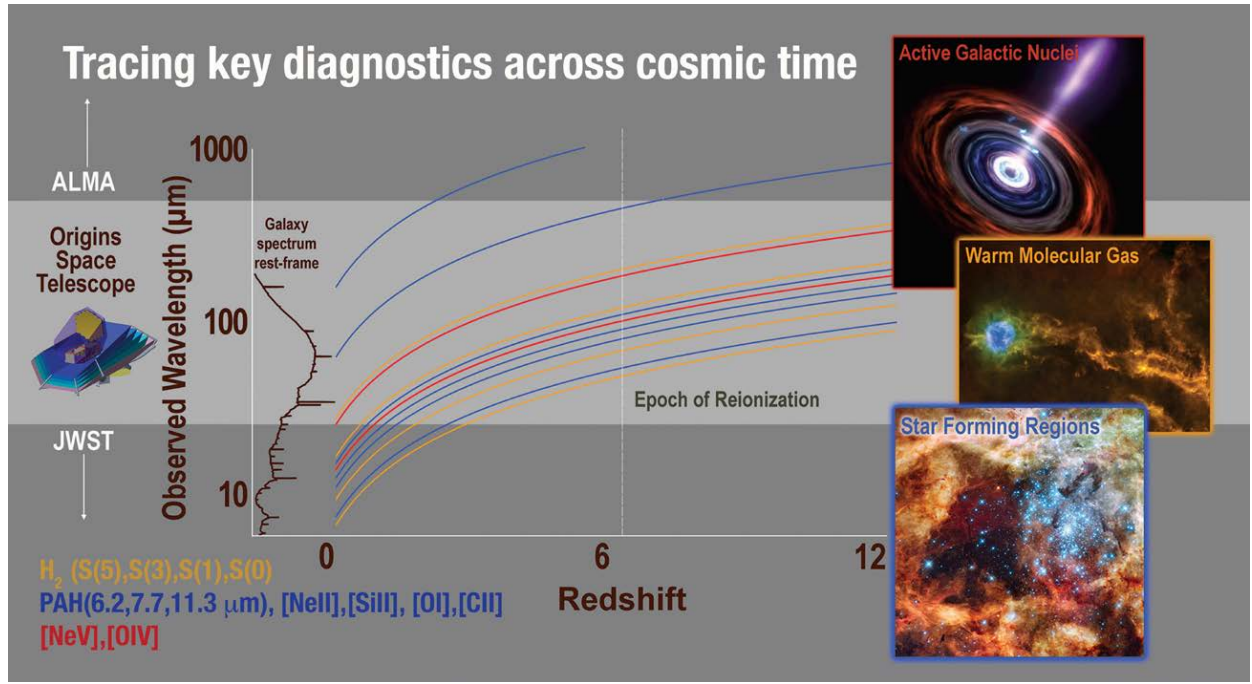

**Figure 1-2:** OST measures key infrared emission line tracers of the star formation rate, black hole accretion rate, and physical conditions of the ISM in galaxies.

In star-forming galaxies, PAHs also supply the photo-electrons that excite gasses in the transition regions between the ionized and neutral ISM. The brightest fine structure lines in the far-infrared, [CII] 157 and [OI] 63, mostly arise from these transition regions, and together with far-infrared thermal dust emission, are a major gas coolant. Other bright, far-infrared lines, [OIII] 57 and [NII] 122, trace the most-embedded HII regions surrounding young stars. Therefore, the fine-structure and dust features in the rest-frame mid- and far-infrared directly constrain the energy balance in the ISM. By quantifying temperatures, densities, and radiation fields characteristic of the gas surrounding dense molecular cores where stars form, OST will reveal the physical conditions in thousands of star-forming galaxies to z~6. In addition to mapping the numbers and luminosities of star-forming galaxies as they build to the peak SFRD at z~2-3, OST is able to chart, for the first time, how the ISM evolves inside galaxies over cosmic time. This can then be used to determine luminosity-weighted physical conditions in the cold neutral, warm neutral, and warm ionized media in thousands of galaxies at z > 4.

A large, low-resolution spectroscopic survey would, for the first time, provide redshifts, BHAR, and SFRs of thousands of high-z galaxies with $L_{IR} < 10^{12}$ L¤. Deep follow-up spectroscopy at higher resolution (R~3000) of representative samples could be used to probe gas dynamics, molecular gas temperatures and masses, shock tracers, and more complete diagnostics of the ISM. For example, a 1500 hr low-resolution survey covering 5 square degrees could reach line flux levels of $5 \times 10^{-21}$ W m$^{-2}$ detecting the PAH and bright fine structure lines in 500 LIRGs at z~5 and 10000 ULIRGs at z=1-2.

## First Dust and the Rise of Metals and Dust Over Cosmic Time

Metal production by stars is a key element of cosmic evolution, since metals and dust are not only important conduits for light emission but also key elements in the dissipation of energy in the ISM. OST will quantify the dust enrichment history of the Universe, uncover its composition and physical conditions, reveal the first cosmic sources of dust, and probe the properties of earliest star formation.

Astrophysical dust comprises less than one-hundredth of one percent of the baryonic mass of the Universe, yet approximately one-half of all energy radiated by stars and accreting black holes over its history has been reprocessed by dust to long wavelengths. Dust shrouds the most intense regions of black hole accretion and star formation, and as it is comprised of condensable species like C, Si, Fe, and O, is a sensitive tracer of the abundance of heavy elements. Studying the broad spectral dust features in the mid-infrared (3–30 μm) is the only way to provide critical and unique information on dust origin, composition, and properties like grain size distribution, as well as to track the underlying gas metal content. Due to the very different expected dust compositions and mass yields of low metallicity PopII and PopIII stars, dust signatures can directly probe the properties of the earliest phases of star formation.

At z>5, only a sensitive far-IR platform like OST can provide access to the rest-frame mid-infrared (MIR) emission of galaxies, which is crucial for assessing the content and conditions of the dusty interstellar medium (ISM) at the earliest star-forming epochs. These wavelengths are almost entirely obscured by Earth's atmosphere. Additionally, since the cosmic microwave background (CMB) temperature ($T^{CMB}$) increases linearly with redshift above z~5, dust temperatures of many galaxies will approach $T^{CMB}$, rendering differential detection in the Rayleigh-Jeans (R-J) tail of their dust spectral energy distributions (SEDs) increasingly difficult. With high intrinsic luminosities and direct or stochastically-heated temperatures of hundreds of Kelvin, the dust features emitting in the rest frame MIR (PAHs, silicates) can be readily observed above this warm CMB background.

OST utilizes the unique power of infrared fine-structure emission lines, which are not susceptible to dust extinction, to trace the rise of metals from the first galaxies across all of cosmic time utilizing relative abundance indicators to minimize systematics. The present day Universe is rich in metals heavier than helium that enable efficient cooling of gas in the ISM to form stars, assemble planets, and make the building blocks of life-as-we-know-it. The Universe did not start in this state – metals had to build up over time with successive generations of stars. However, the rate of metal buildup and how it depends on galaxy properties (e.g., stellar mass) and environment is unknown. Extensive studies of metallicity locally and out to z~1 indicate a factor of 10 decline in metallicity, while knowledge beyond z~1 is much more limited, although dusty galaxies have been detected at z>6 [e.g., Watson et al., 2016]. JWST uses red-shifted optical lines to trace changes in metallicity to z~6, and OST will extend this range to trace the evolution of metals out to z~10 with mid- IR lines (i.e., ([NeII]+[NeIII])/([SIII]+[-SIV]) from tens of thousands of galaxies to build a statistically-significant sample. Using MRSS, OST will easily survey 10-100 s of square degrees to measure the rise of metals and dust as part of the same surveys to study the co-evolution of star formation and blackhole evolution over cosmic time.

## Feedback from Stars and AGN

Feedback is the action of radiative and mechanical processes on the accretion of gas and star formation. How feedback occurs not only can potentially regulate star formation, it also determines the state of the interstellar medium in a galaxy, and the rate at which the interstellar medium is enriched with products of stellar evolution. Feedback likely plays a key role in establishing the shape of the galaxy mass function, the mass-metallicity relation, and the heating and enrichment of the intergalactic medium. Feedback from both star formation and AGN must play a large role in regulating the growth of galaxies and black holes. There is abundant evidence for superwinds and outflows in local and distant (z=1-3) galaxies [Fischer et al. 2010; Sturm et al. 2011, Veilleux et al. 2013, Gonzalez-Alfonso et al. 2017].

The Far-IR provides a unique window to probe the multi-phase nature of feedback, which occurs in atomic and molecular gas. By measuring key parameters of shocked gas and dust, as well as the cool ISM, OST will produce a nearly complete picture. Mid- and far-IR spectroscopy provides a unique window on the physical conditions in the multi-phase ISM in galaxies experiencing powerful feedback. OST can directly observe the role of feedback in quenching galaxies, derive the wind mass, kinetic energy, and mass outflow rates, and correlate these with key galaxy properties (AGN or SB power, environment, mass, age, etc.).

OST will measure $H_{20}$ and [OI] and OH features in absorption against the source continuum, spectrally separating the outflowing gas over a wide range of redshifts. Galaxy outflows have been traced in the diffuse ionized gas in nearby galaxies for decades. The cooler phases of these outflows, however, may dominate the mass budget. Far-infrared absorption features, which trace neutral and molecular material, are probes of this cool phase, first measured in a handful of nearby AGN and ultra-luminous infrared galaxies with Herschel. These unambiguously trace negative feedback on the cold molecular ISM [e.g., Sturm et al., 2011]. These measurements will allow OST to reconstruct the cosmic history of outflows.

OST will use measurement related to the strength and velocity offset of P-Cygni profiles in the FIR OH lines to estimate the speed and mass of the molecular outflow and correlate with AGN and SB strength (using the far-IR flux and mid-IR diagnostics). Broad wings (1-few x 1000 km/s) on the mid-IR fine structure lines will indicate high-velocity ionized gas as well as the ionization mechanism in the different velocity components. The $H_2$/PAH flux ratios will be compared to PDR models to isolate systems with strong $H_2$, indicative of shock heating. Samples of order 100 galaxies will need to be measured because the feedback is expected to be "bursty" and dependent on the coupling of stellar and AGN power to the ISM.

### Understanding the Epoch of Reionization and Probing the Formation of Primordial Galaxies

Reionization is a major event in the evolution of the universe, about which little is currently known, including when it occurred. Dwarf galaxies, which are low in mass, metals, and luminosity, were the likely sources of reionizing photons. To understand reionization sources and processes, it is essential to probe galaxies at z>6 during the Epoch of Reionization (EoR). Tomographic line intensity mapping is a powerful means of observing the dwarf galaxy population during EoR. The power spectrum analysis of intensity fluctuations in spectral line emission is sensitive on large scales to the aggregate intensity of all galaxies in a given volume—including intrinsically faint sources that may otherwise remain below sensitivity thresholds for individual detection.

OST's wide-area spectroscopic surveys with large instantaneous bandwidth will enable high-fidelity power spectrum measurements of rest-frame IR fine structure lines. The [OIII] 52 μm line is expected to be a robust Star Formation Rate (SFR) indicator based on observations of [OIII] 88 μm in low-metallicity dwarfs (assuming [OIII] 88 μm /[OIII] 52 μm ~ 1, as is appropriate for typical HII densities nH+ ~ 100 $cm^{-3}$ in these galaxies). Surveying 1 $deg^2$ in the 400-μm MRSS channel, OST can achieve signal-to-noise ratios (SNRs) of up to ~1,000 on the total [OIII] 52-μm power spectrum (including shot noise) and up to SNR ~ 100 on the linear clustering power at k~0.1 h/Mpc, which is used to extract the mean line intensity. Contamination from brighter interloper line emission, [CII] 158 μm at emitted redshift zem = 1.5, can be removed effectively by masking individual >5σ detected sources. The masking depths for each considered OST survey area are above the spectral confusion limit ($10^{-22}$ W $m^2$). Where masking is not sufficient to remove interloper emission on a source-by-source basis, OST provides additional methods for identifying contaminating foregrounds [Lidz et al., 2016; Cheng et al., 2016].

OST will probe the birth of galaxies through warm $H_2$ emission. A key to understanding formation of the primordial galaxies and evolution of large-scale structure is tracing the cooling of gas into dark matter halos. Emission from pure rotational $H_2$ lines redshifted into the far-IR can provide a direct measurement of the strength of cooling and dissipation of mechanical energy in the gas as it collapses before the onset of star formation. By analogy with well-studied local shock-dominated systems, rest-frame lowest-lying pure-rotational $H_2$ transitions can directly trace the energy deposition into cool (typically $120 < T < 1000K$) molecular gas from low-velocity shocks and turbulence. The rotational $H_2$ lines (e.g., 0-0 S(1), S(3), S(5)) directly trace energy dissipation in the first collapsing structures. For the redshifted universe ($4 < z < 15$), rotational $H_2$ lines fall squarely in the far IR (32 - 455 μm) and cannot be practically observed by any current or planned facility. Gas accumulating in dark matter halos may also show inhomogeneity in metal enrichment; measuring the relative strength of $H_2$ versus the generally brighter metal fine-structure cooling would be a direct observational step toward testing models of early galaxy formation.

**High-z Galaxy Surveys**

To accomplish the mission's high-z galaxy survey goals, OST needs to make spectra and continuum measurements of millions of galaxies that sample the red shift range and galaxy type. To utilize ancillary data from prior work, OST builds upon the legacy of multi-wavelength high-z galaxy surveys by the current great observatories, Hubble Space Telescope (HST), Chandra X-ray Observatory, and Spitzer Space Telescope (Spitzer), and continued by every great observatory in space and on the ground. JWST will push these deep field studies to detect the first galaxies. However, to understand the total process of galaxy formation and co-evolution with black holes requires an observatory like OST that will enable gazing even further back in time (**Figure 1-1**).

OST's extragalactic surveys will adopt the "wedding cake" galaxy survey approach, surveying a large, shallow area (e.g., Stripe 82), plus perform an ultra-deep survey in a smaller area (e.g., Cosmos Evolution Survey (COSMOS) field). A shallow survey of the entire Stripe 82 area (~300 deg$^2$) with FIP would take ~70 hours of observatory time. Performing a MRSS spectroscopic survey of the entire COSMOS field (~1 deg2) would take 500 hours and of the entire North Great Observatories Origins Deep Survey (GOODS) field (~0.02 deg$^2$) would take ~60 hours. These large surveys support OST's multiple science goals for galaxy formation and co-evolution with black holes.

Wide-Field Infra-Red Survey Telescope (WFIRST) plans surveys over hundreds of square degrees that will provide quiescent stellar mass assembly, and galaxy morphologies, which signpost galaxy assembly processes (i.e., mergers, secular processes, emergence and buildup of bulges and disks) on kpc scales. With their complementary information providing a complete picture of all relevant baryonic processes on all relevant scales, combined WFIRST and OST surveys would be the most powerful possible diagnostic of galaxy assembly and large-scale structure.

### 1.1.2 How did we get here? The Ingredients of Life and Habitable Worlds

> OST will trace the elements that support life from the interstellar medium, to planet-forming disks and from there to potentially habitable planets.

How are volatile elements, carbon (C), nitrogen (N), and the hydrogen (H) in water, which are critical for life-as-we-know-it, delivered to habitable planets? This was an extremely inefficient process for the Earth. For example, only one in a million of the available nitrogen atoms made it to Earth to form our atmosphere, and the nitrogen that did was likely carried by organics. Is it similarly difficult to deliver water and organics to typical Earth-sized planets in the habitable zone around other stars?

- OST will reveal the complete history of the formation, evolution, and potential existence of biospheres using unique tracers of water, organics, and nitrogen-bearing species that dominate the infra-

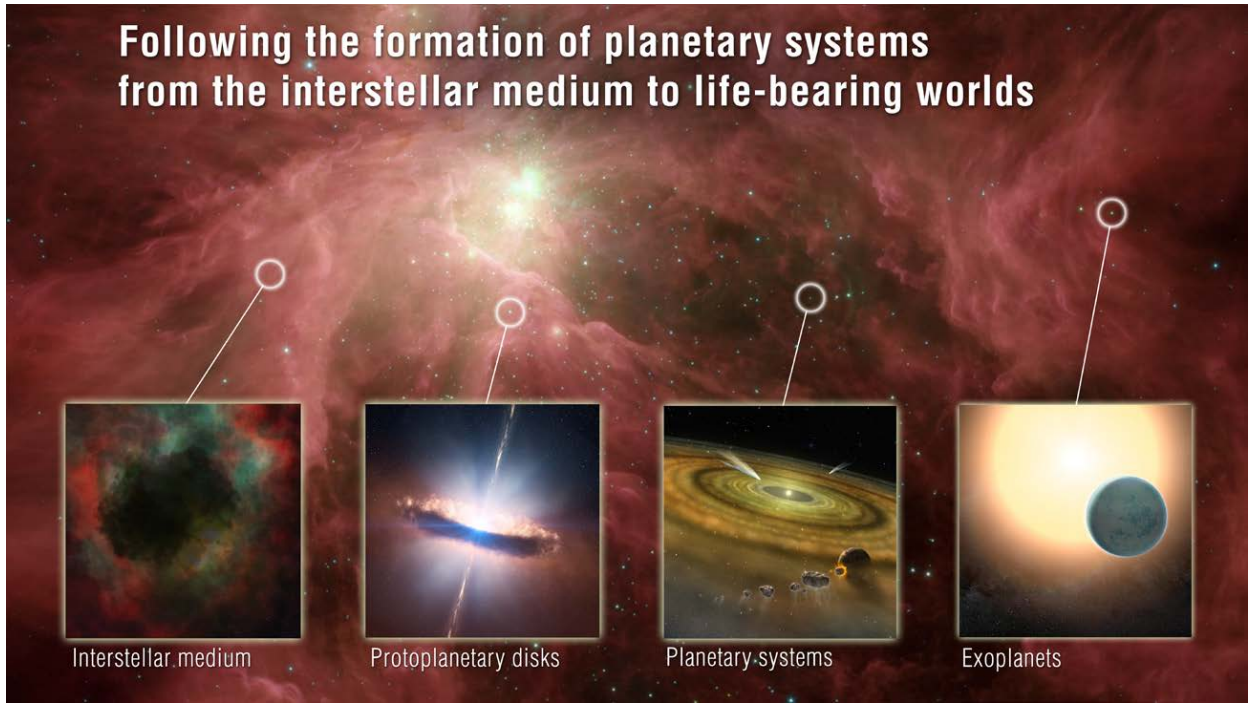

**Following the formation of planetary systems from the interstellar medium to life-bearing worlds**

Interstellar medium     Protoplanetary disks     Planetary systems     Exoplanets

**Figure 1-3:** The tremendous sensitivity from the mid- to far-infrared offered by OST opens new windows through which to explore the trail of life's ingredients, especially $H_2O$, from planetary system origins to inhabited worlds.

red wavelength regions (5-600 μm) (**Figure 1-3**). OST will build a census of water in planet-forming disks around stars of all masses. OST will obtain velocity-resolved spectra of water lines, tracing, for the first time, the full range of gas temperatures in thousands of planet-forming disks around stars of all masses. This will enable scientists to determine the mass and distribution of water as a function of stellar mass, and establish the degree to which planets are seeded with water during formation. OST will thus enable an understanding of planet formation and habitable planetary conditions.

- Determine the total amount of primordial planet-forming gas around all stars: OST will use the ground-state HD line at 112 μm to obtain precise measurements of the total gas mass available for the formation of planets around thousands of stars of all masses. This will fully determine the efficiency of planet formation as a function of stellar mass by comparison to exoplanet demographics.
- Trace the origin of water on Earth and in our Solar System: OST will determine the D/H ratio in hundreds of comets, providing, for the first time, a full statistical sampling of this critical fingerprint for the origin of water on Earth. OST uses low-lying lines of $H_2O$ and HDO to determine the D/H ratio with high precision. This large sample will set stringent constraints on cometary delivery of Earth's water, explore heterogeneity of D/H within cometary reservoirs (Oort Cloud, Kuiper Belt), and transform our understanding of the origin of water in the solar system.
- Trace water and other volatiles back to their interstellar origins: OST will follow the water from the interstellar medium to young protoplanetary disks, revealing how planetary systems form.
- Determine the mass and composition of dust in exoplanetary systems: OST will image the dust in hundreds of debris disks and spectroscopically determine the dust composition using unique fingerprints of PAHs, silicates, and large grains.
- Find and characterize the outer small bodies in our Solar System: OST far-infrared imaging will discover thousands of new Kuiper Belt and trans-Neptunian objects.

**Statistical Constraints on Life's Ingredients in Planet-Forming Disks: The Trail of Water**

Most of the mass that incorporates into planetesimals is found in the disk midplane. OST's spectrometers will observe a pristine ice feature at 47 µm and two processed ice features at 43 and 62 µm. The bulk ice mass and degree of thermal processing measured by the ice feature allow the team to quantify the potential proto-planet habitability. The 62-µm feature has only been detected at a handful of bright, flared protoplanetary disks [e.g., McClure et al., 2015] due to low peak/continuum ratio. The much stronger 43 µm feature, which is an excellent proxy for the ice/dust mass ratio, is not available to Spitzer or Herschel, but is covered by OST.

It is not known whether protoplanetary disks are universally able to seed their planets with water and other volatile species critical to the origin of life-as-we-know-it. To measure the abundance and distribution of water in gas and dust actively being incorporated into planetesimals and planets requires access to far-infrared lines of water vapor. The 3-200 µm spectra of typical protoplanetary disks are rich in emission lines from molecular gas, with cooler gas generally being traced at the longer wavelengths (**Figure 1-4**).

Thus, to understand the full diversity of planet-forming disks and their connection to exoplanetary composition and availability of water to potential biospheres, it is critical to obtain an infrared spectroscopic survey of the water content in a large sample of protoplanetary disks around stars of all masses and for a wide variety of environments. Spitzer observations suggest there are strong chemical differences in disks around stars of different masses, with water being abundant around solar-mass stars, but much less pronounced around low-mass and intermediate-mass stars [Pontoppidan et al., 2010; Pascucci et al., 2009]. A full census will place the Solar Nebula into broad Galactic context. Is water equally common in disks around low-mass and high-mass stars, near young OB star clusters in environments, and in low-metallicity regions? Is our solar system chemically typical or unusual [Pontoppidan et al., 2014; Bergin et al., 2015]?

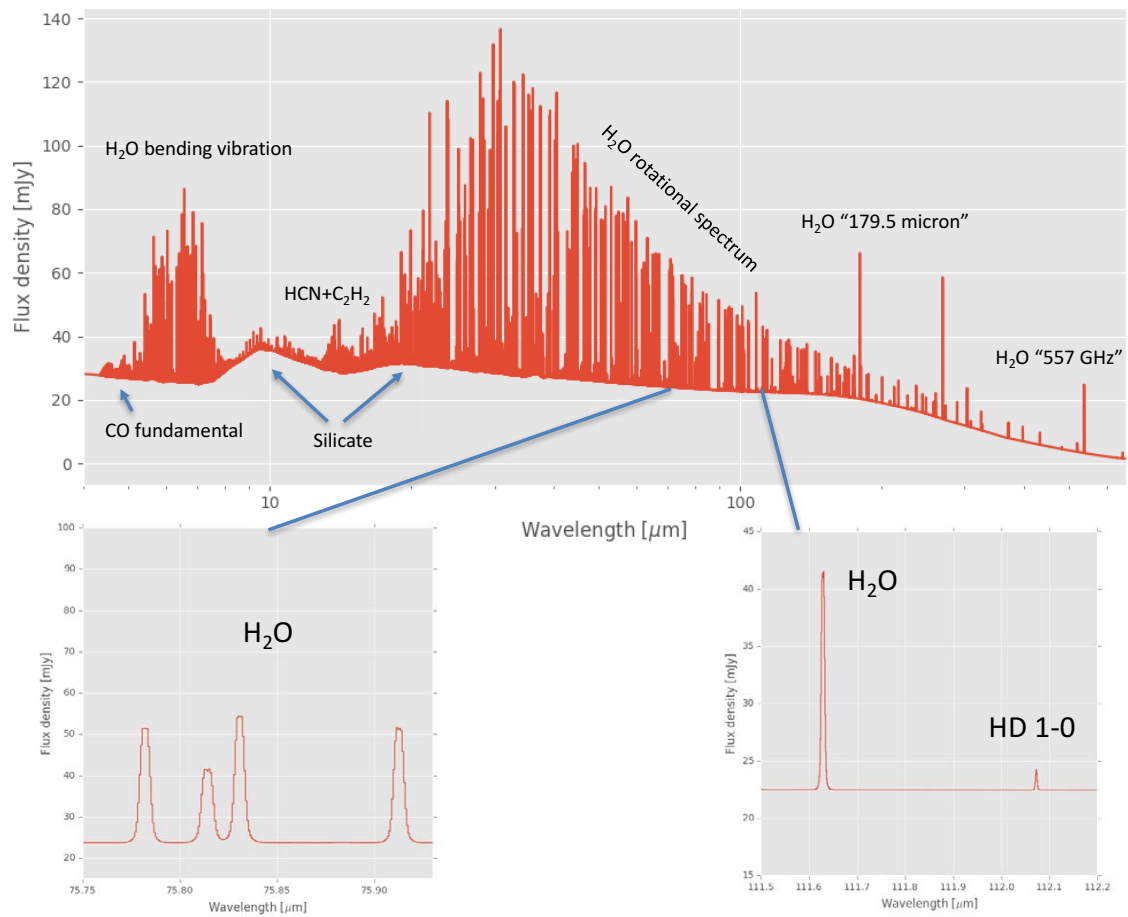

**Figure 1-4:** A disk model spectrum from Pontoppidan et al. [2018], rendered at 4-660 µm and at a uniform 3 km s⁻¹ spectral resolving power showing major spectral features. OST provides diagnostic molecular lines, including the HD 1-0 line, and the $H_2O$ lines, to quantify the gas mass and the location of water in proto-planetary disks.

Herschel was able to detect water from the planet-forming region (1-10 AU) in a few disks, but was sensitivity-limited [Riviere-Marichalar et al., 2012; Podio et al., 2013; Blevins et al., 2016]. And at the end of the Herschel mission, the ability to observe water vapor from gas with temperatures in the 50-500 K was essentially lost. The water at these temperatures uniquely traces the planet- and comet-forming regions at 1-50 AU, including the location of the snow line. Even ALMA is limited, as it can observe only a handful of transitions of $H_2^{18}O$ that are difficult to detect. To characterize water requires going into space, and tracing its evolution requires following the mass reservoir in cool planet-forming gas (20-300 K). JWST traces part of the water story in hot gas (>300 K). However, only OST is able to probe water emission around and beyond the snow-line, peering into the giant planet and comet formation zones. OST is designed to have full access to the complete array of transitions that trace water vapor at all temperatures, and therefore the true distribution and history of water as an ingredient for habitable worlds.

The broad wavelength range needed to efficiently observe water vapor, which is readily feasible with OST, also includes strong transitions from many other chemically important molecular species, including HDO and $NH_3$. These can generally be observed using the same requirements as those needed to observe water.

Additional observations at the highest spectral resolution ($R \sim 10^6$) of the lowest-lying water lines, such as the ground-state ortho and para lines at 1113 and 557 GHz, are essential to measuring the gas phase water abundance in the outer, coldest portions of the disk, corresponding to the Sun's Kuiper Belt (10-30 K). Such observations can only be achieved with a heterodyne instrument on a large far-infrared space telescope. Herschel-HIFI was able to detect such cold water in two disks, TW Hya [Hoger-heijde et al., 2011] and DG Tau [Podio et al., 2013], whereas OST will be able to detect cold water in up to hundreds of disks. Such observations will be highly complementary to observations of warmer water found nearer to the star in the planet-forming region around the snowline. These velocity-resolved observations will make it possible to use the Keplerian rotation to derive the distribution of water vapor in the outer disk.

## Measuring the Mass of Planet-forming Matter

Protoplanetary disk mass is the fundamental quantity that determines planet formation. Estimates of disk masses are complicated because the molecular properties of its dominant constituent, molecular hydrogen, make it unemissive at temperatures in the 10-30 K range, which characterizes much of the disk mass. Bergin et al. [2013], using Herschel, detected the fundamental rotation transition of HD at 112 μm emitting from the TW Hya disk. The atomic deuterium abundance relative to $H_2$ is well known and the lowest rotational transition of HD is a million times more emissive than $H_2$ for a given gas mass at 20 K. It is therefore possible to convert HD emission to $H_2$ gas mass. Due to Herschel's limited lifetime, the only other deep HD observations obtained were in six disks, with the result of two additional detections [McClure et al., 2016].

Understanding the lifetime of disk gas requires a survey of hundreds of stars in stellar groups and clusters with known ages, which will provide a stringent constraint on gas mass as a function of time. An OST HD emission survey, scheduled in conjunction with an $H_2O$ in disks survey, will provide missing insight into the gas masses of planet-forming disks to determine the timescales of giant planet formation, understand the constraints on gas available for super-Earths/mini-Neptunes, and set needed limits for disk dynamical models. Disk mass knowledge also breaks the degeneracy between disk mass and chemical abundance. This information is crucial, as ALMA is now providing resolved images of gas tracers (e.g., CO, etc.) that do not directly trace the gas mass like HD 112 μm can. For example, without the Herschel detection of HD in TW Hya, it would have been inferred from readily-accessible gas tracers (e.g., CO, HCN, etc.) that the gas mass is low, whereas it is now believed abundances of tracers are low, which reveals the early formation of a planetary system [Favre et al., 2013]. Thus, there is tremendous synergy between a future far-IR facility and ground-based instruments. An OST survey will build upon SOFIA/HIRMES (scheduled for flight in 2019), which will observe HD and $H_2O$ in a handful of bright disks, but is incapable of providing a large or complete survey. Among all existing or planned IR missions, only OST can survey significant numbers of disks and measure their gas content and ability to form planets.

**Cometary D/H Ratio as a Fingerprint for the Origin of Water on Earth**

Cometary (Halley, Hale-Bopp) water shared with the Earth had a level of deuterium enrichment well above the estimated value of the proto-Sun and interstellar medium [Ceccarelli et al., 2014] (**Figure 1-5**), and this D/H ratio could be used as a fingerprint to trace the source of the Earth's water. Based on comparing the mineralogy of Earth and asteroid belt, it is believed Earth formed within the nebular snow-line [e.g., van Dishoeck et al., 2014]. Hence, embryonic Earth would have contained little to no water. Instead, water is posited to have been delivered from the outer parts of the asteroid belt and solar system via dynamical interactions, perhaps with the giant planets [e.g., Raymond et al., 2014]. Prior to Herschel, only three Oort cloud comets had D/H measurements, and there were no existing measurements in the short period comets (Jupiter family) that originate in the Kuiper Belt. Based on Herschel and Rosetta observations, it is now known that the D/H ratio in these cometary reservoirs is not uniform, and there is some heterogeneity with an unknown origin.

With its 9-m aperture, OST provides the capability, for the first time, of measuring the D/H ratio in all short and long period comets that approach the Sun during the mission's nominal 5-year lifetime. This sample would comprise ~290 Oort cloud comets and ~200 short-period Jupiter family comets. Compared to existing observations (**Figure 1-5**), this is a factor of ~50 increase in sample size. These observations will enable interpretation of D/H in comet formation zones, with invaluable information about the fossil record of the early history of our solar system, movement of small bodies, and diversity of the early chemical record. This is needed to better understand the D/H ratio and its use in tracing the origins of life-fostering water on Earth.

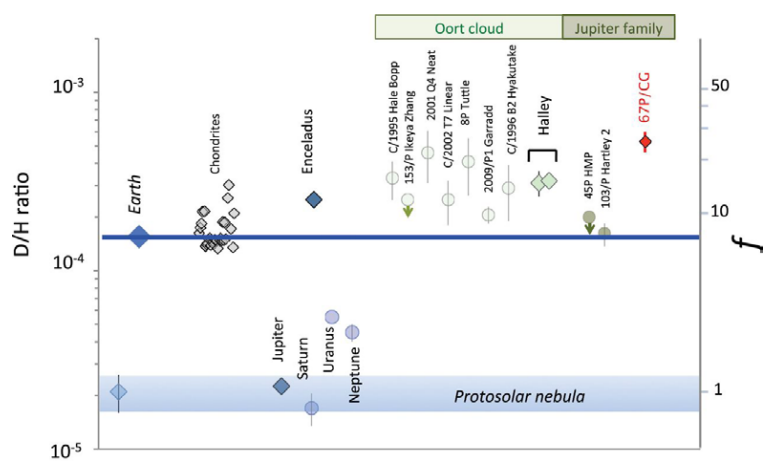

**Figure 1-5:** The current state-of-the-art in D/H measurements in the solar system [from Altwegg et al. 2015] is limited by sensitive measurements of $H_2O$ and HDO. OST provides definitive D/H ratio statistics within the Oort Cloud and Kuiper Belt, critical cometary reservoirs that will reveal if comets are the source of Earth's water.

**A New Frontier in the Water Trail: ISM to Young Disks**

The origin of planetary water is directly tied to its formation in the interstellar medium (ISM). The path of water begins in dense molecular clouds, where water forms. Some of the interstellar water cycles back to diffuse molecular gas and UV-irradiated photodissociation regions, where it is destroyed, while some is incorporated into protostars and eventually into planet-forming disks and planets. Tracing the abundance and state of water through these transitions is necessary to understanding our interstellar origins.

OST will play a critical role in tracing the early path of water from the ISM into young protostellar disks through its unique access to the lower rotational transitions of the water molecule and its isotopologues in the far-IR. With the high sensitivity provided by its cold, 9-m telescope and Heterodyne Instrument's extremely high spectral resolution capabilities, OST will be a transformational tool for tracing the origins of water in the ISM.

Studying the state of water across a wide range of interstellar environments is necessary to understanding the transitions between gaseous and solid phases, as well as the overall formation and abundance of water. Far-IR water lines also provide diagnostics of the overall ISM: the rich THz emission spectrum of excited $H_2O$ allows accurate determination of the total column density [e.g., Emprechtinger et al., 2013], and helps to untangle multiple regimes along the line-of-sight. Observations of water absorption toward continuum sources have proven to be a rich source of information about the chemistry of $H_2O$, as well as physical conditions in diffuse ISM regions [e.g., Flagey et al., 2012]. Thus, OST will explore water emission at all stages of star formation, as well as in planet formation and on planets. This starts with the very cold (~10 K) pre-stellar cores, where Herschel detected water vapor [Caselli et al., 2012]. With its improved sensitivity, OST measures not only water content during star and planet formation, but its D/H ratio. While OST and other observatories build a census of water on other worlds, OST will uniquely provide critical information about how common water is and how it originated.

**Debris Disks and Exoplanetary Systems**

Debris disks, consisting of dust replenished by collisions of leftover planetesimals and cometary activity, are tenuous. They often have a structure analogous to the minor body belts in our solar system, and include asteroid- and/or Kuiper-belt-like components. Their large surface area makes these disks detectable through infrared thermal emission or optical scattered light, providing insight into unseen minor-body populations and their underlying planetary architecture. Their resemblance to the solar system (exo-asteroid and exo-Kuiper-belt analogs) provides a constraint on models of how our solar system formed and evolved.

OST's sensitivity improvements also enable detections of faint, second-generation gas in debris disks. It is believed these molecular gas (CO) and atomic species (CI, CII, OI, some metals) are released in collisions between comets in debris belts [Kral et al., 2016] or falling evaporating bodies similar to outgassing comets in our solar system [Kiefer et al., 2014]. Far-infrared fine structure lines, like [CII] (157 µm) and [OI] (63 µm and 145 µm), are essential to inferring the composition of exo-comets [Kral et al., 2017]. OST observations will not only detect these lines in many debris disks (compared with only three Herschel detections), but also provide high angular resolution maps of nearby systems to better determine the location of the gas, its composition, and formation mechanisms (collision, outgassing). This has great synergy with OST's detailed studies of comets in our solar system.

**Distribution of Minor Bodies in Exoplanetary Systems**

Asteroids and comets (i.e., our "debris disk") are an integral part of our solar system. These small bodies give life by delivering water and take it away through occasional extinction events. Their lo-cations and orbital structure provide strong constraints on the solar system's history, a story that includes giant planets' migration, capture of Jupiter's Trojans, and the Late Heavy Bombardment of the terrestrial planets. Due to current detection methods, most known exoplanetary systems are biased toward the inner zones of mature systems: radically different from our solar system in ways that make them unlikely abodes for intelligent life. Giant or ice-giant planets positioned at large orbital distances not only foster terrestrial planet formation, but also serve to shield habitable terrestrial planets from small bodies. Systems with planets at wide orbits, analogous to Jupiter, during the critical first few hun-dred million years of evolution are virtually unexplored. Since small bodies in exoplanetary systems are inevitably perturbed by the presence of planets, the debris distribution can also reveal hidden planets, especially low-mass ice giants below current detection limits. Therefore, resolved debris disk images offer an effective method to study the formation and mi-gration history of low-mass ice giants from youth to the age of the solar system. OST's unprecedented far-IR spatial resolution can resolve ~50% of the known debris systems by more than ten resolution elements, and more than ~80% by two resolution elements. This will provide valuable insight into the formation and migration history of giant planets at wide orbits and provide a ground truth for testing planet formation and migration theories.

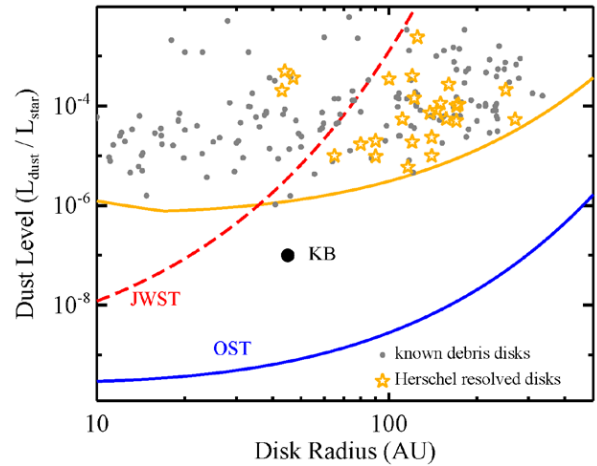

**Figure 1-6:** OST's unprecedented far-IR resolution and great sensitivity enable a complete census of true Kuiper-belt (KB) analogues among nearby stars, placing our own into context. Detection of circumstellar dust around stars, based on measurements by Herschel and Spitzer. The expected sensitivities (lines based on unresolved photometry with JWST (22 µm), Herschel (100 µm), OST (60 µm) without systematic noise from stars and calibration. In addition to sensitivity, resolving the dust structure from the star is the only robust way to detect faint levels of KB analogues.

Although all other stars host some degree of a debris disk, only the brightest 20% are currently detectable, and the rank of our solar system in the remaining 80% is unknown, as is how this rank relates to our solar system's history. True Kuiper-belt analogues (~45 AU planetesimal belt with peak emission ~60 µm [e.g., Vitense et al., 2010]) can only be detected by imaging, as they are too faint (1%) relative to their host stars to detect as an IR excess. OST's finer resolution and two orders of magnitude improvement in sensitivity compared to Herschel enable a complete census of true Kuiper-belt analogues among nearby stars, putting our own solar system in context (**Figure 1-6**).

## Outer Solar System Populations

The Solar System's Trans-Neptunian Objects (TNOs) and their related populations are the small solar system bodies residing between 30 and 50 AU. TNOs, a collection of dynamically-variegated subpopulations, including Centaurs and Scattered-Disk Objects (SDOs), as well as "cold" (low-inclination, eccentricity) and "hot" (high-eccentricity) classical Kuiper Belt populations (KBOs) [Gladman et al., 2008], comprise the majority of our Solar System's small bodies. These minor planets are the reservoir of comets that routinely visit our inner solar system and cloud the distinction between asteroids and comets. They are primordial material, unmodified by the evolution of the solar system after their emplacement and are the sources of volatile materials to the inner solar system. TNO and outer solar system small body size distributions, down to scales of tens of km or a few km [Dones et al., 2015 and references therein], test the formation scales of planetesimals and evolutionary models of the early solar system.

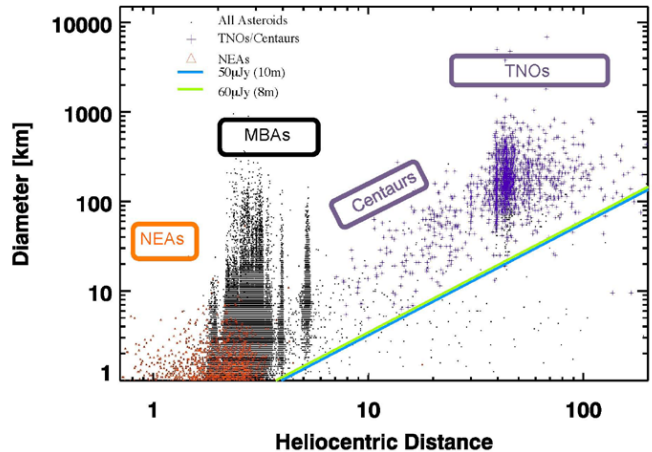

**Figure 1-7:** OST imaging surveys of the ecliptic plane will reveal new small bodies and characterize ones found by LSST via the far-infrared emission from these cool outer solar system bodies. Current knowledge is limited and OST's sensitivity limits provide the opportunity to discover thousands more of these objects.

OST's FIR imaging capability enables characterization of known TNOs and discovery of thousands of new ones, down to scales of a few km (**Figure 1-7**). Small bodies typically vary in surface reflectivity from a few percent to a few tens of percent, a factor of 5 or more, which significantly affects optical measurements of the reflected light, and prevents accurate assessment of their size based on reflected light. These reflected-light based diameters can be off by factors of 2 or more. OST FIR surveys detect emitted light and will provide reliable sizes from the flux, within ~15% of the true diameter on average [c.f., Mainzer et al., 2011].

Objects at TNO distances are best detected at wavelengths near 110 mm, close to or longward of their thermal peak. Shorter (~70 μm) and longer (~200 μm) wavelengths better constrain sizes and temperatures. A 9-m OST can reach a 5σ sensitivity of ~55 μJy in ~14 sec of integration; four repeated observations over the course of several days enables detection of the TNOs, which move in their orbits. Repeated observations boost confidence in an object's detection and enable identification of its orbit type. In 200 hours, OST can cover a field of ~200 square degrees and ~20,000 objects, nearly 7 times the number of presently known TNOs, and more than 200 times the number of measured TNO diameters. These measurements will complement optical LSST and WFIRST measurements and extend JWST, SOFIA, and ALMA TNO characterizations into longer wavelengths.

### 1.1.3 Are We Alone?: The Search for Biosignatures in Habitable Worlds

> OST has the stability and precision required to detect signs of life on rocky planets orbiting within the habitable zones of nearby M dwarf stars. Going beyond the reach of current and planned observatories, OST will enable comparative exoplanetology at the terrestrial level.

For the first time in human history, our species will have the means to answer one of our longest-standing questions: "Are we alone?" Only recently have planet-hunting programs, including TRAnsiting Planets And PlanetesImals Small Telescope (TRAPPIST), MEarth Project, and Kepler/K2, confirmed the first Earth analogues – Earth-size planets orbiting within the habitable zones of their host stars. However, these planets' potential for life-as-we-know-it will only be determined once their atmospheres have been characterized with ultra-stable, high-precision instruments. Using a purpose-built spectrometer, OST will constrain the presence of biosignatures in nearby potentially-habitable planets orbiting mid-to-late M dwarfs.

The transit technique is a promising method for detecting and characterizing the atmospheres of Earth-size planets in the habitable zone of their host stars [de Wit et al., 2016]. During primary transit, the planet is detected passing in front of its host star, which unambiguously determines the planet's size and, when combined with a mass measurement, its bulk density. These two fundamental parameters are the essential first step toward identifying rocky, potentially habitable worlds and neither value can be constrained using other techniques. By leveraging these constraints, OST will focus on planets that are definitively rocky and likely to have volatile-rich atmospheres to search for biosignatures.

Distinguishing potentially habitable planets from those that might have inhospitable environments requires characterizing a planet's atmosphere. During transit, OST will measure the wavelength-dependent change in planet radius (i.e., its transmission spectrum) as light from the host star passes through its atmosphere. This method has been used to successfully detect absorption by atoms and molecules in the atmospheres of hot, Jupiter-size planets [e.g., Charbonneau et al., 2002; Sing et al., 2016]; OST will do the same for temperate terrestrial exoplanets. Moreover, the limiting effect of noise from star spots and stellar granulation on the transmission spectrum is minimized in the mid-infrared [Trampedach et al., 2013; Rackham et al., 2018]. During conjunction or secondary eclipse, when a planet passes behind its host star, OST will measure the planet's thermal emission. The emission from a planet in the habitable zone, as well as the planet-to-star flux ratio, peaks in the mid-infrared. The emission spectrum constrains the absolute abundances of molecular species as well as the temperature profile (vs. altitude) of the planet's dayside atmosphere [e.g., Kreidberg et al., 2014; Line et al., 2016]. For example, Earth's emission spectrum shows our planet has a temperature inversion; the strongest part of the $CO_2$ band at 15 μm, which forms at higher altitudes, is seen in emission rather than absorption.

Outside of secondary eclipse, thermal phase-curve measurements probe variation in emitted flux over an entire orbit, thus providing fundamental information about the radiative, chemical, and advective processes taking place in the planet's atmosphere as a function of planet longitude. When probed spectroscopically, these properties can be measured at different altitudes and, therefore, can be mapped in two dimensions [Stevenson et al., 2014, 2017]. For Earth-size planets observed with OST, spectroscopic phase-curve measurements could readily distinguish between different atmospheric scenarios. Is the planet an airless body, or does it have an appreciable atmosphere? If so, what is its composition and density? Ultimately, these high-fidelity measurements would enable a truly global view of the temperature structure of these habitable-zone planets.

Ground- and space-based surveys have begun to reveal the prodigious number of planets orbiting M-dwarf stars [Dittmann et al., 2017; Gillon et al., 2017]. Kepler has already shown that M dwarfs host Earth-size planets in numbers exceeding those around hotter stars like our Sun, and that a significant fraction of these planets orbit in these cooler stars' compact habitable zones [Dressing and Charbonneau, 2015; Mulders et al., 2015]. Currently, four sub-15 pc transiting HZ exoplanets are known, including TRAPPIST-1 d, e, and f [Gillon et al., 2017] and LHS 1140 b [Dittman et al., 2017]. Several additional non-transiting planets are also known, including Proxima-b and Ross 128b. By the mid 2030s, TESS [Sullivan et al., 2015], MEarth [Berta-Thompson et al., 2013], TRAPPIST [Gillon et al., 2016], SPECULOOS [Search for habitable Planets Eclipsing Ultra-cOOl Stars; Gillon, 2017], and future surveys will have systematically searched all nearby M-dwarf stars and discovered more than a dozen transiting habitable-zone exoplanets. A projection of the Sullivan et al. [2015] sample for TESS – based on the Kepler statistics, and thus only on early M dwarfs – suggests ~8 temperate rocky planets (<300 K, <1.5 $R_E$) around M stars less than 0.25 $M^\odot$. Furthermore, the ground-based project SPECULOOS [Gillon, 2017] is anticipated to detect an additional 25 HZ planets, a few of which will be within 15 pc, for a total of at least a dozen quality targets. However, this number could be a factor of ~2 too low, as the consideration of compact multiple planetary systems implies a revised TESS planet yield for early M dwarfs 50% higher than that of Sullivan et al. [Ballard, 2018]. Additionally, immediate discovery of the TRAPPIST-1 system among the first 50 stars in the survey suggests the frequency of HZ planets around mid-to-late M dwarfs could be higher. Therefore, a number closer to ~2 dozen seems quite plausible.

M dwarfs differ from solar-type stars in important aspects, including the mass of their initial protoplanetary disks [Pascucci et al., 2016], duration of their pre-main sequence phase, UV and X-ray emission relative to their total luminosities, and strength of tides in their habitable zones. The effects of these characteristics have been modeled, but there are many uncertainties and unsupported assumptions in these models, and little or no data [Tarter et al., 2007; Scalo et al., 2007, Shields et al., 2016]. A common concern is that the extreme UV environment and high stellar luminosity during the pre-main sequence phase may preclude terrestrial planets orbiting M dwarfs from having water or being habitable. However, recent mass–radius–composition models based on new mass estimates of TRAPPIST-1 [Grimm et al., 2018] suggest HZ planets TRAPPIST-1f and g probably contain substantial (≥50 wt%) water/ice [Unterborn et al., 2018]. Additionally, M dwarf planets may be tidally-locked, with permanent daysides and nightsides, presenting a unique environment in which to explore habitability. For these planets, the temperature difference on the day and night sides of planets [e.g., Joshi et al., 1997] and slow rotation may even increase the HZ limit due to cloud coverage on the dayside [Kopparapu et al., 2017]. Since M dwarfs are the most ubiquitous stars in our galaxy, accounting for at least 75% of all stars, the "typical" life-bearing planet may orbit such a star. Furthermore, the closest potentially-habitable planets outside of our solar system likely orbit these cool stars. Still, as of yet, it is unknown if these planets are Earth-like or habitable. OST can assess if these planets are capable of supporting life.

The presence of reducing molecules (such as methane) in an oxidizing atmosphere (ozone or nitrous oxide) is a fundamental indicator of life (**Figure 1-8**); thus, taken together, these molecules are referred to as biosignatures [Kaltenegger, 2017]. The proposed wavelength range (5 - 25 μm) for the OST Mid-infrared Imager, Spectrometer, Coronagraph (MISC) contains several biosignature combinations, and includes strong spectral features from water and $CO_2$ that can further constrain a planet's climate. Additionally, acquiring data with R~300 at λ ≥ 18 μm permits reduction of light curve systematics introduced by stellar variability.

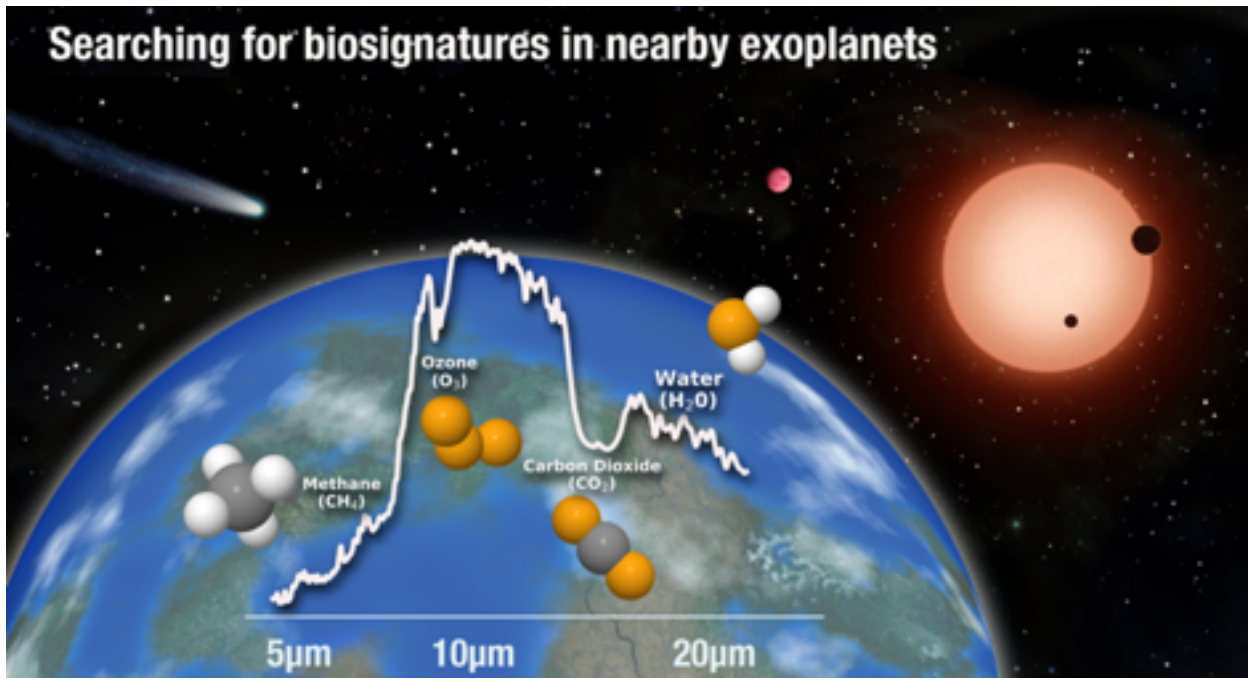

**Figure 1-8:** Using its mid-infrared transit spectrometer, MISC, OST will search for bio-indicators ($H_2O$, $CO_2$) and biosignatures ($O_3$, $CH_4$) in nearby exoplanets to determine if they are habitable.

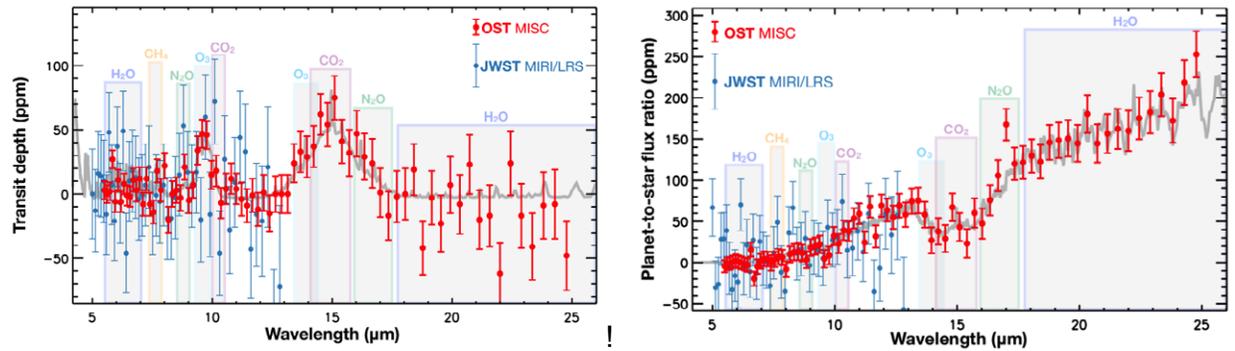

**Figure 1-9:** OST's MISC transit spectrometer channel will characterize habitable planets in search of signs of life. Model transmission (A) and emission (B) spectra of TRAPPIST-1e (0.92 $R_E$, 250 K) with realistic uncertainties from JWST and OST when the system is moved to a distance of ~4.2 parsecs ($K_{mag}$ = 8) show OST's improvement in precision and wavelength coverage over JWST.

Model transmission and emission spectra of TRAPPIST-1e with simulated data show OST can achieve a lower noise floor and, hence, higher precision than JWST (**Figure 1-9**). Simulated JWST MIRI/LRS data assume an optimistic noise floor of 30 ppm and, even after 30 visits, it does not achieve the necessary precision to definitively detect methane or adequately constrain the dayside temperature of this 250 K planet. Due to its lower estimated noise floor (<5 ppm), OST/MISC not only measures thermal emission, but also constrains the atmospheric abundances of molecules, including such as carbon dioxide ($CO_2$, 15 μm), water vapor ($H_2O$, 6.3 and $\lambda \geq 18$ μm), ozone ($O_3$, 9.6 μm), methane ($CH_4$, 7.7 μm), and nitrous oxide ($N_2O$, 17 μm).

OST will perform a survey of the nearest habitable-zone exoplanets and, for the first time, empirically constrain the fraction of planets orbiting M dwarfs that might harbor life. Early estimates based on the Sullivan et al. [2015] sample suggest a 9.1-m telescope OST could detect $CO_2$ and $O_3$ in at least 30 and 20 rocky planets, respectively, at 5σ confidence [Gaidos et al., in prep]. Kepler has successfully determined the frequency of habitable-zone, Earth-size planets, $\eta_e$. OST, in turn, is poised to take the next step in answering the question "Are we alone?" by using proven techniques to constrain the atmospheric temperatures and constituents from the nearest transiting terrestrial worlds.

### Synergies with JWST and Future Ground-Based Facilities

JWST will perform a reconnaissance of the most promising rocky exoplanets transiting mid-to-late M-dwarfs, primarily in transmission at near-infrared wavelengths [Morley et al., 2017]. However, it is unlikely JWST's MIRI/LRS (5 - 12 μm) will have the stability and precision necessary to obtain reliable mid-infrared spectra of rocky HZ planets [e.g., Greene et al., 2016], let alone detect biosignature gases such as $O_3$ and $CH_4$. Beyond 12 μm, JWST/MIRI is limited to broadband photometry. The recently-selected ESA exoplanet spectroscopy mission ARIEL (Atmospheric Remote-sensing Exoplanet Large-survey) will focus on a transiting exoplanet phase space that has little overlap with OST, focusing on a large sample size of hotter and larger planets with a 1-meter telescope [Zingales et al., 2017]. From the ground, the next generation of extremely large telescopes will search for $O_2$ using high-resolution spectrographs [Snellen et al., 2013; Rodler et al., 2014]. At the point OST is launched, there will be some knowledge about the targets of interest, such as whether or not a planet has an atmosphere or contains an indeterminate amount of $O_2$, but the picture will be incomplete. OST will determine the atmospheric compositions and thermal structures of dozens of potentially habitable planets, thus opening the door for comparative exoplanetology of rocky worlds.

**Our Solar System in Context: Detection and Spectra of Cool Transiting Gas Giants**

The last decade of exoplanet science has shown there is a broad continuum of worlds, from sub-Earths to super-Jupiters. For giant planets with hydrogen-dominated atmospheres, which range from ~10 to ~1000 Earth masses, even post-JWST our understanding of these atmospheres across a wide range of temperature will be elusive. The best-characterized JWST targets will be far hotter than the solar system's cold gas giants. OST's wavelength coverage and sensitivity will enable transformational science in the chemistry and dynamics of temperate giant planets. In these atmospheres, the abundances of $CH_4$ and $NH_3$ probe metallicity, non-equilibrium chemistry, and the strength of vertical mixing [Zahnle and Marley, 2014]. OST can probe the rich photochemistry expected for these atmospheres, via production of HCN, $C_2H_2$, $C_2H_4$, and $C_2H_6$, beyond 5 μm. OST can also advance our understanding of the physics and chemistry of clouds, which are currently a major source of uncertainty in models of these atmospheres. A number of cloud species known or expected to impact the spectra of these planets over a wide temperature range have Mie scattering features in the mid-infrared [Wakeford and Sing, 2015]. Additionally, OST's eclipse maps and phase curves will provide an unprecedented view of dynamics and circulation in cool atmospheres, as a bridge to the solar system's gas giants.

**Exoplanet Coronagraphy with OST**

With estimated contrast ratios of $10^{-4}$ to $10^{-5}$, OST/MISC's coronagraph module will be able to directly image and characterize the atmospheres of gas giant planets in wide orbits. JWST will search for Saturn-mass planets in young systems (~100 Myr) that host high-mass ($2M_J$) planets detected through ground-based imaging. These Saturn-mass planets will be excellent targets for characterization with OST. Additionally, because the drop in planet-to-star contrast with planet temperature (and therefore planet age) is less extreme in the mid-IR than in the near-infrared (where current ground-based surveys operate), middle-aged stars (~300-1000 Myr) typically passed over by high-contrast imaging surveys become viable targets for detecting Jovian exoplanets.

### 1.1.4 General Astrophysics Opportunities

> Starting in the first year of operations, OST will be available to all astronomers via competitively reviewed proposals.

Enabled by its proposed cold, large telescope and improved detectors, OST provides a tremendous discovery space that will facilitate breakthrough discoveries in all areas of astronomy through its dramatic increase in spectral line sensitivity (up to $10^4$) (**Figure 1-10**). OST's spectral resolution enables broadband imaging, medium resolution spectroscopy, and high-resolution spectroscopy. The three science themes driving the design could be accomplished in the first year of operations. Most OST observations will be driven by competitively-reviewed proposals selected from the 2030s astronomical community, which is likely to propose additional study concepts based on knowledge at the time. The strength of a large general observatory mission like OST is that it expands the science of the time during which it operates, not the science outlined in the mission's early planning phases.

### 1.2 Science Traceability Matrix

> High priority science, selected in partnership with the astronomical community through a competitive proposal process, drove selection of the cooled, large-aperture telescope design.

At the outset of the study, the OST team reached out to the community and solicited proposals detailing interesting science questions and the types of measurements needed to answer them.

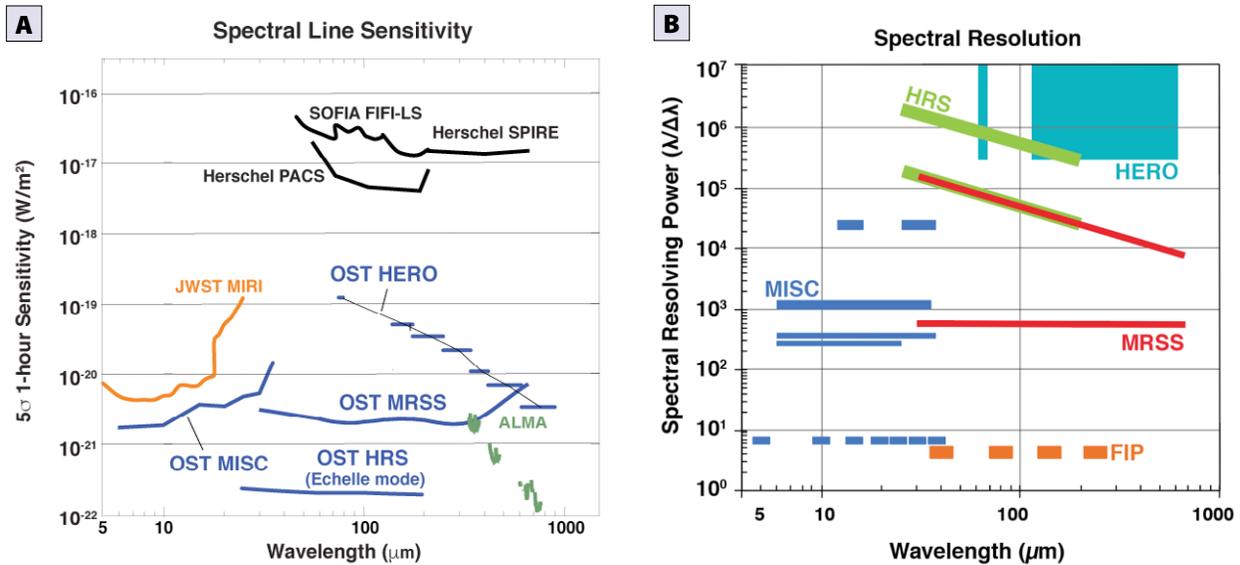

**Figure 1-10:** OST spectral line sensitivity (A) is up to 4 orders of magnitude better than previous missions. The spectral resolution range (B), from broadband imaging to very high spectral resolution, resolves the kinematics of water in protoplanetary disks.

Proposals were reviewed and ranked by the voting STDT members. The complete list of proposals received, ranked by science, is provided in **Table 1-1**. The highest ranked proposals guided selection of a large-aperture telescope architecture, rather than an interferometer, because the most interesting questions required high-sensitivity measurements, but not high angular resolution. The OST team made this architecture decision in August 2016. The team remained open to receiving science proposals throughout the course of the study (four proposals arrived after final ranking was determined). All proposals are available on the OST website at: http://origins.ipac.caltech.edu/.

The team selected the top 14 science proposals as the basis for defining OST Mission Concept 1. Requirements derived from these top proposals are captured in the OST Science Traceability Matrix (**Table 1-2**).

**Table 1-1:** Proposals Submitted to the OST Study for Mission Concept 1

| Submitted Science Proposal Title |
| --- |
| The Rise of Metals |
| Biosignatures of Transiting Exoplanets |
| The First Dust |
| Water Content of Planet-Forming Disks |
| Connection Between Black Hole Growth and Star Formation Over Cosmic Time |
| Direct Detection of Protoplanetary Disk Masses |
| Birth of Galaxies During Cosmic Dark Ages |
| Galaxy Feedback from SNE and AGN to z~3 |
| Survey of Small Bodies in the Outer Solar System |
| Direct Imaging of Exoplanets |
| Star Formation and Multiphase ISM at Peak of Cosmic Star Formation |
| Thermo-Chemical History of Comets and Water Delivery to Earth |
| Galaxy Feedback Mechanisms at z<1 |
| Water Transport to Terrestrial Planetary Zone |
| Frequency of Kuiper Belt Analogues |
| Feedback on All Scales in the Cosmic Web |
| Magnetic Fields and Turbulence – Role in Star Formation |
| The EBL (extra-galactic background light) with OST |
| Determining the cosmic-ray flux in the Milky Way and nearby galaxies |
| Episodic Accretion in Protostellar Envelopes and Circumstellar Disks |
| Formation and History of Low-Mass Ice Giant Planets |
| Find Planet IX |
| Fundamental of dust formation around evolved stars |
| The dynamic interstellar medium as a tracer of galactic evolution |
| Probing magnetic fields with fine structure lines |
| Ice/Rock Ratio in Protoplanetary Disks |
| Role of Environment in Galaxy Evolution |
| Galaxies at Reionization |
| Large-Scale Structure – Crucial FIR Link |
| Obscured AGN |
| Regulating the Multiphase ISM |
| Planetary Origins and Evolution of the Solar System |
| Gas and Comets in Exoplanetary Systems |
| Stochastic vs. secular accretion in forming stars |
| Cooling Power of Molecular Gas in Star-Forming Regions |
| Star Formation Efficiency Outside the Milky Way |
| Comparative Climate and Thermal Evolution of Giant Planets |
| Stellar Mass Buildup in galaxies over cosmic time |
| MIR spectroscopy of rest frame optical lines to estimate metallicity over cosmic time |
| Dust Processing in Outer Planetary Systems |
| Gas in debris disks |

**Table 1-2:** OST Science Traceability Matrix

| OST Science Theme (and NASA Roadmap Science Goal) | Science Objectives | Science Requirements | | Instrument Requirements | | Mission Functional Requirements (Top-Level) |
|---|---|---|---|---|---|---|
| | | Science Observable | Measurement Requirement | Technical Parameter | Technical Requirement | |
| 19, Rise of Metals, Dust, and the First Galaxies<br><br>Trace the dust and metal enrichment history of the early Universe. Find the first cosmic sources of dust, and search for evidence of the very earliest stellar populations forming in pristine environments. | Trace the rise of metals and (a) determine the evolution in metallicity from z=1 to z=3 to 0.1 deg down to $10^{10} L_{sun}$; (b) determine the cosmic metal abundance Ωmetals from z=0 to z=8 to 0.1 deg accuracy in 8 redshift bins; and (c) measure the multiple phases of the ISM to infer the physical phenomena that regulate SF efficiency at the peak of cosmic star formation at z=1–3. | z=1–3 relative metallicity tracer: [NeII]12.8, [NeIII]15.6, [SIII]18.7, [SIV]10.5;<br>z=0–8 relative metallicity tracer: [OIII] 52+88μm, [NIII] 57μm ;<br>cooling and heating of the ISM through [OI], [OIII], [NII], [CII].<br><br>A multi-tiered survey, with a wider tier of ~10 deg², with sensitivity down to $10^{11} L_{sun}$ galaxies at z=3 and $10^{12} L_{sun}$ galaxies at z=8.<br>Aim is a sample with >10000 galaxies with IR luminosity >$10^{11} L_{sun}$ at z=1–3 | Rest-frame mid and far-IR spectral mapping to select z=0 to 8 galaxies | Wavelength range | 20-800 μm | M-1: Observing strategies: TBD<br><br>M-2: Launch Window: TBD<br><br>M-3: Mission Life to complete Science Requirements: 5 years<br><br>M-4: Desired Mission Life or Extended Mission Life: 10 years<br><br>M-5: Observatory will orbit L2<br><br>M-6: Observatory will accommodate five instruments |
| | | | Identify galaxies in a tiered spectral mapping survey | Spatial resolution | 5 arcsec at 200 μm | |
| | | | Measure line flux densities of identified galaxies | Spectral line sensitivity | $1 \times 10^{-21}$ W m⁻² (driven by the MIR lines) | |
| | | | | Spectral Resolving power | $\lambda/\Delta\lambda = 500$ | |
| | | | | Field of Regard | | |
| | | | | Square Degrees | | |
| 9, Tracing the signatures of life and the ingredients of habitable worlds | Determine the mass (with HD as a tracer of $H_2$), abundance and spatial distribution of water vapor inside and outside of the snowline in 1000 protoplanetary disks with +/- 20 degrees of the Galactic plane around young stars across the stellar mass range. | Spectra of rotational emission lines from water with upper level energies of 100-2000 K. Targets are 1000 proto-planetary disks out to distances of 500 pc, including the high-mass star forming region in Orion, and brown dwarfs with masses from 0.05 to 10 $M_{sun}$. | Measure fluxes of multiple (>10) emission lines spanning the full range of upper level energy. | Wavelength range | 28-200 μm | |
| | | | | Bandwidth | 1000 km/s (for HD at 112 μm) | |
| | | | | Spatial resolution | $\theta_{slit} < 2''$ at 30 μm | |
| | | | Spectrally resolve lines beyond the snow line to independently determine the emitting area of the detected gas (using Keplerian line profiles). | Spectral resolving power | $\lambda/\Delta\lambda > 25,000$ | |
| | | | | Spectral line sensitivity | $10^{-21}$ W/m² (5σ) | |
| | | | | Field of Regard | | |
| | | | | Square Degrees | | |
| 27, Rise of Metals, Dust, and the First Galaxies<br><br>Trace the dust and metal enrichment history of the early Universe. Find the first cosmic sources of dust, and search for evidence of the very earliest stellar populations forming in pristine environments. | Mid-IR dust features measure small-grain cooling, provide unambiguous redshift, and discriminate among various underlying stellar populations, and system metallicities. | Measure the rest-frame 7-25 μm PAH spectra of 1000 galaxies identifiable with JWST/WFIRST, uniformly sampling the 0.5 to 1.5 billion-year cosmic time interval (z=4 to 10), with a range of luminosities extending down to $3\times10^{11} L_{sun}$ (modern-universe equivalent mid-IR luminosity). | High-quality mid-IR spectra of galaxies, appearing as point sources.<br>Measure PAH flux densities with sensitivity down to $3\times10^{11} L_{sun}$ | Wavelength range | 35 mm < λ <275 mm | |
| | | | | Spatial resolution | $\theta_{slit} < 5''$ at 100mm | |
| | | | | Spectrometer Sensitivity | $1\times10^{-20}$ Wm⁻² (5σ) in R=50 bin. | |
| | | | | Spectral Resolving power | $\lambda/\Delta\lambda > 50$ | |
| | | | | Calibration / gain accuracy across spectrometer. | 3% relative calibration in R=50 bins across the spectrometer. | |
| | | | | Field of Regard | | |
| | | | | Square Degrees | | |
| 14, Tracing the signatures of life and the ingredients of habitable worlds<br><br>Explore whether planets around other stars could harbor life (NASA Roadmap Plan) | Search for bio-signatures and measure climates in nearby terrestrial super-Earth exoplanets through transit emission spectroscopy. | Secondary transit emission spectrum and non-transiting thermal phase curves of rocky planets, selected by 2025 transit and radial velocity surveys including all-sky TESS. | Spectrally resolve $CO_2$, $N_2O$, $O_3$ and $H_2O$ absorption features in the emission spectrum | Wavelength range | 9 μm < λ <40 μm | |
| | | | | Spatial resolution | $\theta_{slit} < 2''$ at 10 μm | |
| | | | | Spectral Resolving power | $\lambda/\Delta\lambda > 500$ | |
| | | | Separate thermal emission from the planet from the host star | Photometric sensitivity | 1 μJy at 10 μm (5σ) | |
| | | | | Photometric precision | 25ppm | |
| | | | | Transit monitoring cadence | ~5 minute | |
| | | | | Field of Regard | | |
| | | | | Square Degrees | | |
| 4, Tracing the signatures of life and the ingredients of habitable worlds<br><br>Trace the trail of water from interstellar clouds, to protoplanetary disks, to Earth itself in order to understand the abundance and availability of water for habitable planets. | Establish the water content from ISM to protoplanetary disks and establish the dominant formation mechanisms of water in interstellar clouds, the processes that set the grain-surface and gas-phase abundances and the physical state of water as clouds collapse to form stars and protoplanetary disks. Measure velocity field within dense, collapsing cores to improve modeling of the process of star-formation. | Using observations of 120 clouds and cores, imaged over an area of 5 arcmin2, character-ize the $H_2O$ abundance, temperature, velocity structure with ob-servations of low energy (tracing cool gas) $H_2O$ spectral lines and isotopologues in quiescent clouds, shocked regions, and in different stages of the col-lapse of interstellar clouds to form stars. | Spectrally resolve lowest rotational lines of ortho and para $H_2O$ plus isotopologues at 509-557 GHz and 1107-1113 GHz | | 509-557 and 1107-1113 GHz (590-539 and 270 μm) to cover key $H_2O$ lines and isotoplogues) | |
| | | | | Spectral Resolving power | I/DI = 5 x 10⁻⁶ (0.06 km/s to resolve velocity structure) | |
| | | | | Bandwidth | < 20 GHz | |
| | | | | Angular resolution | 30″ (to resolve star forming clouds in the Milky Way) | |
| | | | | Spectral Line sensitivity | 4 mK in 0.1 km/s @ 600 GHz = 1.6 x10⁻²¹ W m⁻² for 6 m telescope | |
| | | | | Field of Regard | | |
| | | | | Square Degrees | | |

| OST Science Theme (and NASA Roadmap Science Goal) | Science Objectives | Science Requirements | | Instrument Requirements | | Mission Functional Requirements (Top-Level) |
|---|---|---|---|---|---|---|
| | | Science Observable | Measurement Requirement | Technical Parameter | Technical Requirement | |
| 21, Understanding the co-evolution of BH and galaxies over cosmic time "How did we get here?" | Measure sta-formation rate (SFR) and blackhole accretion rate (BHAR) in IR-selected $10^{11}L_{sun}$ and $10^{12}L_{sun}$ galaxies to determine the cosmic SFR density and BHAR density for z > 2-3 | Large area map to build large sample of 2 < z < 7 LIRGs and ULIRGs with sensitive spectra covering the mid-IR PAH and bright fine-structure lines. At the lowest redshift bins, want to be sensitive to sources with SFR > 10-20 $M_\odot$yr-1 | Spatial-spectral mapping of 5 sq. degree area to generate samples and for each source, measure line fluxes and equivalent widths of key dust and gas diagnostic lines: 6.2-12.7 μm PAHs, and [OIV] 26, [NeII] 12, [NeIII] 15, [NeV] 14. (At z=5 a $10^{12}$ $L_\odot$ source should have a [NeII] line flux ~6x$10^{-21}$ W m-2). 5 deg2 results in at least $10^4$ sources, with at least 500-1000 galaxies at z > 5. | Wavelength range | 15-300 μm | |
| | | | | Spatial resolution | 5 arcsec at 100 μm | |
| | | | | Spectral Resolving power | l/Dl >300 (maximize PAH and fine structure line detections) | |
| | | | | Spectral line sensitivity | 3x$10^{-21}$ W m-2 (5σ) | |
| | | | | Field of Regard | | |
| | | | | Square Degrees | | |
| 26, Rise of Metals, Dust, and the First Galaxies Trace the dust and metal enrichment history of the early Universe. Find the first cosmic sources of dust, and search for evidence of the very earliest stellar populations forming in pristine environments. | Spectroscopic imaging of gas cooling via rotational transitions of molecular hydrogen over the epoch of reionization to cosmic dark ages at z > 9. Directly detect H2 molecular gas cooling at 4 < z < 15 $10^9$ to $10^{10}$ M_sun dark matter halos, through magnification provided by gravitational lensing. | Survey of lensing caustics around 30 a priori known massive lensing clusters to identify lensed H2 cooling in early galaxies. Obtain line detections in at least three rotational emission lines. | Detect at least 3 H2 rotational lines from S(0) (28 microns) and S(5) (6.9 μm) lines, over 4 < z < 15. Sensitivity to reach and identify $10^{10}$ solar mass halo at z=7 (per current H2 models), including 10x lensing boost. | Wavelength range | 32 μm < λ <455 μm | |
| | | | | Spatial resolution | $θ_{pix}$ < 5" at 100 μm | |
| | | | | Spectral line sensitivity | 2x$10^{-22}$ Wm-2 (5σ) | |
| | | | | Spectral Resolving power | λ/Δλ > 500 | |
| | | | | Spectrometer relative calibration accuracy | 3% from channel to channel | |
| | | | | Field of Regard | | |
| | | | | Square Degrees | | |
| 18, Understanding the co-evolution of BH and galaxies over cosmic time | Measure the feedback from SNe and AGN on the molecular and atomic ISM in galaxies from z=1-3. Determine ubiquity and importance of outflows for SFRD drop since z=3. Use the OH absorption to derive a mass outflow rate. Compare to the SFR from [NeII] and [NeIII] lines to derive depletion timescales. For sources with AGN, compare outflow rates and velocities with AGN power from [OIV]. | Pointed spectra of 2000 infrared-selected galaxies from the general extragalactic survey. Detect absorption at the 10% level at z=3 or 1/20 line flux of [NeII] at z=3. Spectral resolving power is set to separate absorption and emission in OH profiles (100 km/s level) and detect broad line wings in lines with intrinsic widths of 200-300 km/.s. | Ability to detect OH emission and absorption, with the requirement to detect 79 μm line at z=3, Perform spectral fits to OH and fine-structure lines to get velocities of blue-shifted absorption (OH) and line wings (fine-structure lines). Measure line fluxes, profiles of bright fine structure lines - [NeII] [OIV], [NeIII] and H2 | Wavelength range | 20-350 μm | |
| | | | | Spatial resolution | 5" at 150 μm | |
| | | | | Spectral Resolving power | λ/Δλ =3000 | |
| | | | | Spectral line sensitivity | 1x$10^{-21}$ W m-2 (5σ) | |
| | | | | Field of Regard | | |
| | | | | Square Degrees | | |
| 29, Tracing the signatures of life and the ingredients of habitable worlds | Determine the distri-bution of D/H values in outer Solar system comets to reveal the thermo-chemical his-tory and determine their role in delivery of water to the early Earth. Determine both the mass content of water and the dis-tribu-tion. | Measure the HDO/H2O ratio in 100s of comets. | Spectrally resolve H2O and HDO lines | Wavelength range | 548, 557, 988, 995, 1107, 1113 GHz (H2O & isotopologues) 509, 600, 894, 919, 1010 GHz (HDO) | |
| | | | | Spectral Resolving power, resolve lines for DV~1.3 km/s | λ/Δλ= 2 x $10^5$ | |
| | | | | Bandwidth (to cover both isotopes simultaneously) | 1 GHz | |
| | | | | Spectral Line sensitivity | 2 x $10^{-21}$ W m | |
| | | | | Moving Target tracking | 60 mas/s (based on statistics of JWST at 30 mas/s) | |

| OST Science Theme (and NASA Roadmap Science Goal) | Science Objectives | Science Requirements | | Instrument Requirements | | Mission Functional Requirements (Top-Level) |
|---|---|---|---|---|---|---|
| | | Science Observable | Measurement Requirement | Technical Parameter | Technical Requirement | |
| 7, Revealing the interplay between stars, black holes, and interstellar matter over cosmic | Measure the role of magnetic fields and turbulence in star formation, connecting Galactic-scale ISM physics to proto-stellar cores | Measure the dust emission and polarization from the peak of the dust SED. | 1000 deg² photometric and polarization mapping and photometry across the dust SED peak | Wavelength range | 200 μm <l<500 μm | |
| | | | | Spatial resolution | $\theta_{res}$<2" at 100 μm | |
| | | | | Sensitivity to high-dynamic range targets | Dynamic range 1000. | |
| | | | | Polarization capabilities | 0.1% in linear and circular polarization, ±1° in pol. Angle | |
| | | | | Photometric mapping | 1μJy at 250 μm (5σ) | |
| | | Trace turbulent shocks and directly measure both energy injection scales and energy losses in the neutral ISM. | Detailed (high spatial and spectral resolution) maps of spectral lines in particular of [CII], and the CO Spectral Line Energy Distribution (CO SLED). | Spatial resolution | $\theta_{res}$<2" at 100 μm | |
| | | | | Spectral resolution | l/Dl > 3x10⁶ | |
| | | | | Spectral line sensitivity | $3 \times 10^{-19}$ W m⁻² | |
| | | | | Field of Regard | | |
| | | | | Square Degrees | | |
| 5, Reveal the interplay between stars, black holes, and interstellar matter over cosmic time. | Characterize the dominant mechanism of feedback (AGN and/or star-formation) and the amount of material ejected in outflows across a population of 1000 z<1 galaxies with varying properties, morphologies and AGN activity. | (a) Map extraplanar dust emission in continuum or PAH; (b) Measure H₂O, [OI], and OH lines in absorption to identify outflowing gas; (c) Map extraplanar gas in emission to measure the ejected mass | Resolve and map dust continuum, PAH emission, and H₂, HD, [CII] and molecular ions around galaxies to measure mass outflow rates | Wavelength Range | 10-500 μm | |
| | | | | Angular resolution | 5" at 100 μm | |
| | | | | Continuum Sensitivity | 50 μJy (5σ) | |
| | | | | Surface Brightness Sensitivity | $1 \times 10^{-12}$ W/m²/sr | |
| | | | Determine speed of outflowing neutral gas by resolving P-Cygni line profiles. | Spectral Line Sensitivity | $6.6 \times 10^{-22}$ W/m² | |
| | | | | Spectral Resolving power | λ/Δλ = 10⁴ | |
| | | | | Bandwidth | 1 GHz | |
| | | | | Field of Regard | | |
| | | | | Square Degrees | | |
| 30, Tracing the signatures of life and the ingredients of habitable worlds | A wide 1000 sq. degree area survey on the ecliptic plane to detect and catalog Trans-Neptune Objects (TNOs) and related Outer Solar System Small Bodies, measuring 1000s of small bodies (7-20 km) to inform us about the early history of the solar system, and how its composition has evolved over the time. | Determine the temperature and size distribution of 1000s of TNOs, on the ecliptic plane. | Image a large area of the sky (>1000 square deg.) and capture >1,000 TNOs (desire multiple wavelengths). Cadence: 4 times over a year over 1,000 sq. deg. with a cadence of few days to weeks. | Wavelength range | 75 < λ < 250 μm | |
| | | | | Instantaneous field of view (to track motions) | 14 x 14 arcmin | |
| | | | | Continuum Sensitivity | 50 μJy at 125 μm (5σ) | |
| | | | | Angular Resolution | 3" at 125 μm | |
| | | | | Field of Regard | | |
| | | | | Square Degrees | | |
| 16, Tracing the signatures of life and the ingredients of habitable worlds | Characterize (measure composition, temperature and size) Jupiter analogs (1 M_Jup planets at 10 AU) around nearby stars. | Mid-infrared emission spectra obtained in direct imaging, targeting strong features from NH₃, H₂O and CO₂, as well as their effective temperature. | Use coronagraphic high contrast spectro-imaging to directly image Jupiter analogs around within 10 pc | Wavelength range | 9-40 μm | |
| | | | | Spatial resolution | <0.25" at 10 μm | |
| | | | | Spectral Resolving power | λ/Δλ > 500 | |
| | | | Obtain spectra of Jupiter analogs to detect spectral molecular bands | Photometric sensitivity | 1 μJy at 10 μm (5σ) | |
| | | | | Coronagraphic contrast | 10⁶ @ 0.5" | |
| | | | | Field of Regard | Over the course of the mission = 4π sr. | |
| | | | | Square Degrees | | |

## 1.3 Design Reference Mission

A design reference mission (DRM) can be used to define the operations of an observatory and act as a demonstration of an observatory's science capability. The team used science proposals (**Table 1-1**) derived for the mission architecture to create a DRM for OST Mission Concept 1. Science proposers converted their science proposal ideas into observing proposals using the instrument suite for Mission Concept 1. Each proposer estimated necessary observing time needed to accomplish the goals in their submitted proposal using the instrument performance sheets for OST Mission Concept 1.

**Table 1-3** shows the top 14 programs' requested science hours. The total requested science hours for all 14 programs, as proposed, is 2.5 years. Assuming 50% efficiency for the observatory, the total time required to complete all 14 proposed science programs is 5 years, the nominal mission lifetime. The OST observatory program will be governed by a time allocation process, with the final program a combination of the large comprehensive programs outlined in this proposal and smaller programs focusing on specific objects and targets. Nevertheless, demand from so many potential exciting programs demonstrates that this observatory will generate amazing results over its lifetime. Programs beyond the current top 14 proposals would be desired, and the OST Mission Concept 1 can accommodate a diverse set of programs beyond this initial potential set.

A DRM is also useful for defining an observatory's operational modes. Several key proposed programs (e.g., programs 1 and 9) require large area sky surveys, and the observatory will need to have the basis for accomplishing necessary maneuvers to fulfill these types of observations. The observatory also needs stability to enable transit spectroscopy (i.e., program 2). Transit target locations are in all quadrants of the sky, requiring an annual field of regard that covers the entire sky.

**Table 1-4** outlines the usage per instrument mode for all programs submitted (**Table 1-1**) showing that a realistic, broad observatory program utilizes the full capability of concept 1. **Table 1-5** uses only the top 14 (**Table 1-3**). The instrument usage provides a rough instrument prioritization as a starting point for Mission Concept 2 instrument suite.

**Table 1-3:** The top 14 proposed science programs were used to inform the OST Concept 1 DRM.

| Rank | Proposal Title | Hours | Instruments |
|---|---|---|---|
| | **Design Reference Mission for Mission Concept 1** | | |
| 1 | The Rise of Metals (* totals include studies in proposals 3, 7, 8, 11) | 3000 | MRSS – survey R~500 |
| 2 | Biosignatures of Transiting Exoplanets | 4000 | MISC, transit channel, R~500 |
| 3 | The First Dust (*) | Studies incorporated into Rise of Metals | |
| 4 | Water Content of Planet-Forming Disks | 4795 | HRS, HERO |
| 5 | Connection Between Black Hole Growth and Star Formation Over Cosmic Time | Folded in with Rise of Metals | |
| 6 | Direct Detection of Protoplanetary Disk Masses | 4181 | MRSS (R~43,000), HERO |
| 7 | Birth of Galaxies During Cosmic Dark Ages (*) | Studies incorporated into Rise of Metals | |
| 8 | Galaxy Feedback from SNE and AGN to z~3 (*) | Studies incorporated into Rise of Metals | |
| 9 | Survey of Small Bodies in the Outer Solar System | 8000 | FIP |
| 10 | Direct Imaging of Exoplanets | 1000 | MISC, coronagraph |
| 11 | Star Formation and Multiphase ISM at Peak of Cosmic Star Formation (*) | Studies incorporated into Rise of Metals | |
| 12 | Thermo-Chemical History of Comets and Water Delivery to Earth | 120 | HRS, HERO |
| 13 | Galaxy Feedback Mechanisms at z<1 | 1100 | MRSS +FTS, MISC imaging |
| 14 | Water Transport to Terrestrial Planetary Zone | 66 | HERO |
| | **Total Science Hours for Top 14 Proposals** | **22081 Hours** | **2.5 Years** |
| | **Total Science Hours (Top 14) Assuming 50% Observatory Efficiency** | **44162 Hours** | **5.0 Years** |

**Table 1-4:** Instrument usage from all the submitted programs (see Table 1-1) shows the wide variety of instrument capability desired by the astronomical community.

| Instrument Mode | # Programs | % Programs | # of Hours 22079 Hrs (~2.5 Yrs) | % Hours | 1/Rank of Highest Ranked Prop * %Program |
|---|---|---|---|---|---|
| FIP | 13 | 25.5 | 1635 | 7.4 | 2.8 |
| FIP Survey | 6 | 11.8 | 885 | 5.0 | |
| FIP Pointed | 5 | 9.8 | 250 | 1.1 | |
| FIP Polarimetry Short | 1 | 2.0 | 250 | 1.1 | |
| FIP Polarimetry Long | 1 | 2.0 | 250 | 1.1 | |
| MRSS | 14 | 27.5 | 8581 | 38.9 | 27.5 |
| MRSS 30-660 Scanning Survey | 7 | 13.7 | 4300 | 19.5 | |
| MRSS FTS | 4 | 7.8 | 4181 | 18.9 | |
| MRSS 30-140 Pointed | 2 | 3.9 | 100 | 0.5 | |
| MRSS 140-660 um pointed | 1 | 2.0 | 0 | 0 | |
| HRS | 5 | 9.8 | 4647 | 21.0 | 2.45 |
| HRS Pointed High Resolution | 4 | 7.8 | 4535 | 20.5 | |
| HRS Pointed Moderate Resolution | 1 | 2.0 | 112 | 0.5 | |
| HERO | 11 | 21.6 | 2316 | 10.5 | 5.4 |
| HERO Scanning | 2 | 3.9 | 1126 | 5.1 | |
| HERO Polarization | 0 | 0 | 0 | 0 | |
| HERO Pointed Double Beam Switch | 2 | 3.9 | 1100 | 5.0 | |
| HERO Pointed Freq Swtich | 1 | 2.0 | 0 | 0 | |
| HERO w/Off | 3 | 5.9 | 32 | 0.1 | |
| HERO Stare | 1 | 2.0 | 18 | 0.1 | |
| HERO Raster Map | 1 | 2.0 | 30 | 0.1 | |
| HERO Scanning | 0 | 0 | 0 | 0 | |
| HERO Freq. Scanning | 1 | 2.0 | 10 | 0 | |
| MISC | 8 | 15.7 | 4900 | 22.2 | 7.85 |
| MISC MIR Transit | 1 | 2.0 | 4000 | 18.1 | |
| MISC Coronagraph | 1 | 2.0 | 50 | 0.2 | |
| MISC Coronagraph Spectroscopy | 1 | 2.0 | 50 | 0.2 | |
| MISC Imaging | 3 | 5.9 | 650 | 2.9 | |
| MISC Low Res Spectroscopy | 2 | 3.9 | 150 | 0.7 | |
| MISC Medium Res Spectroscopy | 0 | 0 | 0 | 0 | |

*Note: Assumptions about observing hours were extrapolated for proposals that did not provide estimated observing time.*

**Table 1-5:** Instrument usage for the top 14 proposals.

| Instrument Mode | # Programs | % Programs | # of hours 17536 hrs 2.0 years | % Hours | 1/rank of highest ranked prop * %program |
|---|---|---|---|---|---|
| FIP | 2 | 11.1 | 300 | 1.7 | 1.2 |
| FIP Survey | 1 | 5.6 | 300 | 1.7 | |
| FIP Pointed | 1 | 5.6 | 0 | 0 | |
| FIP Polarimetry Short | | 0 | 0 | 0 | |
| FIP Polarimetry Long | | 0 | 0 | 0 | |
| MRSS | 7 | 38.9 | 7281 | 41.5 | 38.9 |
| MRSS 30–660 Scanning Survey | 5 | 27.8 | 3300 | 18.8 | |
| MRSS FTS | 2 | 11.1 | 3981 | 22.7 | |
| MRSS 30-140 Pointed | 0 | 0 | 0 | 0 | |
| MRSS 140-660 um pointed | 0 | 0 | 0 | 0 | |
| HRS | 1 | 5.6 | 4295 | 24.5 | 1.4 |
| HRS Pointed High Resolution | 1 | 5.6 | 4295 | 24.5 | |
| HRS Pointed Moderate Resolution | 0 | 0 | 0 | 0 | |
| HERO | 4 | 22.2 | 1060 | 6.0 | 5.6 |
| HERO Scanning | 0 | 0 | 0 | 0 | |
| HERO Polarization | 0 | 0 | 0 | 0 | |
| HERO Pointed Double Beam Switch | 1 | 5.6 | 1000 | 5.7 | |
| HERO Pointed Freq Swtich | 0 | 0 | 0 | 0 | |
| HERO w/Off | 1 | 5.6 | 12 | 0.1 | |
| HERO Stare | 1 | 5.6 | 18 | 0.1 | |
| HERO Raster Map | 1 | 5.6 | 30 | 0.2 | |
| HERO Scanning | 0 | 0 | 0 | 0 | |
| HERO Freq. Scanning | 0 | 0 | 0 | 0 | |
| MISC | 4 | 22.2 | 4600 | 26.2 | 11.1 |
| MISC MIR Transit | 1 | 5.6 | 4000 | 22.8 | |
| MISC Coronagraph | 1 | 5.6 | 50 | 0.3 | |
| MISC Coronagraph Spectroscopy | 1 | 5.6 | 50 | 0.3 | |
| MISC Imaging | 1 | 5.6 | 500 | 2.9 | |
| MISC Low Res Spectroscopy | 0 | 0 | 0 | 0 | |
| MISC Medium Res Spectroscopy | 0 | 0 | 0 | 0 | |

# 2 - SCIENCE IMPLEMENTATION

> OST is the next logical step in space astrophysics, with an international partnership providing a powerful instrument suite that will enable the community's search for life in the cosmos, fulfill a longstanding quest to learn how Earth acquired its life-supporting conditions, and extend the legacy of JWST with information needed to complete our understanding of galaxy formation and evolution.

The Origins Space Telescope (OST) offers the next leap forward in far-infrared exploration with a large 9.1 m aperture, actively-cooled (~4 K) telescope that covers a wide span of the mid- to far-infrared spectrum. Paired with the mission's five instruments (**Section 2.2**), OST allow scientists to survey large swaths of sky in three dimensions and make detailed spectroscopic measurements to address several of the community's highest science priorities. OST's colder telescope temperature, larger-format arrays, and complement of instruments with greater detector performance than those on any previous infrared mission enable these new measurements. The OST Concept 1 instruments meet all science cases and requirements outlined in **Section 1**, and instrument requirements were flowed down from the science objectives and measurement requirements.

The OST Concept 1 payload consists of the Optical Telescope Element (OTE) and five state-of-the-art spectrometers and imagers, which cover the mid-to-far infrared wavelengths (5 to 660 μm) and approach fundamental performance limits (**Table 2-1**). To ensure telescope emission is lower than the sky background, the OTE is cooled to ~4 K. The cooling system combines passive cooling provided by a five-layer deployed sunshield with active cooling provided by four high Technology Readiness Level (TRL) mechanical cryocoolers operating in parallel. Housed in the protective Instrument Accommodation Module (**Figure 2-1**, **Section 3.2.1**), OST's five instruments, MRSS, HRS, FIP, MISC, and HERO (**Section 2.2**), are clustered near and upstream of the telescope focal plane.

## 2.1 Optical Telescope Element

The OST Optical Telescope Element (OTE) (**Figure 2-2**) collects light during science operations and delivers it to the focal surface, where each instrument receives the signal and processes it further.

The OTE interfaces with the Instrument Accommodation Module (IAM) (**Figure 2-1**), which houses several OTE components, including the secondary, tertiary, and Field Steering Mirrors (FSM), and the entire instrument suite.

**Table 2-1:** OST's five state-of-the-art spectrometers and imagers cover the mid-to-far infrared wavelengths.

| Instrument | Organization | Specifications | Operation Modes |
|---|---|---|---|
| Mid-IR Imager Spectrometer Coronograph (MISC) | JAXA NASA ARC | Bandpass: 5 to 38 μm Spectral Resolution: 300, 1000, 20000 Section 2.2.2 | Imager; Low Res Spec; Medium Res Spec; High Res Spec; Coronograph; Transit |
| Mid-Res Survey Spectrometer (MRSS) | NASA JPL | Bandpass: 30 to 660 μm Spectral Resolution: 500 (Grating), 40000 (FTS) Section 2.2.3 | Multi-band; Wide Field Survey; Pointed |
| High-Res Spectrometer (HRS) | NASA GSFC | Bandpass: 25 to 200 μm Spectral Resolution: 100,000 at 50 μm Section 2.2.4 | High Resolution with Fabry-Perot Etalons |
| Far-IR Imager/ Polarimeter (FIP) | NASA GSFC | 7.5 x 15 arcmin FOV FIR Imaging (40, 80, 120, 240 μm) Spectral Resolution: ~15 Section 2.2.5 | Polarization Short; Polarization Long; Continuum-Multi-band |
| Heterodyne Instrument (HERO) | Europe [a Consortium] | Bandpass: 63 to 66 μm (111 − 641) Spectral Resolution: 1 x 10[7] Section 2.2.6 | Multi-beam High-resolution Spectroscopy |

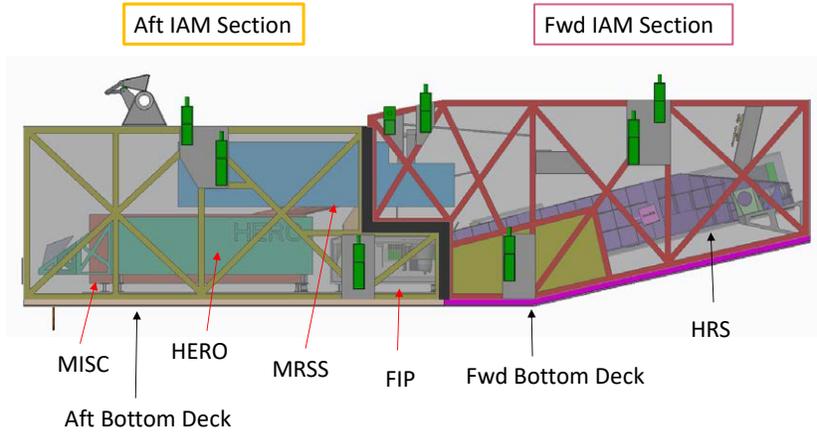

**Figure 2-1:** The five OST instruments are housed in the Instrument Accommodation Module (IAM) module (Section 3.2.1), which provides the necessary power, cooling, and mechanical operations enabling the science observations.

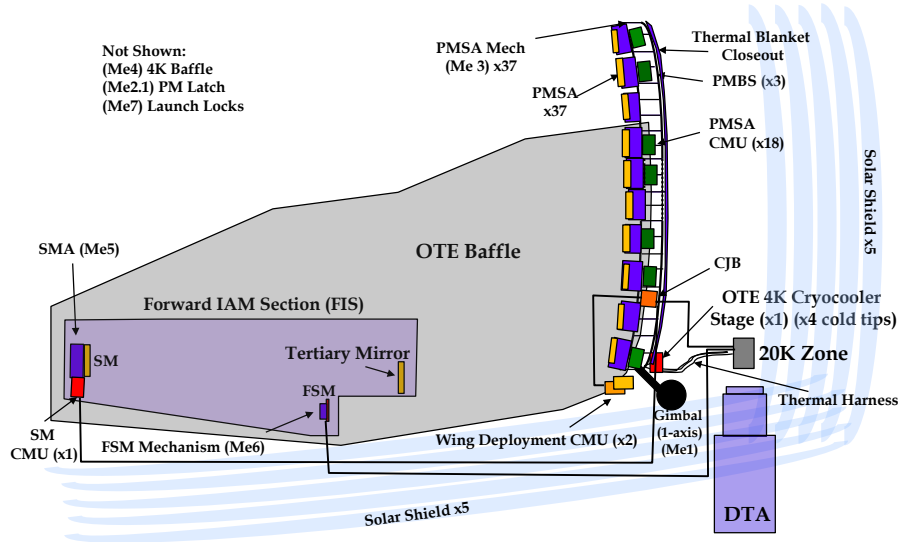

**Figure 2-2:** The OST Optical Telescope Element (OTE) and instruments are cooled to 4 K. A surrounding baffle, also cooled to 4 K, prevents stray light from entering the field of view. The OTE uses high TRL mechanisms from JWST.

**Table 2-2:** Telescope system driving requirements flow from the mission's Science Goals and instrument specifications.

| Driving Science Goals | Required Instrument Specifications | Telescope Specifications |
|---|---|---|
| Tracing the signatures of life and the ingredients of habitable worlds<br><br>Unveiling the growth of black holes and galaxies over cosmic time<br><br>Charting the rise of metals, dust, and the first galaxies<br><br>Characterizing small bodies in the Solar System | **Mid-IR Imager Spectrometer Coronagraph (MISC)**<br>Bandpass: 5 μm to 38 μm<br>Spectral Resolution: 300, 1000, 20000 | **FOV:** 15 x 25 arcmin<br><br>**Wavelength:** 5 to 660 μm<br><br>**Spatial Resolution:** Diffraction limited at ~30 μm (MISC instrument diffraction limited at 5 μm with the deformable mirror)<br><br>**Telescope Design:** 9.1 meter off-axis (unobstructed view), three-mirror anastigmat with field steering mirror, 37 hexagonal segments each 1.277 m flat to flat<br>**Operating Temperature:** 4K; 500mK thermal stability<br><br>**Pointing Requirements:**<br>Knowledge: 30 mas (MRSS inertial point)<br>Control: 44 mas<br>Jitter: 22 mas RMS (at MISC; telescope rqmt TBD)<br>Scan Rate: 100 arcsec/sec (for MRSS and FIP large survey) |
| | **Mid-Res Survey Spectrometer (MRSS)**<br>Bandpass: 30 to 660 μm<br>Spectral Resolution: 500 (Grating), 40000 (FTS) | |
| | **High-Res Spectrometer (HRS)**<br>Bandpass: 25 to 200 μm<br>Spectral Resolution: 100,000 at 50 μm | |
| | **Far-IR Imager/Polarimeter (FIP)**<br>7.5 x 15 arcmin field-of-view<br>FIR Imaging (40, 80, 120, 240 μm)<br>Resolution: ~15 | |
| | **Heterodyne Instrument (HERO)**<br>Bandpass: 63 to 66 μm (111 − 641)<br>Spectral Resolution: 1 x 10[7] | |

### 2.1.1 Telescope Requirements and Specifications

The team developed the telescope system's driving requirements, and optical specifications flowed-down from the OST science goals and instrument specifications (**Table 2-2, Table 2-3**).

Providing an unobstructed 9.1 m aperture view was the driving telescope-packaging requirement. The OST field of view (FOV) requires a three-mirror-anastigmat (TMA) design to provide sufficient performance at the 30-micron design wavelength. The 4 K operating temperature prevents thermal emission from exceeding the sky background. Only one instrument (MISC) operates below 30 μm; a deformable mirror internal to MISC corrects mid-spatial frequency wavefront errors so that the MISC has diffraction-limited performance down to 5 μm.

**Table 2-3:** OST Concept 1 telescope first-order specifications.

| Parameter | OST Concept 1 |
|---|---|
| Aperture Size | 9.1 m |
| f-number | f/12.8 |
| Effective Focal Length | 116 m |
| Field-of-view | 25 x 15 Arcmin |
| Waveband | 5 - 660 μm |
| Operating Temp. | 4K |
| Optical Performance | Diffraction limited at λ = 30 μm |
| Design Form | Three mirror anastigmat Unobstructed (off-axis pupil) |

### 2.1.2 Optical Layout

Like JWST, which is also a TMA, the OST telescope is composed of four mirrors – three with optical power (the primary, secondary, and tertiary mirrors) plus a field-steering mirror (FSM) that provides a higher-order wavefront aberration correction (**Figure 2-3**). The hexagonal f/1.29 primary mirror is partitioned into 37 hexagonal segments, and measures 9.1 m tip-to-tip. Between the secondary and tertiary mirrors, the rays form an internal image (Cassegrain focus), which allows the tertiary mirror to image a real exit pupil. This real image includes baffling to help reject stray light. The FSM, placed at the exit pupil of the telescope, actively tilts to control the FOV directed into each instrument. The image surface of the telescope is concave, with the center of curvature located at the FSM surface effectively making a locally-telecentric system for each field point, where each chief ray is normal to the curved image surface. This configuration prevents defocusing when the FSM is tilted.

The team developed an Optical Image Performance Budget for the telescope that details wavefront error allocation, thermal stability, line of sight jitter, and reserve for each of the five instruments and telescope.

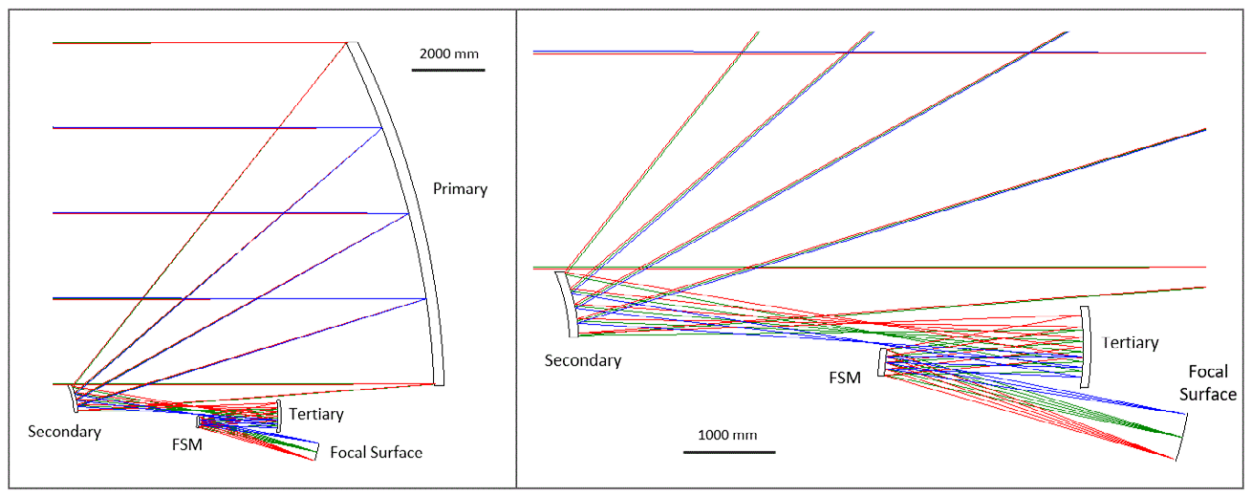

**Figure 2-3:** OST Concept 1 telescope layout includes three mirrors with optical power and a field-steering mirror. The design provides the required FOV for all OST instruments.

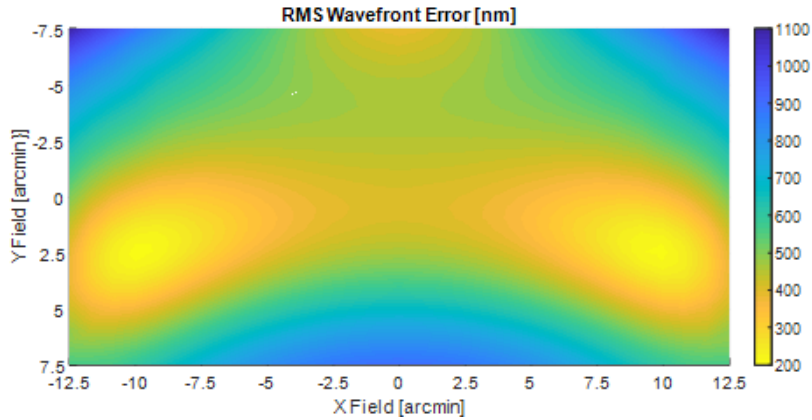

**Figure 2-4:** Evaluation of telescope's RMS wavefront error in nanometers as a function of FOV position.

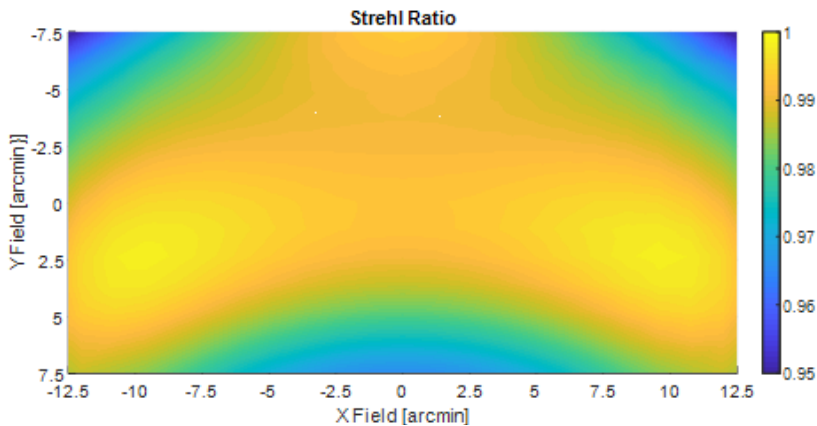

**Figure 2-5:** Evaluation of telescope's Strehl ratio at a wavelength of 30 µm as a function of FOV position.

### 2.1.3 Imaging Performance

To meet Science requirements, OST is diffraction-limited at a wavelength of 30 µm. The team used two optical performance metrics to evaluate this requirement: 1) Root-mean-square wavefront error (RMSWE) over the FOV, and 2) Strehl ratio over the FOV.

RMSWE: Maintaining less than $0.07\lambda$ wavefront error across the FOV is a common diffraction-limited performance standard. For a design wavelength of 30 µm, this corresponds to less than 2.1 µm RMSWE. The OST optical design meets this criterion (**Figure 2-4**).

Strehl Ratio: A diffraction-limited design corresponds to a Strehl ratio of greater than 0.8. **Figure 2-5** shows this requirement is met across the entire FOV.

### 2.1.4 Instrument Field of View Allocations

The OST telescope is designed to have a rectangular 15 x 25 arcminute full FOV. Each instrument is allocated a portion of that field, with empty space reserved for mounting structures (**Figure 2-6**).

### 2.1.5 Stray Light Analysis

Baffles are incorporated into the telescope design to block potential light ray "skip paths." OST's Instrument Accommodation Module (IAM) supports and contains the five instruments, as well as elements of the telescope, except for the primary mirror. A triangular opening in the top of the IAM clips unwelcome rays, eliminating skip paths. The design includes a notional baffle location within the IAM, around the telescope's Cassegrain focus. Because the FSM is located at the exit pupil of the telescope, an image of the primary mirror forms there. To further mitigate stray light, the design incorporates a baffle around the FSM, as was successfully implemented on JWST (**Figure 2-7** C).

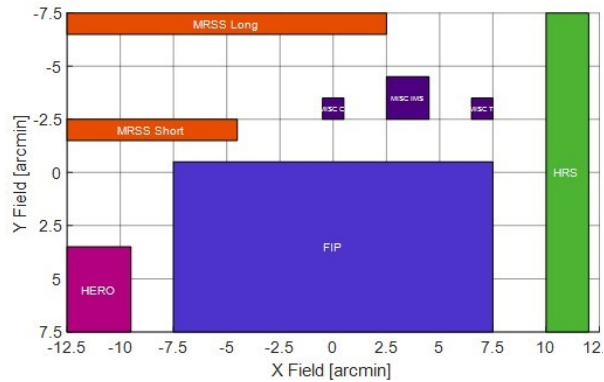

**Figure 2-6:** FOV allocations for the Concept 1 instruments provide unimpeded views for all instruments. FIP, has the largest FOV; MRSS and MISC have multiple channels.

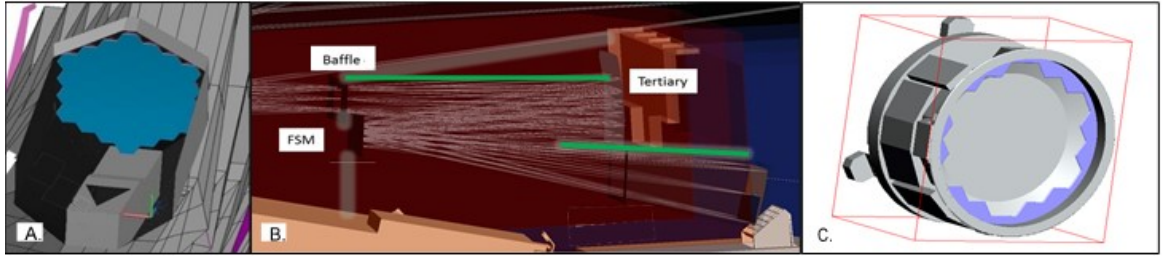

**Figure 2-7:** A) The IAM box protrudes away from the primary mirror and holds the secondary as well as the instruments. B) Baffles (notional) within the OST IAM are used to limit skip paths. C) Model of the baffle structure implemented on JWST.

### 2.1.6 Telescope Trade Studies

The OST team explored a number of telescope designs before settling on the final design architecture based on this analysis. The investigations included trade studies between two- and three-mirror telescopes, on- and off-axis pupils, non-rotationally symmetric pupils, and systems utilizing a primary mirror with a spherically-shaped surface (as opposed to a conical or aspheric surface).

The primary trade study consisted of four different telescope configurations, A through D, as summarized in **Table 2-4**. The study purpose was to identify what combination of powered mirrors and pupil bias would offer the best fit for the science objectives and become the working telescope design. Optical layouts for each option are shown in **Figure 2-8**. Each of these four designs identifies the primary mirror

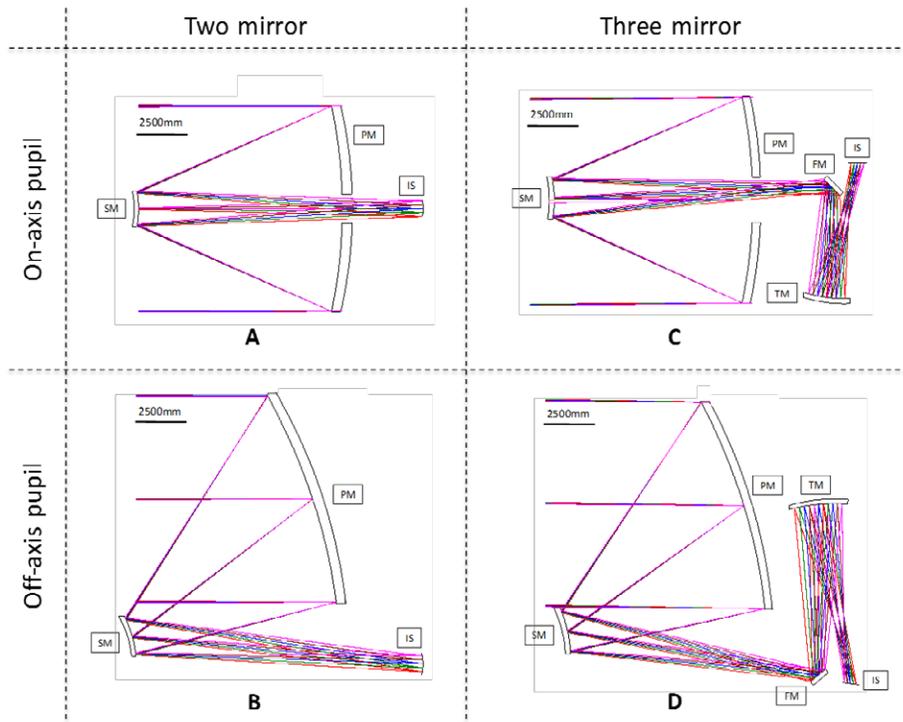

**Figure 2-8:** The team studied four telescope optical configurations, including two- and three-mirror solutions.

(PM), secondary mirror (SM), and image surface (IS), tertiary mirror (TM), and fold mirror (FM), as appropriate. In all four design cases, the image surface is curved to better correct field curvature, an optical aberration caused by a difference in best focus with field. As expected, the image surface curvatures for the three-mirror designs (C, D) are less than that for the two-mirror designs (A, B). This is because the additional powered mirror is better able to spread the optical power over a greater number of mirrors, therefore "flattening" the field.

**Table 2-4:** The OST team studied four alternative telescope architectures.

| Configuration | No. of Mirrors | Pupil Bias | Field Bias |
|---|---|---|---|
| A | Two | On-axis | On-axis |
| B | Two | Off-axis | On-axis |
| C | Three | On-axis | Off-axis |
| D | Three | Off-axis | Off-axis |

## Spherical Primary

The final study considered a spherical-surface (as opposed to aspheric or conical) primary mirror. This design would enable identical surface shape segments to compose the primary mirror, simplifying the design, and mitigating cost and risk. This design would ease fabrication and testing processes compared to a design requiring each segment or subset of segments to have a unique surface prescription.

Historically, spherical mirrors have been used in telescope designs due to their relative ease of manufacture and test as compared to aspheric surfaces. The major fault of spherical surfaces is that they introduce optical aberration into the telescope, most notably spherical aberration. Spherical aberration is defined as a difference in best focus based on the position of the ray within the aperture. Essentially, the rays striking the center of the optic experience a difference optical power than those at the edge of the optic, resulting in a blurred image. The simplest way to correct this error is to utilize a parabolic mirror, which has the property of focusing all the rays from a single on-axis field point to a single point. Unlike a spherical surface, a parabola does not have a constant local curvature across its aperture, which is why it can bring all the rays to the same focus. For these reasons, spherical mirrors are not often used today in telescope design, especially when image quality is of high concern.

The potential cost-savings of the segmented spherical primary mirror were considered beneficial enough to warrant a design study. For this purpose, a 4500 mm-diameter three-mirror anastigmat (TMA) design was attempted, utilizing the same 15 x 25 arcmin full FOV and design wavelength of 30 μm as the final OST Concept 1 design. As expected, the optical design proved to be quite challenging due to the greater amount of aberration induced by the primary mirror. The optical design is shown in **Figure 2-9**. This design is an initial design used to explore the solution space and does not include all constraints as the final Concept 1 telescope design, including matching the FSM position to the location of the exit pupil. Adding these constraints will only make the design process more challenging. Lockheed Martin optical designers attempted a spherical mirror concept in a separate internal study and arrived at similar results. Based on these two independent studies, the spherical primary was not pursued for Concept 1. The team expects this idea presents an opportunity for greater benefit as the number of individual segments composing a mirror increases, and the studies could be applied to another mission concept, depending on its architecture and needs.

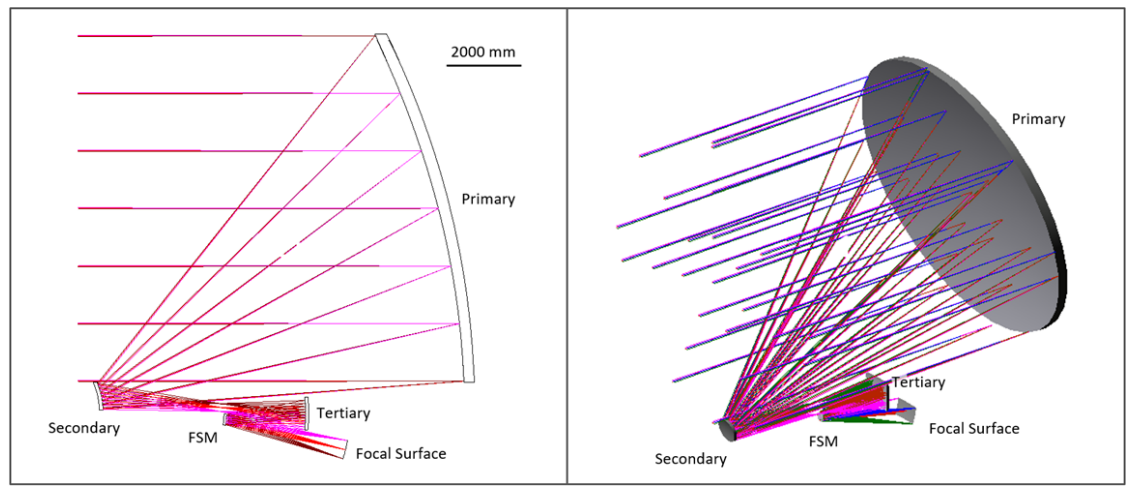

**Figure 2-9:** A notional TMA configuration was used to study pros and cons associated with a spherical primary mirror. For OST, the benefits (mirror segments with identical optical prescriptions) were found to be outweighed by disadvantages (greater residual wavefront error).

### 2.1.7 Telescope Mechanical System Design

The 9-m diameter Primary Mirror (PM) is stowed with the mirror-side facing the top and sides of the IAM for launch (**Figure 2-10A**). The PM's on-orbit position is achieved by first rotating the PM wings out, then rotating the entire PM 75 degrees using the four-bar linkage mechanism that interfaces the PM to the top/back surface of the IAM. A PM light baffle is stored on the two sides of the IAM in troughs (not pictured) under the PM's stowed wings (**Figure 2-10B**). The baffle is secured to the edges of the PM wings, the sides of the IAM, and to the front of the IAM by two "swing arms." The stowed PM assembly, having a first mode of 28 Hz stowed and weighing ~5700 kg, is made up of 37 aluminum hex mirror segments. Each segment attaches to the 0.5-m deep 6061 aluminum backplane via three brushless DC motors with flexures in-line to prevent motor shaft bending. The motors pro-vide tip, tilt, and piston adjustment for each mirror segment. The deployed PM has a first mode of 2.0 Hz. PM backplane structure strength was sized using a FEM with 12 g forces applied in each of three axes. The PM backplane is made of I-beams and channels. The PM faces the IAM primarily to avoid rotating the PM 180 degrees once it is rotated off the IAM. Securing the PM to the IAM with launch locks is more difficult in this configuration, but still possible. Stowing the PM facing toward the IAM also protects the mirror surfaces.

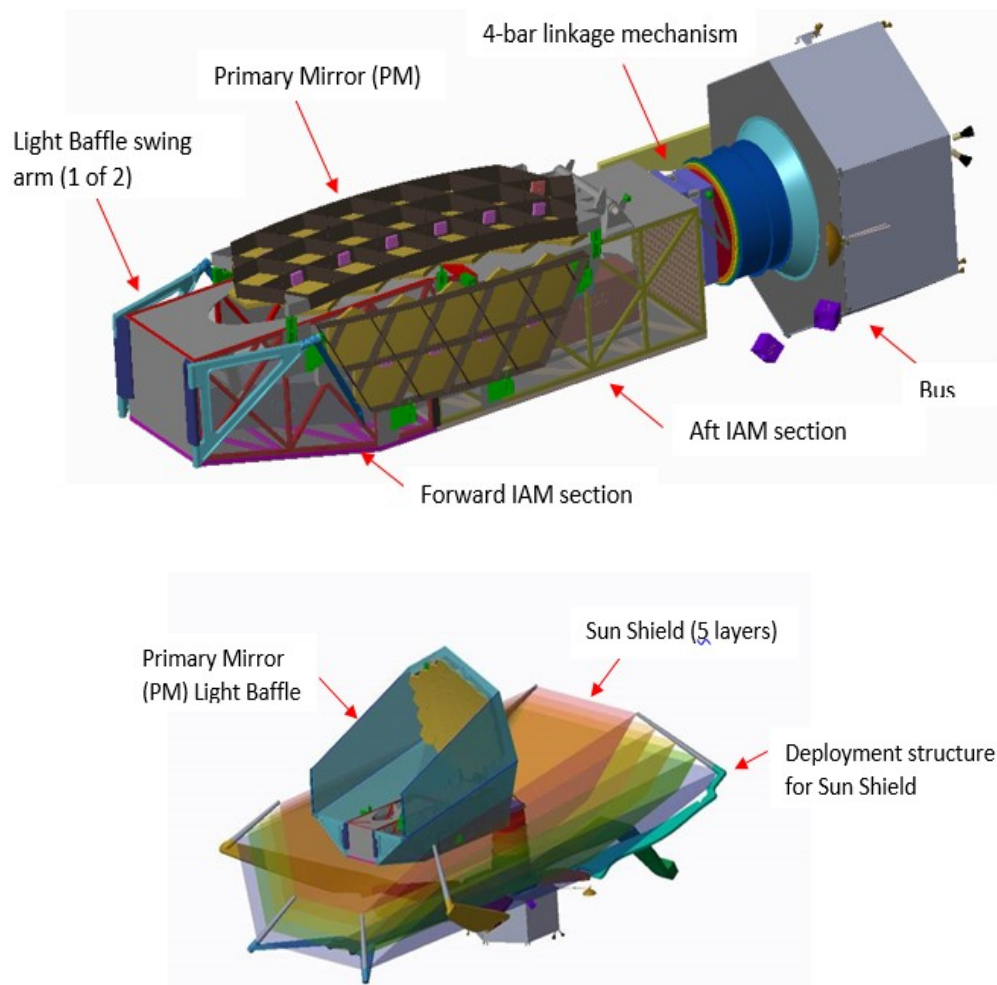

**Figure 2-10:** The OST in stowed (A; top) and deployed (B; bottom) configurations. This concept relies heavily on deployment mechanisms, but deployment is feasible in principle.

### 2.1.7.1 Telescope Deployment and Mirror Mechanisms

**Figure 2-11** shows the PM wing deployment sequence. Each wing is rotated by a common bearing hinge line consisting of a motor-driven hinge system on one end of the wing, and a follower hinge system on the opposite side of the same wing. Because each wings holds nine mirror segments and the mirrors face inward toward the IAM in the stowed position, the wing deployment drive system and hinges must be on the non-optical side of the mirrors. Therefore, each wing deployment system is packaged on the backside (backplane) of the support structure that holds the 37 mirror segments.

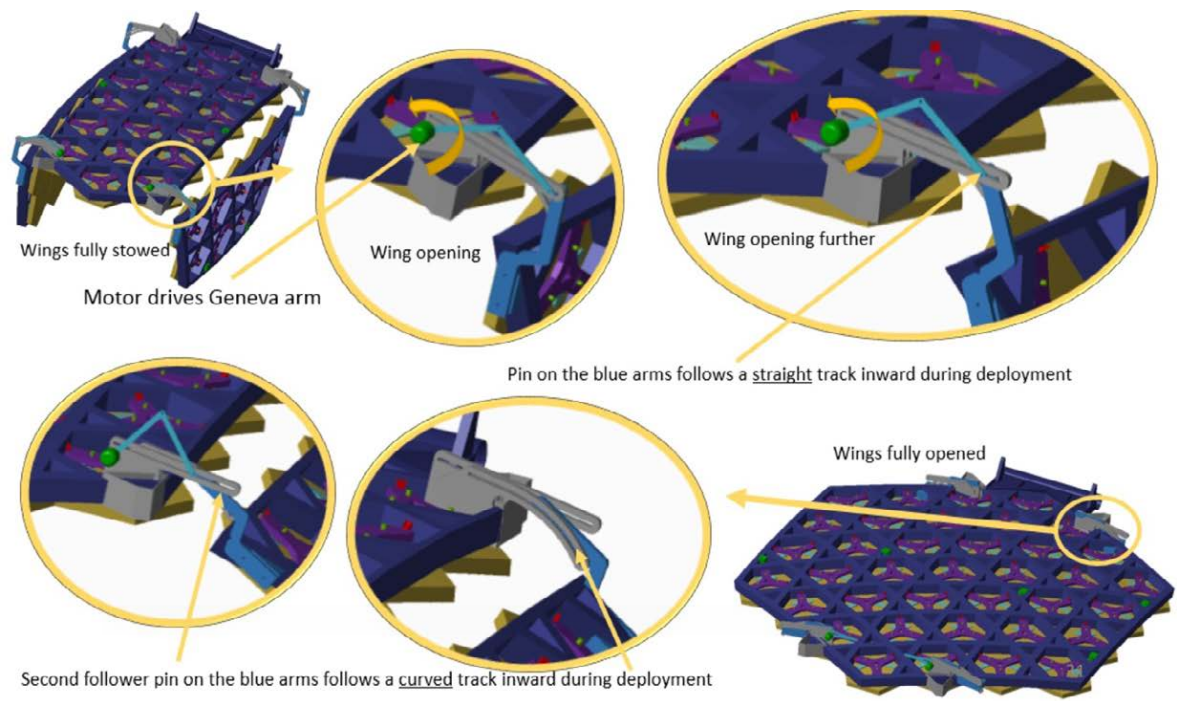

**Figure 2-11:** Details of the primary wing deployment sequence.

**Figures 2-12** to **-15** show the ~90 degree deployment sequence of the primary mirror wings. Two separate actuators are used to drive each wing hinge axes from stowed to deployed in a controlled sequence.

**Figure 2-16** shows the details of the individual mirrors for the OST observatory and how they are precisely moved once deployed and on orbit. Each mirror segment had three linear actuators spaced ~120 degrees apart on the back side of each primary mirror segment. There is support structure in between the mirrors and the backplane that allows relative mirror motion via activating and controlling the three actuators. The actuators allow for micron level tip-tilt-piston motions of each mirror segment.

**Figure 2-17** shows the deployment of the primary mirror system and backplane assembly deploying away from the Instrument Accommodation Module (IAM). This is accomplished by a actuator driven bar linkage system that allows good mechanical advantage for deploying this large assembly, which

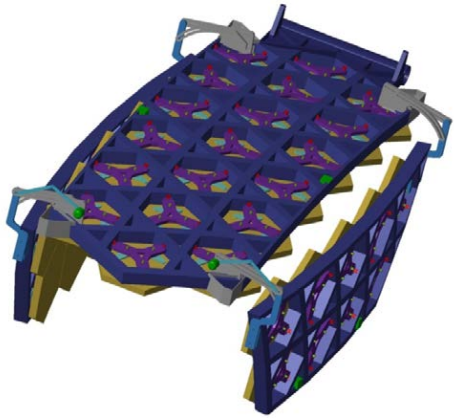

**Figure 2-12:** Fully Closed Wings of the Primary Mirror and Backplane assembly.

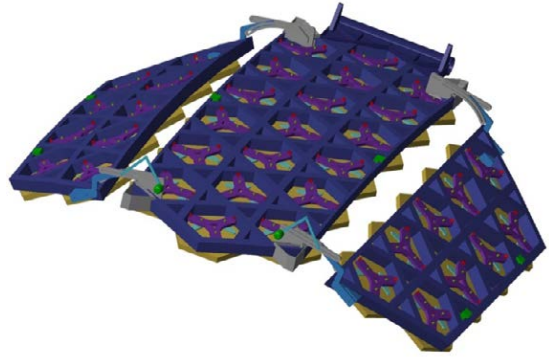

**Figure 2-13:** Beginning to opening wings for the primary mirror and backplane assembly.

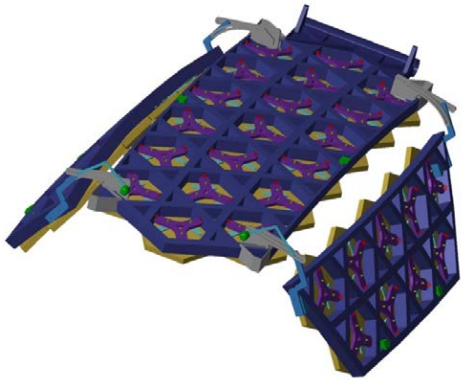

**Figure 2-14:** Continuing to open wings for the primary mirror and backplane assembly.

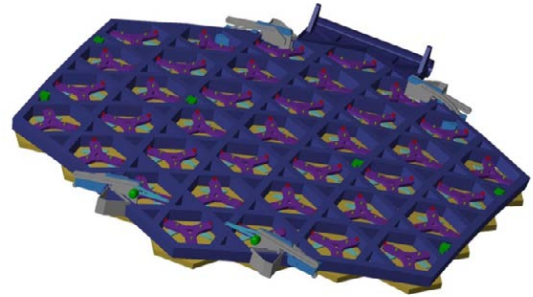

**Figure 2-15:** Fully Deployed wings for the primary mirror and backplane assembly.

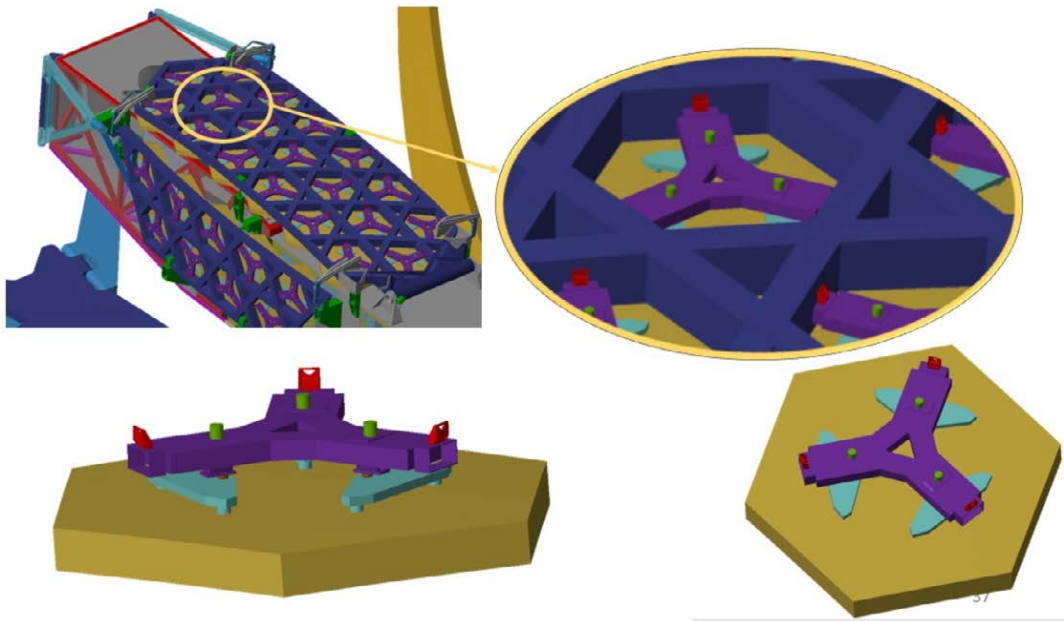

**Figure 2-16:** Mirror Adjustment Mechanisms for the mirror segments for the primary mirror and its attachment to the backplane structure for OST.

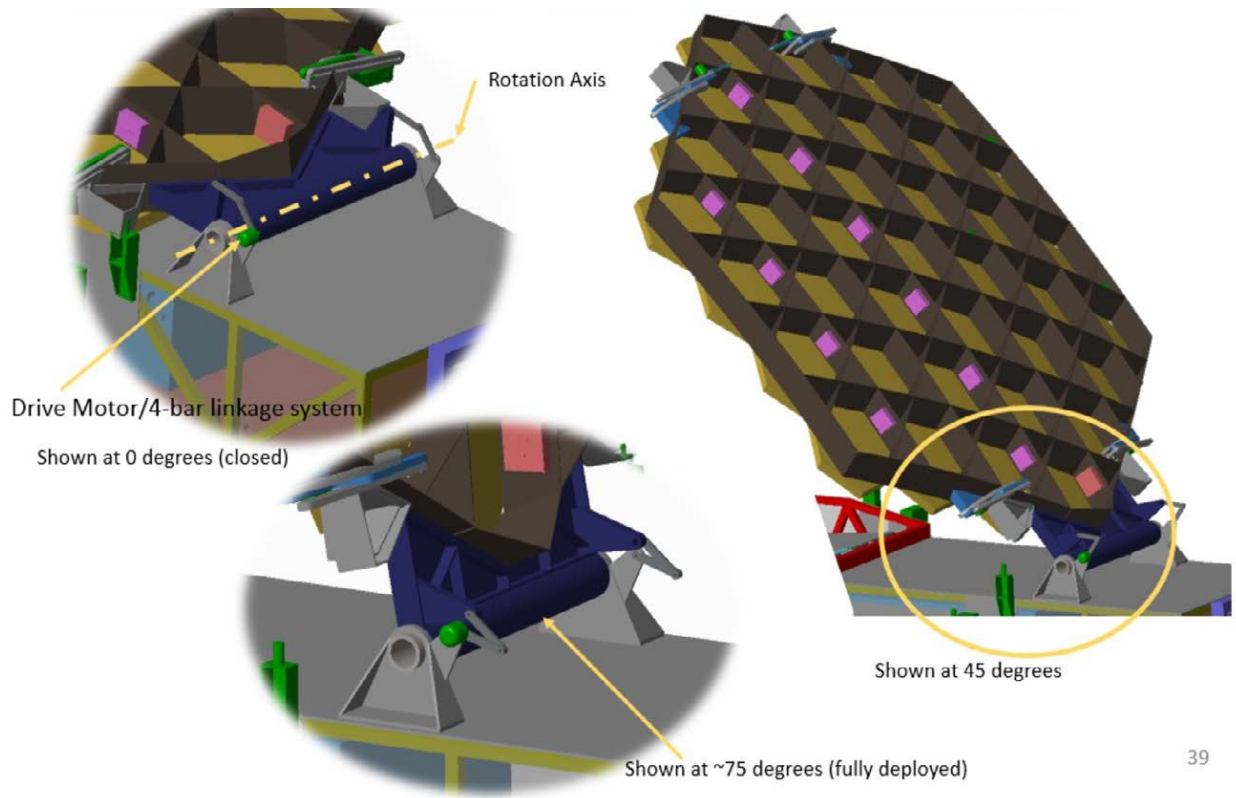

**Figure 2-17:** Primary Mirror (PM) Assembly Deployment away from the IAM structure.

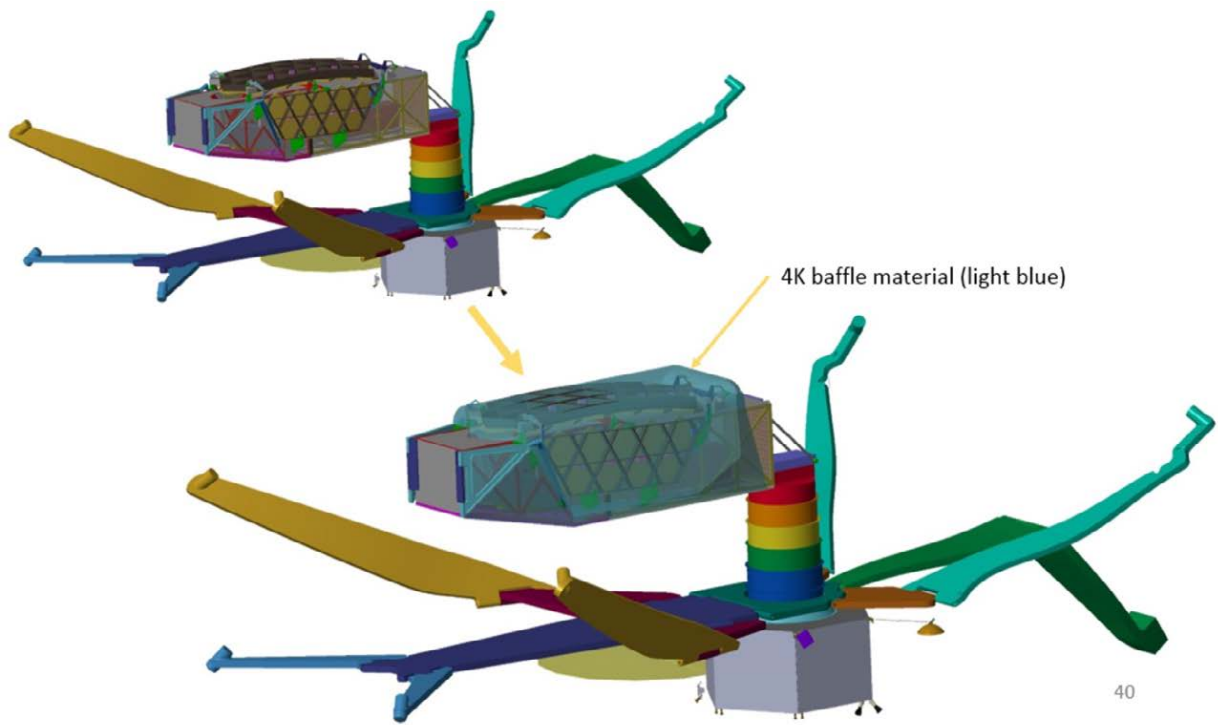

**Figure 2-18:** Primary Mirror Assembly Deployment also showing 4K Baffle Deployment (Stage 1).

locks in place once fully deployed for greater deployed stiffness. The angle of rotation for this motion is about 75 degrees. The 4 K baffle system is not shown deploying for clarity.

**Figures 2-18** through **-26** show the 4 K baffle thermal material affixed to the primary mirror backplane, wings, and the IAM structure in the stowed state and its deployment sequence. As the PM wings deploy, the flexible 4 K baffle material is deployed simultaneously, as shown in the sequence. Two hinge arms and finger booms are attached to the IAM and release to allow the 4 K baffle to fully deploy the baffle material around the primary mirror assembly.

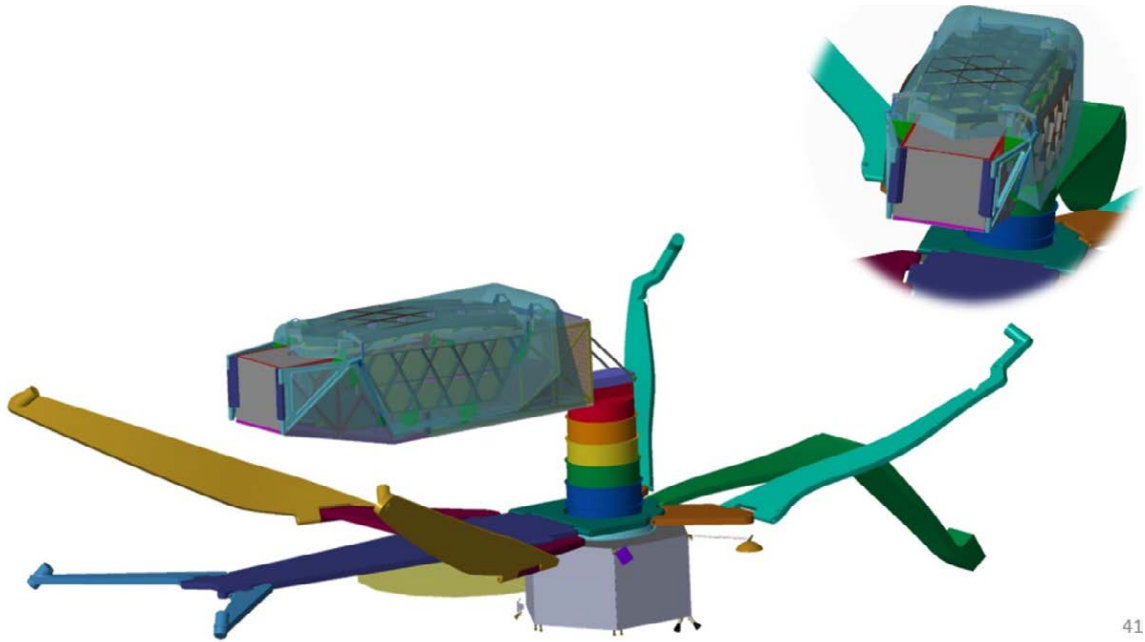

**Figure 2-19:** PMA Deployment showing 4K Baffle and mirror deploying together (Stage 2).

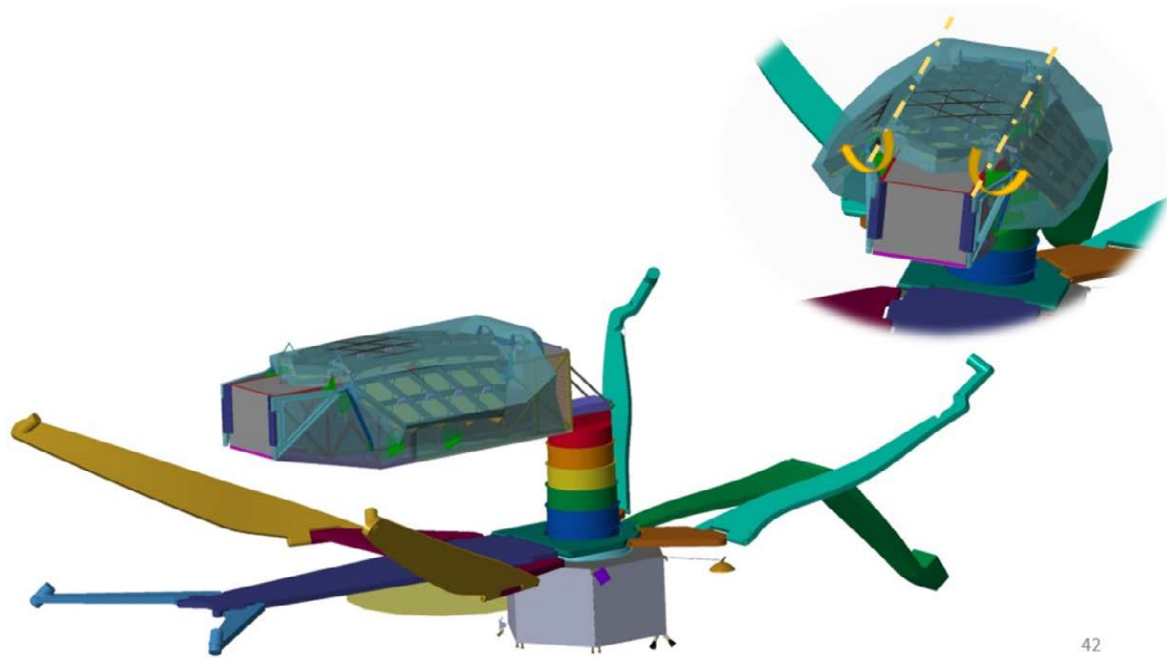

**Figure 2-20:** Primary Wings with 4K Baffle partial deployed (Stage 3).

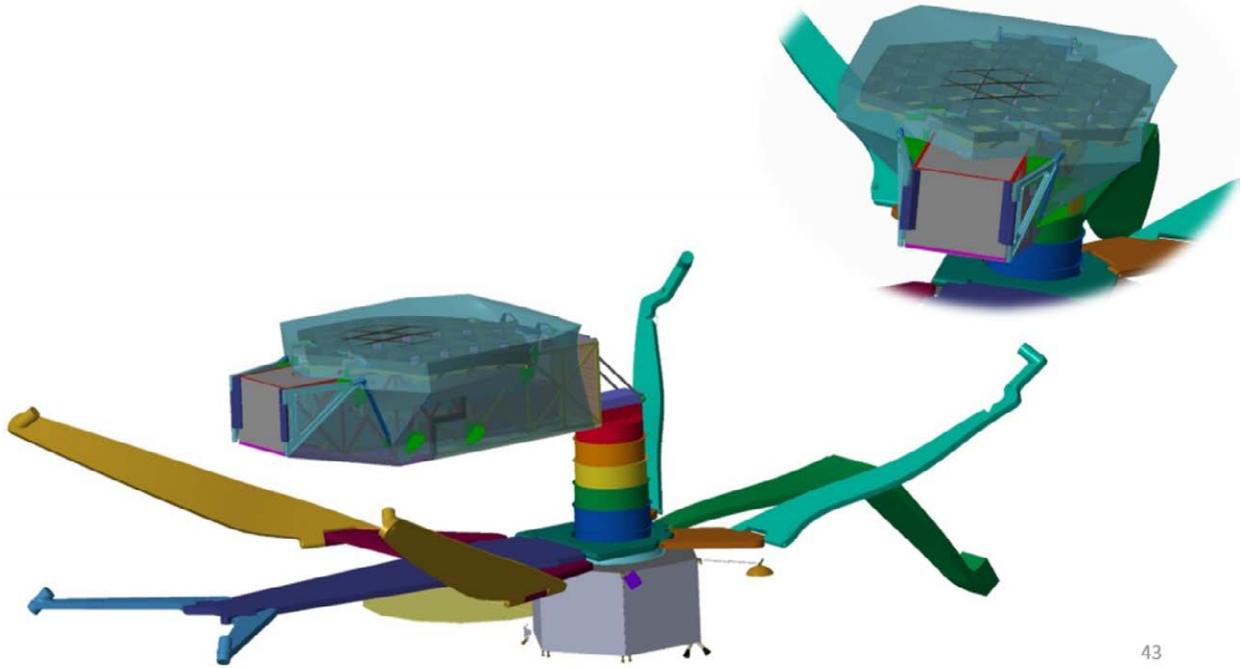

**Figure 2-21:** Primary Wings and 4K Baffle begin deployment (Stage 4).

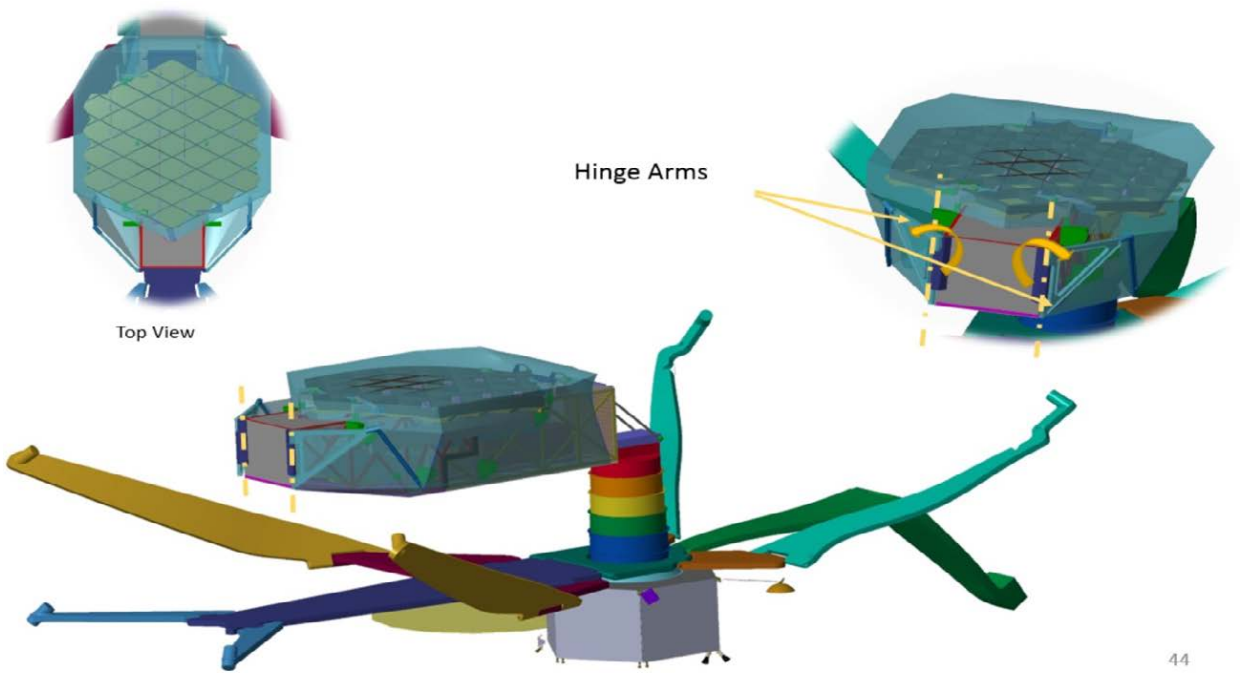

**Figure 2-22:** Hinge Arms Begin Deployment (Stage 5).

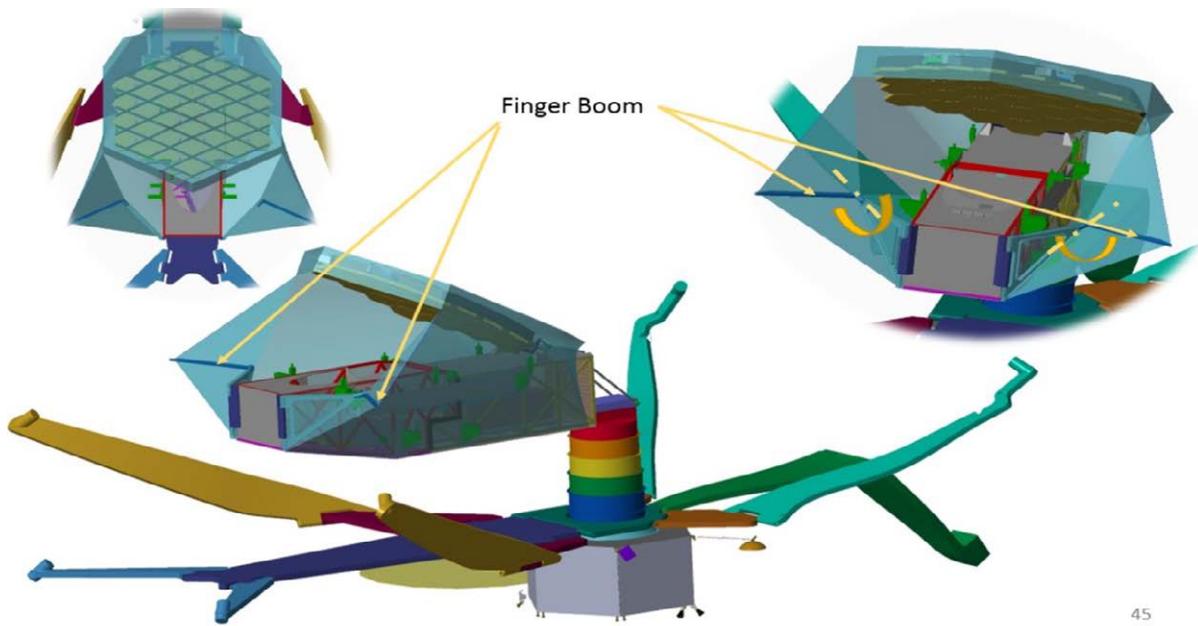

**Figure 2-23:** Mirror and Finger Boom Begins Deployment (Stage 6).

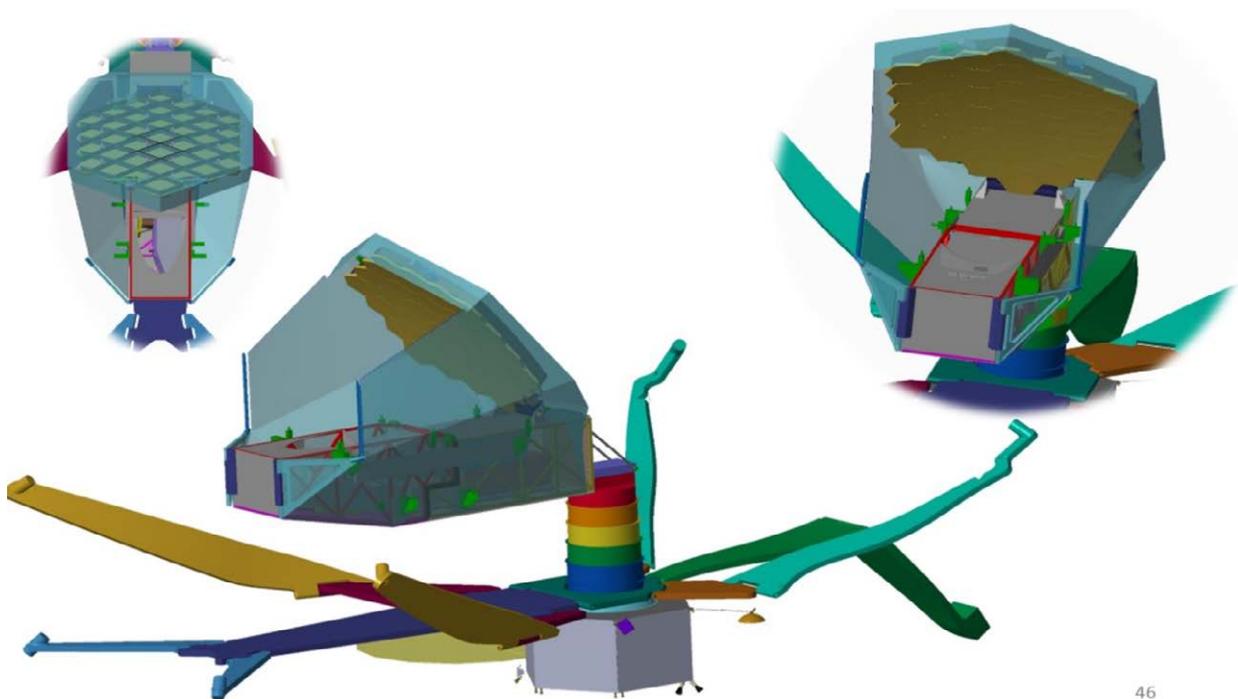

**Figure 2-24:** Mirror Deploys and Finger Booms Lock in place (Stage 7).

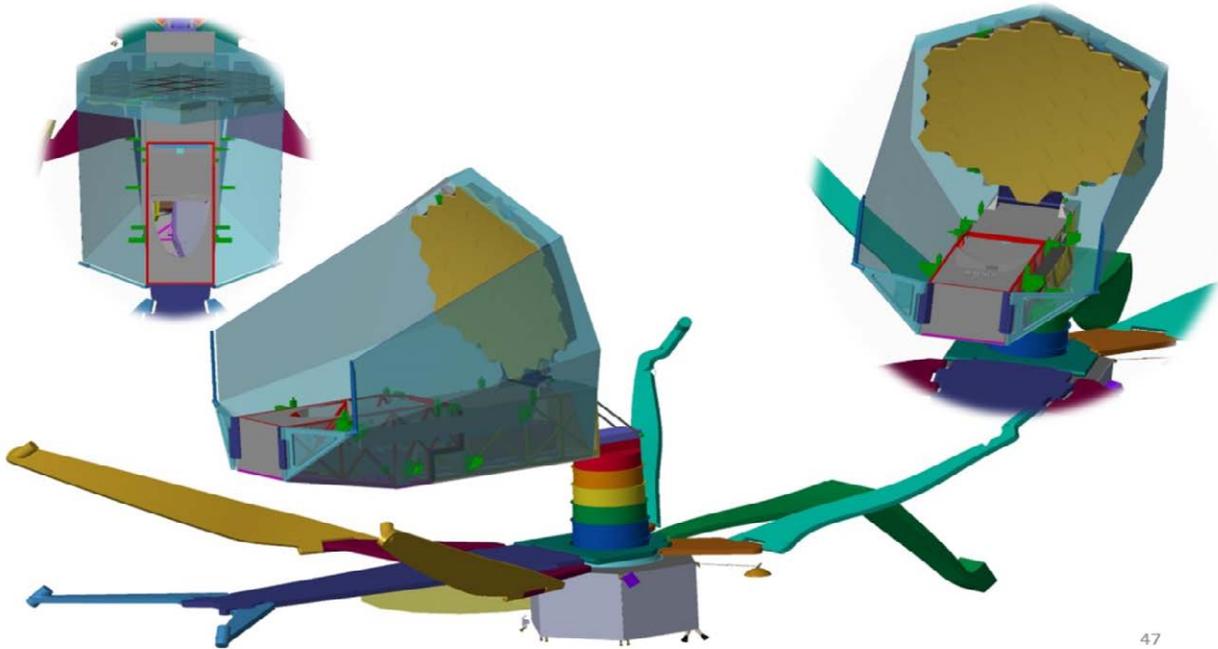

**Figure 2-25:** Mirror Finishes deploying and Hinge Arm Continues (Stage 8).

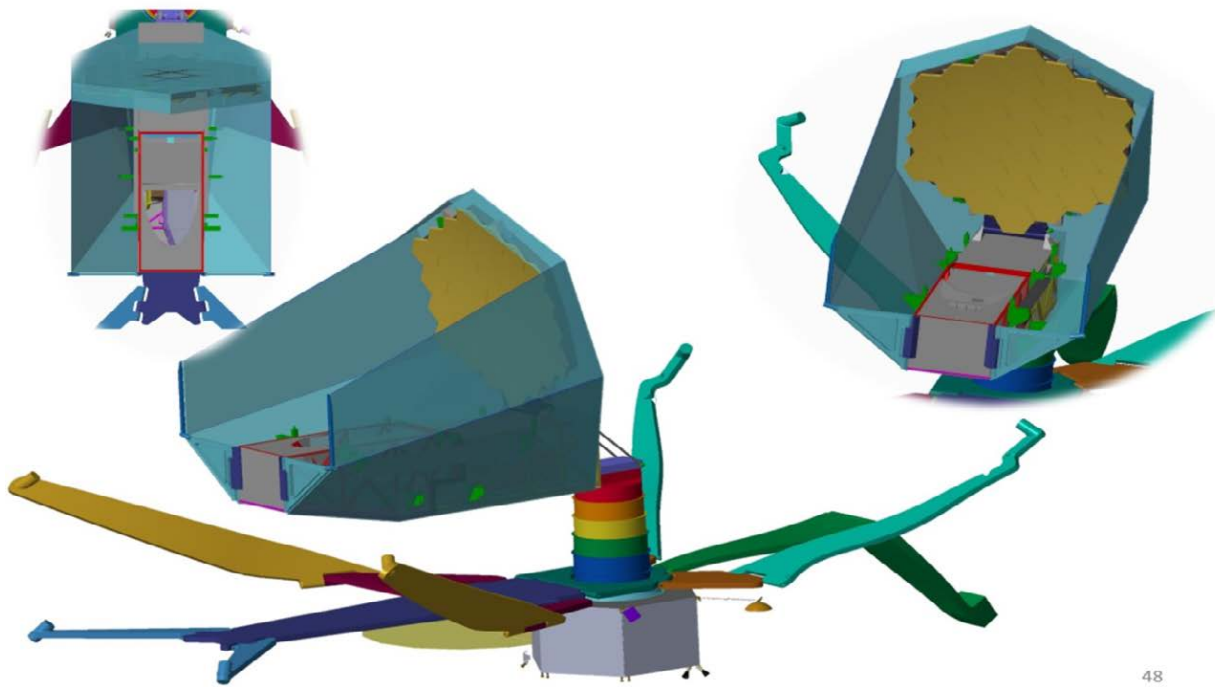

**Figure 2-26:** Primary Mirror Assembly and 4K Baffle Deployment completed (Stage 9).

### 2.1.8 Number of Mirrors

The ability of a telescope to correct optical aberrations is evaluated based on a map of Strehl ratio over the FOV. The OST team evaluated four designs for the Concept 1 mission and ran Strehl ratios for each, then plotted the Strehl ratio over the same 1° x 1° full FOV (**Figure 2-27**). As expected, imaging performance is always superior at longer wavelengths and at the center of the FOV. Based on this analysis, as shown in **Figure 2-6**, the team allocated the center portion of the telescope's FOV to the instruments with the strictest wavefront performance specifications and/or those operating at the shortest wavelengths (often one and the same).

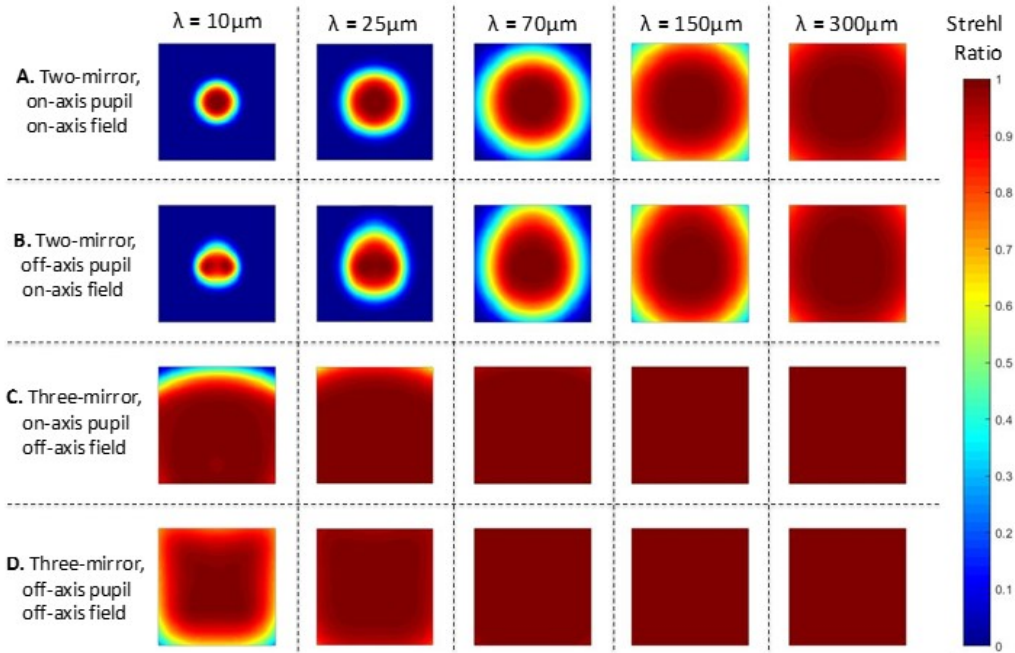

**Figure 2-27:** An analysis of Strehl ratio as a function of FOV for four potential OST telescope designs enabled the team to select the best optical configuration and place the instruments with the strictest performance requirements in the most suitable locations.

While a TMA offers superior imaging over a larger FOV when compared to a two-mirror design, the advantage is not without trade-offs. Compared to a two-mirror design, a TMA often requires fabricating, integrating, and testing two additional mirrors (the powered tertiary as well as a flat fold mirror). A fold flat mirror is often required to ensure the telescope's focal surface is in a convenient location for interfacing with the instruments. And if a steering mirror is required to provide pointing stability (as was the case for JWST), this fourth mirror becomes a necessity. However, it is very difficult to include a steering mirror in a telescope with two powered mirrors, as the steering mirror must be placed at the telescope's exit pupil. TMAs are a useful solution for relaying the pupil to a convenient location for a steering mirror. If pointing stability is an issue, but the design does not include a steering mirror, pointing may also be provided in the individual instruments, although this significantly increases the mass and size of most instruments.

Pursuing a two-mirror or three-mirror design essentially becomes a trade of cost versus performance. Based on analysis that identified improved imaging performance over a larger FOV and better pointing stability afforded by the steering mirror, the OST team selected a TMA design for Concept 1.

### 2.1.9 Aperture Bias

Once the three-mirror-anastigmat (TMA) baseline was set, the team next considered if the aperture stop (located on the primary mirror) should be on- or off-axis. Off-axis telescopes are designed so the secondary mirror never obstructs light incident on the primary mirror, which maximizes the amount of light reaching the detector. Because maximizing throughput facilitates meeting OST's science objectives, the team selected the off-axis/unobstructed telescope design.

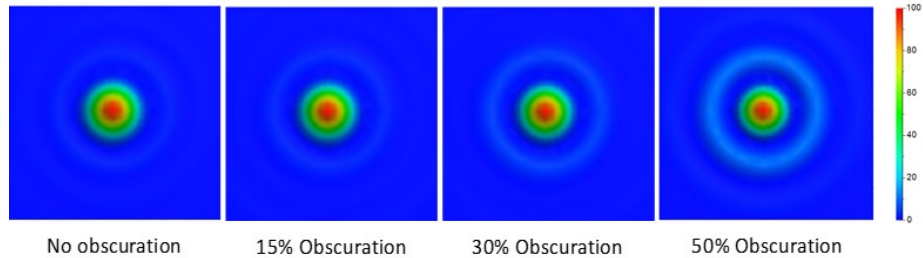

**Figure 2-28:** Comparison of PSFs for increasing values of linear obscuration for an on-axis telescope. OST's primary mirror is unobstructed (no obscuration), giving the telescope optimal image performance (lower sidelobe structure) in the PSF.

Another consideration is how pupil bias affects the telescope's point spread function (PSF). The PSF, or impulse response, is one metric for determining the image quality of an optical system. The image formed can be mathematically calculated for a given optical system by convolving the object being viewed with that system's PSF, which changes with FOV location and wavelength. For perfect stigmatic imaging, the PSF should be a delta function; however, in reality, the PSF is extended due to the effects of diffraction. For a system with a circular aperture, the PSF is an Airy disk. By increasing the size of the obscuration, more energy ends up in the side lobes of the Airy disk pattern (**Figure 2-28**). The 80% encircled energy diameters (for the on-axis field point at a wavelength of 30 μm) are 0.49, 0.69, 0.90, and 1.05 mm for the no obscuration, 15, 30, and 50% obscuration cases, respectively. Spreading out the PSF is undesirable, as it dissipates the energy contained in the image of a single field point, further deviating the telescope from stigmatic imaging. In addition to the effect of the size of the obscuration, the mounting structure (spider) of the secondary mirror also causes diffraction of light and reduces overall throughput to the imaging surface.

To maximize telescope throughput and minimize the power in the sidelobes of the PSF, the OST team pursued an unobstructed, off-axis aperture design. Combining this concept with the advantages of a TMA design vastly improved image quality over a wide field of view with the ability to have an FSM, the team decided upon a three-mirror, unobstructed design. This corresponds to the 'three mirror, off-axis pupil, off-axis field' design for which the Strehl ratio is shown in **Figure 2-27**.

Decentered mirrors that compose off-axis aperture telescope systems used to be far more complicated and costly to fabricate and test compared to centered mirrors. However, in the last 10 years, the technology to manufacture and test such mirrors has become cheaper and better understood. All segments (except the center one) of a segmented mirror, such as the mirror successfully implemented on JWST, are examples of decentered mirrors. In addition to JWST, decentered mirrors have been used in a variety of programs and missions, including JWST, New Horizons, Cassini, and Hubble instruments, providing heritage processes for manufacturing and testing this type of mirror. Additionally, since OST operates at long wavelengths, the polishing requirements for surface figure and surface roughness are much coarser than they would be for an equivalent telescope operating in the visible spectrum, easing the fabrication process. Based on all of these factors combined, a TMA off-axis aperture telescope provides the greatest benefit, in terms of optical performance, to the OST mission and astrophysics observation.

### 2.1.10 Aperture Shape

Telescope apertures are commonly circular. To give OST the largest collecting area possible that would fit into a given fairing, the team explored a number of alternative mirror shapes, most notably ellipses of differing aspect ratios. Ultimately, the team determined a circular primary mirror was the best solution based on design complexity, and that a symmetric PSF better supports the mission's sci-ence objectives. A non-rotationally symmetric aperture will result in a non-rotationally symmetric PSF. For an elliptical aperture, the effect of this is to essentially have different f-numbers and therefore, different angular resolutions in directions orthogonal to one another. The dependence of the PSF on aperture shape for circular and elliptical apertures of different aspect ratios is shown in **Figure 2-29**. **Figure 2-30** shows an analogous comparison of PSFs for both square and rectangular primary mirrors. Such shapes were not explored in depth for Concept 1, having the same issue of an asymmetric PSF that the elliptical apertures do, but to a larger degree.

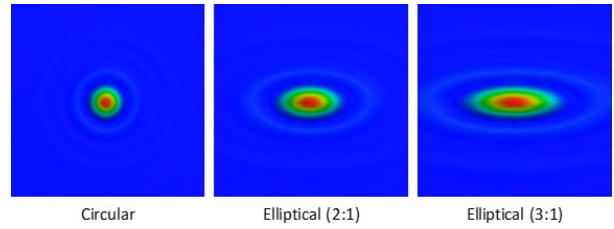

**Figure 2-29:** PSFs for circular and elliptical aperture shapes.

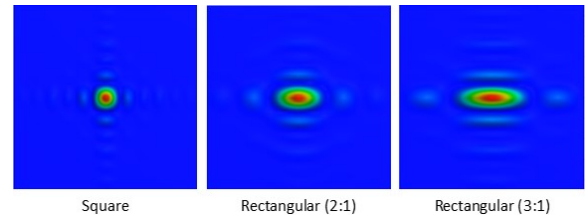

**Figure 2-30:** PSFs for square and rectangular aperture shapes.

Due to its large size and required deployments, the primary mirror could not be a monolith. Instead, it is comprised of mirror segments. The team decided it best to utilize a previously-explored technology in the form of JWST-like hexagonal segments which, when tiled together, would result in a primary mirror with an overall hexagonal shape. While a circular telescope aperture would offer a cleaner PSF, the team found utilization of a design with heritage preferable to a new segmentation scheme. Using hexagonal segments for OST allows the use of a tested deployment scheme: the "wings" of the primary mirror are folded for launch, like those of JWST. To preserve symmetry in the PSF, OST's 37 hexagonal primary mirror segments were laid out in the closest approximation to a circle.

## 2.2 Science Instruments

### 2.2.1 Overview

> The five Concept 1 instruments were designed to meet all the science goals and enable a rich General Observer program.

OST Concept 1 is equipped with five instruments (**Table 2-5**). For this phase, the OST team studied each instrument in sufficient detail to assess its viability, expected performance, capability to satisfy science requirements, technical challenges, technology readiness, and cost. These instrument concepts are products of a single design cycle, and based on priorities for OST Concept 2, the designs will be further developed prior to the Decadal Survey, including to reduce the instruments in mass, size, and power consumption.

**Table 2-5:** OST's instrument suite is the product of an international team effort.

| Instrument | Contributing Organizations | Wavelength Range (μm) | Observing Modes |
|---|---|---|---|
| Mid-Infrared Imager Spectrometer Coronagraph (MISC) | JAXA, NASA ARC | 5 - 38 | • Imaging, spectroscopy<br>• Coronagraphy ($10^{-6}$ contrast)<br>• Transit Spectrometer (<10 ppm stability) |
| Medium Resolution Survey Spectrograph (MRSS) | JPL | 30 - 660 | • Multi-band spectroscopy |
| High Resolution Spectrometer (HRS) | NASA GSFC | 25 - 200 | • High-resolution, high-sensitivity spectroscopy |
| Far-infrared Imager and Polarimeter (FIP) | NASA GSFC | 40, 80, 120, 240 | • Broadband imaging<br>• Field of view 2.5' x 5', 7.5' x 15'<br>• Differential polarimetric imaging |
| Heterodyne Receiver for OST (HERO) | Europe [a Consortium] | 63 - 66, 111 - 641 | • Multi-beam high-resolution spectroscopy |

### 2.2.2 Mid-Infrared Imager Spectrometer Coronagraph (MISC)

> The dual-purpose MISC serves as a science instrument and telescope guide. With its specially-designed transit spectrometer channel and coronagraph, MISC enables all OST exoplanet science.

#### 2.2.2.1 MISC Science Traceability

MISC is essential to OST's search for biosignatures in the atmospheres of exoplanets via ultra-stable (<10 ppm on timescales of hours to days) spectrophotometric observations of primary and secondary transits. MISC measures exoplanet atmosphere constituents such as ozone, methane, and water with spectroscopy in the 5-18 μm (R~100) and 17-25 μm (R~300) bands.

MISC coronagraphic mode directly images and characterizes exoplanets, such as Saturn and Jupiter analogs, and ice giant planets at ice-melting temperatures (~300 K). To achieve this objective, the MISC coronagraph is designed to provide $10^{-7}$ contrast at 0.5" (~2λ/D) at 10 μm.

MISC mid-IR imaging in the 5-40 μm band will be used to study episodic accretion in circumstellar debris disks, and to support the biosignature observations, while spectroscopy with resolving power ranging from ~$10^2$ to $10^4$ in the MISC spectral range will support prioritized observing campaigns in all of OST's science themes.

#### 2.2.2.2 MISC Instrument Description

MISC has separate optical modules dedicated to imaging, spectroscopy, and coronagraphy. The MISC Imager and Spectrometer (I&S) module offers: 1) a wide field imaging (3' x 3') and low- resolution spectroscopic capability with filters and grisms for 6-38 μm; 2) a medium- resolution (R~1,000) Integral Field Unit (IFU) spectroscopic capability for 9-38 μm (with a goal of 5-38 μm); and, 3) a high-resolution (R~25,000) slit spectroscopic capability for 12-18 and 25-36 μm. The MISC Coronagraph (COR) module covers 6-38 μm. The special Densified Pupil Spectrometer (TRA) module provides R~100-300 exoplanet transit and emission spectroscopy from 6-26 μm with very high spectrophotometric stability. As the shortest wavelength focal plane imager, the MISC I&S module is also used for focal plane guiding, including as needed for the other OST science instruments. Tip-tilt mirrors and deformable mirrors in the MISC I&S and COR modules improve pointing stability and wavefront performance beyond that provided by the OST telescope.

As a result of a trade-off study on coronagraphic methods among a PIAA coronagraph [Guyon et al., 2014], 4QPM coronagraph, 8OPM coronagraph [Murakami et al., 2016], Vortex coronagraph, and a binary pupil mask (BPM) coronagraph [Enya et al., 2010], the MISC team selected the PIAA coronagraph as the baseline for MISC COR. The corongarph inner working angle (IWA) and outer working angle (OWA) are derived from the direct imaging and characterization of true exoplanet analogs of Jupiter and Saturn as well as ice giants. The coronagraph must detect Jupiter at 3.4 λ/D with a contrast of 2 x $10^{-6}$ at 15 μm and at 2.1 λ/D with a contrast of 3 x $10^{-5}$ at 24 μm. The IWA is smaller than ~3 λ/D at 15 μm and ~2 λ/D at 24 μm. The same science objective requires the capability to detect Saturn at 6.1 λ/D at 15 μm with a contrast of ~1 x $10^{-7}$ and at 3.8 λ/D at 24 μm with a contrast of 3 x $10^{-6}$. The OWA is larger than 7 λ/D at 15 μm and 4 λ/D at 24 μm. Although the PIAA coronagraph method requires its own independent module within the MISC instrument, it achieves the highest contrast at a small inner working angle with the highest throughput among the various methods considered. The 4QPM and 8OPM methods do not necessarily require an independent module and can add a coronagraphic capability to the I&S module with a relatively minor impact on its optical and mechanical design. However, as is the case for the Vortex coronagraph, they also require very stringent pointing accuracy and stability and, above all, fabrication of the phase masks that achieve a homogeneous capability in a wide mid-infrared wavelength range would require further technical development.

MISC employs densified pupil spectroscopy, a newly-studied method for transit spectroscopy [Matsuo et al., 2016] in which the aperture is subdivided, yielding multiple independent spectra on the focal plane. Robust averaging reduces susceptibility to low-order wavefront error, cosmic ray hits, and defective pixels, greatly improving MISC's spectrophotometric performance. The science image is not disturbed by minor telescope pointing jitter or wavefront error. Intra- and inter-pixel sensitivity variations are mitigated by spreading the light over multiple pixels. A number of reference pixels are also employed for calibration of potential detector gain fluctuations. Focusing on the fact that the detector plane is optically conjugated to the pupil plane for this system, a pupil mask on the plane where the densified pupil is formed completely blocks any thermal astronomical light incoming into residual pixels (i.e., reference pixels) except for the science pixels. The variation of the detector bias common to the entire detector can be traced by a number of the reference pixels and may be reduced to a random component. With the aid of a prototype instrument testbed during instrument development, the team will develop an optimal calibration technique using the reference pixels.

### 2.2.2.2.1 General Instrument Operating Principle and Heritage

MISC's wavelength coverage and capabilities partly overlap with those of JWST/MIRI, SOFIA/FORCAST, SOFIA/EXES, SPICA/MCS, SPICA/SCI, SPICA/SMI, TAO/MIMIZUKU, Spitzer IRS, and TMT/MICHI, providing its component and subsystem heritage and technical maturity (**Table 2-6**).

The entrance apertures to MISC's three modules occupy three different regions of the OST telescope focal plane. Within each module, beam splitters subdivide the overall spectral range. This configuration enables separate optimization of the optics and detectors used for transiting exoplanet spectroscopy and coronagraphy, enhancing OST's utility in these roles.

Other than guiding the OST focal plane using the MISC I&S imager, one MISC mode is active at a time. MISC's seven distinct modes are all intended for pointed observing. A mode using the TRA module is dedicated to transiting spectroscopy. Four modes use the I&S module for either imaging: long-slit, low-resolution spectroscopy; medium-resolution IFU spectroscopy; or high-resolution slit spectroscopy. Two modes use the COR module, one for imaging, the other for coronagraphic spectroscopy.

MISC employs detectors and amplifier chips similar to those used extensively in previous space missions, including Spitzer, AKARI, and JWST, and SOFIA instrumentation with some variation (**Section 3.3.2.2.5** & **Section 3.3.2.4**). These devices operate at cryogenic temperatures.

### 2.2.2.2.2 MISC Optical Design

The I&S, COR, and TRA module ray traces (**Figure 2-31**, **Figure 2-32**, and **Figure 2-33**) reflect the MISC instrument's design maturity.

**Table 2-6:** MISC has component and subsystem heritage from a long string of successful missions, including JWST/MIRI.

| Item | Subsystem/ Component | Heritage | TRL |
|---|---|---|---|
| Deformable Mirror | C | SPICA/SCI | 3 |
| Tip-Tilt Mirror | C | SPICA/SCI, JWST/ NIRCAM, TAO/MIM- IZUKU | 3 |
| 2k x 2k Si:As, 2k x 2K Si:Sb (*) | C | JWST/MIRI, SPICA/SMI | 4-5 |
| Binary Pupil Mask | C | SPICA/SCI | 4 |
| Beamsplitter, Bandpass Filters (Multi-layer Interference Filter) | C | SPICA/MCS | 5 |
| SiC Mirrors | C | AKARI, JWST/NIRCAM | 4 |
| Image Slicer | S | SPICA/MCS, TMT/ MICHI | 4 |
| Immersion Grating (12-18 μm) | C | SPICA/MCS | 4 |

*Detector stability (i.e., long-term fluctuation of gain generated in detector and readout electronics) must be significantly improved to be employed in MISC transit spectrometer module.

A potential alternative detector suitable for the MISC transit spectrometer module is "super-conducting nanowire detector", which will be studied in Concept 2.

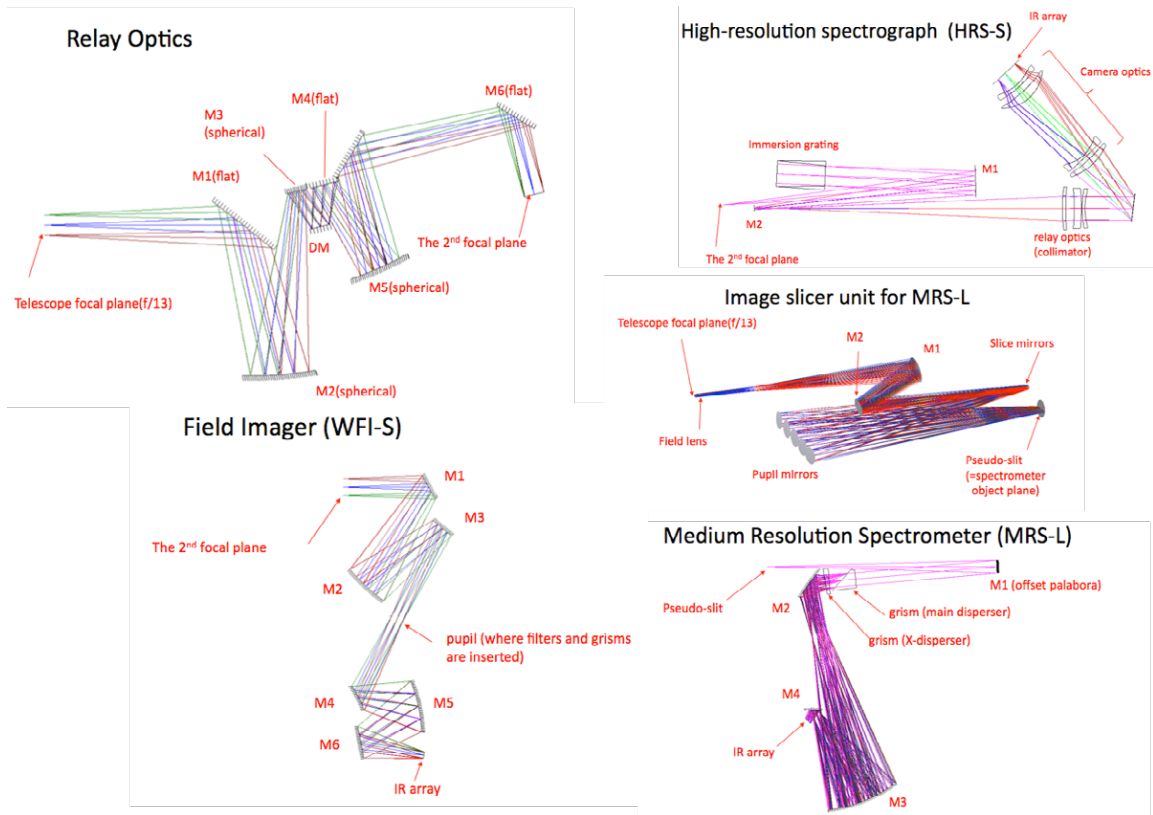

**Figure 2-31:** The MISC I&S module optical design is based on heritage from designs developed for the mid-infrared instrument on the SPICA mission. A set of Offner fore-optics includes a deformable mirror and a tip–tilt mirror to compensate for telescope aberrations and image motion. A beam splitter at 18.5 μm sends the light into short and long wavelength optical paths, while slit mirror changers in each path direct the light to imaging cameras, medium-resolution, or high-resolution spectrometers.

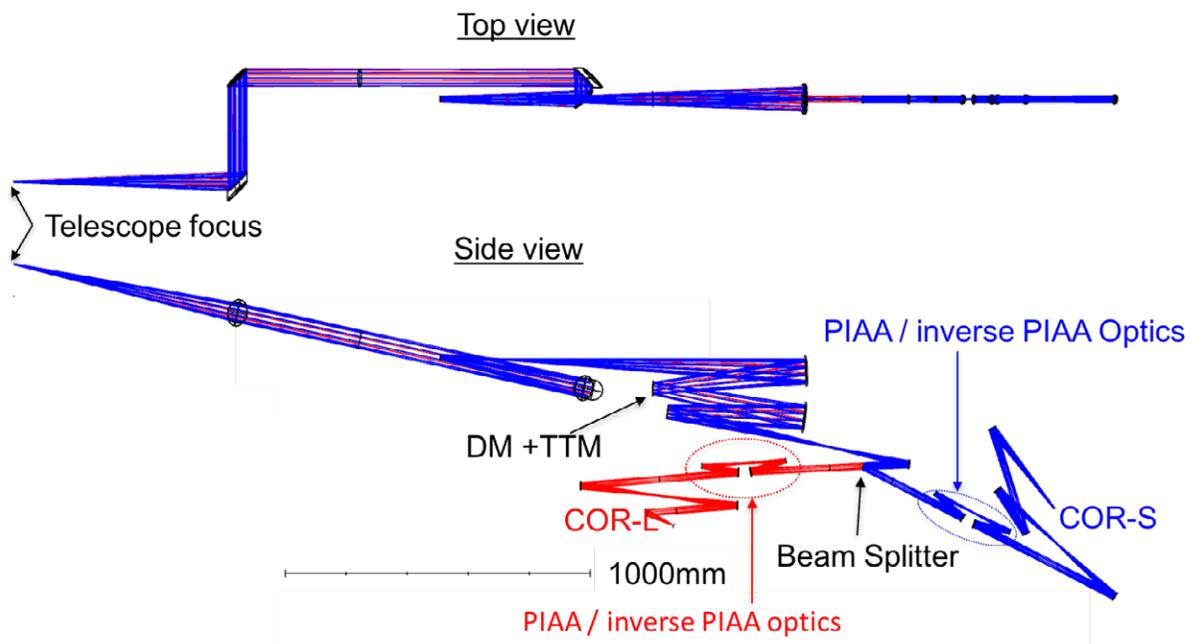

**Figure 2-32:** The MISC PIAA coronagraph channel optical design includes deformable and tip-tilt mirrors in its fore-optics to achieve the required image stability. A beamsplitter located at 15.5 μm sends the light down short and long wavelength paths.

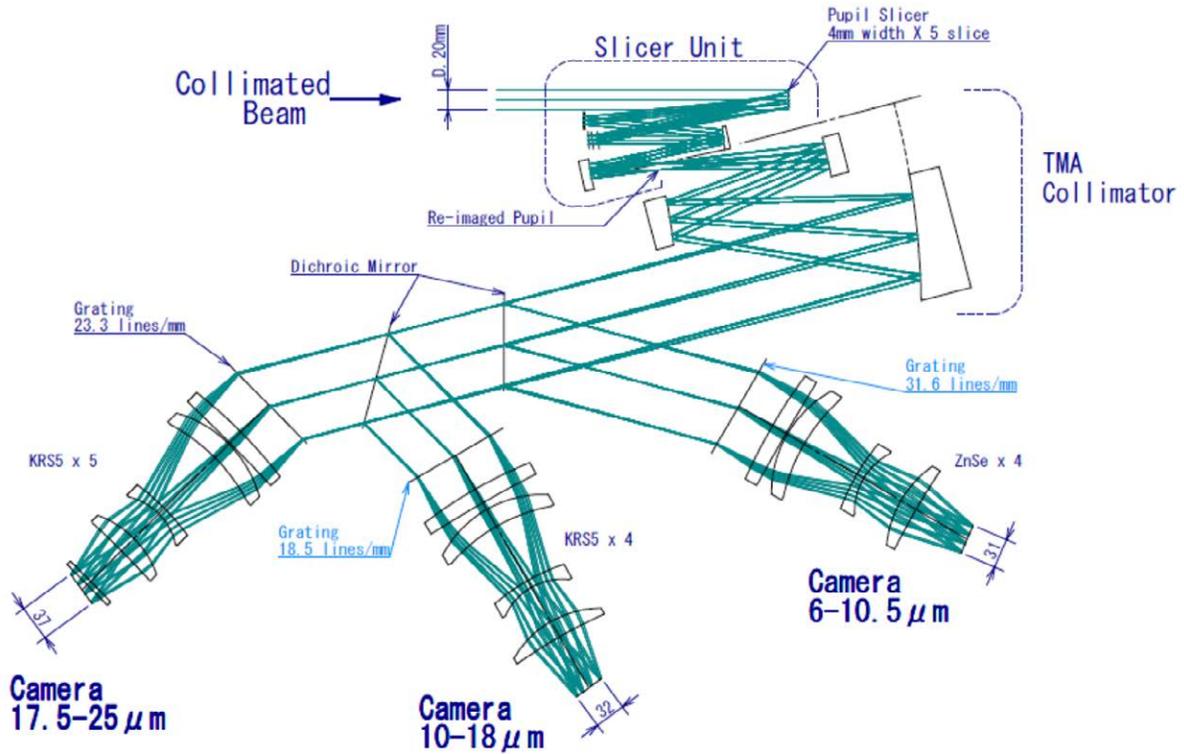

**Figure 2-33:** The MISC transit spectrometer channel employs a new densified pupil design described in Matsuo et al. [2016] and incorporates beamsplitters located at 10.2 μm and 17.2 μm that feed three detector arrays.

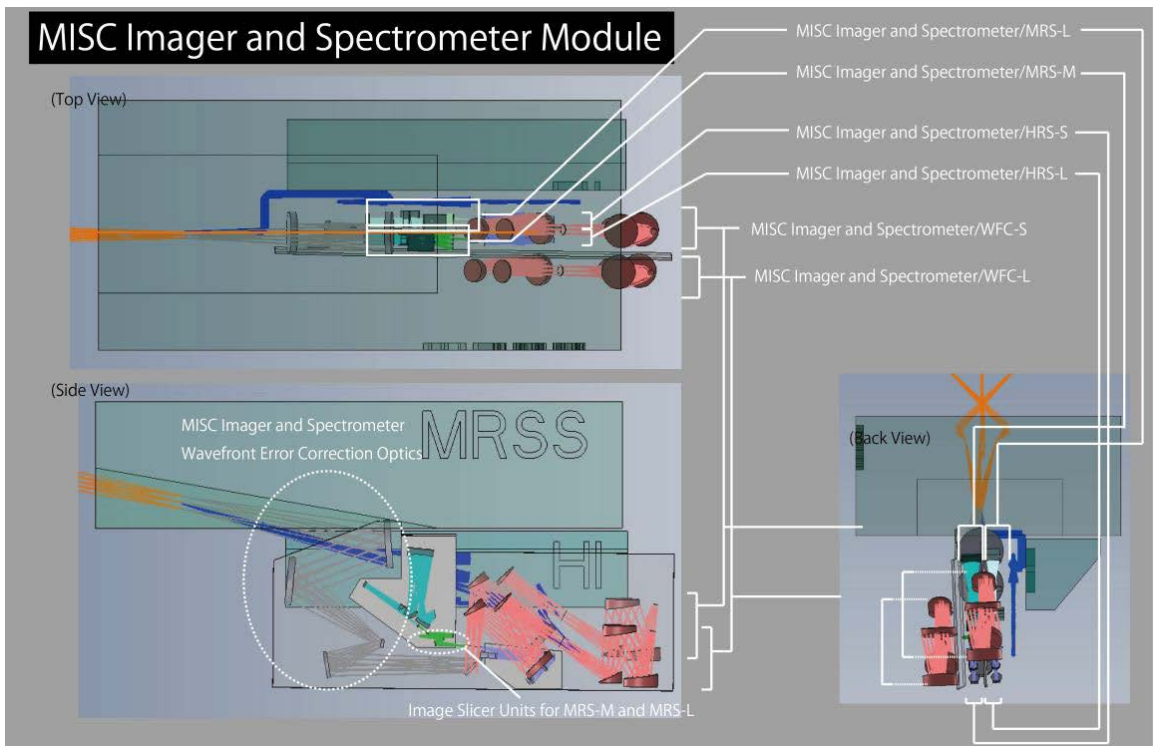

**Figure 2-34:** The MISC I&S module optical elements and mounting structures are aligned on a base plate, which accounts for a significant fraction of the total mass. To reduce mass, the mirrors, mirror support structures, and base plate are SiC.

### 2.2.2.2.3 MISC Mechanical Design

Three-dimensional (3D) solid models of the I&S and TRA modules (**Figure 2-34** and **Figure 2-35**) are also indicative of the maturity of the MISC instrument design.

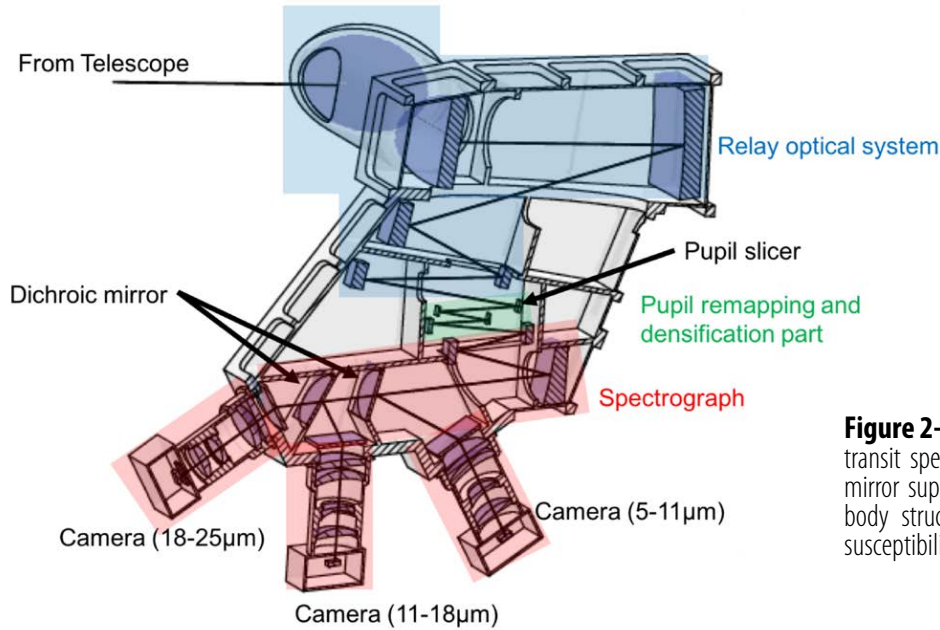

**Figure 2-35:** Structural design of the MISC transit spectrometer, in which the mirrors, mirror support structures, base plates, and body structures are Al 6061, minimizing susceptibility to thermal deformation.

### 2.2.2.2.4 MISC Mechanisms

The MISC instrument has eleven mechanisms. The I&S module includes a Deformable Mirror (DM); a tip-tilt mirror (TTM); a Slit Mirror Changer; and two filter wheels, one for the Wide-Field Imager (WFI)-S (short-wavelength channel) and one for the WFI-L (long-wavelength channel). The COR module includes a DM; a TTM; and two slit wheels, one each for COR-S and COR-L. The TRA module has no mechanisms. Since the OST telescope is diffraction-limited at >30 μm, the MISC I&S and COR modules provide wavefront error correction internally using DMs and TTMs, achieving diffraction-limited image performance at >5 μm for sources within a 3'x3' FOV. The COR DM, with 32 x 32 actuators, excludes speckles in regions up to 16 λ/D from the PSF peak [Takahashi et al., 2017]. The slit mirror changer for I&S has four 2-inch positions for slit mirrors. It is used to switch observing modes between WFI-S and WFI-L, medium-resolution spectrometer MRS-M and MRS-L, and high-resolution spectrometer HRS-S and HRS-L. The I&S filter wheel assemblies are comprised of triple wheels and each wheel has twelve 1.25-inch positions for band-pass filters and/or grisms. The COR filter wheel assemblies have double wheels and each wheel has six 1-inch positions for band-pass filters and/or grisms. The slit wheel in COR is a single wheel with six 0.5-inch positions for slit masks. These masks are used for low-resolution spectroscopy.

### 2.2.2.2.5 MISC Detection Subsystem

All MISC detectors are a similar Si Blocked Impurity Band (BIB) design bonded to Si readout multiplexers. Two doping agents, As and Sb, are employed to provide good detective quantum efficiency over the MISC 5-38 μm wavelength range. Required MISC detector development is a modest factor of 2 increase in the detector dimension format and multiplexer work to increase the detector stability for the TRA module. The multiplexer is an enabling OST technology (**Section 2.2.2.4**).

At the time of this writing, the team is developing a detector technology roadmap. The roadmap will acknowledge that substantial effort will be required to advance from the state-of-the art, particularly where Sb is used as a dopant. Alternative detector types, such as bolometers, will also be considered.

### 2.2.2.3 MISC Estimated Performance

To estimate MISC sensitivity, the team applied assumptions, including:

Exposure Time: The longest exposure time is 300 sec for the Si:As 2k×2k detectors (30 μm/pix; Raytheon) and 600 sec for the Si:Sb 2k×2k or 1k×1k detectors (18 μm/pix; DRS), taking into account the number of pixels affected by cosmic ray events, with an assumed event rate $5×10^4$ $m^{-2}$ $sec^{-1}$ at Sun-Earth L2 [Swinyard et al., 2004].

Detector Performance: Detector Dark Current is assumed as 0.06 $e^-$ $sec^{-1}pixel^{-1}$ for the Si:As detectors and 0.8 $e^-$ $sec^{-1}pixel^{-1}$ for the Si:Sb detectors. The readout noise (reduced up to ¼ by means of Fowler-16 sampling) is assumed to be 10 $e^-$ for the Si:As detectors and 25 $e^-$ for the Si:Sb detectors. Saturation Full Well is assumed to be $2.5 \times 10^5$ $e^-$ for the Si:As detectors and $1.0 \times 10^5$ $e^-$ for the Si:Sb detectors.

Background: Zodiacal emission is modeled as a greybody at 268.5 K normalized to 80 MJy/sr at 25 μm for the high-background case, and as a greybody at 274.0 K normalized to 15 MJy/sr at 25 μm for the low-background case.

Predicted MISC measurement capabilities are provided in **Section 2.2.9**.

### 2.2.2.4 MISC Enabling Technology

MISC has some enabling technology for which no heritage has been identified (**Table 2-7**). Ultra-stable detection technology is critical to the gain in mid-IR transiting exoplanet spectroscopic detection capability necessary for OST. The Densified Pupil Spectrometer includes science and reference pixels, and a key to improving the detection system is learning how to use the reference pixels. The team will build a subsystem prototype testbed to optimize this critical calibration technique.

**Table 2-7:** MISC component and subsystem enabling technology and current maturity.

| Description | Subsystem/Component | TRL |
|---|---|---|
| PIAA Coronagraph Module | S | 3 |
| 8-Octa Phase Mask for MIR (8 – 36 μm) | C | 2 |
| Immersion Grating (25 – 38 μm) | C | 2 |
| Densified Pupil Spectrometer | S | 3 |

### 2.2.3 Medium Resolution Survey Spectrograph (MRSS)

> MRSS offers simultaneous low-resolution spectroscopy over the entire 30-660 μm band, enabling spectral surveys of high-z galaxies. A moderate resolution mode can enhance detection of lines in galaxies and planet forming disks.

### 2.2.3.1 MRSS Science Traceability

MRSS measures the rest-frame mid- and far-IR spectral lines essential to charting the cosmic history of galaxies, enabling a new understanding of the interplay between stars, black holes, and the interstellar medium. To access the full history of the universe since Reionization requires superlative sensitivity and uninterrupted coverage of the far-IR, which MRSS provides. The spectral lines are approximately uniformly distributed in log λ, and access to a large range of redshifts is imperative. Spectral resolving power R ~ 500 is sufficient for these observations, as the goal is to measure integrated line fluxes. Pointed observations are used to follow up on interesting known sources (i.e., based on JWST, WFIRST, ground-based mm/sub-mm surveys), but the OST team also prioritized blind spatial-spectral surveys, for which a scanning mode is required. These requirements set the basic instrument architecture.

Higher resolving power (R ~ several thousand) is required to enhance the line-to-continuum ratio and enable OST's characterization of protoplanetary disks. This is particularly necessary in the mass-tracing HD line at 112 μm and in the OI 63 μm line, which carries much more gas flux per column density and is therefore better suited to characterizing older gas-poor disks. The desire to understand planet habitability development drives the requirement to measure low-lying far-IR water lines and distinguish between gas interior and exterior to the snow line in these disks and, in turn, a requirement for R ~ 25,000 at wavelengths as long as 180 μm.

### 2.2.3.2 MRSS Instrument Description

The heart of MRSS is a suite of six R=500 wideband grating spectrometer modules which combine to cover the full 30-660 μm range instantaneously. The detector system is designed to maximize sensitivity to line detection, and each grating couples a long slit on the sky, so that MRSS can be used for both pointed observations or spectral mapping.

MRSS also includes a a Fourier-transform interferometer, which can be brought into the beams in front of the base grating backends using a pair of sliding mirrors on a stage, providing resolving power or order $10^4$.

MRSS observes point sources or makes small maps (~0.3 deg) with the observatory in a staring mode, for which FSM provides spatial modulation. This mode requires pointing control to 0.1 arcsec to provide photometric precision. Alternatively, MRSS can be used to conduct wide-field surveys. In this mode, the entire observatory scans in a manner similar to that used with Herschel, and pointing is reconstructed, obviating the need for precise pointing control. The challenge is to cover large fields (up to tens of degrees, with good recovery on degree scales) fast enough to beat low-frequency systematics, while still being able to capture information on the short wavelength beamsize at frequency ≤100 Hz. A scan speed of ~60 arcsec per second satisfies these requirements.

### 2.2.3.2.1 MRSS General Instrument Operating Principle and Heritage

MRSS provides outstanding spectral survey capability. The full 30-660 μm wavelength range is divided into six bands, spaced uniformly in log λ. Each band covers a 1:1.72 bandwidth, and can be processed efficiently with a single first-order grating that covers the full band instantaneously with no moving parts. This approach is similar to that employed in the Spitzer Infrared Spectrograph's (IRS) four grating modules [Houck et al., 2004] (**Figure 2-36**). Each grating module couples only one polarization with high efficiency, specifically the mode with perpendicular to the slit length and parallel to the direction of spectral dispersion.

The slits for Bands 1, 2, and 3 overlap in a single field, and those for Bands 4, 5, and 6 overlap in a second (**Figure 2-37**). This improves the speed of obtaining full-band spectra of individual targeted sources in follow-up mode. For each sub-field, light is routed to the three spectrometers using a polarizing grid beamsplitter to split the light into its two linear polarizations. In one polarization arm, a dichroic is used to further split the light into two bands, while the other arm couples the third band through a polarization rotator.

High-resolution mode is implemented using a pair of identical Martin-Puplett Fourier Transform spectrometers (FTS), one each for the long-wave and short-wave slits. Similar interferometers have flown successfully on COBE/FIRAS, Cassini/CIRS, and Herschel/SPIRE.

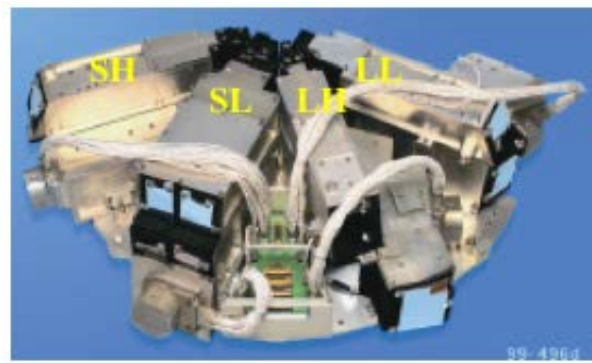

**Figure 2-36:** The Spitzer IRS module performed its spectroscopy with no moving parts by sending the light to four grating modules. MRSS adopts a similar approach.

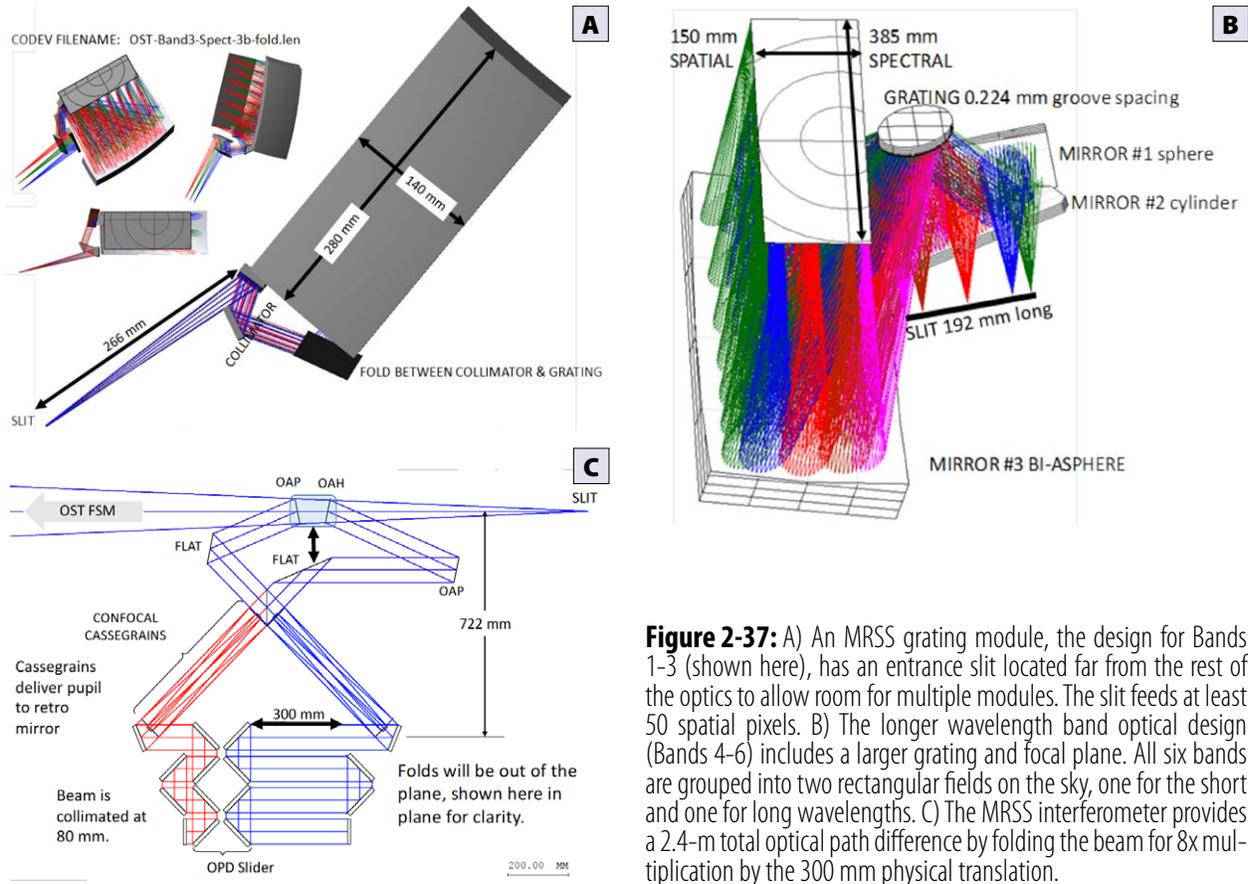

**Figure 2-37:** A) An MRSS grating module, the design for Bands 1-3 (shown here), has an entrance slit located far from the rest of the optics to allow room for multiple modules. The slit feeds at least 50 spatial pixels. B) The longer wavelength band optical design (Bands 4-6) includes a larger grating and focal plane. All six bands are grouped into two rectangular fields on the sky, one for the short and one for long wavelengths. C) The MRSS interferometer provides a 2.4-m total optical path difference by folding the beam for 8x multiplication by the 300 mm physical translation.

#### 2.2.3.2.2 MRSS Optical Design

The optical building blocks of the MRSS instrument all have scientific heritage. All of the grating spectrometer modules are simple bolt-and-go aluminum structures with no moving parts, similar to the Spitzer infrared spectrograph shown in **Figure 2-37**. Five of the six MRSS grating spectrometer modules are simple R=500 first-order grating systems. Light is routed to the six spectrometer backends with a combination of polarizing grids and dichroics, configured in a converging beam so each grating module has unique access to the focus to define a slit entrance to a light-tight enclosure. For the short-wave slit (Bands 1-3), this is possible in the converging beam from the telescope. For the long-wavelength slit (Bands 4-6), light passes through the telescope focus and then is relayed from the crowded focal-plane area to provide more space for the larger grating spectrometers.

As needed, an interferometer is engaged by inserting two mirrors to intercept the converging beams from the telescope to the gratings. The same grating backends are used for detection. Through careful design and pupil reimaging, the interferometer subsystem adds no appreciable aberrations. With the interferometer in place, each spectral channel from the grating measures an interferogram, and the Fourier transforms of the interferograms are stitched together to create the full high-resolution spectrum.

#### 2.2.3.2.3 MRSS Mechanical Design

The MRSS grating modules are enclosed in light-tight machined aluminum (Al) assemblies supporting diamond-turned Al mirrors (**Figure 2-38**). They are thus homologous and not susceptible to thermal deformations (e.g., no focus shift when cooling). These modules are cooled to 2-4 K, depending on the wavelength band.

#### 2.2.3.2.4 MRSS Mechanisms

MRSS mechanisms are associated with the FTS subsystem. A small sliding stage inserts two mirrors into the telescope's beam, engaging the interferometer in a small field. A 2.4-m total optical path difference through the interferometer is obtained by folding the light, resulting in an 8x multiplication factor on the physical translation of 300 mm. This approach, with a dual-sided stage and folded optical path, is similar to that used in the Herschel SPIRE instrument, albeit with a longer scan length.

#### 2.2.3.2.5 MRSS Detection Subsystem

MRSS adopts the strongest detector technology available. All the low-temperature detector types require about the same support infrastruc-ture, in terms of temperature, space, cabling, and support electronics. There is room for progress in the multiplexing readouts. As currently conceived, MRSS uses detector arrays based on RF or microwave micro-resonators. In this scheme, pioneered by groups developing kinetic inductance detectors (KIDs) [Day et al., 2003; Zmuidzinas, 2012], each detector is coupled to a micro-resonator and incident photons shift the frequency of the resonator. If the resonators have sufficiently high quality factor (Q), then 1000 or more can be arrayed in frequency and accessed with a single readout circuit. Digital processing and ACD/DAC operations are performed in a custom silicon application-specific integrated circuit (ASIC).

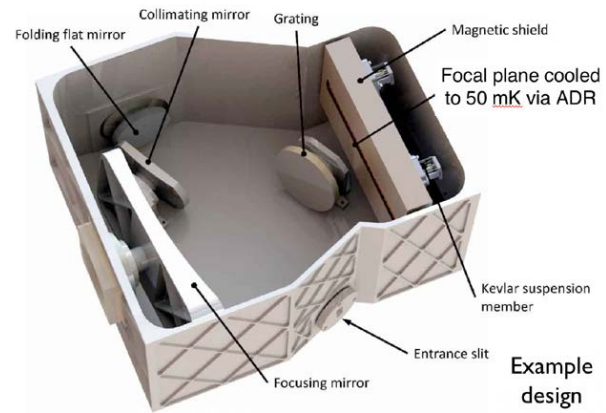

**Figure 2-38:** The MRSS grating modules have no moving parts. In Bands 1 – 3, the focal plane assembly drives the size, not the grating.

#### 2.2.3.3 MRSS Estimated Performance

To estimate MRSS sensitivity (**Section 2.2.9**), the team assumed detectors with sensitivity (Noise Equivalent Power, NEP) of $2 \times 10^{-20}$ W/$\sqrt{\text{Hz}}$ in array formats up to 50 kilopixels would be available (TRL 5 by PDR).

Because the same grating backends are used for detection in the high-resolution mode, the line sensitivity in this mode is comparable to that in the base grating mode.

#### 2.2.3.4 MRSS Enabling Technology

Detectors are the most important enabling technology for MRSS, and for OST in general. OST's superlative sensitivity with MRSS is largely due to its cryogenic temperatures and advanced detectors. To satisfy MRSS science goals, its detector requirements include:

- A per-pixel NEP of $3 \times 10^{-20}$ W/$\sqrt{\text{Hz}}$ or lower, with a goal of $2 \times 10^{-20}$ W/$\sqrt{\text{Hz}}$, for which the detector noise is sub-dominant to astrophysical background photon noise.
- Six detector arrays, each with at least 50 spatial by ($500 \times ln(1.7) = 270$) spectral pixels, 13,000 pixels total per array, with a goal of up to $200 \times 270 = 50,000$ pixels per array.
- High absorption efficiency from 30 to 660 μm, allowing for changes in design for optimum performance at various wavelengths.

NASA is currently investing in the maturation of Kinetic Inductance Detectors (KIDs) and Transition Edge Sensor (TES) bolometers, which have potential to satisfy these requirements, as do the more exotic but less mature Quantum Capacitance Detectors (QCDs) (**Section 4.1**).

### 2.2.4 High Resolution Spectrometer (HRS)

> HRS is OST's most sensitive high resolution spectrometer and will spectrally resolve the water and HD lines of forming planetary systems.

#### 2.2.4.1 HRS Science Traceability

HRS is integral to OST's ability to characterize planet-forming circumstellar disks, derive the distribution of water, and understand habitable planet formation. HRS is optimized to provide the high spectral resolving power required to Doppler-resolve the mass-tracing HD J = 1 → 0 112.8 μm and ground-state para $H_2O$ 179 μm lines, providing an unmatched high-sensitivity survey of protoplanetary disks. The far-IR spectral region contains water lines representing a full range of excitation temperatures, allowing HRS to trace gaseous water over the entire protoplanetary disk. The HERO instrument (**Section 2.2.6**) will make complementary observations of the ground state ortho water line at 538 um. By resolving these important lines, and assuming Kepler-like rotation, the distributions of total mass and gas-phase water can be derived. HRS sensitivity enables detailed studies of Solar-mass protoplanetary systems beyond 1 kpc, ensuring access to the full range of systems and characterization of the processes that drive the production of Earth-like planets.

#### 2.2.4.2 HRS Instrument Description

HRS, an echelle spectrometer, simultaneously provides the required resolving power and high sensitivity. The sensitivity per spectral channel is maximized because only photons in the selected bandwidth contribute to noise. Optimal sensitivity requires detectors that can reach the background limit (**Section 2.2.4.3**), but excellent performance is still achieved with detectors at the current state-of-the-art. Overall simultaneous spectral coverage is set by the number of detector elements, which is only limited by budget.

##### 2.2.4.2.1 HRS General Instrument Operating Principle and Heritage

HRS is an echelle spectrometer with a grating cross disperser, providing resolving power R ~$10^5$ x (50 μm /λ) in the far-IR. The grating allows simultaneous sampling of a contiguous spectral region where the bandwidth of each resolution element is set by the echelle resolution. A high-resolution Fabry-Perot interferometer (FPI) is used in series with the echelle to provide resolving power as high as R ~$10^6$, with the echelle order-sorting the high-resolution, high-order etalon. The instrument has two operating modes: echelle-only mode, which provides R~$10^5$, and high-resolution (R ~$10^6$) mode when the FPI is introduced. The high-resolution mode can be used to observe a single on-axis point source, making it well-suited for follow-up measurements. In echelle mode, the wavelength region is selected by choosing an input slit and order-sorting filter, and adjusting the angles of the echelle grating and cross-dispersing grating. A single echelle is used over the full spectral range, but three cross-dispersion gratings are required to cover the full range. A selection mirror diverts the input beam to switch between modes. The selection mirror can divert internal spectral and intensity sources into the light path for calibration. To provide spectral stabilization at better than a part in $10^7$ and facilitate spectral radiometric calibration, HRS uses quantum cascade lasers (QCLs) and thermal sources like those being developed for SOFIA/HIRMES.

The HRS etalon design is derived from Cornell University devices being developed for HIRMES. The etalons, which have larger reflector spacing and are about twice the diameter of those for HIRMES, can be redily fabricated.

The HRS focal plane is comprised of arrays of detectors optimized to about a single octave of spectral coverage. This provides full sampling of the dispersed images and efficient coupling to the detectors. Since the highest sensitivity detectors often have limited dynamic range, the HRS team allowed for the possibility of including detectors for operation at higher power levels; OST's increase in sensitivity compared to prior missions is so large the entire brightness range is discovery space. Only a subset of detectors acquires science data at any given time, so only these relevant detectors operate, saving significant power.

### 2.2.4.2.2 HRS Optical Design

HRS optical design is typical of a conventional echelle spectrograph; system packaging is the primary challenge. High resolving power requirements and the need to minimize length drives the high angle echelle (**Figure 2-39**). The linear layout was driven by the available space; it is optically convenient, but not as mechanically efficient as a more compact layout. The design provides diffraction-limited performance over the full spec-tral range, and the diamond turned aluminum surfaces offer near-ideal performance in the far infrared. The Fabry-Perot module is mounted below the optical bench and is inserted into the optical path using a selector mirror.

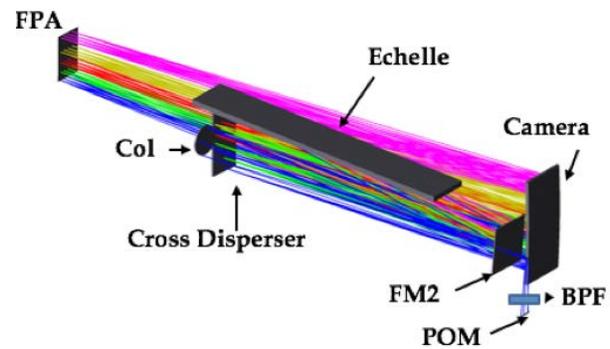

**Figure 2-39:** The HRS pick-off mirror (POM) diverts light from the telescope focal plane into the instrument to its focus on the spectrometer input slit. The light is collimated (Col) and illuminates the echelle. The light is reflected from the echelle to FM2, then to the cross disperser. The camera lens then focuses the cross-dispersed beam onto the focal plane. The etalon and its optical path is under the optical bench. When selected, it serves as a prefilter for the echelle spectrograph.

### 2.2.4.2.3 HRS Mechanical Design

For this early design phase, GSFC's Instrument Design Lab (IDL) studied a lightweighted all-aluminum instrument, which was not optimized for mass or packing efficiency. The design provides a stable system, so good system cooling stability is expected with careful annealing. The grating was modeled as a monolithic, lightweighted structure. The study provided a preliminary mass, indicated that HRS could be built with conventional material, and indicated no technology advances are required for the mechanical system or assemblies. Additional studies are needed to optimize layout and structure, should consider carbon fiber elements, and will benefit from JWST heritage.

Given HRS's large optical surfaces and modest wavefront error requirements, lightweight optical structures with active control could significantly decrease instrument mass. Modest developments in this area could result in significant savings.

### 2.2.4.2.4 HRS Mechanisms

To enable coarse HRS echelle spectrograph configuration for observations at particular wavelengths, the observer selects a slit that matches the diffraction-limited beam of the telescope, an order sorting filter, and a cross dispersing echelle to direct the desired echelle order to the required detector. The fine adjustment requires tilting the echelle to direct the line of interest to the center of the detector array. The required mechanisms include: Slit Wheel (rotary), Filter Wheel (rotary), Cross Disperser Exchanger/Rotator (rotary with precision control), Echelle angle adjustor (jack screw), and FPI/Calibration Selector (rotary). The slit and filter wheels have heritage from JWST/NIRSpec. The Cross Disperser Exchanger/Rotator must execute large-angle motions to select gratings and provide high-resolution continuous adjustment for wavelength tuning. HIRMES will demonstrate a similar mechanism on SOFIA. Finally, the echelle requires precision angle control over a limited angle range. The planned mechanism uses a flexure and jackscrew to control the angle. This will require routine engineering development; none of the angular position or stability requirements fall outside the range of previously demonstrated mechanisms.

Since HRS employs high-sensitivity detectors, the motors and frictional elements in the mecha-nisms must be kept at low temperatures to avoid self-emission. Sub-Kelvin cooling is provided by an Adiabatic Demagnetization Refrigerator (ADR), a descendent of the ones developed for Astro-H and now being built for its successor XARMS, and being developed for Athena. The ADR design has not been established, the thermal, mechanical, and magnetic designs will be based on those with previous X-ray flight heritage.

### 2.2.4.2.5 HRS Detection Subsystem

The detector assemblies will be derived from those developed for HIRMES and those being developed by SRON for Athena. These assemblies provide the required performance under launch loads, assure adequate immunity from vibrational heating in operation, and provide adequate magnetic shielding in Earth's field for ground testing and adequate rejection of spacecraft and instrument-generated fields in flight.

### 2.2.4.3 HRS Estimated Performance

The HRS team estimated the instrument's sensitivity assuming different levels of detector sensitivity, a noiseless detector, such as an efficient photon counter, a case with NEP $1 \times 10^{-20}$ W/$\sqrt{\text{Hz}}$, and a case with NEP ~ $1 \times 10^{-19}$ W/$\sqrt{\text{Hz}}$. Overall instrument efficiency was estimated at 15% (including quantum efficiency of the detector) using HIRMES as an analog. A detailed HIRMES model predicts 12% efficiency for a system with more lossy optical elements, so 15% is conservative. Sky brightness was assumed to be the ecliptic pole, as measured with COBE's DIRBE and FIRAS instruments.

### 2.2.4.4 HRS Enabling Technology

High-sensitivity detector arrays are a critical enabling technology for OST, and HRS in particular. While high-quality science is achievable using current state-of-the-art ($1 \times 10^{-19}$ W/$\sqrt{\text{Hz}}$) detectors, very large improvements in observing speed can be achieved for the faintest objects by developing detectors that can reach the photon shot noise limit (~$3 \times 10^{-21}$ W/$\sqrt{\text{Hz}}$). Such an increase in sensitivity would significantly increase the advantage the HRS has over alternatives, such as an FTS. If such high-performance detectors are developed, an instrument like HRS is required to reap the benefits of their sensitivity. Any development progress that increases the size of the detector array and can be operated within the available OST resources is of great importance.

### 2.2.5 Far-infrared Imager and Polarimeter (FIP)

> FIP will perform wide-field imaging and polarimetry observations, enabling deep field images of high-z galaxies and detailed studies of nearby star formation regions.

### 2.2.5.1 FIP Science Traceability

FIP allows OST to survey large areas for continuum emission from dust, survey solid bodies in the outer solar system, and measure dust polarization, yielding information about the direction and strength of the interstellar magnetic field. With astrophysical background-limited sensitivity and angular resolution unprecedented in a far-IR telescope, FIP on OST enables studies of the first dust in the universe, properties of the dust and magnetic fields in our own and nearby galaxies, and probes the origins of our solar system.

### 2.2.5.2 FIP Instrument Description

FIP performs large-area survey imaging in four far-IR wavelength bands ranging from 40-240 μm and two modes: Total Power Mode and Polarimetry Mode. In Total Power Mode, FIP quickly reaches the confusion limit at longer wavelengths (32 ms at 240 μm; 9 s at 120 μm), but allows integrations of unprecedented depth and is unlikely to reach the confusion limit in the 40-μm band. The FIP team optimized the design to meet the maximum achievable optical efficiency in each band and provide simultaneous wavelength coverage while requiring only a moderate FOV from the telescope.

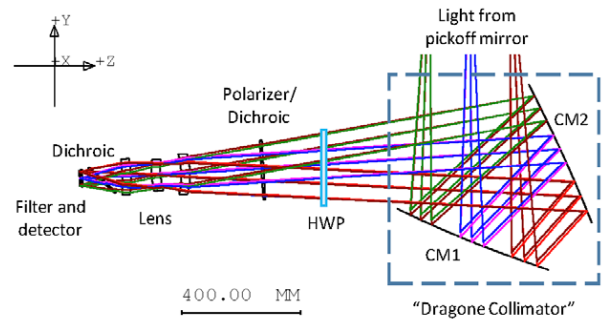

**Figure 2-40:** The FIP optical layout has heritage from SOFIA/HAWC+.

### 2.2.5.2.1 FIP General Instrument Operating Principle and Heritage

FIP is similar to the SOFIA/HAWC+ instrument (**Figure 2-40**), which has been performing successful science flights since December 2016. In Total Power Mode, FIP observes in four bands (40, 80, 120, 240 μm) simultaneously. In Polarimetry Mode, FIP operates as a differential polarimeter, simultaneously covering two bands (40 and 80 μm or 120 and 240 μm). FIP provides diffraction-limited images (1.1 x (λ/40 μm) arcsec) and covers an instantaneous FOV of 2.5' x 5' at 40 and 80 μm and 7.5' x 15' at 120 and 240 μm.

FIP has two orthogonal arms, each with one 'blue' and one 'red' channel. Light between the two arms is split into the two observing modes. In Total Power Mode, a dichroic separates the incoming radiation into a "long band," which includes the 120 μm (blue), and 240 μm (red) bands and a "short band," which includes the 40 μm (blue) and 80 μm (red) bands. In Polarization Mode, a half-wave plate modulator is brought into the beam and the dichroic is replaced by a polarizer. One optical arm then receives the horizontal (h) polarized radiation, while the other receives the vertical (v) polarized radiation. The red and blue channels in both arms are then identical (either "long" (l) or "short" (s)), enabling differential polarimetry simultaneously in two bands. The center band frequencies are chosen, so each detector operates in resonance at nearly 100% optical efficiency in the l and h frequency bands.

### 2.2.5.2.2 FIP Optical Design

Light is directed into the FIP instrument box by a pickoff mirror located near the telescope focal surface. Once inside the instrument box, a Dragone collimator comprising a pair of free-form (polynomial surface) mirrors (CM₁, CM₂), is used to collimate the light and correct aberrations. Depending on the mode of operation, the collimated beam either traverses or bypasses a half-wave plate (HWP) before passing through a dichroic or polarizer. The path depends on the waveband of interest, with a total of three possible bands: 1) short, 40 – 80 μm, 2) long, 120 – 240 μm, and 3) the full band. If either the short or long band is used, the light passes through one of two HWPs (one for each band) and then through a polarizer. If the full band is used, the light does not pass through an HWP and is incident upon a dichroic rather than a polarizer. In both cases, light is split into two orthogonal paths: reflection and transmission.

The rays from these two beam paths are imaged by separate lens systems. In each case, the converging light is split in two again before passing through a filter and onto a detector plane, bringing the total number of beam paths (and detectors) to four. Like the HWP and polarizer/dichroic, the lens system switches in and out during operation depending on the waveband. The system uses an f/7 or f/2 lens for the short and long bands, respectively, bringing the total number of possible beam paths to eight. Although each of the four channels has two possible lens configurations, for either lens, the detector does not change. The HWP, dichroic/polarizer, and lenses are all switched in and out through mechanisms.

**Table 2-8:** FIP mechanisms have heritage from SOFIA/HAWC+.

| Mechanism | Requirement | Design Concept |
|---|---|---|
| Half-wave Plate Flip | Three positions, open center<br>Element diameter: 150 mm<br>Life: 5 yr, 22 k cycles + test<br>Duty cycle: low | Superconducting stepper motor drives crank between toggle positions, rotating element housings around axis parallel to optical axis; stop motor at center position to achieve open condition |
| Half-wave Plate 1 Rotation and Half-wave Plate 2 Rotation | Rotate HWP at 60 rpm<br>Read position to +/-0.1 deg<br>Life: $10^7$ revolutions<br>Duty cycle: high | Superconducting brushless DC motor and resolver to rotate HWP and sense its position |
| Polarizer/Dichroic Flip | Two positions<br>Element diameter: 150 mm<br>Life: 5 yr, 22 k cycles + test<br>Duty cycle: low | Superconducting stepper motor drives crank between toggle positions, rotating element housings around axis parallel to optical axis |
| Lens Flip 1 and Lens Flip 2 | Two positions<br>Element diameter: 150 mm<br>Elements move simultaneously<br>Life: 5 yr, 22 k cycles + test<br>Duty cycle: low | Superconducting stepper motor drives crank between toggle positions, translating linear carriage |

### 2.2.5.2.3 FIP Mechanical Design

The FIP structure (**Figure 2-41**) is aluminum. Mechanical parts include exterior flexure mounts, a housing frame, thin enclosure, optical bench, and brackets that hold the three main mirrors. The design also includes mechanical components associated with the internal optics and detector assemblies.

### 2.2.5.2.4 FIP Mechanisms

FIP's six mechanisms (**Table 2-8**) are well understood and characterized.

### 2.2.5.2.5 FIP Detection Subsystem

The FIP uses membrane-suspended, close packed Transition Edge sensors (TES) and a fre-quency domain multiplexer, the µWave SQUID, which can handle FIP's large pixel counts. µWave SQUIDs have been demonstrated, but currently do not achieve the channel density desired for low-power readout. If the density cannot be enhanced to meet OST needs, the team proposes using a hybrid scheme, which combines Code Division Multiplexers with µWave muxes. All elements for this scheme have been demonstrated. The pixel dimensions of FIP's four detector arrays are 120 x 240 at 40 µm, 60 x 120 at 80 µm, 120 x 240 at 120 µm, and 60 x 120 at 240 µm.

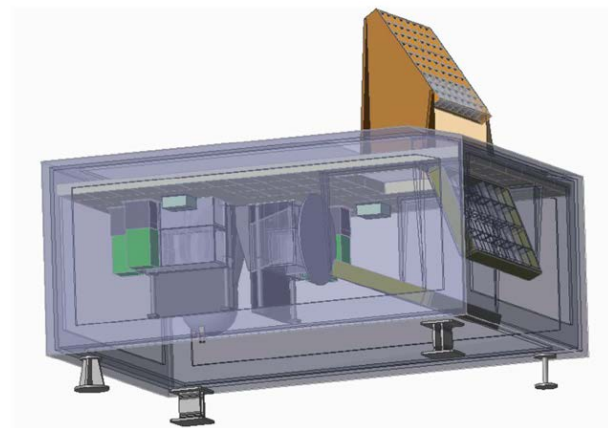

**Figure 2-41:** The FIP mechanical design is derived from HAWC+. Mechanical enclosures for the warm electronics are located in the spacecraft bus.

### 2.2.5.3 FIP Estimated Performance

The predicted FIP measurement capabilities are provided in **Section 2.2.9**.

### 2.2.5.4 FIP Enabling Technology

The technology needed for FIP is already well within reach. Existing single-pixel laboratory detectors already meet the requirements for background-limited far-IR imaging in space. A group at SRON developed low-noise (NEP ~ $10^{-19}$ W/√Hz) TES bolometers on suspended SiN membranes for the SAFARI instrument, with pixel pitch 800 µm, comparable to the detector requirements for FIP. Suzuki [2015] described attaining this sensitivity in a device with time constant <1 ms. Although this device had no absorber, other devices with absorbers have achieved similar sensitivity. GSFC is developing phonic filters to provide even better thermal isolation of suspended TES detectors, potentially enabling another order of magnitude improvement in sensitivity [Rostem et al., 2016], surpassing the FIP requirement. The GSFC team has demonstrated kilopixel Backshort Under Grid (BUG) detector arrays with readout multiplexers in a relevant environment on SOFIA.

### 2.2.6 Heterodyne Receiver for OST (HERO)

### 2.2.6.1 HERO Science Traceability

> HERO offers the highest spectral resolution of OST's spectrometers and will resolve the detailed physics of water lines in the ISM and planet forming disks.

The HERO instrument brings high spectral resolving power ($R > 10^6$) to the OST instrument suite, enabling spectral line profile measuring observations of the far-IR water lines and retrieval of the water distribution in a large sample of protoplanetary disks. HERO can also be used to measure water emanating from small bodies in the solar system. Thus, OST observers will be able to follow the trail of water from the interstellar medium to developing habitable planets.

By accessing spectral lines across a broad frequency range, HERO observes many transitions of water vapor and its isotopologues, including the low-lying transitions that trace cold gas. It also offers high spectral resolution access to the 112 μm line of HD, a tracer of molecular gas, a variety of excited CO transitions, and key diagnostic fine structure lines of oxygen, ionized carbon and nitrogen, which can be used to probe the flow of energy through the interstellar medium. Many targets are expected to present complicated line profiles, with a combination of narrow (<1 km/s) and broad (≤10 km/s) emission features, or narrow (<0.5 km/s) or broad (~10 km/s) absorption dips, which result from a complex interplay of gas kinematics with line photon propagation through the emitting and absorbing medium.

HERO's large array sizes offer excellent mapping capabilities absent from the previous generation of heterodyne instruments on space-borne platforms. This capability can be used to map filamentary structures and molecular clouds on square-degree scales, providing a new view on star formation.

### 2.2.6.2 HERO Instrument Description

To meet the scientific requirements, HERO covers the wavelength range 111 - 641 μm (in six bands) continuously, and the [OI] line separately at 63 μm with a range that allows for the redshifts of nearby galaxies. As OST goals require mapping, HERO employs focal plane arrays that allow the observer to obtain, instantaneously, many spectra at 2Fλ spacing. HERO provides 2x16 pixels in each of the two orthogonal linear polarizations in the low-frequency bands, and 2x64 pixels in the high-frequency bands. Thus, a similar area can be mapped at low and high frequencies with a diffraction-limited beam. The receiver pixels are spaced equidistantly in a square array. When observing unpolarized objects, the data obtained from both polarizations can be co-added to reduce the noise by $1/\sqrt{2}$, but HERO also can be used to make polarization measurements.

### 2.2.6.2.1 HERO General Instrument Operating Principle and Heritage

HERO is a passive heterodyne receiver system. In a heterodyne receiver, the "radio" frequency (RF) signal from the sky is compared to that of an artificial monochromatic source, the local oscillator (LO). The RF and LO signals are mixed, and a signal at the beat frequency, or intermediate frequency (IF), is created. The system allows the LO to be selected so the intermediate frequency is constant and convenient to amplify using commercial components. A nominal heterodyne receiver system consists of optics, a receiver front end with a mixer and LO, an IF chain for signal amplification and processing, and a receiver backend containing a spectrometer.

HERO builds on a rich heritage of heterodyne instruments for space and near-space, including ODIN/SMR, SWAS, Rosetta/MIRO, and more recently, the Heterodyne Instrument for the Far-Infrared (HIFI) on Herschel, as well as advanced designs for JUICE/SWI, and the Galactic/Extragalactic Ultra-long Duration Balloon Spectroscopic Terahertz Observatory (GUSTO). HERO exploits technology developed for HIFI, including mixers, LOs, IF, backends, mirrors, control electronics, and observing modes. HERO uses very sensitive cryogenic SIS and HEB mixers that also contributed to the success of Herschel/HIFI.

HERO uses the most sensitive mixer technology available, approaching the quantum limit. It uses SIS mixers at frequencies below 900 GHz (wavelengths >333 μm), and HEBs at higher frequencies. The LO reference signal is provided by amplifier-multiplier chains for all bands, including the [OI] band at 63 μm.

### 2.2.6.2.2 HERO Optical Design

The HERO optical system is divided into four parts: fore-optics, receiver modules, cold LO optics, and warm LO optics (**Figure 2-42**). The common fore optics section picks off the signal from the focal plane and adjusts the focal length for compactness (**Figure 2-43**).

The Offner relay allows the selection of the receiver module for observations. While one module sees the target, the others see different locations on the sky. Each receiver module has two focal plane arrays to measure the two linear polarizations. Each receiver module also contains two refocusing mirrors and a facetted field mirror, which matches the sky signal to the pixels of the focal plane array.

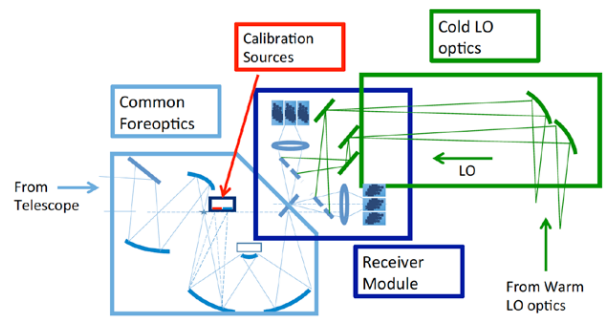

**Figure 2-42:** Schematic illustration of the HERO optical design.

Because waveguide and cables losses would be too great at HERO frequencies, the LO signal is transported via free space from the spacecraft bus to the receiver modules, traversing a distance of ~6 m. This signal must penetrate the sunshield. To minimize the number and size of the holes in the sunshield and retain some redundancy, a focusing mirror forms a very slow beam and the LOs are superimposed onto two channels, one for each polarization. Compensating optics keep the optical path between LO and mixers constant, and the path aligned.

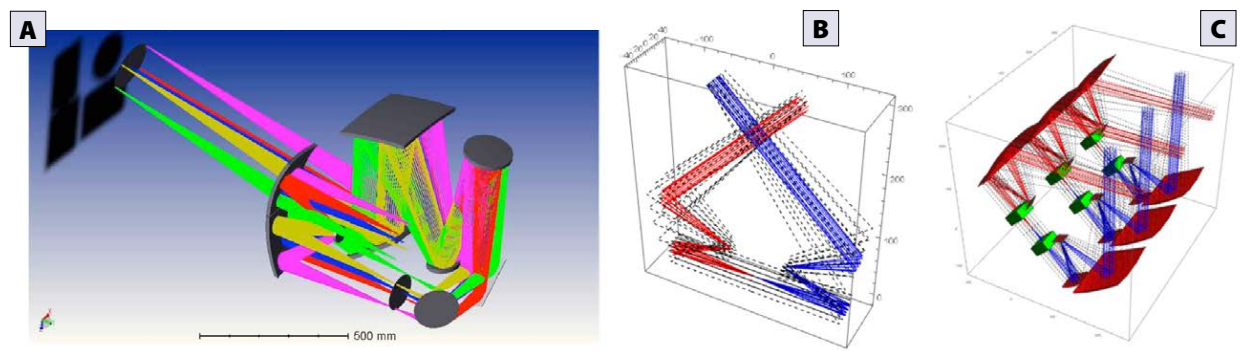

**Figure 2-43:** HERO is designed for compactness. A) Ray tracing of the foreoptics. B) One of six receiver modules. C) Stacked receiver modules. The signal passes through an Offner relay, which 1) projects the astronomical target onto one of six different focal plane arrays, 2) provides focal-plane chopping between source and off-source, 3) switches between the sky and calibration loads, and 4) compensates for telescope jitter.

### 2.2.6.2.3 HERO Mechanical Design

HERO mechanical structures work in three thermal environments: a 4 K zone in the IAM, a 20 K structure for the IF amplifiers, and the warm structure in the spacecraft bus. Structures in the latter zones house position-insensitive common RF/DC electronics.

In the 4 K zone, the OST mechanical structure supports the mirrors, receiver frontend, and calibration loads. These components are held in the correct position to within a few μm (for optical efficiency) and are vibration-sensitive. The mechanical subsystem consists of a light-weighted aluminum structure (with thermal contraction taken into account), as it is thermally-conductive and allows for easy matching with the IAM structure and diamond-turned aluminum mirrors. The mirrors and their supports also serve as the unit frames.

The individual HERO optical units (fore-optics, receiver module, LO distribution, LO pick-off mirror) are connected by a frame-like structure that provides maximal stiffness with minimal weight. The HERO focal plane unit is an open structure since stray light is not a concern for a heterodyne instrument. The team will verify the design using Finite Element Analysis (FEA) and there is enough flexibility to modify the design, if necessary.

### 2.2.6.2.4 HERO Mechanisms

The HERO mechanisms are based on proven, high-TRL designs.

- The Offner relay includes a two-axis motor on the mirror and a 2D cryogenic chopper mirror. Similar mechanisms were used in Herschel-HIFI and SPIRE, and an extended low-power motor is currently being developed within the SPICA consortium in Heidelberg.
- The calibration load shutters are closed when HERO is not operating to minimize stray radiation into the other instruments. Light-tight cryogenic shutter mechanisms have been studied at ETH in Switzerland for SPICA's SAFARI instrument.
- When HERO is not operating, a door closes off the LO path through the sunshield. This door, designed by NASA, uses permanent magnets to hold the open or shut position without current. Coils around the permanent magnets can be activated to demagnetize the permanent magnet and allow the door to shut or open. An additional motor ensures door movement. In case of failure, a non-explosive device permanently closes the door.
- To take advantage of mechanisms with flight heritage, the LO path compensation mechanism employs flex-pivots and voice-coil actuators in a pendulum construction.
- A tip-tilt mirror provides LO beam alignment, and the actuation motor has heritage from Herschel.

### 2.2.6.3 HERO Estimated Performance

HERO performance parameters are summarized in **Section 2.2.9**. The receiver sensitivities are calculated assuming double sideband receiver noise temperatures ranging from 40 K in the lowest frequency band to 500 K at the highest. These noise temperatures are very conservative extrapolations from today's state-of-the-art. Resolving powers of $10^{10}$ have been demonstrated in the laboratory, but are not required for astronomy, so HERO has a spectral resolution of up to $10^7$. HERO provides diffraction-limited angular resolution.

### 2.2.6.4 HERO Enabling Technology

HERO relies heavily on HIFI technology, and a less ambitious 4-pixel instrument based on this heritage could be built today. The HERO design takes recent technological progress into account and surpasses HIFI in three key aspects: 1) HERO uses arrays of mixers rather than individual mixers, 2) HERO can observe at wavelengths down to 63 μm, and 3) HERO has a larger instantaneous IF bandwidth.

HERO uses compact mixer arrays with integrated amplifiers. European funding (Radionet, CNES)

supports research on heterodyne arrays, as does NASA through ROSES grants. GUSTO, a NASA long-duration balloon mission, has 8-pixel arrays at 1.4, 1.9, and 4.7 THz. HERO exploits recent developments in low-power SiGe LNAs. Estimated HERO power consumption is based on projected advances in CMOS technology for backends. JPL is conducting R&D work on backends, and R&D applications have been submitted to CNES. HERO uses arrays of multiplier-amplifier chains like those being developed at JPL.

HERO uses high-frequency receivers that exist on balloons (STO$_1$, STO$_2$) and SOFIA (GREAT), with one exception: at 4.7 THz, an amplifier-multiplier chain, rather than a Quantum Cascade Laser (QCL), provides the LO on OST, as QCLs require significant cooling power. Projects including JUICE, ESA funded R&D, and NASA APRA and SAT studies are paving the way to 4.7 THz LO chains with enough power at 4.7 THz to pump a mixer.

HERO's 8 GHz IF bandwidth is already attained with SIS mixers, but require HEB mixers with matching performance. ESA's Radionet project supports research on wideband HEBs. Cryogenic LNAs also need further development to achieve the combination of low noise, low power consumption, and wide bandwidth needed for HERO.

HERO's engineering challenge is the long distance (~6 m) between the LO and mixer. The team proposes building a prototype to demonstrate the path-length compensation and tip-tilt alignment mechanisms.

## 2.2.7 Risk Management Approach

OST is envisaged as a NASA Class A mission. Accordingly, all OST instruments are designed with redundancy and cross-strapping wherever feasible. This is particularly important for MISC, as the imaging array in the imager and spectrometer module also serves as the focal plane and guiding array for the observatory. Each of the three MISC modules has its own warm electronics box, each with two redundant sets of warm electronics. Within each electronics box the single board computers and the rest of the warm electronics are cross-strapped. All moving mechanisms in the MISC cold areas have dual coils/actuators, with each coil/actuator capable of being driven by the A or B warm electronics sides. Although the focal plane arrays do not have dual redundancy, their multiplexer readouts are arranged so sections of the arrays fail gracefully, as on JWST.

All HERO sub-systems are fully or internally redundant. If a single pixel fails, the others are not affected. Should an entire focal plane array or one of the two LO source units fail, there is still redundancy, as each band consists of two focal plane arrays/LO source units. There are two LO channels, and each contains all six bands. To mitigate these single points of failure, the Offner relay and sunshade door use high-TRL, proven, reliable mechanisms.

## 2.2.8 Partnership Opportunities

The MISC and HERO studies were led by a JAXA team and a French CNES team respectively, and OST's instruments plug into the IAM in modular fashion. The OST team envisions international participation in the OST mission, with these instruments contributed by NASA's international partners. JWST serves as a model for the potential partnership.

Partnership opportunities also exist at the component level. Europe has invested heavily in far-IR detectors, and Japan has extensive cryocooler experience.

Based on feedback and current international interest in the OST mission concept from the science community, OST's science goals and discovery space will attract Guest Observers from the global community. The OST team will work with NASA to establish a Guest Observer program that supports international participation.

## 2.2.9 Summary

OST offers unprecedented angular resolution, 2.7 (λ/100 μm) arcseconds, over its entire wavelength range, 5 – 660 μm, except where ground-based submillimeter interferometers peer through atmospheric windows.

OST sensitivity to an unresolved spectral line from a point source with each of the four spectrometers (**Figure 2-44**) is unprecedented and will open a new era of exploration. OST's far-IR instruments, MRSS and HRS, are estimated to be three to four orders of magnitude more sensitive than any far-IR instrument flown to date. At mid-IR wavelengths, MISC modestly improves on JWST's sensitivity due to OST's greater light collecting area, and is significantly more sensitive than JWST at wavelengths >20 μm due to OST's colder temperature. HERO's sensitivity is close to the quantum limit in all bands.

The FIP instrument provides estimated Noise Equivalent Flux Density (NEFD) 3, 6, 9, and 18 μJy √s at 40, 80, 120, and 240 μm, respectively. The imaging sensitivity is unprecedented and provides enormous discovery space for new phenomena. FIP reaches the confusion noise floor in 32 ms at 240 μm, 9 s at 120 μm, and 1 hour at 80 μm, but should never reach the confusion limit at 40 μm. The confusion limit at 40 and 80 μm enables deep field surveys of galaxies at high z.

A wide range of spectral resolving power is available with the OST instruments (**Figure 2-45**), leading to a discovery space far more powerful than Herschel and SOFIA.

The MISC coronagraph covers 6 – 38 μm, achieving $10^{-7}$ contrast at 0.5" from the central star. The MISC transiting spectroscopy module covers 5 – 26 μm and is estimated to achieve 5 ppm spectrophotometric stability, enabling a survey of exoplanet atmospheres for biosignatures. The unprecedented stability of the transit channel will allow breakthrough exoplanet science.

Collectively, OST's instruments open a vast discovery space to exploration and enable the science community to answer profound astrophysical questions (**Section 2**) spanning topics from the first stars to life.

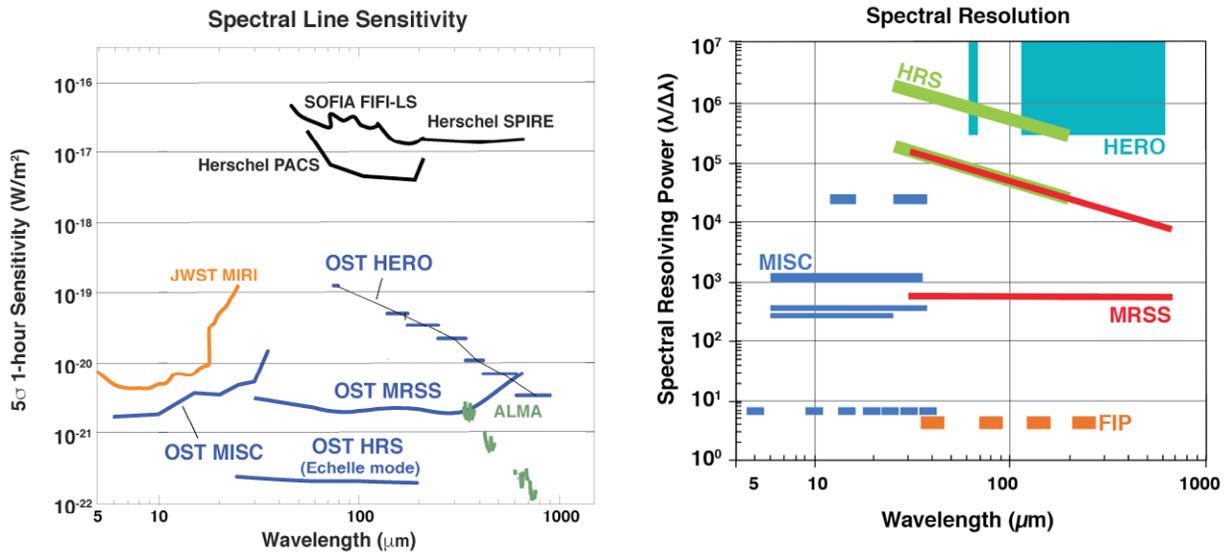

**Figure 2-44:** OST is orders of magnitude more sensitive than past far-IR telescopes, providing new understanding of galaxies in the young universe and development of habitable conditions during planet formation, while opening up a vast new discovery space.

**Figure 2-45:** The OST instrument suite includes broadband imagers and moderate and high-resolution spectrometers, providing measurements optimized to detect biosignatures in exoplanet atmospheres, follow the water trail from the ISM to habitable planets, and extract information from the dominant cooling and diagnostic spectral lines from galaxies out to high redshifts.

## 3 - MISSION IMPLEMENTATION

Origins Space Telescope (OST) offers the next step in astrophysics far-infrared exploration with a large 9.1-m aperture, actively-cooled telescope (~4 K) covering a wide span of the mid-to far-infrared spectrum. Its spectrographs will enable three-dimensional surveys of the sky to discover and characterize the most distant galaxies, exoplanets, and the outer reaches of our Solar System. The observatory's science measurements are enabled by instrument designs that have greater detector performance, large format arrays, and cold telescope temperatures than any previous mission. OST Concept 1 meets all science cases and requirements in **Section 2**. All mission requirements flowed down from the science objectives, measurement requirements, and instrument performance requirements.

### 3.1 Mission Architecture and Overview

OST Concept 1 consists of a sky background-limited telescope for the mid-to-far infrared wavelengths, 5 to 660 microns, state-of-the-art spectrometers and imagers for these wavelengths, a thermal system to passively and actively cool the telescope and instruments to 4 Kelvin (K) and below, and a spacecraft that provides the necessary attitude control, power, radiators, subsystems control, and data collection and transmission.

To keep telescope emission lower than the sky background, the telescope is cooled to ~4 K. The cooling system is a combination of passive cooling provided by a five-layer deployed sunshield, and eight high technology readiness level (TRL) mechanical cryocoolers in parallel. The multilayer thermal shield, similar to that used on the James Webb Space Telescope (JWST), is positioned between the telescope and Sun/Earth/Moon. The layers are reflective in the Sun-ward direction, but effectively black in the perpendicular direction, allowing radiative cooling to deep space. The sunshield layers are designed so any layer only sees its next neighboring shields or the 4 K baffle. In addition to the 4 K baffle, the cryocoolers provide multi-stage intermediate cooling for the structure and wires, spanning the 300 K to 4 K temperatures of the spacecraft to telescope/instruments. Notionally, these cooling intercepts are at ~70 K, ~20 K, and ~4 K. An innermost baffle at ~4-4.5 K blocks stray radiation from the sunshield from reaching the telescope and instruments. Multiple stages of cooling, resulting in an achievable heat load (200 mW) at

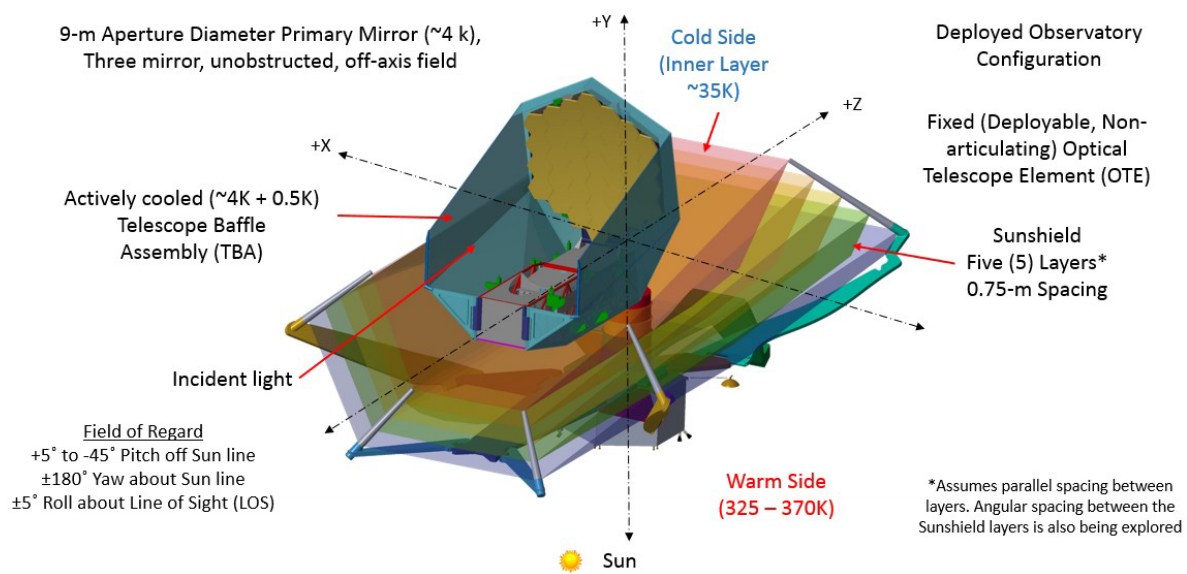

**Figure 3-1:** The OST Concept 1 observatory has a 9.1 m diameter off-axis telescope with five science intruments.

4 K, are provided by thermally-conductive materials (*i.e.,* aluminum) kept at low temperature so the heat is easily transported to the cryocoolers with minimal temperature gradient (<100 mK).

Contamination, especially from water, is curtailed by avoiding the use of composites on the cold side of the sun shields. Any off-gassing from the spacecraft and other warm portions of the observatory has no line of sight view to the cold optical and instruments, but will condense on the cold sunshield layers

Instruments are clustered near and upstream of the focal plane. The warm instrument electronics are located in the warm spacecraft instead of in a warmed box on the cold side of the sunshield. OST mitigates long harness lengths by using cold/low-dissipation amplifiers.

OST observatory (**Figure 3-1**) key elements are the 9-meter primary mirror, Instrument Accommodation Module (IAM) that houses five science instruments, cryocoolers, 4 K baffle, Deployable Tower Assembly (DTA), sunshield, and spacecraft bus.

The OST top-level observatory block diagram shows the major system elements (Instrument Accommodation Module (IAM), spacecraft bus), which are modular designs that improve serviceability and facilitate I&T.

The OST orbit is the thermally-advantageous Sun-Earth Lagrange Point 2 (L2). The observatory's Field of Regard (FoR) is +5˚ to -45˚ pitch off sun line, ±180˚ yaw about sun line, and ±5˚ roll about Line of Sight (LOS). OST's FoR is the same as JWST's and provides a consistent ~40% view of the sky. Any point on the sky is viewable at least twice per year for ~50 days.

**Mission Orbit Design**

OST Concept 1 is deployed at L2 in a combination Quasi-Halo (an imperfectly-periodic Halo) and Lissajous orbit. OST requires maintaining an orbit about L2 for five years, and carries sufficient consumables for ten. To achieve L2, the OST Launch Vehicle (LV) must provide a C3 of -0.55 to -0.75 km²/sec² to achieve the correct energy for the outbound trajectory.

The OST orbital design includes a requirement to keep the observatory out of shadows cast by the Earth and Moon, and an angular requirement was placed on the orbit to meet FoR requirements and limit sunshield size. To ensure stray light from the Moon does not impact the OST optics, the angle between the Earth to the L2 line and the Moon to OST vector must be no larger than 31°. This angle effectively forces OST's L2 orbit to fall within a L2-Earth-Vehicle (LEV) angle no larger than 16.7° off the Earth-to-L2 line. Additionally, the OST mission orbit at SEL2 must maintain an LEV angle of 4° or larger to ensure it does not pass through Earth's shadow.

### 3.1.1 First Decision: Architecture Trade Study

The team researched two architecture paths for a large mid- to far-infrared space observatory – a large single-aperture design and an interferometer – from which the Science and Technology Definition Team (STDT) intended to select a single architecture to pursue for the full study. The STDT chairs approached selection by requesting scientific proposals, enabling the team to evaluate the science community's greatest demand and the most intriguing science potential. **Table 1-1** shows all proposals used in the selection process and the top 14 proposals used to make the architecture decision.

**Section 3.1.1.1** is a report from the external facilitator, Dr. Thomas Greene (NASA Ames), who coordinated the decision process. This initial report includes draft parameters and performance characteristics of large single aperture and interferometer architectures. At this earliest stage of the concept study, the project's working name was Far-IR Surveyor (FIRS), adopted from the NASA Astrophysics Roadmap. To better reflect the selected science path, the study was later renamed Origins Space Telescope (OST) after an August 2016 meeting during which the team selected to focus on shorter wavelength ranges (~5 microns).

### 3.1.1.1 Far-IR Surveyor STDT Mission Concept Decision and Process

August 22, 2016

Dr. Thomas Greene, Study Facilitator

The Far-IR Surveyor science and technology definition team (FIRS STDT) met at NASA's GSFC on August 16 – 17 2016 and came to a consensus on the mission concept architecture and size. The group followed the process developed in May 2016 and described in **Section 3.1.1.2**. Briefly, the process consisted of four major components:

1. Establish performance requirements needed to achieve the mission's highest priority scientific objectives, separated into 'Musts' and 'Wants.'
2. Develop several possible mission concepts utilizing single filled aperture and interferometric architectures. Determine how well each concept performs in executing the planned science program.
3. Involve the entire STDT in a face-to-face evaluation of how well each mission concept meets the requirements derived from their selected science program. Identify potential risks in each viable concept.
4. Choose the best overall mission concept based on these considerations. The GSFC-based FIRS study office will feature this concept in a detailed study plan submitted to NASA in late August.

The study co-chairs (A. Cooray, M. Meixner) developed a process for identifying and ranking the science programs that would drive the mission requirements. They led the STDT in writing and ranking science cases (**Table 1-1**), and the 14 highest ranked cases were used to define the mission performance requirements. The STDT determined two baseline performance requirements were useful for discriminating between concept options when evaluating their ability to perform the selected science program: 1) Spatial resolution must be better than 1.24" at $\lambda$ = 50 $\mu$m wavelength, and, 2) The total set of proposed observations must be completed in under 5 – 10 years integration time (not accounting for observing efficiency).

T. Roellig led the Mission Concept Working Group (MCWG), which was tasked with determining the mission performance needed to complete the science program and developing viable mission concepts. A total of four mission concepts were developed: small and large versions of each architecture option. Option 1 was a 5-m filled aperture telescope, Option 2 was a 15-m filled aperture telescope, Option 3 was a 3.5-m dual aperture interferometer with up to a 20-m baseline, and Option 4 was a 6.5-m dual aperture interferometer with up to a 50-m baseline. **Figure 3-2** shows how the science proposals drive the telescope aperture size requirement. M. Bradford developed performance information (**Appendix 3**) for Options 1 and 2, while D. Leisawitz developed performance information (**Appendix 3**) for Options 3 and 4. During the August 16 – 17 STDT meeting, T. Roellig led the MCWG in determining the total integration time required for completing the selected science program ('Must' requirement #2) for each concept option (**Figure 3-2** and **Appendix 3**). M. Bradford and D. Leisawitz computed these quantities for each of the concept options.

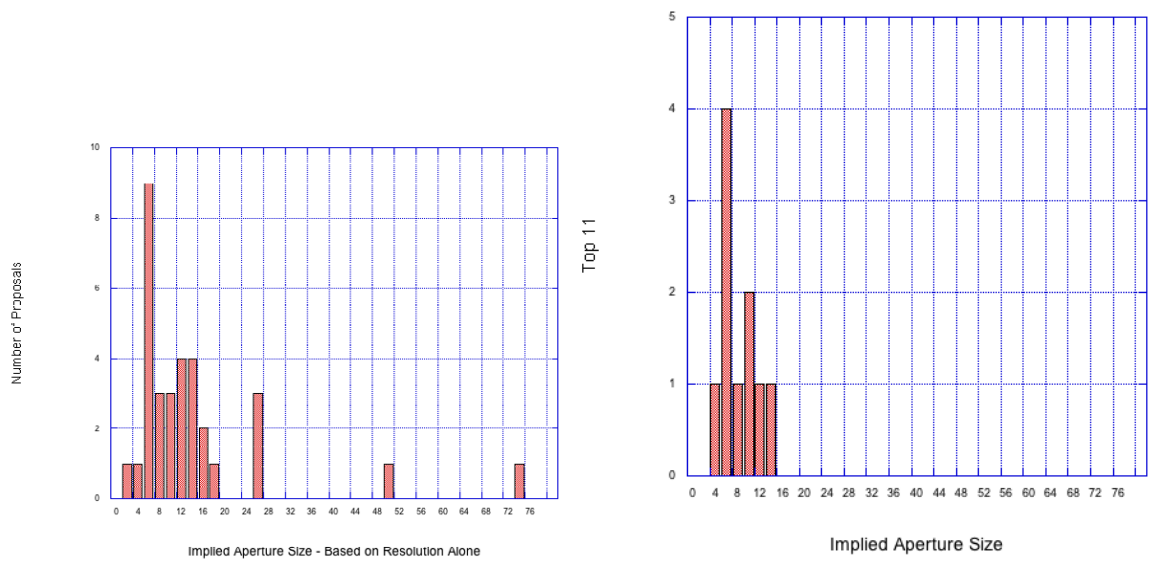

**Figure 3-2:** Summary graphs of angular resolution requirements for the collected science proposals, for A) all submitted proposals and the B) top ranked proposals. Analysis indicated that a single large aperture would service the majority of the proposed science cases.

At the STDT meeting on August 17, T. Greene led the STDT in evaluating the suitability of the different mission concepts in meeting the mission requirements. This included assessing absolute performance relative to the two 'Must' requirements and relative performance in additional desired areas ('Wants'). The STDT identified Option 2, the 15-m filled aperture, as the configuration with the best overall performance. The STDT also identified several risks (cryogenic testing, launch vehicle availability) associated with this aperture, and therefore judged that a filled aperture telescope with diameter between 10 and 15 meters was the best overall choice. The assessment of each concept in the 'Must,' 'Want,' and risk categories is shown **Figure 3-3**.

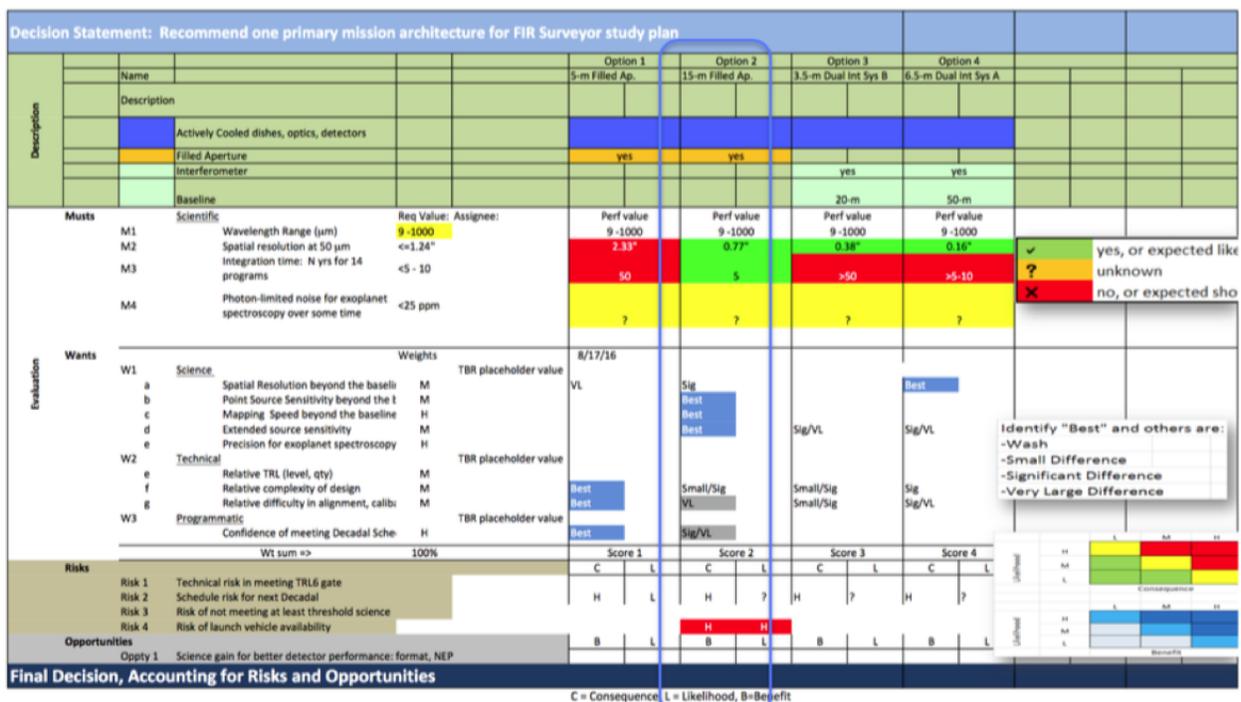

**Figure 3-3:** The assessment of each concept in the 'Must,' 'Want,' and risk categories.

### 3.1.1.2 Far-IR Surveyor Architecture Assessment Process, Guidelines, and Schedule

May 12, 2016
Prof. Asantha Cooray and Dr. Margaret Meixner, Community Chairs
Dr. Thomas Greene, Study Facilitator

This document describes the process for recommending a basic mission architecture concept for the FIR Surveyor study plan due August 26, 2016. The study schedule does not allow a full quantitative analysis of well-developed competing architectures, so this is a preliminary assessment. Further mission requirement and architecture development will likely require refining this assessment in FY 2017.

1. **Establish Science Requirements:** The Community Chairs will lead the FIRS STDT in establishing science objectives and the consequent science requirements. The FIRS STDT will concur with the process used and must agree that the requirements are sufficiently quantitative to be useful for defining the mission architecture concept. This may require establishing some performance requirements that are derived from science requirements.

2. **Program Requirements:** The Community Chairs and study scientist will work with the greater STDT in establishing and communicating overall mission readiness dates (e.g., TRL 6 by date 1 to allow launch by date 2) and other program requirements.

3. Architecture Definition: The Community Chairs will lead the STDT in developing, defining, and documenting (to some extent) basic concepts for one or more FIRS architectures likely to meet the science and performance requirements and need-by dates.

4. Weighting the Science Requirements: With assistance from the Community Chairs, the facilitator will lead the STDT in establishing a grade scale and assigning numeric weights to each of the science and performance requirements. The grading scale for each requirement will include the minimum performance level that any architecture must pass, as well as the value of exceeding that level. Each requirement will be weighted such that the weights sum to 100 for all requirements.

5. **Architecture Assessment:** The facilitator will lead the STDT in a quantitative evaluation of how well each architecture meets the requirements using the grading scale and weights established in Step 4. Ability to meet need-by dates and system risks will also be assessed. Assessment of risk probabilities and likelihoods and creation of resulting risk matrices may or may not be required. It is desirable for the STDT to reach consensus in the evaluation of each criterion. However, dissenting opinions during the evaluation process will be noted, and dissenters will be given the opportunity to provide information supporting their viewpoints.

   The process will be deemed successful when the STDT has reached consensus on a single executable FIRS mission architecture concept or else has significantly narrowed the range of concept options. Consensus is defined as every STDT member agreeing that the process was followed fairly. In this case, it is expected that all STDT members will support the group's decision even if some members do not like the outcome. If this is not possible, then dissents will be documented and included in STDT reports.

6. **Schedule:** The Community Chairs desire the initial FIRS concept assessment process, as defined above, to be completed by August 26, 2016 and included in the study plan due on that day. To achieve this, Steps 1 - 3 must be completed prior to the second STDT meeting in August. Steps 4 and 5 need to be completed no later than the end of that meeting. Steps 1 - 3 can be conducted in parallel to some extent, and Step 1 needs to be completed before starting Step 4. It would be useful to make some progress on Step 4 before the August meeting, perhaps by an initial canvassing of the STDT. The remainder of the assessment/selection process is to be determined depending on the outcome of the August meeting. The initial August assessment will likely be refined before mid-FY2017.

7. Other Personnel: An architecture-neutral person will record the activities of Steps 4 and 5 during the August meeting, including STDT votes and dissents.

### 3.1.2 Mission System Requirements and Traceability

All driving OST mission functional requirements trace directly to science and instrument measurement requirements. The Mission Traceability Matrix (**Appendix 3**) links the design to the science drivers.

The top-level mission requirements are:

- **Instrumentation** (observatory to accommodate five instruments):
    - Medium Resolution Survey Spectrometer (MRRS) – JPL
    - Hi Res (Far-IR) Spectrometer (HRS) – GSFC
    - Heterodyne Instrument (HI) – CNES
    - FIR Imager/Polarimeter (FIP) – GSFC
    - MID-IR Imager Spectrometer/Coronagraph (MISC) – JAXA
- **Instrument Wavelength Coverage:** 6 to 600 microns
- **Mission Duration:** 5 years nominal, consummables up to 10 years
- **Telescope Size:** 9.1-meter diameter (37 hexagonal segmented mirrors)
- **Telescope Type:** Three mirror anastigmat, unobstructed, off-axis field
- **Observatory Operating Temperature:** ~4 to 4.5 K
- **Observing Strategies:** Each instrument operates in succession
- **Data Download:** Observatory transmits science data to Science Operations Centers (MOC, SOC)
- Observatory Sericing and I&T: The instruments are designed in modules to enable servicing and to allow flexibility during I&T.
- Orbit: Sun-Earth $L_2$, selected for continuous observing (no eclipses) and thermal stability
- Redundancy: Fully redundant
- Launch Vehicle: Space Launch System (SLS) 8.4 m fairing
- Launch Date: September 1, 2035
- Optical Telescope Assembly (OTA) Temperature: 4 K, actively cooled using eight cryocoolers
- Instrument Redundancy: Instruments include appropriate internal redundancy, as appropriate
- Instrument Electronics: Located in the spacecraft bus (warm side), separated from the cold portion of the instruments located in the IAM behind the OTA on the cold side of the observatory
- Instrument Cooling: ~4 K for the instruments and an intermediate ~20 K zone for HEMTs
- ACS System Guidance: MISC instrument includes a guider
- Consumables: Sized for 10 years

### 3.1.3 System Implementation

**Sunshield**

OST sunshield layers are spaced further apart at the core (750 mm compared to 10 mm apart) than in the outermost layers. This additional spacing provides enhanced cooling of the critical core region. The initial OST design has heritage from JWST and also uses five sunshield layers. The OST team plans to conduct a study that will evaluate what impact eliminating one layer would have on the cryocoolers and overall heat load.

**4 K Cryocoolers**

The mission's 4 K requirement is met using multiple high-TRL cryocoolers. The planned cryocoolers provide cooling at ~70 K and ~20 K in addition to 4 K, allowing more efficient heat removal at higher temperatures.

**Thermal Budget for 4 K Region**

The initial thermal budget at 4 K is 100 mW for the telescope including dissipation, thermal radiation, and thermal conduction through harnesses and support structure, and 100 mW for the instruments, including dissipation and heat rejection from the sub-Kelvin coolers. The current thermal budget is provided in **Table 3-1**.

**Table 3-1:** OST Thermal Budget

| Component | Heat Dissipation: mW |
|---|---|
| Radiation | 42 |
| Harness | 15 |
| Structure | 30 |
| Instrument | 113 |
| Total | 200 |

## 4 K Baffle

The 4 K telescope and instruments are surrounded by a 4 K baffle assembly that eliminates stray radiation from the relative warm (~35 K average) inner sunshield layer. This baffle represents the boundary of the 4 K zone and comprises the primary telescope, IAM, and a simply-deployed flexible conductive membrane between the two, consisting of high purity aluminum foil in a sandwich of two aluminized Kapton layers. This type of material has been well characterized and has heritage from the JWST/MIRI shield.

### 3.1.4 OST Optical Performance Budget

An initial OST Optical Performance Budget (OPB), is shown in **Figure 3-4**. OST optical design input parameters are provided in **Table 3-2**.

The telescope portion of the OPB is shown in **Figure 3-4**. No addition reserves are held at the system level and the team has made initial reserve allocations for the telescope (21%), design (5%), thermal stability (1%), and line-of-sight (LOS)/jitter (73%).

**Table 3-2:** Optical Design Input Parameters

| Telescope Design | INPUT | Units | Comment |
|---|---|---|---|
| Aperture | 9.03 | meters | Circular primary mirror, unobstructed |
| Full FOV X | 25 | arcmin | |
| Full FOV Y | 15 | arcmin | |
| Wavelength (design) | 30 | microns | We consider this to be defining the WFE. Not the spectral coverage |
| Angular Resolution | 1.10 | arcsec | |
| Pointing Error Fraction | 0.10 | fraction of angular resolution | |
| Pointing Error Jitter | 109.64 | milli-arcsec | Total LOS accuracy |
| Area of Primary Mirror | 64.04 | meter2 | |
| Telescope F/# | 12.8 | | |
| Telescope Focal Length | 115.6 | meters | |
| Telescope Plate Scale | 1.784 | as/mm | |
| Telescope WFE | 2255 | nm rms at 0.8 Strehl | |
| Telescope WFE | 0.075 | waves | 0.05 |
| Telescope WFE | 13.3 | inverse fraction of a wave | 0.0655 |

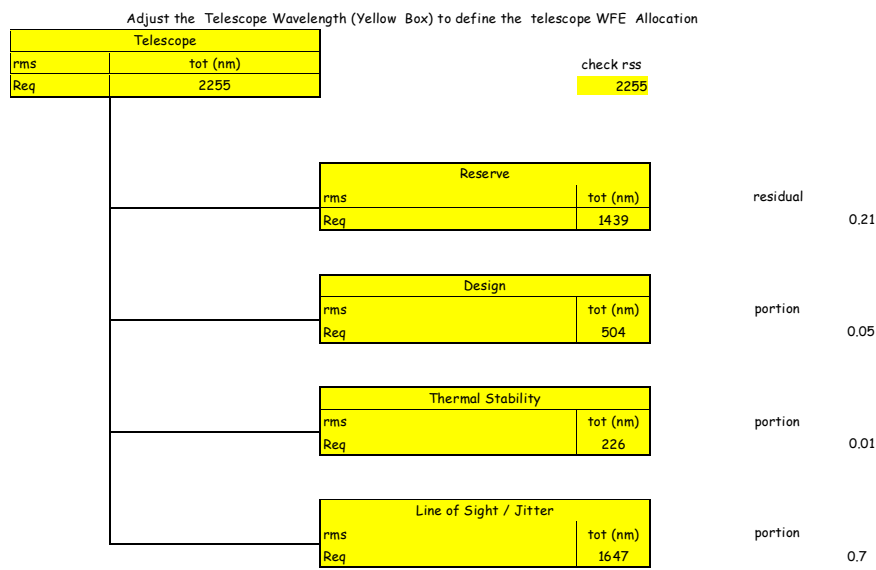

**Figure 3-4:** The OST Telescope Optical Performance Budget shows initial wavefront error.

OST Wave Front Error (WFE) allocations are generous compared to the state-of-the-art for this architecture (*i.e.,* JWST margins are reserve (45 nm), design (50 nm), thermal stability (50 nm), and LOS/jitter (7 mas)) [*Lightsey et al., Optical Requirements Allocation for the James Webb Space Telescope, Proc. SPIE Vol. 5487, 2004*]). These differences are consistent with the difference between JWST's diffraction limit of 2 microns versus OST's 30 microns. It allows the OST team to incorporate potential cost reduction, risk reduction, and resource reallocation to non-optical systems. There is also ample margin/reserve to reallocate within the OST breakouts for future design changes, if necessary.

The telescope WFE is 2,255 nm rms at 0.8 Strehl, which allows for three degrees-of-freedom (DoF) for each of the primary mirror segments. Based on this, OST does not require an additional mirror radius of curvature actuator or central actuator. Therefore, only three actuators – providing tip, tilt, and piston actuation – are required for each OST mirror segment, simplifying the design and reducing cost and risk.

The team made system-level allocations for all five instruments, and each was allocated reserve, design, thermal stability, and LOS/jitter values.

The initial OST total image motion results are shown in **Appendix 3** and assume a simple breakout. The MISC instrument also provides the guider for the observatory. The telescope dominates the LOS/jitter. The LOS/jitter is a calculation of all image motion terms, which is equivalent to tilt in the wavefront. The principal sources are of jitter are assumed to emanate from the six reaction wheels and/or control moment gyros (CMGs) and eight cryocoolers.

### 3.1.5 Mission Orbit Design

To achieve the science goals described in **Section 3-1**, OST is designed for a Sun-Earth Libration Point 2 (SEL2) orbit.

The Sun-Earth system has five Libration Points (**Figure 3-5**). These points rotate with the Earth as it orbits the Sun, and at these points, the gravitational forces and orbital motion of the spacecraft, Sun/Moon, and Earth interact to create a "stable" environment for a spacecraft.

However, the collinear Libration Points (L1, L2, L3) are not truly stable. These environments are very sensitive to energy, and without routine orbit maintenance, any perturbation will nudge the spacecraft out of its Libration Point Orbit. OST, which will make observations from $L_2$, must perform routine stationkeeping maneuvers to maintain its orbit.

A variety of orbit types can be implemented at the Libration Points. These orbits can vary from large amplitude Halos (similar to JWST), to smaller Quasi-Halos (similar to WFIRST), or smaller Lissajous orbits (similar to DSCOVR). **Figure 3-6** shows a comparison of JWST and DSCOVR in the Rotating Libration Point (RLP) coordinate system.

JWST's orbit is much larger compared to DSCOVR. Although the orbits look somewhat similar in **Figure 3-6's** A, D and B, E panels, the C and F panels show the true difference between a Halo and Lissajous orbit. JWST orbits in a perfectly periodic Halo, whereas DSCOVR's Lissajous grows and expands for several years, and then ultimately begins shrinking. While both Halo and Lissajous orbits have six-month periods, the larger

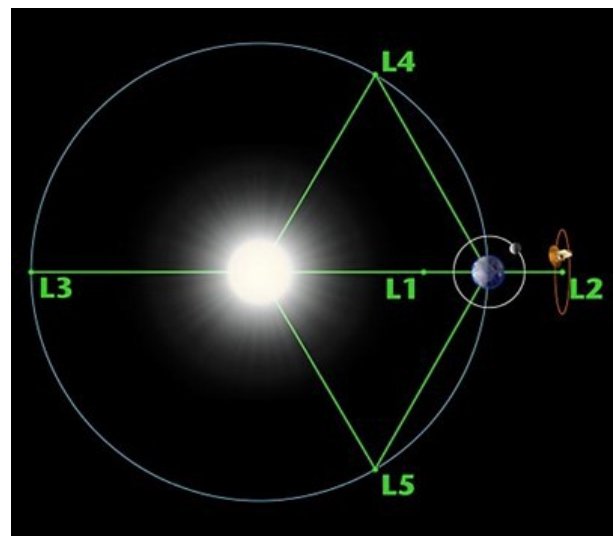

**Figure 3-5:** The five Sun-Earth Liberation Points provide stable locations from which to make observations.

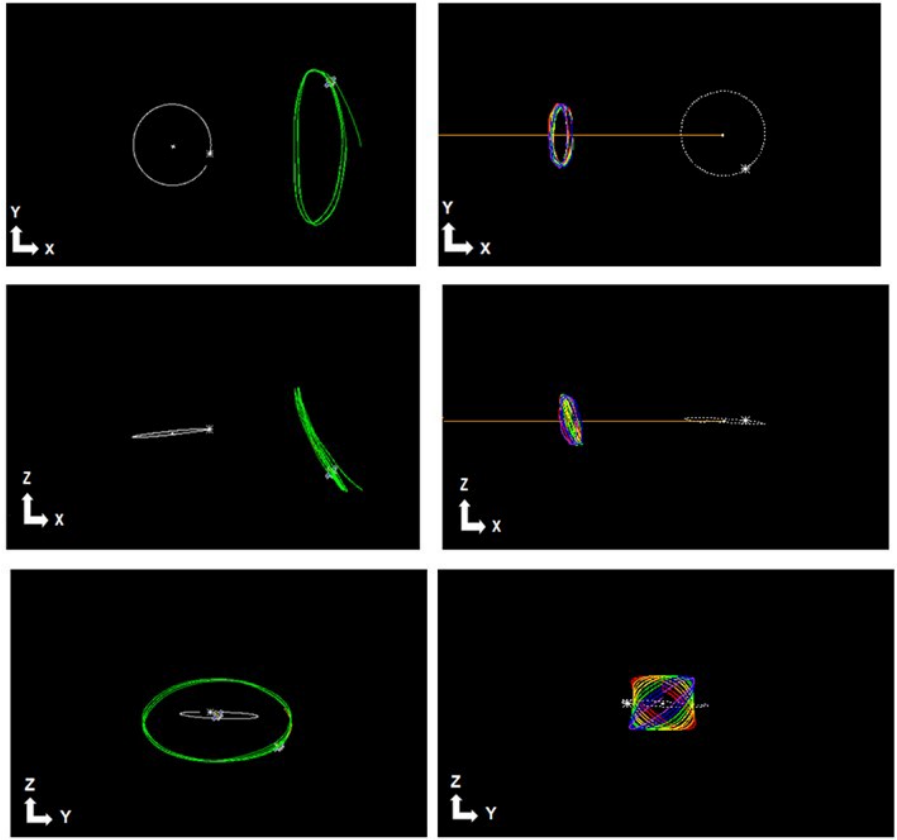

**Figure 3-6:** Comparison of JWST's (A, B, C) and DSCOVR's (D, E, F) orbits in the RLP frame.

Halo orbits about the same crossing points, whereas the Lissajous evolves around them and repeats its cycle. This difference is important because OST uses a combination orbit: a Quasi-Halo (a non-perfectly periodic Halo) and a Lissajous.

Current OST requirements are to maintain an orbit about SEL2 for five years (the observatory carries sufficient consumables for 10 years). To achieve the SEL2 orbit, the Launch Vehicle (*e.g.*, SLS) must be capable of providing a C3 of -0.55 to -0.75 km²/sec² to achieve the correct energy on the outbound trajectory. Also, while orbiting at SEL2, OST must avoid flying through Earth or Moon shadows. Finally, based on the current size and shape of the OST sunshield, an angular requirement was also placed on the orbit. And to ensure stray light from the Moon does not enter the OST optics, the angle between the Earth-to-SEL2 line and the Moon-to-OST vector must be no larger than 31°. This angle effectively forces OST's SEL2 orbit to fall within a SEL2-Earth- Vehicle (LEV) angle no larger than 16.7° off the Earth-to-SEL2 line. **Figure 3-7** shows these angular requirements. Additionally, to ensure it does not pass through Earth or Moon shadows, OST's orbit at SEL2 must maintain an LEV angle of 4° or greater.

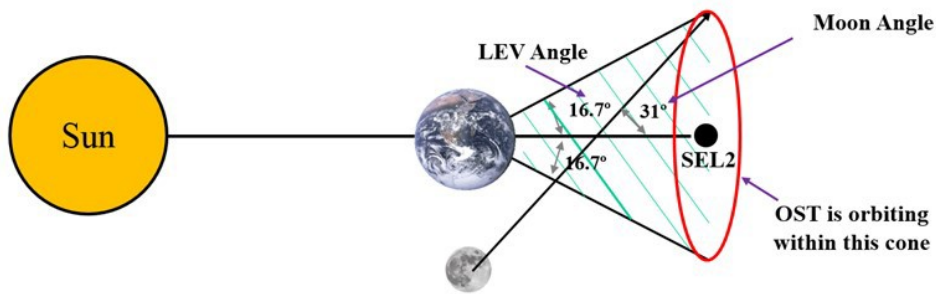

**Figure 3-7:** OST LEV mission orbit angle requirements include avoiding Sun and Moon shadowing.

From a Flight Dynamics perspective, it is much easier to define the OST mission orbit using LEV angular requirements versus size requirements. For example, JWST's orbit is confined to fit within a box that is no bigger than 832,000 km in the RLP Y direction and 500,000 km in the RLP Z direction. By constraining JWST to a box, launch window cases are ultimately thrown out if the mission orbit violates those sizes even by 1 km. By defining OST's orbit to fit within an LEV angle, it opens the launch window to more opportunities to achieve a desired mission orbit that meets science requirements.

**Figures 3-7** and **-8** show OST's orbit for 5 years. OST's orbital period is six months, resulting in two orbits per year. OST inserts into the SEL2 orbit at an LEV of ~4° and opens up to a maximum LEV angle of 17.62°. OST orbits counter-clockwise during its entire mission lifetime.

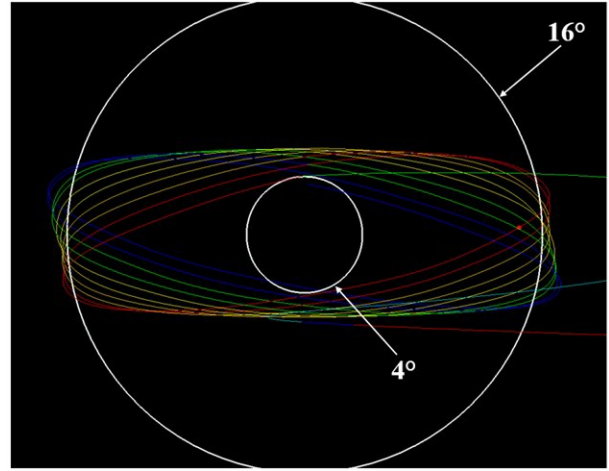

**Figure 3-8:** OST's SEL2 mission orbit for 5 years shows OST's first year is shown in red, second year is orange, third year is yellow, fourth year is green, and fifth year is blue.

The transfer trajectory from Earth to SEL2 and the mission orbit are shown in **Figure 3-9**.

For this orbit design, the STK scenario began with Second Engine Cut-Off 1 (SECO1) state of 01 September 2035 11:52:15.710 UTCG. This SECO1 state represents a typical state provided by the Launch Vehicle (assuming a launch from KSC) used in launch window analysis. SECO1 is shown in **Figure 3-10**.

OST then coasts in a Low Earth Orbit (LEO) of 185 km for ~20 minutes. The Transfer to Insertion Point (TIP) state then imparts a very large ΔV (currently 3.19 km/sec) to OST, putting it on the correct outbound trajectory toward SEL2. The TIP state is shown in **Figure 3-11**.

Because of OST's thruster configuration and sunshield, it faces the same consequence JWST does at launch. In most launches, the LV executes a nearly-perfect TIP burn and the spacecraft only needs to execute a very small (cm/s) Mid-Course Correction (MCC) maneuver 24-31 hours after launch. OST, similar to JWST, does not have thrusters pointed toward the Sun to ensure nothing impinges the sunshield (*i.e.,* no heat hits the sunshield, risking the instruments). This essentially means OST and JWST cannot brake. Therefore, the LV is forced to undershoot/underperform during TIP burn, and the LV imparts less ΔV to OST. Then, OST needs to perform a relatively large MCC (currently 30.42 m/s) at Launch + 24 hours to escape Earth's gravity onto the correct outbound trajectory toward SEL2.

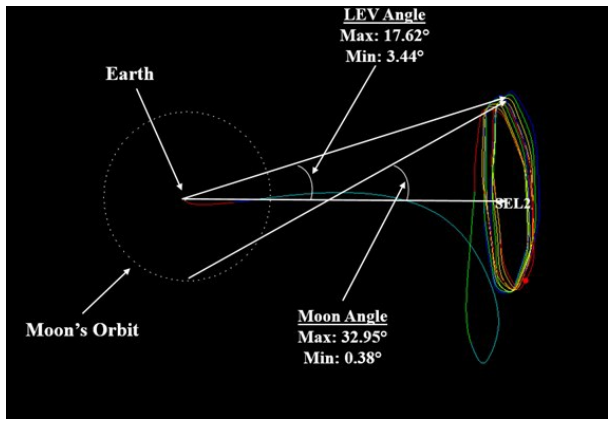

**Figure 3-9:** transfer trajectory and mission orbit for 5 years shown in the RLP XY frame with corresponding LEV and Moon angles shows OST's first year is shown in red, second year is orange, third year is yellow, fourth year is green, and fifth year is blue.

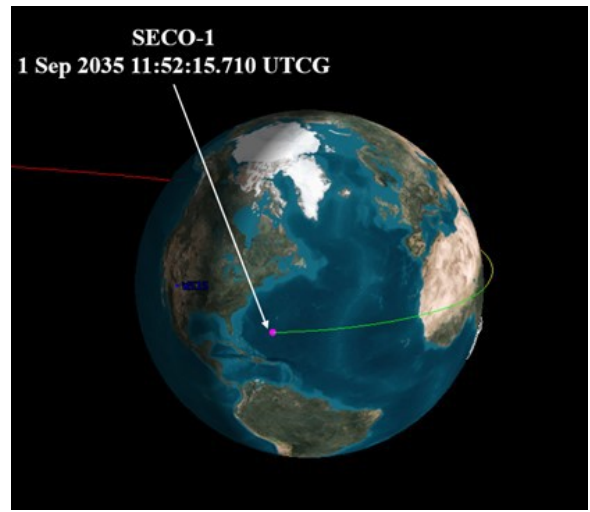

**Figure 3-10:** SECO1 for OST's Launch on September 1, 2025.

During its 113 day transfer to SEL2, OST performs smaller MCC2 (currently 10 m/s) and MCC3 (currently 5 m/s) to help minimize Libration Orbit Insertion (LOI) burn at $SEL_2$. Due to OST's LEV and Moon angle requirements, the resulting orbit is classified as a very small Quasi-Halo. As Libration Orbits shrink, from large Halos to smaller Quasi-Halos, the amount of energy (and therefore ΔV) to get into the orbit increases. For example, JWST is in a very large Halo orbit, which requires a LOI ΔV of 5 m/s. DSCOVR, which is in a very small 4° x 18° Lissajous, required a LOI ΔV of 160 m/s. The LOI ΔV required to attain a 16° Quasi-Halo with the current OST mission orbit and sunshield constraints is 100 m/s. OST's transfer to SEL2 and transfer maneuvers are shown in **Figure 3-12**.

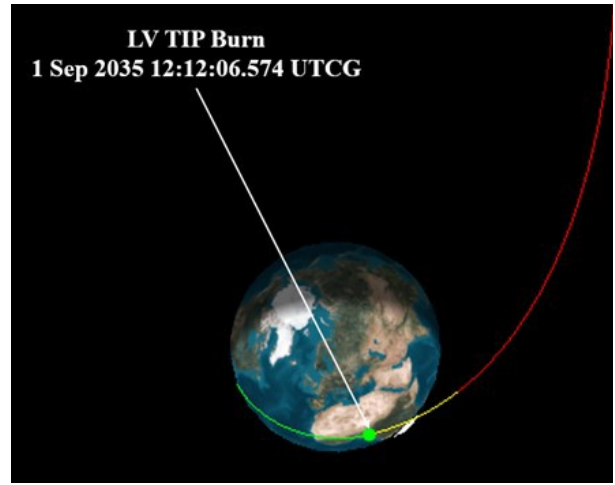

**Figure 3-11:** TIP state for OST's launch on September 1, 2035.

Once OST has completed its LOI maneuver, it begins routine stationkeeping maneuvers, which are required every 21 days to maintain the orbit. Total OST ΔV is provided in **Table 3-3**.

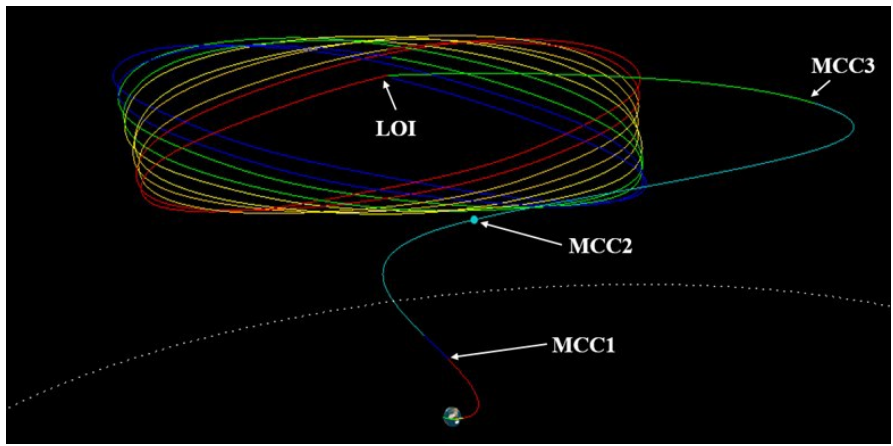

**Figure 3-12:** OST's 113-day transfer to SEL2 with corresponding MCCs and LOI burns show OST's first year is shown in red, second year is orange, third year is yellow, fourth year is green, and fifth year is blue.

**Table 3-3:** Total Mission ΔV

| Maneuver | ΔV (m/s) | Mission Time | Analysis |
|----------|----------|--------------|----------|
| MCC-1 | 30 | L +24h | Dependent on LV performance reliability |
| MCC-1a | 10 | MCC-1 + 5h | Clean up for MCC-1 efficiency |
| MCC-2 | 10 | L +20 days | Clean up for MCC1 and SRP effects after Sunshield deployment; Assist with LOI burn |
| MCC-3 | 5 | L +60 days | Clean up before LOI |
| LOI | 100 | L + 113 days | Insertion into 4x16° LEV Mission Orbit |
| Orbit Maintenance | 45 | Every ~3 weeks after LOI | ΔV for 10 years ~ 3.5 m/s for stationkeeping and 1 m/s for Momentum Unloading per year |
| Total | 200 | | Very conservative, needs to be re-evaluated with full thruster models (pointing and efficiency) in simulation |

## 3.2 Flight System

### 3.2.1 Instrument Accommodation Module

The Instrument Accommodation Module (IAM) (**Figure 3-13**) is the principal OST structure. It contains all five OST instruments, and supports the primary mirror. The IAM assembly is connected to the spacecraft bus via a Deployable Tower Assembly (DTA). **Figure 3-14** shows the IAM in the stowed position. **Figure 3-15** shows the IAM with the instruments.

The IAM supports the science instruments and provides cooling to 4 K. The IAM, in concert with OTE components (*e.g.*, the baffle system and cryo-cooler system located in the bus), must be thermally controlled to provide the cooled environment necessary for instrument and systems operation (**Section X.x**). Careful instrument placement within the IAM allows all five instruments to use the focal surface signal (**Section X.x**) and optimally balances thermal and mechanical factors.

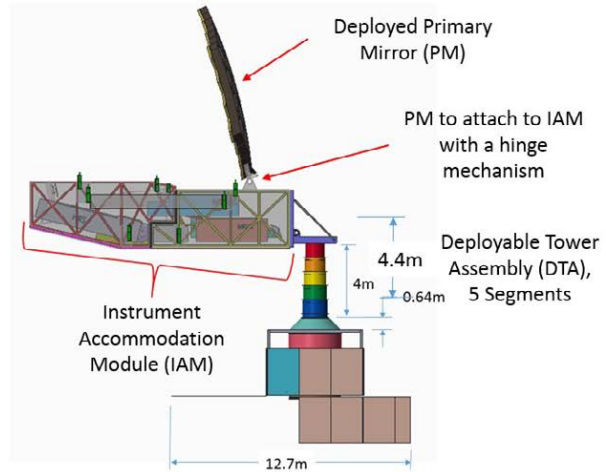

**Figure 3-13:** The IAM is OST'S main structure. It supports the primary mirror, telescope, and houses all five instruments. Sunshade and baffle assemblies are removed for visibility.

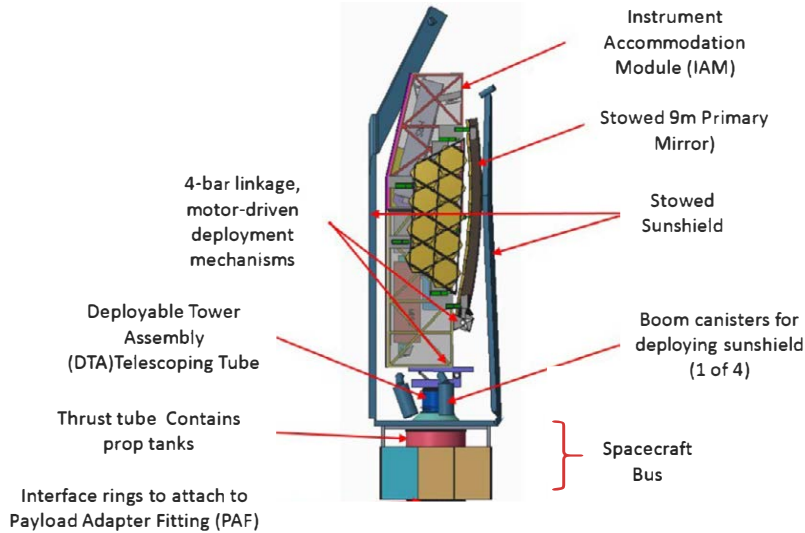

**Figure 3-14:** (Left) The IAM in the stowed configuration can be accommodated in SLS Block 2 (8.4m fariring). The sunshade and baffle assemblies are removed for visibility.

**Figure 3-15:** (Below) The IAM houses all five OST instruments, providing a stable environment and observation platform. The sunshade and baffle assemblies are removed for visibility.

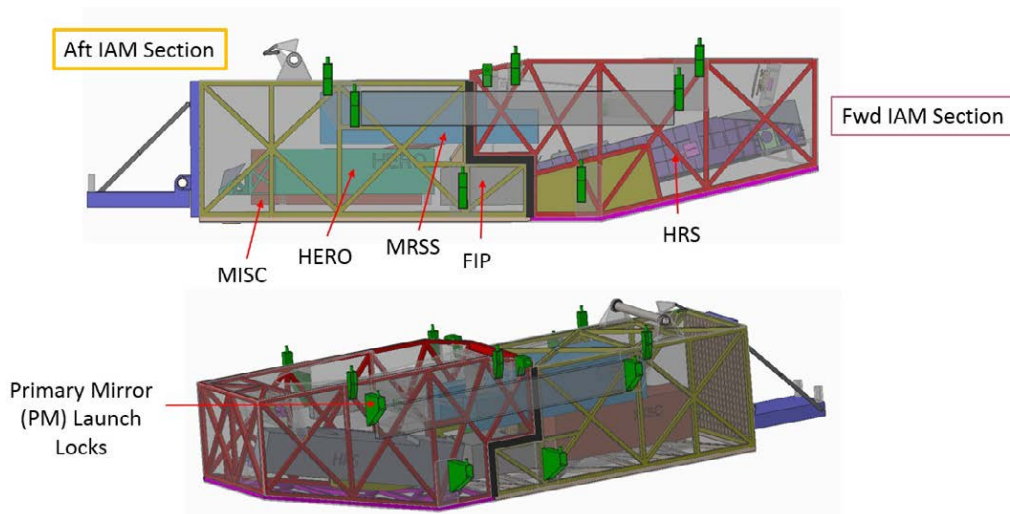

### 3.2.1.1 IAM Requirements

IAM top-level requirements are listed in **Table 3-4**.

### 3.2.1.2 IAM Mechanical System Design

The OST observatory main structural load-path elements consist of the spacecraft bus, Deployable Tower Assembly (DTA), IAM, and IAM rotation mechanism (**Figure 3-16**). The complete OST observatory, including all elements and instruments, is 34,000 kg and has a center-of-gravity (CG) 8 m above the separation system interface to the Space Launch System (SLS).

OST launch loads have a direct path from the IAM to the SLS Payload Adapter Fitting (PAF). This path goes through the IAM rotation mechanism, DTA, small interface cone, the bus's thrust tube, and Spacecraft Separation System (SSS).

The 14-m long IAM has two sections; both are aluminum truss frames covered with aluminum honeycomb closeout panels. The forward IAM section holds the secondary, tertiary, and field steering mirrors and allows the optics to be integrated, aligned, and qualified in the forward section before being integrated with the aft IAM section that houses the five instruments. The five instruments are secured to the IAM using flexures. The IAM also contains light baffles, cryocooler lines for each instrument, and a small door that allows the HERO instrument Local Oscillator light beam to reach HERO from the bus. The IAM interfaces to the DTA via a four-bar linkage mechanism, which deploys the IAM 90 degrees off the DTA once in zero-gravity.

### 3.2.1.3 IAM Deployment

The IAM is deployed once the DTA is fully deployed. A four bar linkage system, powered by a motor/actuator system, deploys the IAM ~90 degrees from its stowed to its deployed configuration.

**Table 3-4:** IAM Top-Level Requirements

| IAM Requirement |
| --- |
| Sahll support mechanically and thermally, and enclose the OST instruments (HERO, MRSS, MISC, FIP, HRS). |
| Shall support Mechanically and thermally, and enclose the OST OTE. |
| Shall provide attachment for the OST baffle assembly. |
| Shall control IAM internal temperature to 4 K (+/-0.5 K). |
| Shall attach to the DTA, which attaches to the spacecraft bus. |
| No internal structure shall interfere with optical ray traces for instrument pickoffs or OTE. |
| Shall support appropriate harnessing for each instrument or OTE component, as required. |
| Shall allow laser optical ports access for the HERO instrument. |
| Shall support instrument cooling to 4 K as performed by the observatory. |

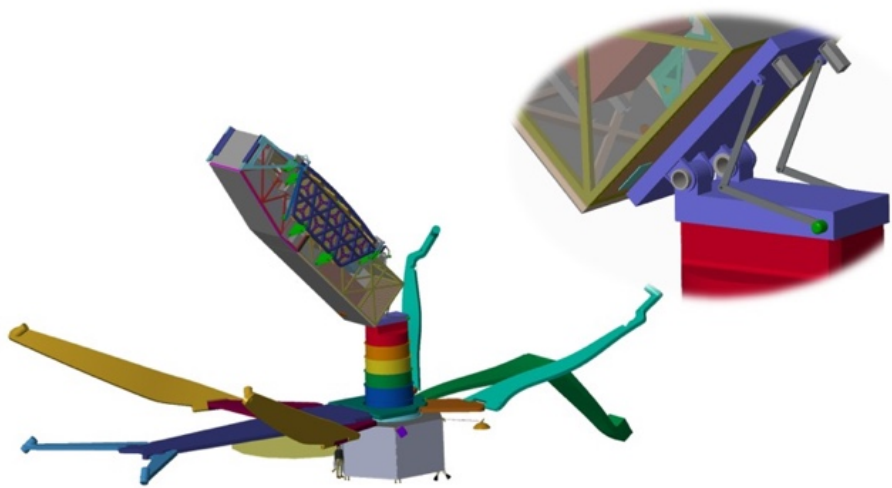

**Figure 3-16:** The IAM continuing to move into the fully deployed position. All motion is slow due to spacecraft control considerations.

### 3.2.2 Spacecraft Bus

The OST spacecraft bus consists of the Avionics, Electrical Power System with a solar array, Propulsion, Thermal, Mechanical Systems including mechanisms, Flight Software, Attitude Control System (ACS), and the Optical and RF Communications Systems. Additionally included in the spacecraft bus are the five instruments' warm electronics, the HERO instrument's Local Oscillator, and all eight cryocoolers. The spacecraft bus interfaces to the DTA. The spacecraft bus, in particular, the warm instrument electronics bay, could accommodate servicing and the propulsion tanks could be refueled in space.

OST spacecraft bus meets all mission level requirements, has robust contingencies and margin, and can be built with existing technology. **Figure 3-17** and **Figure 3-18** show the OST observatory in launch configuration and the spacecraft bus.

The spacecraft bus top-level requirements are to:

- House and provide all necessary spacecraft support subsystems and functions, including Avionics, Electrical Power System, Propulsion, Thermal, Mechanical Systems and Mechanisms, Flight Software, Attitude Control System, Optical, and RF Communications Systems
- House all warm instrument electronics
- House all eight cryocoolers
- Provide a solar array and power system for observatory power
- Provide a thermal radiator system for thermal control
- Permit mission servicing (warm electronics instruments) and propulsion tank refueling

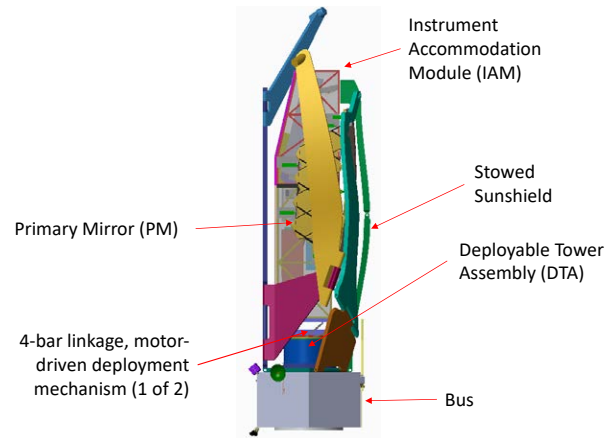

**Figure 3-17:** OST Observatory's launch configuration shows the stowed IAM, sunshield, and OTA.

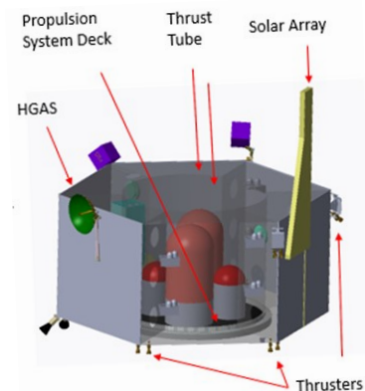

**Figure 3-18:** OST's spacecraft bus hexagonal frame is constructed of aluminum. The bus houses S/C subsystems, warm instrument electronicsm and cryocoolers. The outer panels of the bus are thermal radiators.

### 3.2.2.1 Mechanical Systems

The OST spacecraft bus materials consist of aluminum and fiber composite. The spacecraft 3.5-m diameter thrust tube is made of M55J fiber composite. The frame is made out of aluminum, while the internal panels have aluminum face sheets with aluminum honeycomb core. The six closeout panels are thermal radiators made of aluminum face sheets with aluminum honeycomb core with embedded heat pipes. The top and bottom panels are made of $T_{300}$ fiber composite, while the propellant deck is made of T300 fiber composite face sheets with aluminum honeycomb core.

The spacecraft bus, running at an average temperature of 40° C, carries all spacecraft subsystems, eight cryocoolers, warm instrument electronics, and the HERO instrument's Local Oscillator (light beam generator) (**Figure 3-19**). A 46-m² Orbital/ATK UltraFlex solar array uses a single-axis gimbal. The S-band High Gain Antenna (HGA) and two optical antennas utilize two-axis gimbals.

The deployment tower assembly (DTA) is structurally sized to accommodate the launch mass and CG offset of the IAM/PM system above it. The DTA is secured during launch using eight EBAD pyro-technically actuated launch locks, which use 2.54 cm diameter steel bolts. These launch locks were designed and tested under an Orion Service Module Fairing contract. Each of the five sunshield layers is attached to a DTA canister, so when the DTA deploys, the sunshield is also deployed.

### SLS Fairing, Solar Array, DTA, and Sunshield Deployments

A timeline of all significant deployments after fairing and launch separation is provided in **Table 3-5**. **Figure 3-20** shows the stowed OST telescope in the SLS fairing.

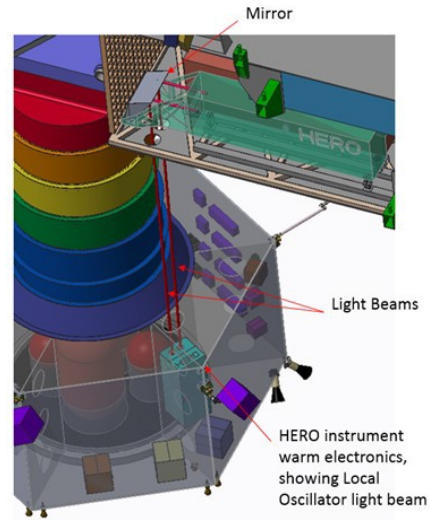

**Figure 3-19:** HERO Local Oscillator light beams run from the spacecraft bus (near the warm electronics) to HERO located in the IAM at 4 K.

**Table 3-5:** OST Observatory Deployment Timeline

| Time after Launch | Event |
| --- | --- |
| 30 minute | Solar Array Deploy |
| hours | Propulsion Checkout |
| 12 hours | Mid Course Correction |
| 1 day | HGA deploy |
| Two weeks | Spacecraft Subsystem checkout |
| Day 3 through day 20 | Telescope and sun shield deployment<br>- Deploy forward and aft sun shade Support structures<br>- Deploy DTA and sun shade & the 4 lower booms, so sunshield basic shape is taken. IAM is vertical.<br>- Deploy the IAM 90°<br>- Release baffle attachments to IAM<br>- Release mirror wing launch locks<br>- Rotate mirror 75°<br>- Deploy wings and baffle<br>- Deploy sun shade 8 upper booms for 5 layer separation - completes sunshade deployment. |
| 21 | Deploy momentum trim tab |
| 21 | Final ACS Calibration |
| 21-35 | Initial Instrument checkout (warm) |
| 30 | LOI maneuver |
| 30 | Cryocoolers on |
| 37 | Cooldown complete |
| 37-50 | Telescope alignment |
| 40-60 | Instrument checkout (cold) and calibration |
| 60-90 | Science Commissioning |
| 90+ | Normal Operations |

### SLS 8.4-m Fairing Deployment Launch Vehicle separation

The 8.4-m SLS fairing is a clamshell deployment system that exposes the observatory prior to separation from the launch vehicle. Thereafter, the observatory is separated from the launch vehicle before observatory deployments begin.

**Figure 3-21** shows the OST configuration on the spacecraft bus with the 8.4-m SLS fairing deployed and in the stowed state. The figure also shows the stowed instrument accommodation module (IAM), stowed DTA, and spacecraft bus. The sunshield has a forward and aft skeletal deployment structure that, once deployed, is the frame for the multi-layer sunshield material. Once this skeletal structure is deployed, the sunshield layers unfurl and separate.

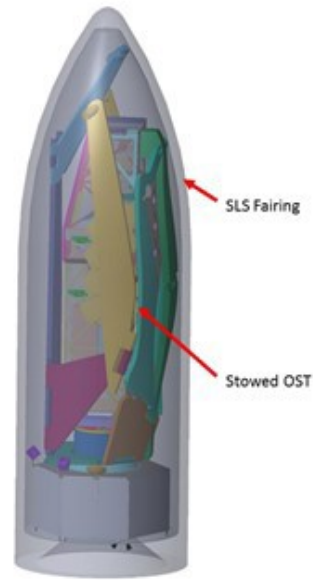

**Figure 3-20:** The OST observatory is well-accommodated in the SLS 8.4-m fairing.

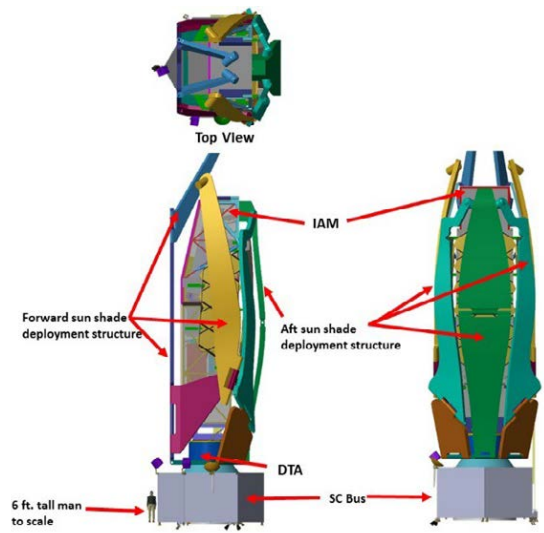

**Figure 3-21:** Top and side views of the observatory with the 8.4-m SLS fairing deployed.

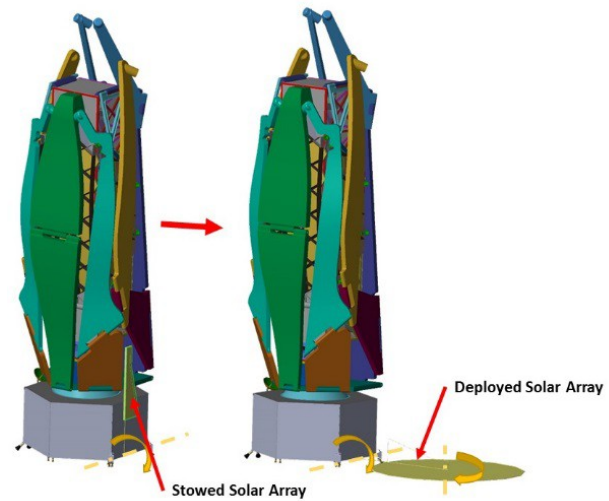

**Figure 3-22:** Solar Array System Deployment.

### Solar Array Deployment

**Figure 3-22** shows the deployment of the solar array system. The solar array is attached to the spacecraft bus side panel via launch locks and a spring loaded hinge boom. It is deployed ~90° when the solar array launch lock releases upon command. Thereafter, the "accordion fan" shape solar array substrate is subsequently released and deploys in a circular pattern.

### High Gain Antenna Deployment

**Figure 3-23** shows HGA deployment. The HGA is attached to the spacecraft bus side panel via launch locks and a spring-loaded hinge boom assembly. The antenna and boom assembly launch locks are released on command to deploy the HGA. **Figure 2-34** shows the OST observatory prior to sunshield deployment.

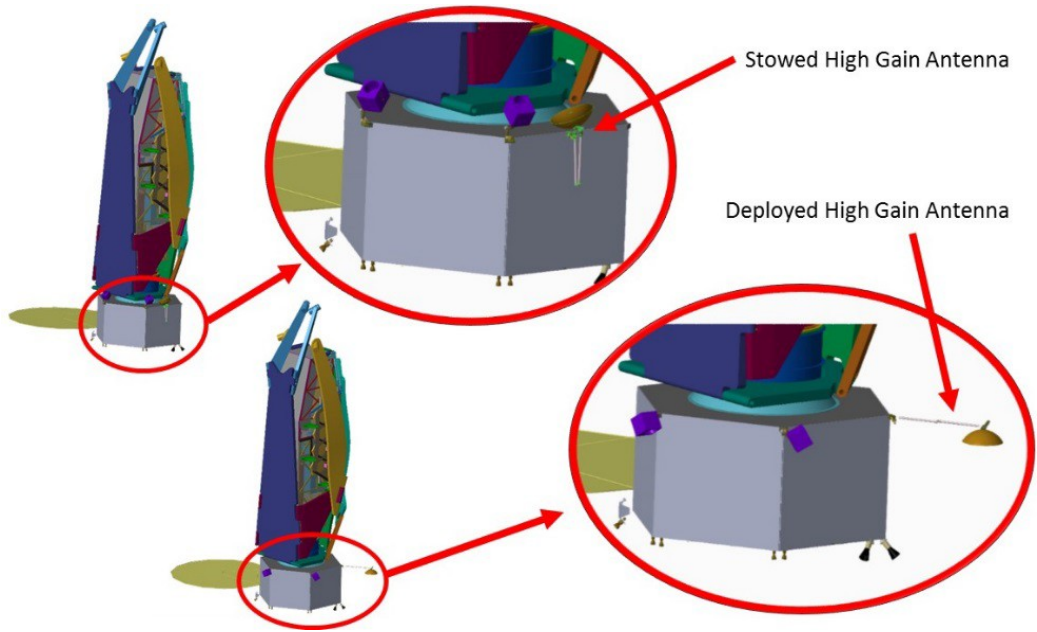

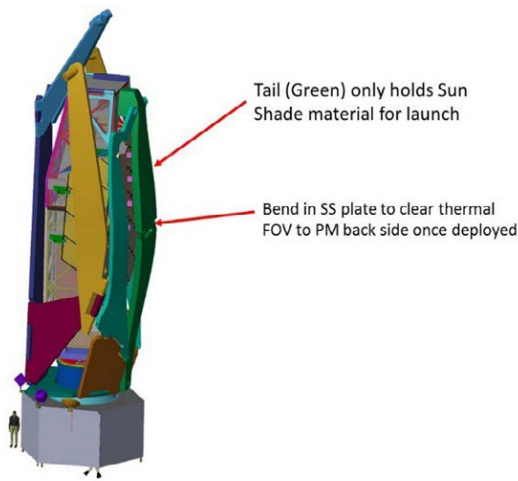

**Figure 3-23:** (Top) OST HGA shows prior and post deployment.

**Figure 3-24:** (Left) Observatory prior to sunshield deployment.

   **Figures 3-25 to -32** illustrate the sunshield forward and aft structural elements, that once deployed, hold the sunshield membrane in its deployed state on-orbit. **Figure 3-25** shows several views of the sunshield fully stowed with the HGA and solar arrays deployed. **Figure 3-26** shows the forward and aft structural arms beginning to swing out, with the "tail" swinging down. **Figure 3-27** shows the forward and aft arms swinging down. **Figure 3-28** shows the side arms swinging away and all arms continuing downward into the deploy state. **Figure 3-29** shows the fore arms down and side arms out. **Figures 3-30** and **-31** show all arms fully deployed, along with the deployed aft "tail." **Figure 3-32** shows the pivot line axes needed for the sunshield structural sections to deploy. These axes are driven by spring/hinge/motor systems to allow positional control of the various sun shade structure parts. The sunshade is not depicted for clarity.

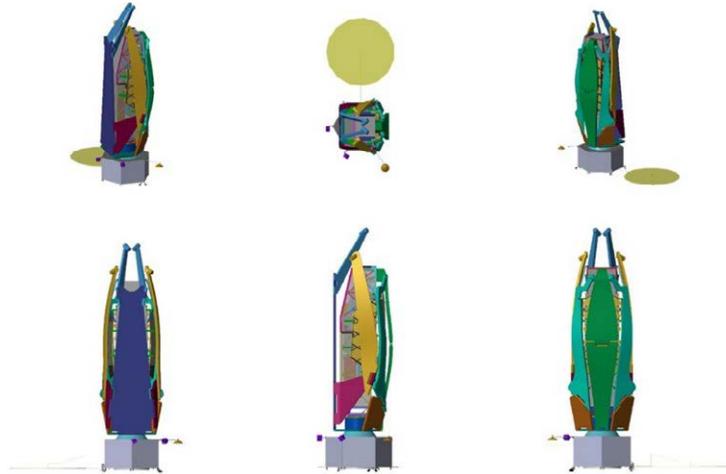

**Figure 3-25:** The fully stowed sun shield with the HGA and solar arrays deployed.

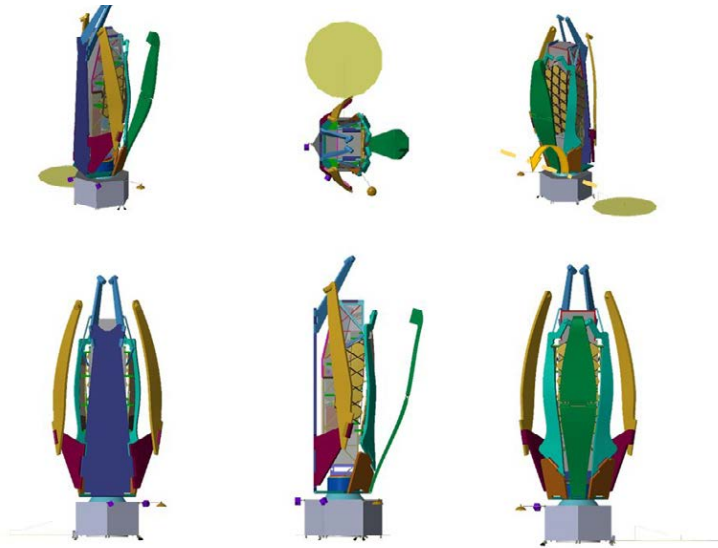

**Figure 3-26:** The fore and aft structural arms swing out and tail swings down.

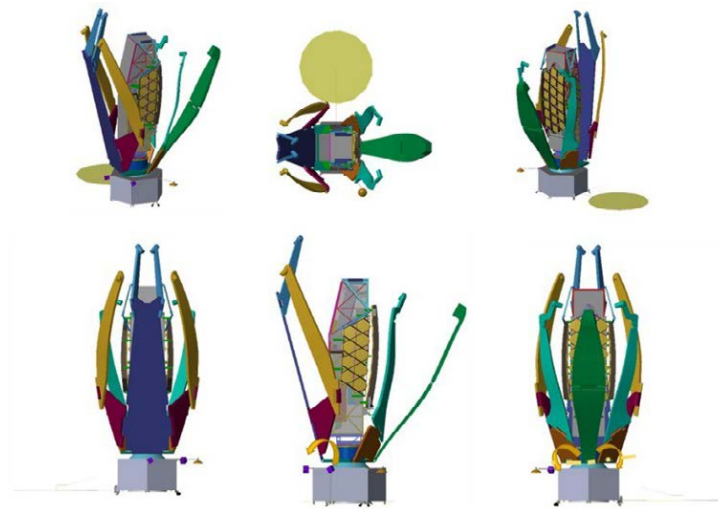

**Figure 3-27:** The front and back structural arms swing down.

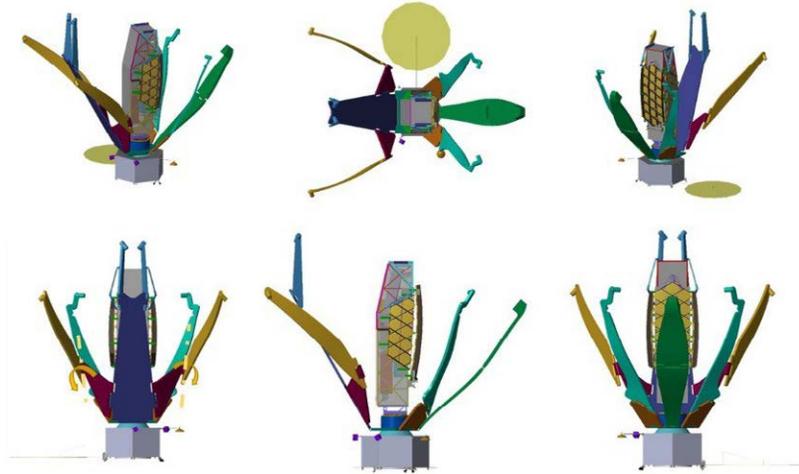

**Figure 3-28:** Side arms swing away and all arms continue downward.

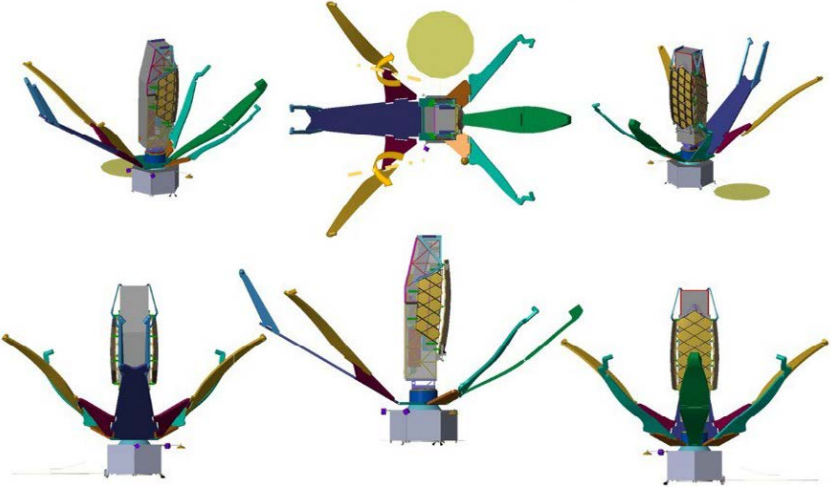

**Figure 3-29:** Front arm down and side arms out.

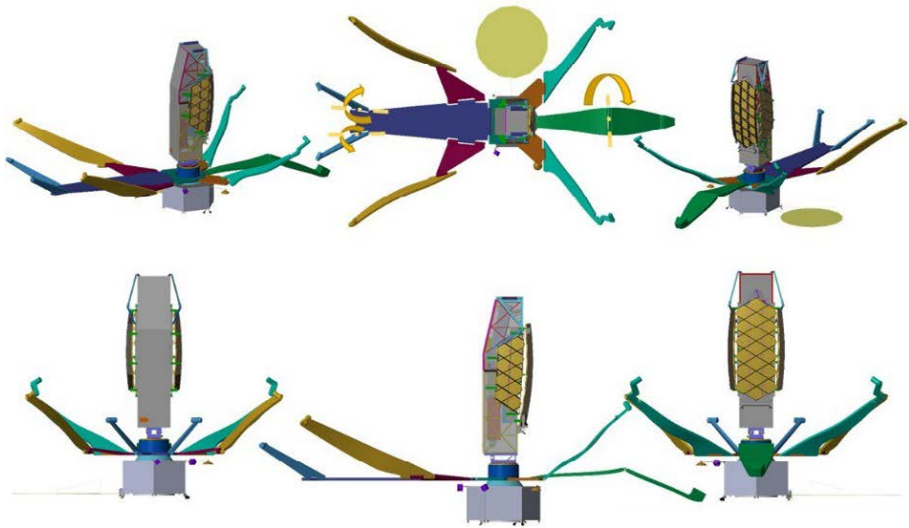

**Figure 3-30:** Front arms deploy and the aft green tail descends.

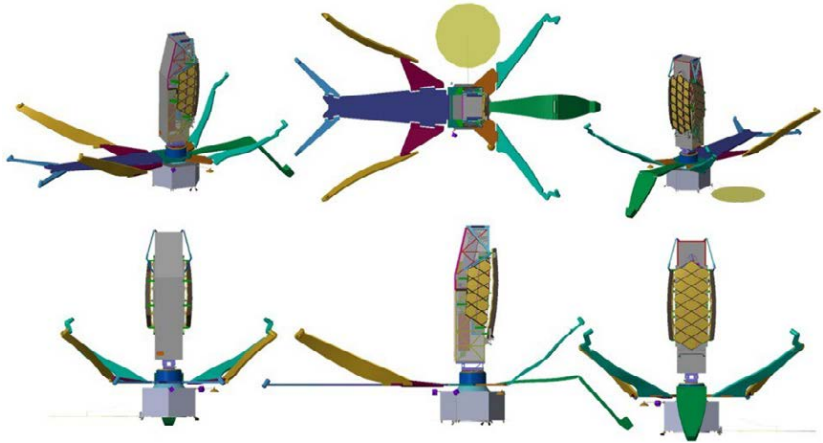

**Figure 3-31:** Arms fully deployed.

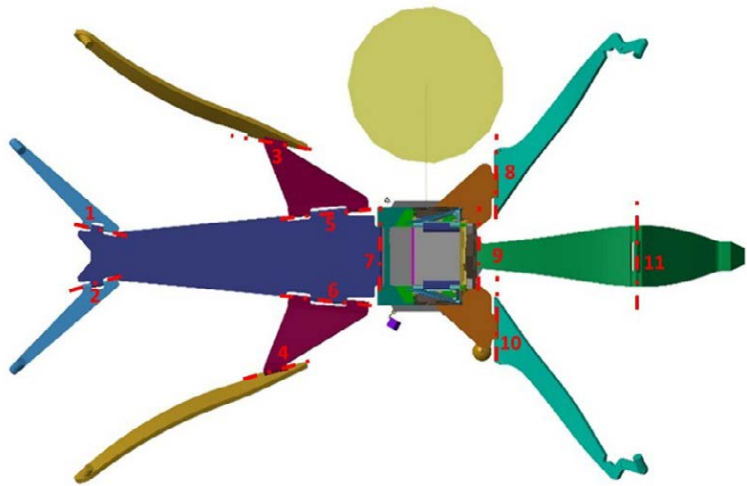

**Figure 3-32:** OST has eleven pivot axes that are spring, hinge, and actuator deployed.

## DTA Deployment

The DTA separates the spacecraft bus, which is relatively warm (~325 to 370 K), from the cold telescope region held at cryogenic temperatures. The DTA is a series of nested structural section tubular cylinders, ~2.5-meters in diameter at the base, that deploy linearly via an internal linear mechanism drive system. As each tubular section deploys and bottoms out at the adjacent tube, it interlocks with the adjacent tube to provide the deployed DTA with greater stiffness and rigidity. The full deployed length of the DTA is ~4.2 meters. **Figure 3-33** illustrates the DTA deployment sequence.

The sunshade multi-layer material becomes visible as the sunshade arms deploy. **Figures 3-34** through **-43** show the full sequence of deploying the five-layer sun shield that thermally-insulates the OST observatory from the sun.

In **Figures 3-34** and **3-35** the shield layer five begins to unfurl out of the carrier arms and is pulled toward each boom extension location.

**Figures 3-35** through **-38** show the sun shade's fifth layer of material continues to tension and pull toward the extension boom locations, which will ultimately separate the sunshade material layers.

Once the layer five shield is close to the layer separation booms, the six booms of the tensioning system extend in a linear deployment, allowing the sunshield material to deploy and separate to its final configuration (**Figure 3-39**).

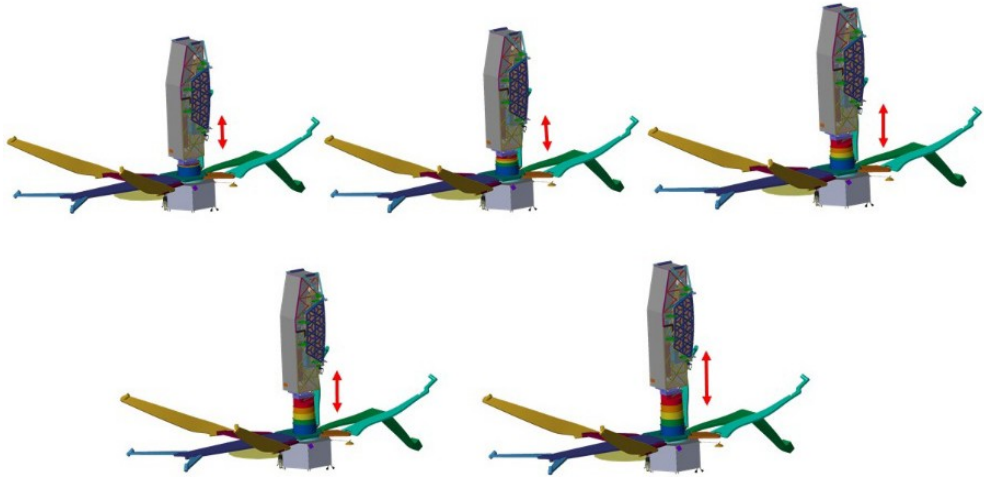

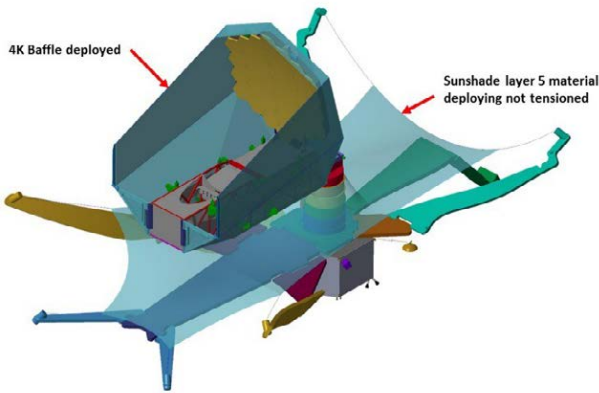

**Figure 3-33:** (top) DTA deployment occurs via an internal linear mechanism. The IAM, primary mirror, and baffle assembly deployments are shown in Section 2.1.

**Figure 3-34:** (left) Sunshield layer five begins to deploy.

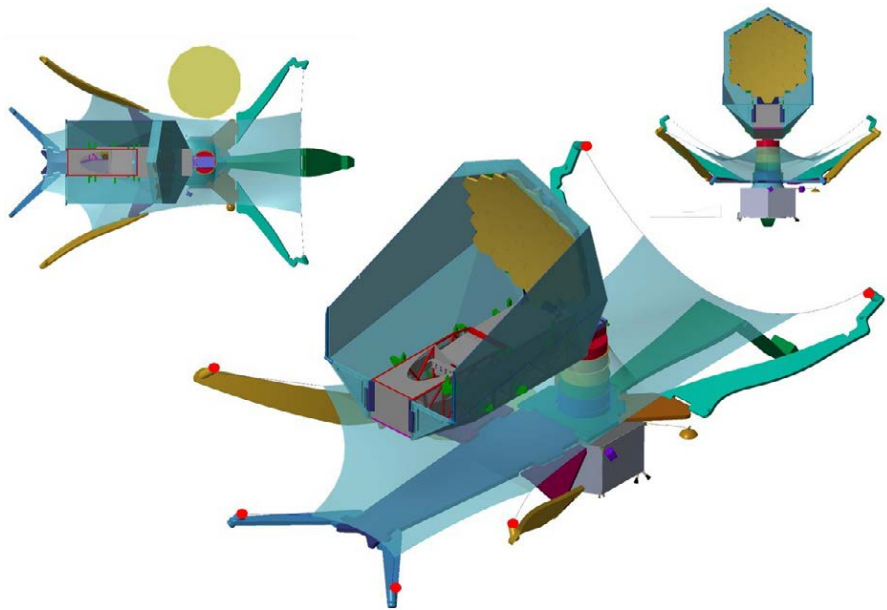

**Figure 3-35:** Sunshield layer five, the closest layer to the arms, unfurls.

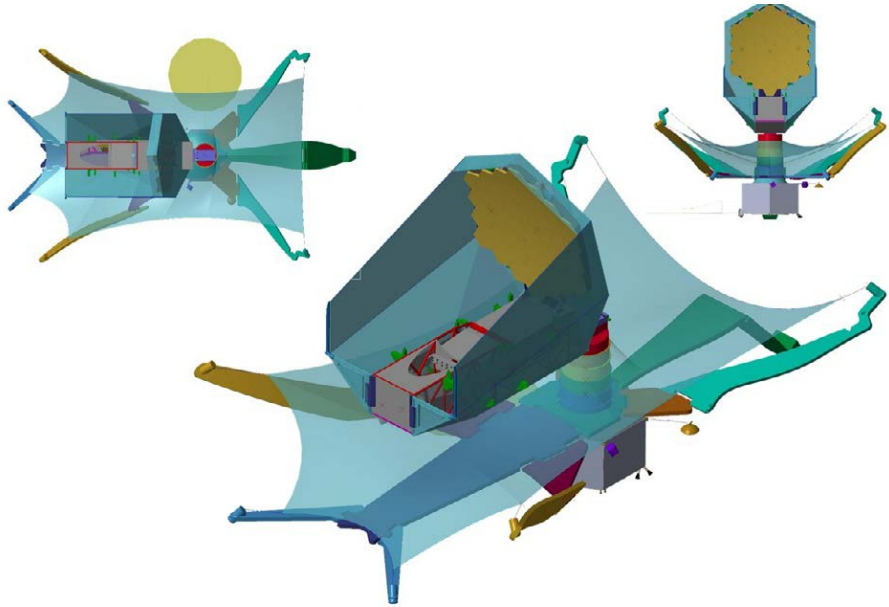

**Figure 3-36:** Sunshield layer five continues to tension and pull toward the extension

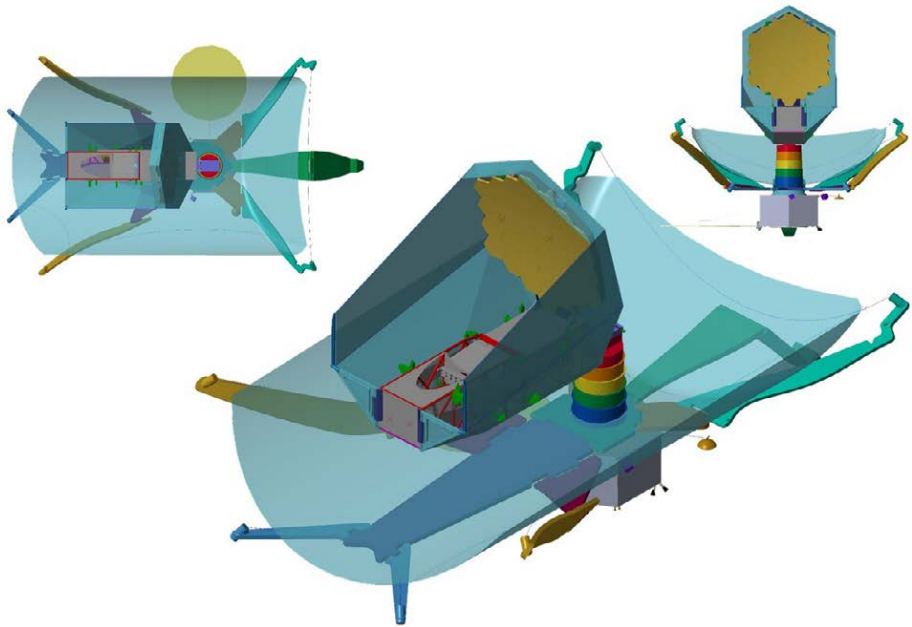

**Figure 3-37:** Sunshield deployment continues with layer five unfurling toward full deployment

   **Figure 3-40** shows layer one, the next to deploy and the layer closest to the primary mirror and the 4 K baffle, being tensioned to its final configuration. All other layers begin similar tensioning steps to reach their final configurations, as shown in **Figures 3-41** through **-43**.
   **Figure 3-44** shows the final deployed and tensioned configuration of the OST sunshield system.

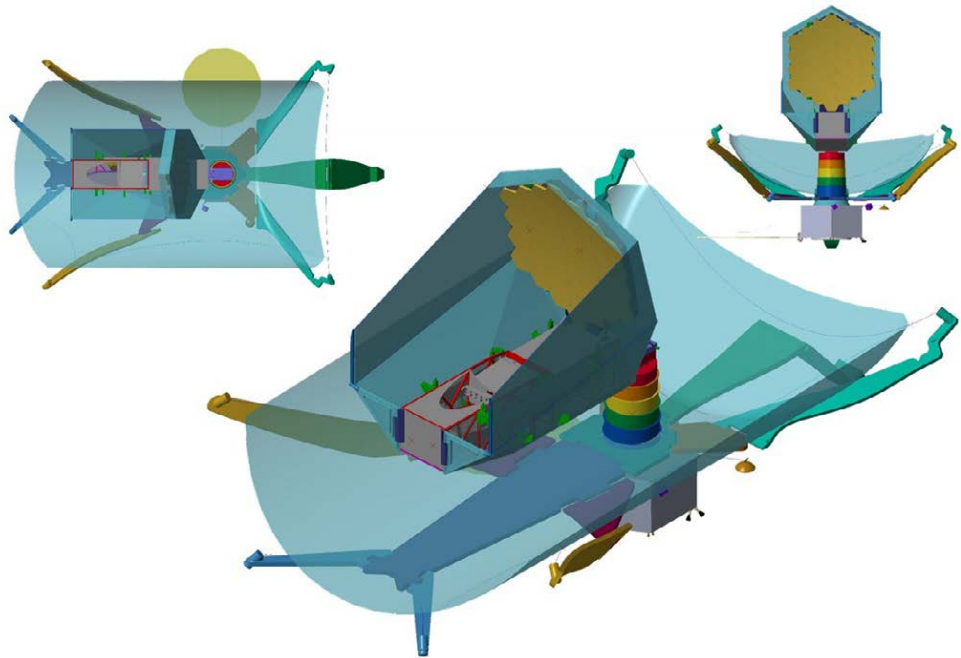

**Figure 3-38:** Sun shield layer five continues deployment, with tension pulling it into position

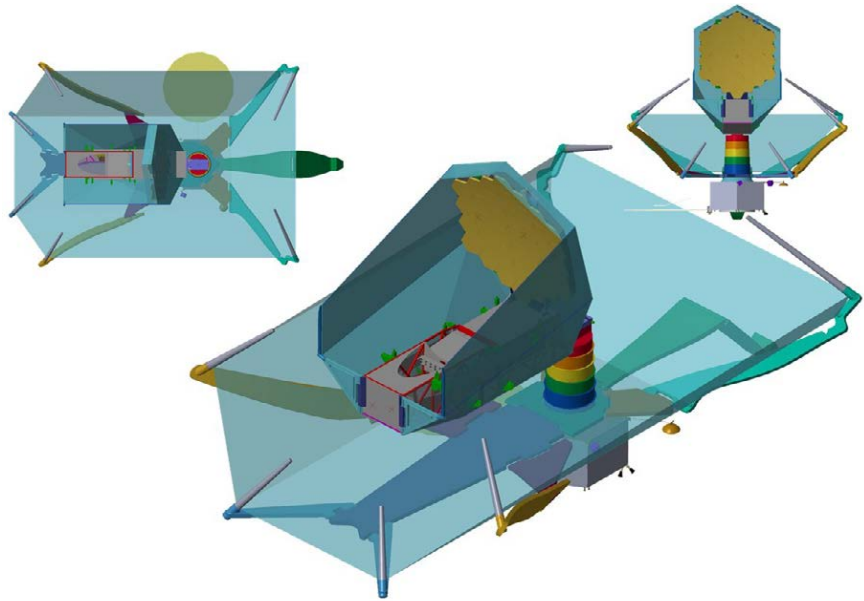

**Figure 3-39:** OST has six booms that deploy once sun shield layer is in position

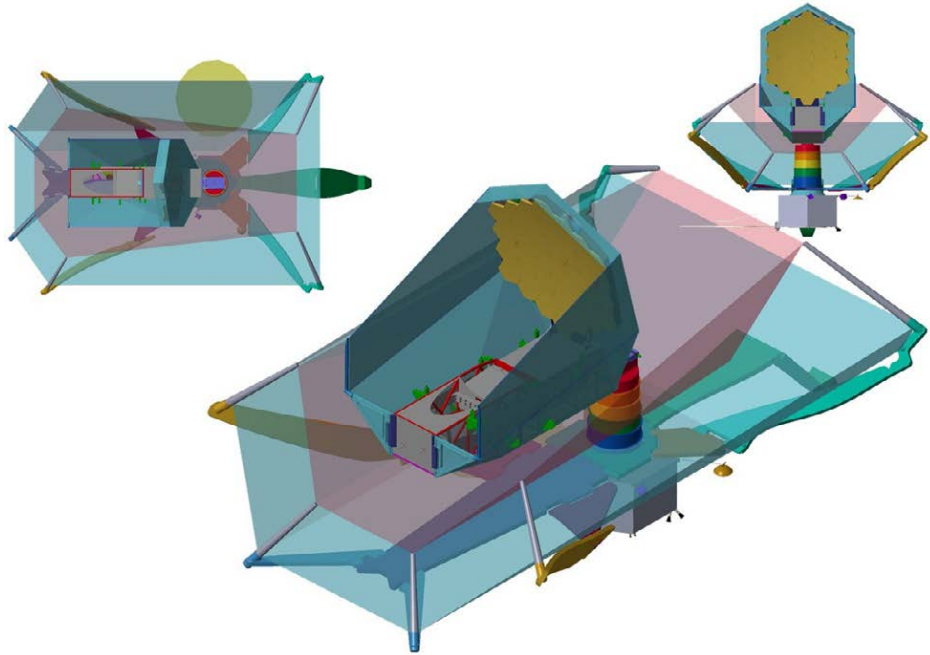

**Figure 3-40:** Sun shield layer tensioned to its final configuration.

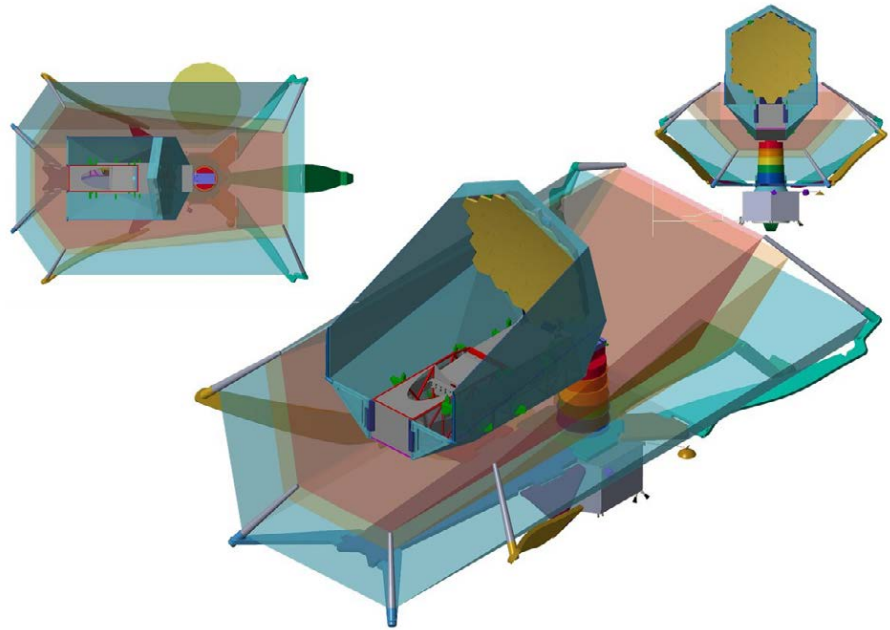

**Figure 3-41:** Sun shield layer two undergoing tensioning.

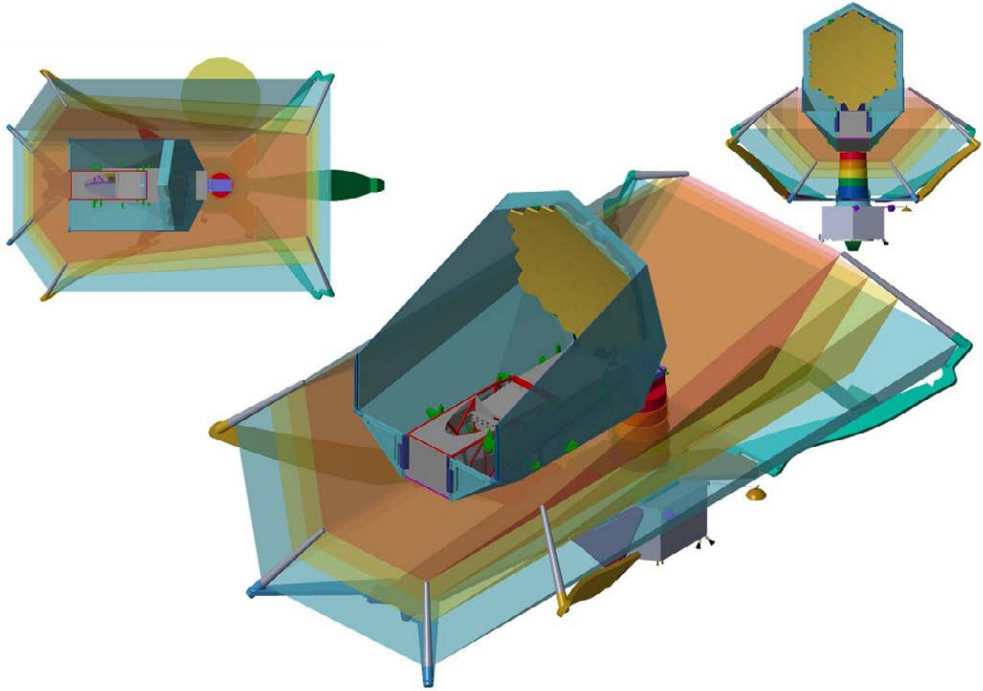

**Figure 3-42:** Sun shield layer three undergoing tensioning.

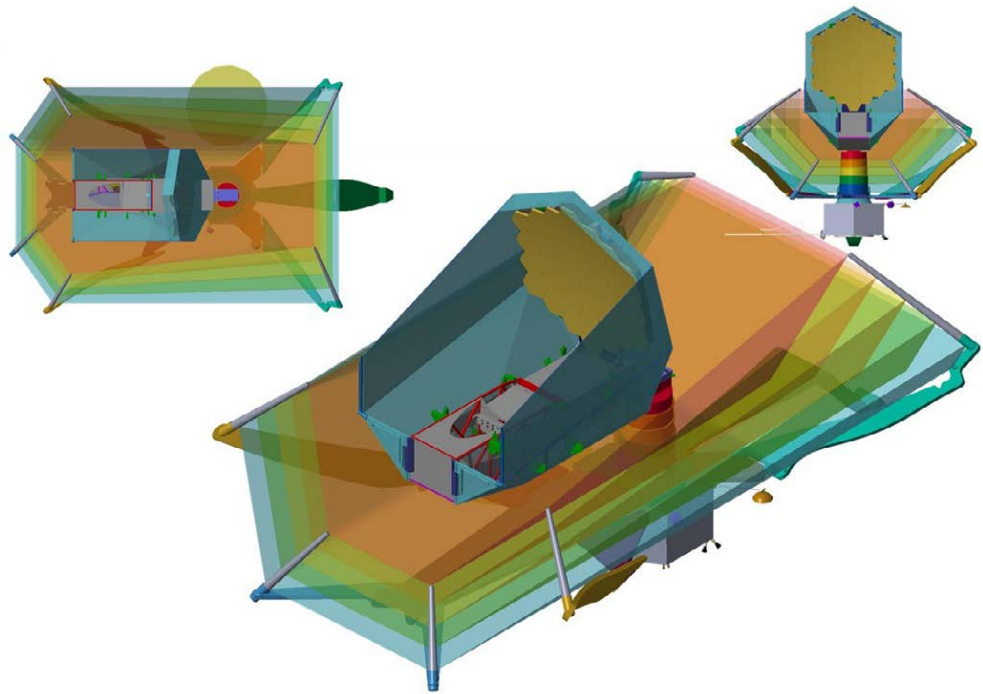

**Figure 3-43:** Sun shield layer four undergoing tensioning.

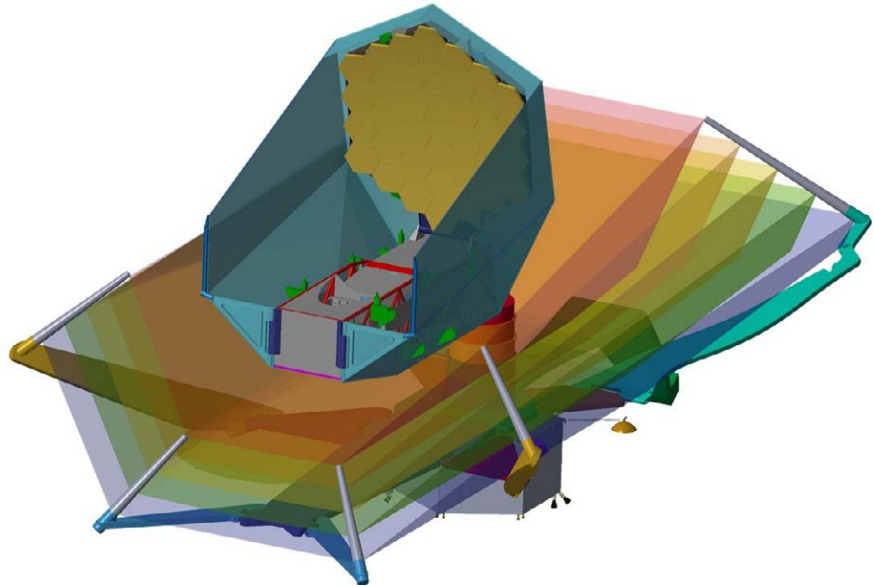

**Figure 3-44:** OST in its final on-orbit configuration with the sun shield fully deployed.

### 3.2.2.2 Electrical Power System

To meet the mission's 8,592 W end-of-life power requirement, the OST EPS is a battery-dominated 28 V DC bus with 42.4m² TJ GaAS solar array, one 35 Ah Lithium Ion battery, and associated PSE for PMAD functions. Sunlight is converted to electrical energy by a single wing DET Solar Array that is tracked via a single axis drive. Solar array power is regulated by the PSE using digital (on/off shunts) and linear shunts. Power is distributed via the PSE to the spacecraft loads via switched and unswitched services, where Solid State Power Controllers (SSPCs) provide over-current protection for switched services and fuses for unswitched services. The battery is connected directly to the power bus, and is only used for the initial launch and cruise phase to L2. Solar Array and battery will be procured in a competitive manner and integrated and tested at an appropriate test site. The PSE can either be designed and built in-house, or acquired via a contractor. All EPS components are TRL 7 or better. The topology and PSE hardware elements have heritage to recent in-house GSFC spacecraft. The EPS block diagram is shown in **Figure 3-45**.

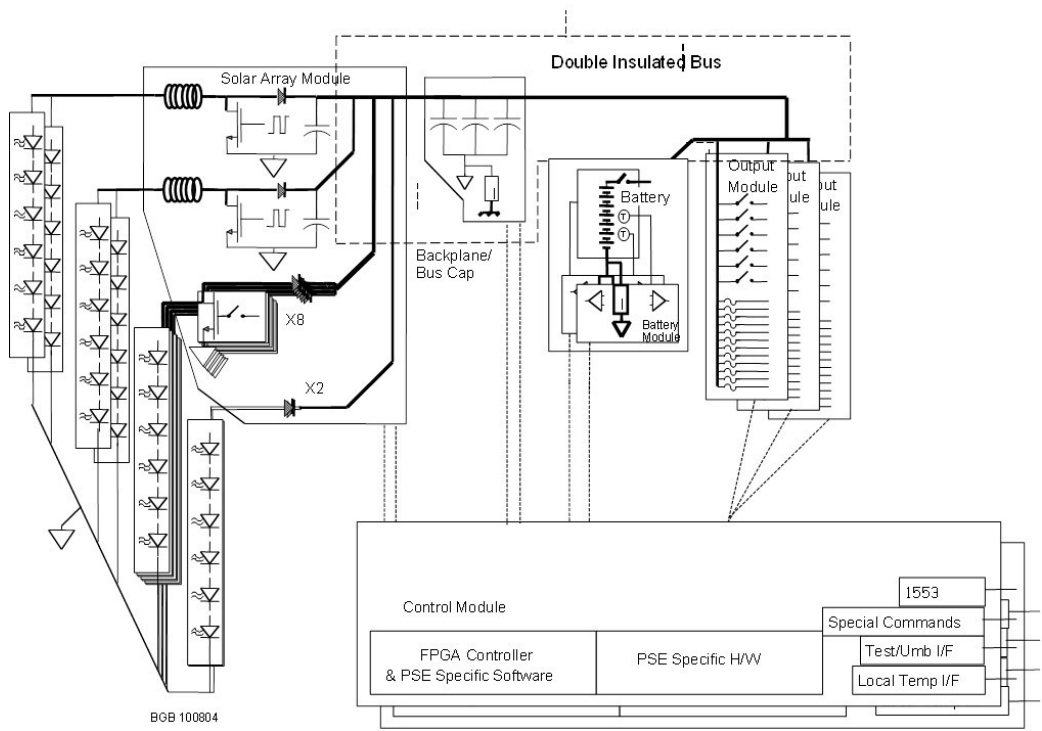

**Figure 3-45:** EPS block diagram shows the control module, 1553 bus, S/A module, battery, and output modules.

The solar array utilizes triple junction GaAs solar cells at 29.5% efficiency. To provide for 8,592 W of power at the end-of-life, a tracking single wing array of 42.4 m² size is required, which will provide a beginning-of-life perihelion power output of 10,795 W. Calibrated standard cell measurements were used for efficiency assumption. Solar cell life degradation over the ten year mission life was based on TJ solar cell radiation degradation data and SPENVIS and EQFLUX model runs for 10 years at the L2 location. Solar array strings are diode isolated with redundant PRTs. The PSE is designed with selective redundancy for reliability; only one PSE is on at a time.

### 3.2.2.3 Avionics

The OST avionics system is a SpaceWire-based design with flexible and modular GSFC's Mustang electronics. The system is fully redundant and comprises Command and Data Handling (C&DH), Attitude Control Electronics (ACE), and a 64-Tbit Solid State Recorder (SSR). **Figure 3-46** shows the system's redundancy with cross-strapping at each subsystem interface. The entire avionics system resides in the warm area of the observatory.

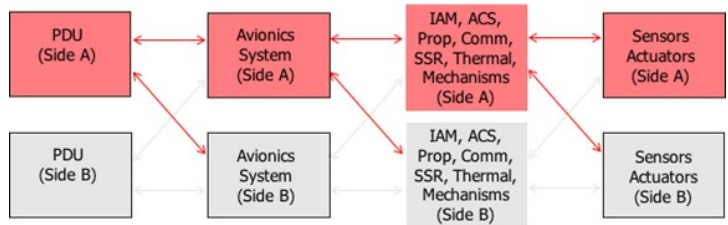

**Figure 3-46:** Avionics system has redundancy/ cross-strapping.

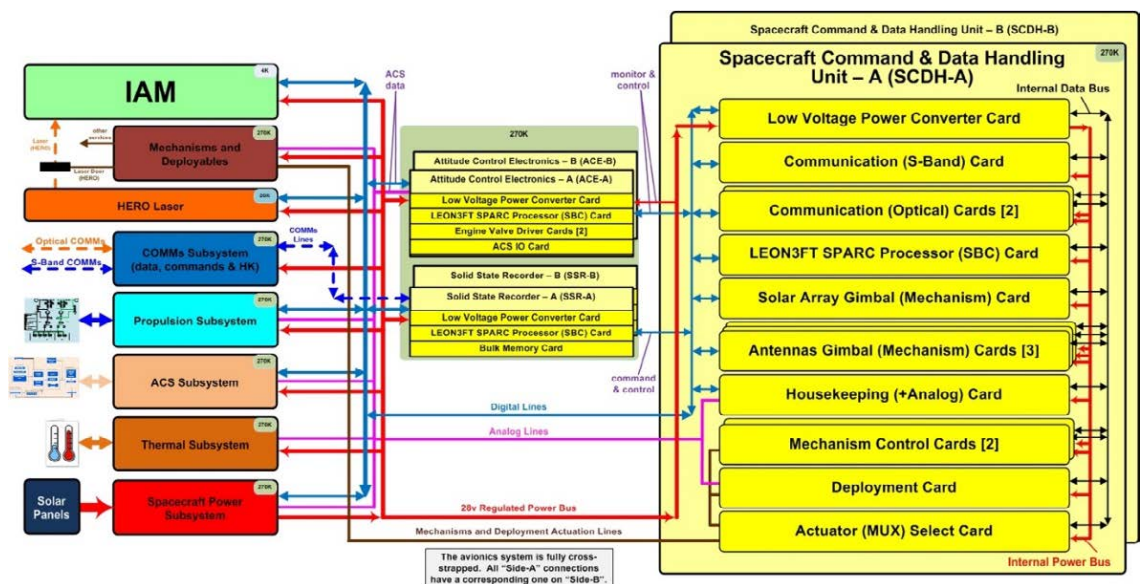

**Figure 3-47:** Avionics block diagram

The C&DH contains all drive electronics for the warm side mechanisms such as sun-shield, high gain antenna, solar array, and Deployment Tower Assembly. It provides SpaceWire data interface to and from the science instrument electronics. ACE controls all ACS components, including the Reaction Wheel Assembly (RWA), CMGs, Star Tracker (ST), Thrusters, SSIRU, and Coarse Sun Sensor (CSS). The avionics system is designed to withstand at least 30 krads of Total Ionizing Dose (TID) at the box level. **Figure 3-47** illustrates the avionics block diagram, showing interfaces of the C&DH to various spacecraft subsystems. **Figure 3-48** shows the overall spacecraft bus module, and the avionics system interfaces to the rest of the OST observatory subsystems.

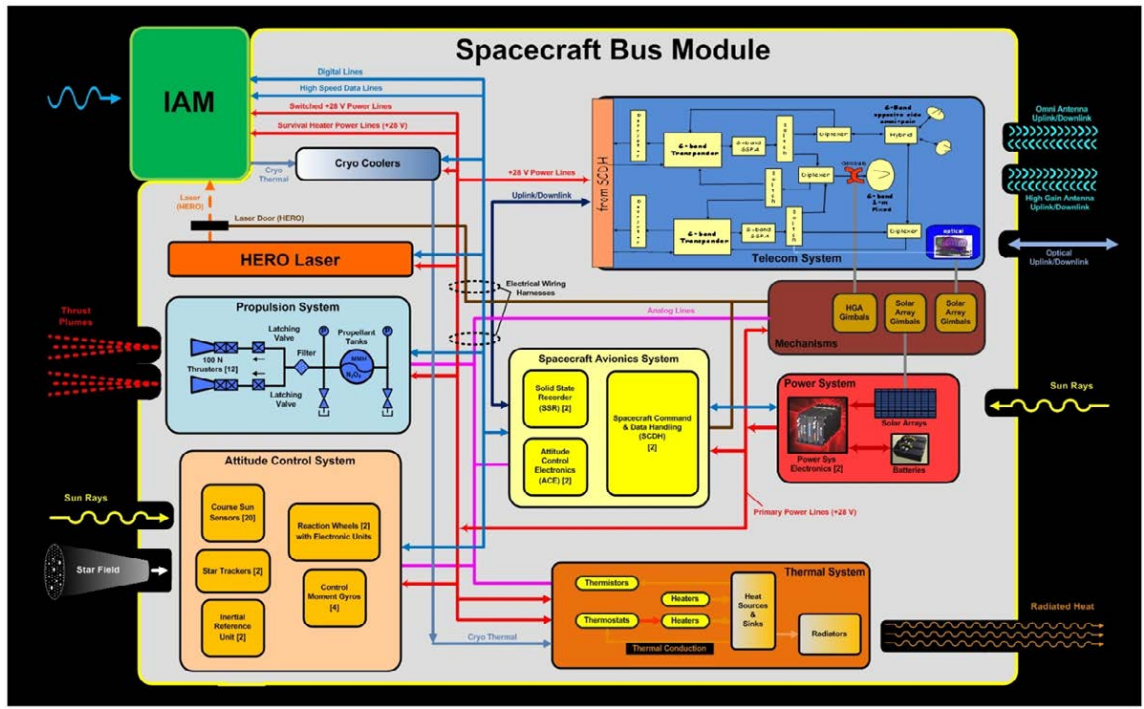

**Figure 3-48:** Spacecraft Bus Module.

### 3.2.2.4 Thermal System

#### Thermal Control Strategy

Three science power dissipation modes seek to minimize the variation and magnitude of dissipated heat loads. Constant conductance aluminum ammonia heat pipes (CCHPs) are used to transport heat to bus external panel radiators. **Figure 3-49** shows heat pipe routing and couplings. Dedicated CCHPs are routed directly to the MRSS base plate and to the cryocooler mounts. Externally-located CCHPs are also used to thermally couple some adjacent bus panels. CCHPs are also assumed to be embedded within external panel radiators to increase radiator efficiency. Component heat loads to each bay were distributed as shown in **Figure 3-50**.

The spacecraft bus structural central cylinder houses the propulsion tanks. MLI is located on the central cylinder to thermally isolate this zone.

#### Deployable Tower Assembly, Thermal Control Strategy

The overall system thermal objective is to minimize the parasitic heat load from the warm spacecraft to the cold telescope. Low conductivity composite materials (M55J_T300 assumed) minimize con-

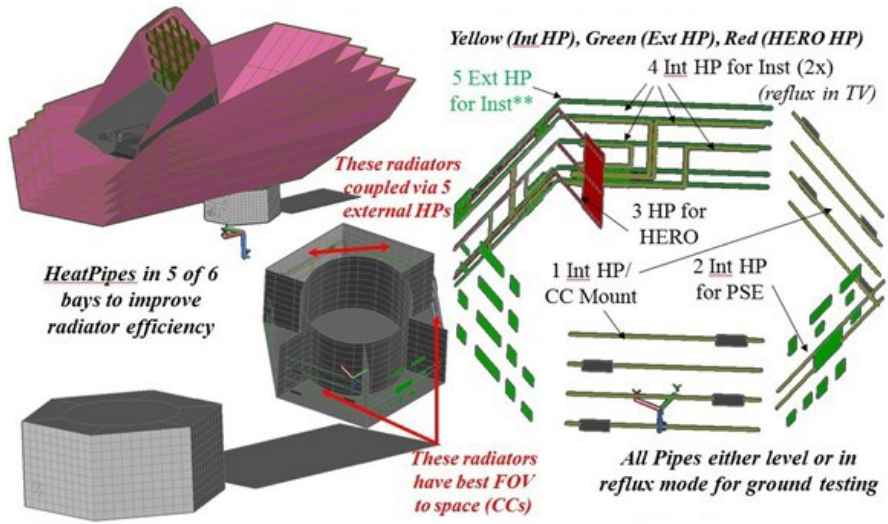

**Figure 3-49:** Spacecraft bus heat pipe routing and couplings showing dedicated CCHPs are routed directly to the cryocoolers.

- -X,-Y Bay: 4 <u>cryocoolers</u> (4 each)
- -X Bay: PSE, CDH, SSR, Battery, ACE
- -X, +Y Bay: 4 <u>cryocoolers</u> (4 each)
- +X,+Y Bay: HRS, FIP
- +X,+Y Radial Panel: HERO
- +X Bay: MRSS, MISC
- +X,-Y Bay: <u>Comm.</u> ACS

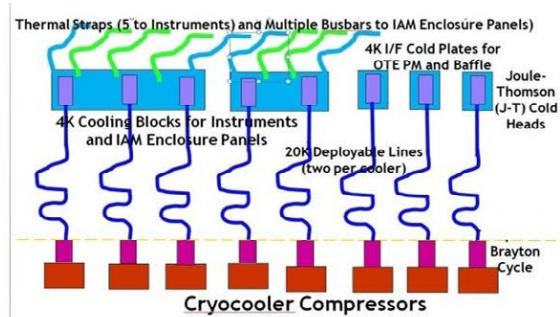

**Figure 3-50:** Distribution of component heat loads to each S/C bus bay.

ductive heat transfer through the DTA to the telescope cold sections. The main structural cylinders are assumed to be 1-cm thick. Significant thermal resistance is assumed between these cylinders at the interfaces where the assembly deploys. MLI thermally decouples the DTA system from the environmental heat loads between Sun Shade Layer 1 and the spacecraft bus. Thermal radiation to the cold sink between sun shade layers is promoted via bare composite (high emittance, 0.80 assumed) DTA surfaces to remove heat from the system. Thermal model results show this strategy to effectively reduce parasitic heat flow to the telescope Cold Zones. Internal to the DTA, low thermal emittance membranes separate each sunshade layer zone. These membranes include a 2-mil high purity aluminum foil for temperature equalization.

The sequence of energy balances show how the parasitic heat load introduced by the spacecraft bus is reduced. The final energy balance of the top disk viewing the Cold Telescope shows its temperature to be ~35 K.

The optimal sunshade geometry minimizes the number of radiation bounces between sunshade layers while satisfying solar and lunar view angle exclusion requirements. The number of layers is an important parameter. Five sunshield layers were used. Improved thermal performance is realized as the spacing between layers is increased to ~75 cm.

### Cryocoolers

OST uses eight cryocoolers, each producing 50 mW of cooling at ~4 K. This 400 mW of cooling capacity gives a factor of 2 margin over the estimated 200 mW of cooling required. Intermediate tem-perature stages at ~20 K and ~70 K have cooling capacities of about 800 mW and 35 W (total for eight coolers) respectively, over and above the heat lift necessary to precool the 4 K stage. The large amount of heat lift available at 70 K allows resistive wire to be used for high current circuits to be used above this temperature, while zero-dissipation HTS leads will be used at lower temperature.

The architecture for the cooling is flexible, allowing multiple possible designs from different companies. The common element of the cryocooler architecture is that the vast majority of the heat is produced at room temperature at the cryocooler's compressors. The final cooler stage reaching ~4 K is located in the cold part of the observatory. Small diameter tubing connects the cold expansion stage to the warm part of the observatory. This tubing is flexible and may be deployed similar to JWST's cryocooler.

### 3.2.2.5 Propulsion

The OST propulsion is a regulated bipropellant system, consisting of twelve engines fed by two independent pressure regulated fluid legs (one for monomethyl hydrazine fuel, one for MON-3 oxidizer). Each leg consists of its own high-pressure gaseous helium feed section and a low-pressure liquid propellant section. The liquid propellant sections each consist of a single Composite Overwrapped Pressure Vessel (COPV) propellant tank with a Propellant Management Device (PMD) and a Propellant Isolation Assembly (PIA). The pressure of the propellant section is maintained by the high-pressure helium feed system, consisting of a COPV pressurant tank and regulator-based pressure control assembly (PCA). Pressure is monitored in each tank via pressure transducers (single for the pressurant tank and primary and redundant for the propellant tanks. Eight fill and drain manual valves in the PS are used for fluid loading, unloading, and component testing (four for each leg)(**Figure 3-51**).

The acceleration limitations of the spacecraft's sunshield constrains the allowable thrust levels of the propulsion system. Given the low TRL-level of the OST sunshield and deployment scheme, acceleration limits from JWST were assumed. The 100-N thrust level is in agreement with the JWST level given the OST Concept 1 CBE mass.

Given the large sunshield design, contamination concerns, off-axis design, and thermal limitation (4 K instrument suite, 35 K sunshield), engine placement and control authority is a challenge. The PS design uses specific engine sets for both the stowed and fully deployed configurations. The AFT eight engines are used for launch vehicle dispersion and tipoff correction with the telescope in the stowed configuration in **Figure 3-52a**. The arrows indicate the approximate thrust vector. The Center of Mass (CM) of the telescope changes significantly from stowed to the deployed configuration. Consequently, the aft engines no longer provide the necessary control once in the deployed configuration. The pair of the aft engines sets (AFT-1, -2) with the side engine banks provide the control authority of OST in the deployed configuration (**Figure 3-52b**). The aft and side engines are canted to an optimum angle off the x-axis, radial outward to maintain this authority while the CM migrates due to propellant consumption. This configuration provides OST with two-axis control, which is all that is necessary given the ACS scheme being implemented. The only maneuvers are delta-V, station keeping, and occasional angular momentum dumping.

The cold temperature requirement of the optics precludes the use of any forward facing engines with their hot plume gases. This constraint, coupled with the large migration of the CM between the stowed and deployed configurations, prevents pure couple momentums from the engines. Angular momentum accumulation due to solar pressure is managed by large momentum wheels along a single axis. Control Moment Gyros (CMGs), also used for all other momentum management, can be positioned along the same axis as the momentum wheels. Therefore, the angular momentum dumping maneuvers are only required along a single axis within the control authority of the engines. As designed, OST can orient itself enough to provide sufficient station keeping maneuvers to counter act solar pressure, while keeping the optics in shadow at L2.

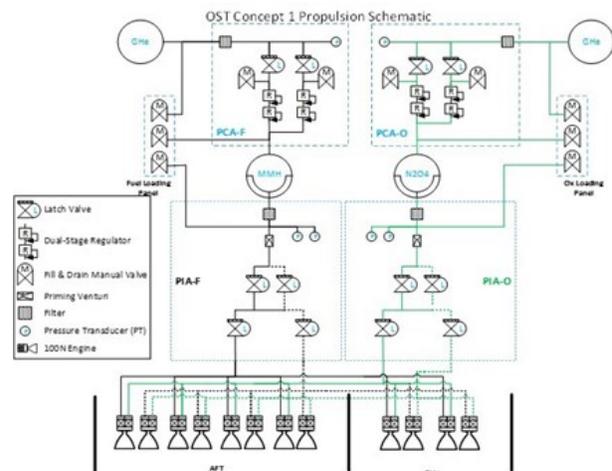

**Figure 3-51:** OST regulated bipropellant propulsion system schematic.

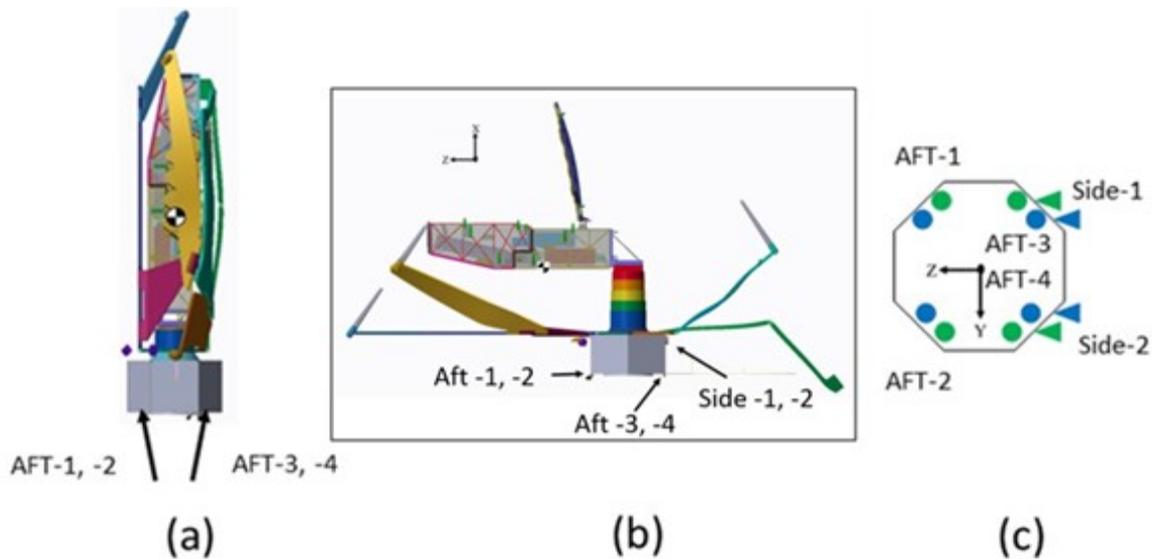

**Figure 3-52:** A) Stowed in launch configuration controlled with eight aft thrusters with thrust vectors. B) OST in the fully deployed configuration controlled via a combination of aft and side thrusters to ensure the SC CM is within the thruster control box. C) Approximate location and layout of thrusters.

All PS components are commercial-off-the-shelf (COTS) and have a TRL level of 9. The fuel and oxidizer masses are 1,718 kg and 2,869 kg, respectively, for a total of 4,588 kg including standard 10% residual propellant mass. The operating pressure of the fuel and oxidizer feed systems and gaseous helium feed system are 300 psia, 300 psia, and 5,000 psia, respectively. The engine baselined for the Concept 1 design is the Aerojet-Rocketdyne R-1E 100N. Assuming a peak oxidizer/fuel mixture ratio for the engine of 1.67 and a 300 psia operating pressure, the specific impulse of the engine is assumed to be a conservative 274 seconds. Redevelopment of this engine is planned to increase its performance, which would significantly reduce the required propellant masses. The size of OST and the bus volume is robust for scaling and repositioning engines, if necessary, during the detailed design phase. The spacecraft conceptual design has ample surfaces and integration points for tanks, engines, valves, and propulsion-related mounting hardware.

### 3.2.2.6 Attitude Control System

The OST science operations profiles impose some challenging requirements on the attitude control system. During inertial pointing of the MISC, the absolute pointing requirement is 44 mas RMS, derived from the requirement to keep the target object within the MISC slit. This requirement is allocated 22 mas to drift and 22 mas to jitter. These requirements drive the requirement for a fine guidance sensor for inertial pointing. During survey operations of the MRSS, scanning rates of up to 100 arcsec/sec are required, with a required absolute pointing accuracy of 0.1 arcsec RMS, which drives the actuator complement.

## ACS Architecture

The size of OST and the range of operational modes require an attitude control system with both fine attitude sensing, high torque authority, and high momentum storage capacity. During science operations with relatively coarse (0.1 arcsec) pointing requirements, the ACS uses star trackers and an inertial measurement unit (IMU) for primary attitude determination. During inertial pointing operations, a fine guidance channel incorporated in the MISC instrument is used to support fine pointing (44 mas RMS). For primary actuation, a set of four control moment gyros (CMGs) provide the high torque authority needed for survey operations. Two high-capacity momentum wheels are dedicated to countering solar radiation pressure torques. The accumulated angular momentum is periodically unloaded using thrusters.

In conjunction with the CMGs, the field steering mirror (FSM) is used to provide fine control of the optical line of sight and to provide some rejection of internal disturbances. The FSM control loop follows tip/tilt commands from the ACS control loop assisted by accelerometer and tip/tilt encoder measurements, and feeds back its tip/tilt to the ACS so that the ACS can keep it centered. Similarly, FSM loop accepts inputs from the MISC tip/tilt steering control loop to keep the MISC tip/tilt mirror (TTM) centered in its operating range. This multi-stage control loop architecture is shown in **Figure 3-53**.

In addition to active vibration rejection by the FSM control loop, vibrations from the CMGs and cryocoolers are further attenuated by mounting them on passive isolators. These isolation mounts are tailored to reject vibrations within a limited frequency range. The cryocoolers and CMGs have limited and tunable operational speeds, which is helpful in tailoring isolators for best effect.

## Technology Readiness

The OST ACS concept makes use of a combination of TRL 9 COTS and optical-mechanical components integrated with the instruments. COTS star trackers, IMUs, and momentum wheels are all available with extensive flight heritage in the deep-space environment. Control moment gyros are less common in astrophysics applications, but are commonly used in Earth observation and reconnaissance applications, and are also available as COTS components. The field steering mirror control is part of the Instrument Accommodation Module (IAM). The MISC fine guidance channel, and MISC tip/tilt mechanism are integral to the MISC instrument.

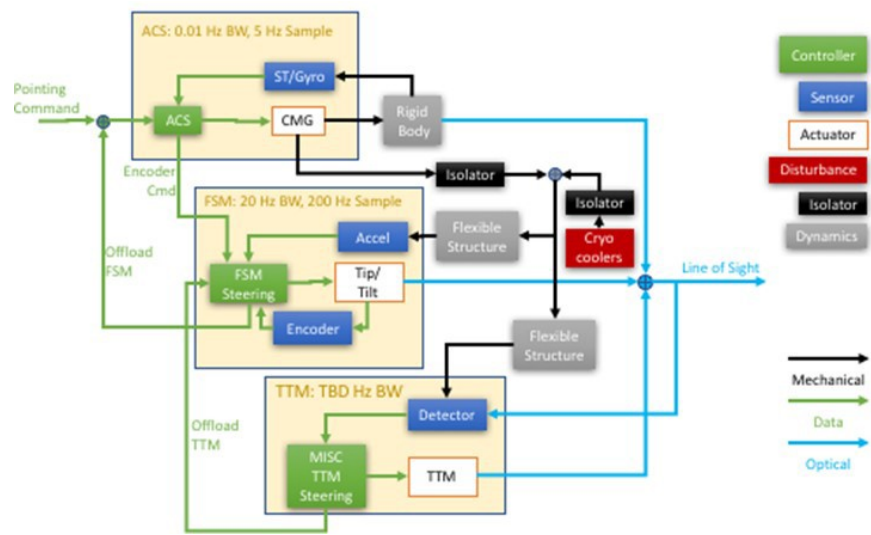

**Figure 3-53:** Multi-Stage Control Loop architecture.

## Pointing Performance

There are two driving cases for pointing performance: inertial pointing and survey operations. Preliminary pointing performance estimates were generated for each case.

During inertial pointing, pointing performance is driven by sensor noise, internal disturbances propagated through the attitude control loop and the flexible structure. The MISC fine guidance channel provides attitude measurements with noise-equivalent angle of 15 mas RMS. Disturbance data from a representative CMG shows tonal components at the rotor spin rate (100 Hz) and several harmonics. The cryocoolers are also modeled as tonal disturbances at 50 Hz. A finite-element structural model has not been developed at this stage, so some flexible modes were modeled as transfer functions. A few modes were placed arbitrarily, with a lowest mode at 0.1 Hz. Additional flexible modes were then placed directly atop the tonal disturbances due to the CMGs and cryocoolers to provide a reasonable worst-case performance estimate. The ACS control loop and FSM loop were modeled. The line-of-sight frequency response was then found and decomposed into drift and jitter using the windowing frequency derived from the MISC exposure time (300 sec). **Figure 3-54** shows the resulting cumulative RMS drift and jitter as a function of frequency. Drift performance (1.5 mas RMS) is dominated by sensor noise. Major contributors to jitter performance (6.2 mas RMS) are the sensor noise and the CMG disturbances, especially the fundamental harmonic at 100 Hz. Structural modes below 20 Hz are effectively suppressed by the FSM loop, at the expense of some amplification of the CMG fundamental. Passive isolators were not modeled for this preliminary analysis, but are held in reserve to mitigate concentrated disturbances if needed. This analysis suggests that the CMG fundamental mode would be a prime candidate for a passive isolator.

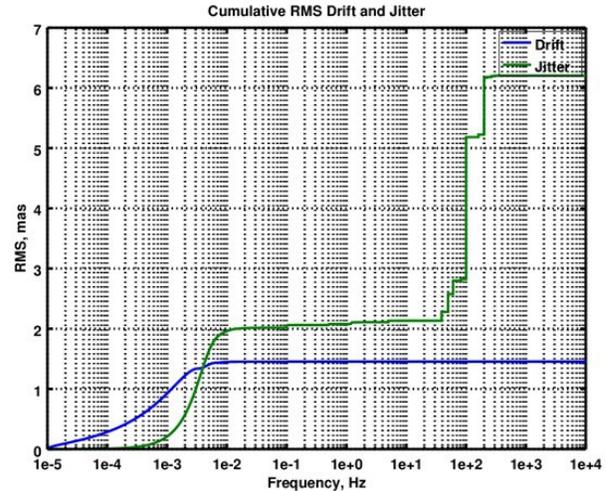

**Figure 3-54:** Drift and jitter performance during inertial pointing shows the resulting cumulative RMS drift and jitter as a function of frequency.

During survey operations, the scanning rates are too high (100 arcsec/sec) for the MISC fine guider channel to be used, but the pointing accuracy requirement is relaxed to 0.1 arcsec RMS. Using performance parameters for a representative off-the-shelf star tracker and IMU, the steady- state attitude accuracy may be estimated. The predicted performance indicates that the requirement of 0.1 arcsec RMS is feasible for a representative star tracker and IMU for a sample rate of 4 Hz or greater.

## Momentum Management

OST presents a very large surface area to the Sun, and the center of pressure is offset from the ob-servatory mass center by ~10 m. Preliminary estimates of angular momentum accumulation due to solar radiation pressure torques are on the order of 2000 Nms/day. Two 250-Nms momentum wheels are baselined, dedicated to storing this angular momentum. Using the full +/- 250 Nms range, they provide 1000 Nms storage capacity, and so need to be unloaded twice per day with thrusters.

To reduce the momentum accumulation rate, the team will investigate options for adjusting the center of pressure. It is expected that moving the center of mass will be difficult due to constraints imposed on the structure by the launch environment. The team will assess feasibility of adding a solar sail (*i.e.*, "trim tab") to modulate the center of pressure. Operational mitigations may also be of some limited use, and will be assessed.

### 3.2.2.7 Communications System

To meet the OST data estimates of ~39 Tbps, Optical Communications was considered as the best choice for the Concept 1 mission for a 2035 launch date. S-band was chosen for telemetry (TLM), command (CMD), and ranging (tracking) (TT&C).

S-band TT&C includes the 18-m at NASA WSC and a NASA contract antenna; 13-m, at South Point, HI. The NASA-contract 13-m at Dongara, Western Australia requires a 2 kWatt klystron upgrade for OST. The DSN would range via the spacecraft bus Omnis in the event of contingency.

The Ka-band backup alternative to Optical requires two sites in addition to WSC, longitudinally separated with 18-m Ka-band capability (possibly a pair of 14-m antennas matrixed together). The S-band system block diagram and Optical Communication System are shown in **Figure 3-55**.

Optical, ~1 Gbps margin at WSC above 20° with no clouds provides 2.7 dB (for sites like Haleakala). WSC 18-m S-band with 2 kbps up, 80 kbps down with ranging, and all margins are positive. South Point, HI (USHI) 13-m with 2 kbps up (2 kW) 20 kbps down while ranging, and all margins are also positive. Dongara could equal this with an upgrade to 2 kW.

Without ranging, 18-m 32 kbps up and 154 kbps down is possible, as well as low rate CMD and TLM with no ranging via S/C Omnis. DSN TT&C with low data rate via S/C Omni is used for contingency backup.

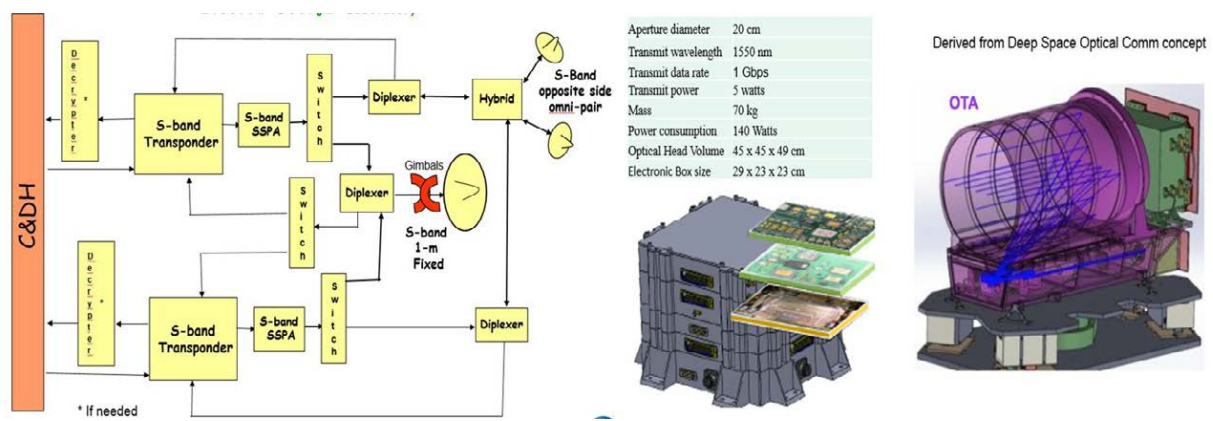

**Figure 3-55:** A) S-Band block diagram. B) Optical Communications System concept.

### 3.2.2.8 Flight Software

OST is a fully redundant mission with cold backups including a hot independent ACE, and it is a fail-safe mission. The OST Flight Software (OFSW) implementation is low risk, based on high heritage code and hardware coupled with a robust flight computing platform. The OFSW includes all the C&DH, ACS, and instrument support software that run on the GSFC MUSTANG DLEON$_3$ computing platform with VxWorks real-time operating system. The ACE safehold, PDU, and SSR software run on three independent GSFC MUSTANG LEON$_3$ computing platforms with RTEMS real-time operating system. Virtually all of the core C&DH FSW will be re-hosted from existing NASA's GSFC Core Flight Executive Software System (cFS) and are common on both the C&DH, ACE, PDU, and SSR processors. The cFS has flown on LRO, LADEE, GPM, MMS, RBSP, NICER, and is being used NASA-wide for future missions.

The cFS was developed specifically for re-use on multiple missions. The cFS are fully tested and are considered as off-the-shelf products at TRL 9. Much of the cFS will only need to be parameterized for the OST mission. Only those OST components unique to the OST system will need to be developed as reengineered components. A heritage analysis has been done and the estimated logical source lines of code reuse is 67% for the C&DH processor, 85% for the ACE processor, 81% for the PDU processor and 67% for the SSR processor (ref to: Source Line of Code Reuse Estimates Table). Standard interfaces (MIL1553, RS422, SpaceWire) and protocols connect the OFSW system with the external environment and provide internal connectivity among the instrument components. The OFSW is highly table-driven and supports memory, table, and FSW image uploads. The Stored Command processor and Limit Checker capabilities provide the backbone functionality for OST operations and fault management. The OFSW supports absolute and relative time tags. Sequences are programmable via table loads. The Limit Checker can be configured to trigger on simple or complex telemetry signatures and can be individually enabled or disabled by command. Unique monitoring and control applications will be developed specifically for OST to provide operational control of the instruments, GNC, and power subsystems. The payload consists of five instruments: MRSS, HERO, FIPS, HRS, and MISC with dedicated instrument avionics. A maximum science data input rate of 350 Mbps goes to the spacecraft SSR via five SpaceWire interfaces.

The GSFC MUSTANG team has ported this rich heritage code to its upgraded avionics. The heritage cFS C&DH software is currently running on the Processor ETU. To this, OST adds only a few mission unique custom applications, including: GNC/safehold algorithm; mission-unique Spacewire and interface manager applications to manage communications to the specific instruments; management of the DTN enabled CCSDS file delivery protocol used for uplink and downlink; telemetry downlink management; ground command ingest and command distribution to subsystems and instruments. OST has healthy onboard processing resources margins (ref to: CPU Utilization Estimates table). The high-speed data rate presents a challenge, therefore, OST proposes using the WFIRST SSR with Delay/Disruption Tolerant Networking (DTN) protocols to handle up to 3.5 Gbps high-speed data downlink. DTN is being infused onto the PACE mission and the conceptual design is available for that system. OST has a requirement to perform high speed data downlink while in science mode. Disturbance caused by optical COMM tracking will require further study. The proposed FSM control has to be modified as follows: addition of an accelerometer and the control algorithm should be performed on the instrument side, not at the spacecraft bus (ideally completed in the ADU). Confirmation is needed whether the ADU has enough processing resource to accommodate the control algorithm. The control bandwidth should be ~200 Hz.

The OFSW will be developed in compliance with NPR 7150.2 using GSFC's CMMI compliant processes. A waterfall development model will transform software requirements into tested flight code. Major coding occurs in three incremental builds: the first provides all core FSW and external/internal interfaces drivers, which is mostly inherited from the cFS library and the ongoing MUSTANG effort. The second build includes mission-unique functionalities, such as science data acquisition/management, power, thermal, and attitude control. The third build includes mission operations and fault management. All requirements for each build of the code will be defined prior to the start of coding. Formal reviews will be held prior to proceeding beyond the following key phases: Requirements Analysis, Preliminary Design, Critical Design, and Acceptance Test. Additionally, all new and modified code receives code inspection by the team and unit tested by the developer prior to build. Once the build code has been integrated into the flight software, the software test team will perform verification tests of the integrated HFSW requirements. No new test is planned for all heritage flight software components except the fact that their functions will be exercised through system test. After all of the builds have been completed, a system level acceptance test will be performed, during which the entire flight software suite will be tested as a total integrated unit. GSFC's Code 582 will maintain HFSW configuration management and discrepancy reporting systems with full NASA's GSFC Quality Assurance insight.

### 3.3 Observatory Thermal Analysis

The 200 mW total 4 K cryo refrigeration budget is a key thermal requirement that drives concept feasibility. The system architecture allocates 100 mW to instrument power dissipation and the remaining 100 mW for everything else. Thermal analyses must show that the telescope, Instrument Assembly Module (IAM), and baffle assemblies can all be cooled to ~4 K within the allocated refrigeration budget. This is largely accomplished by exploiting geometry and low temperature thermal characteristics to effectively eliminate the thermal background.

**Table 3-6** shows the hot case cold zone energy balance and predicted 4K cryo performance as a function of observatory pitch angle. 4K cryo zones are set as boundaries in the thermal model. 4K required cryo-cooling is defined as the heat required to hold 4K based on the steady state solution. Heat distribution inefficiencies are not considered. The data show that the 200 mW requirement is satisfied for the full range of observatory pitch angles.

**Table 3-6:** Cold Zone Energy Balance and Required Cryo-cooling Thermal Model Results

| Pitch Angle to Solar Vector | 5 | 0 | -15 | -30 | -45 |
|---|---|---|---|---|---|
| Required 4K Cryo Cooling | W | W | W | W | W |
| MISC Power Dissipation | 0.1050 | 0.1050 | 0.1050 | 0.1050 | 0.1050 |
| FIP Power Dissipation | 0.0190 | 0.0190 | 0.0190 | 0.0190 | 0.0190 |
| Harness Parasitic Heat Leak | 0.0140 | 0.0140 | 0.0140 | 0.0140 | 0.0140 |
| Baffle Surfaces | 0.0248 | 0.0200 | 0.0272 | 0.0195 | 0.0196 |
| IAM Struct Panels | 0.0263 | 0.0326 | 0.0273 | 0.0345 | 0.0318 |
| PM Segment #28 Cryo Attach | 0.0001 | 0.0001 | 0.0002 | 0.0001 | 0.0001 |
| 4K Required Cryo Cooling | 0.1893 | 0.1907 | 0.1927 | 0.1920 | 0.1895 |

### 3.3.1 Thermal Model Simulation

The Thermal Desktop thermal modeling analytical platform was used to perform these analyses (Version 5.8 Patch 9 - 64 Bit). This platform is built on an AutoCad user interface to facilitate model geometry formation. Monte Carlo statistical ray trace methods are used to calculate absorbed heat fluxes and radiation exchange factors. Conduction couplings are calculated based on physical characteristics and material properties. For each case, the platform generates a SINDA thermal finite difference mathematical model. This tool was used to calculate the thermal network steady state solution for each case considered.

**Table 3-7** summarizes thermal model component power dissipation assumptions, and **Table 3-8** shows the assumed sun shade thermal optical properties. The sun shade properties are based on JWST sun shade material and thermal coatings.

**Table 3-7:** Thermal Model Power Dissipation Assumptions

| | # Units | Power Dissipation Standby W | CBE W | Margined W |
|---|---|---|---|---|
| **Instrument Components On Bus** | | | | |
| CryoCooler | 8 | 2000 | 4000 | 4000 |
| FIP Electronics | 1 | 50 | 394 | 512.2 |
| HRS Electronics | 1 | 50 | 565 | 734.5 |
| HERO Electronics | 1 | 50 | 450 | 585 |
| MRSS Electronics | 1 | 40 | 2086 | 2711.8 |
| MISC Electronics | 1 | 0 | 46 | 59.8 |
| MISC Guider | 1 | 46 | 46 | 59.8 |
| **Bus Components** | | | | |
| Power Supply Electronics (PSE) | 1 | 24.3 | 430 | 559 |
| Battery | 1 | 0 | 0 | 0 |
| Command Data Handling 1(CDH1) | 1 | 70 | 108 | 140.4 |
| Command Data Handling 2(CDH2) | 1 | 20 | 20 | 26 |
| Attitude Control Electronics #1 (ACE1) | 1 | 23.2 | 23.2 | 30.16 |
| Attitude Control Electronics #2 (ACE2) | 1 | 0 | 0 | 0 |
| Solid State Recorder #1 (SSR1) | 1 | 20 | 100 | 130 |
| Solid State Recorder #2 (SSR2) | 1 | 20 | 20 | 26 |
| OTA Electronics | 1 | 0 | 140 | 182 |
| Solid State Inertial Reference Unit (SSIRU) | 1 | 20 | 20 | 26 |
| CMG-G | 1 | 31 | 400 | 520 |
| CMG-E | 1 | 4.03 | 52 | 67.6 |
| Reaction Wheels | 4 | 16 | 16 | 20.8 |
| Telescope ADU | 1 | 39 | 38 | 49.4 |
| SBand Transceiver | 1 | 40 | 40 | 52 |
| SBand Receiver #1 | 1 | 7 | 7 | 9.1 |
| SBand Receiver #2 | 1 | 7 | 7 | 9.1 |

| Power Dissipation Modes | MODE | | |
|---|---|---|---|
| Description | A | B | C |
| | mW | mW | mW |
| **Cold Telescope Zone** | | | |
| MRSS | 105 | 0.0 | 0.0 |
| MISC Science Ops | 0 | 105.0 | 0.0 |
| MISC Guiding Function | 8 | 0.0 | 8.0 |
| HERO | 0 | 0.0 | 70.0 |
| FIP | 0 | 19.0 | 19.0 |
| HRS | 0 | 0.0 | 14.0 |
| | 113 | 124.0 | 111.0 |

**Table 3-8:** Sun Shade Assumed Thermal Optical Properties

| Sun Shade Thermal Optical Properties | Thermal Coating | Solar Alpha | Hemi Emittance | Specularity Infrared Wavelengths |
|---|---|---|---|---|
| Layer 1    Facing Sun | SiO2 on Kapton | 0.467 | 0.723 | 0 |
| Opposite Side | VDA | 0.095 | 0.02 | 0.98 |
| Layer 2    Facing Sun | SiO2 on Kapton | 0.451 | 0.65 | 0 |
| Opposite Side | VDA | 0.095 | 0.027 | 0.98 |
| Layer 3    Facing Sun | VDA | 0.095 | 0.021 | 0.98 |
| Opposite Side | VDA | 0.095 | 0.021 | 0.98 |
| Layer 4    Facing Sun | VDA | 0.095 | 0.021 | 0.98 |
| Opposite Side | VDA | 0.095 | 0.021 | 0.98 |
| Layer 5    Facing Sun | VDA | 0.095 | 0.01 | 0.98 |
| Opposite Side | VDA | 0.095 | 0.01 | 0.98 |

## 3.3.2 Thermal Discussion and Results

### 3.3.2.1 Instrument Accommodation Module (IAM)

**Figure 3-56** shows the IAM thermal block diagram. Eight (8) Cryo Cold heads service three (3) cryo cold blocks . Each Al-1100 cryo block measures 30.48 x 30.48 x 0.3175 cm. High purity aluminum thermal straps thermally couple cold blocks to required locations. A 0.00058 cm (2 mil) high purity aluminum foil is included within black kapton layers which externally covers the IAM structure. High emittance surfaces are required to reduce stray light. As shown in figure, high purity aluminum conduits (0.158 cm thick by 2.54 cm wide) traverse IAM side and front panels to facilitate their cooling to 4K. These conduits are heat strapped to the respective cryo blocks based on proximity and placed in thermal contact with the high purity aluminum foil.

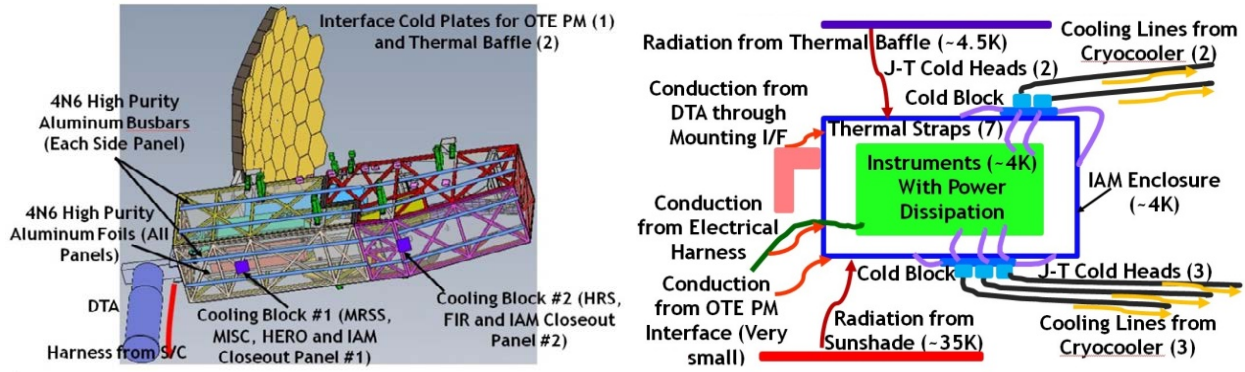

**Figure 3-56:** This Instrument Accommodation Module (IAM) thermal block diagram indicates the study team's attention to detail, in recognition of the importance of an excellent thermal design.

**Table 3-9** describes thermal strap assumptions, and **Figure 3-57** depicts the thermal model internal IAM geometry. IAM temperature results are depicted in **Figure 3-58** as temperature contours for the -15 and -45 degree observatory pitch cases.

**Table 3-9:** Cold Zone Thermal Strap Assumptions

### 4K Thermal Strap Assumptions

| Parameters | Assumptions |
|---|---|
| Number of Foils per Thermal Strap | 330 |
| Foil Thickness, cm | 0.003048 |
| Foil Width, cm | 5.08 |
| Foil Length, m | 1.5 |
| Thermal Conductivity of Foils, Wm$^{-1}$K$^{-1}$ | 2700 |
| Thermal Conductance of Foils, WK$^{-1}$ | 0.92 |
| Thermal Contact Area at Joints with Cold Tip or Cryostat Ring, cm$^2$ | 25.806 |
| Thermal Contact Conductance at Joint with Cold Tip, WK$^{-1}$ | 11.429 |
| Thermal Contact Conductance at Joint with Cryostat Ring, WK$^{-1}$ | 11.429 |
| Overall Thermal Conductance of Thermal Strap, WK$^{-1}$ | 0.792 |

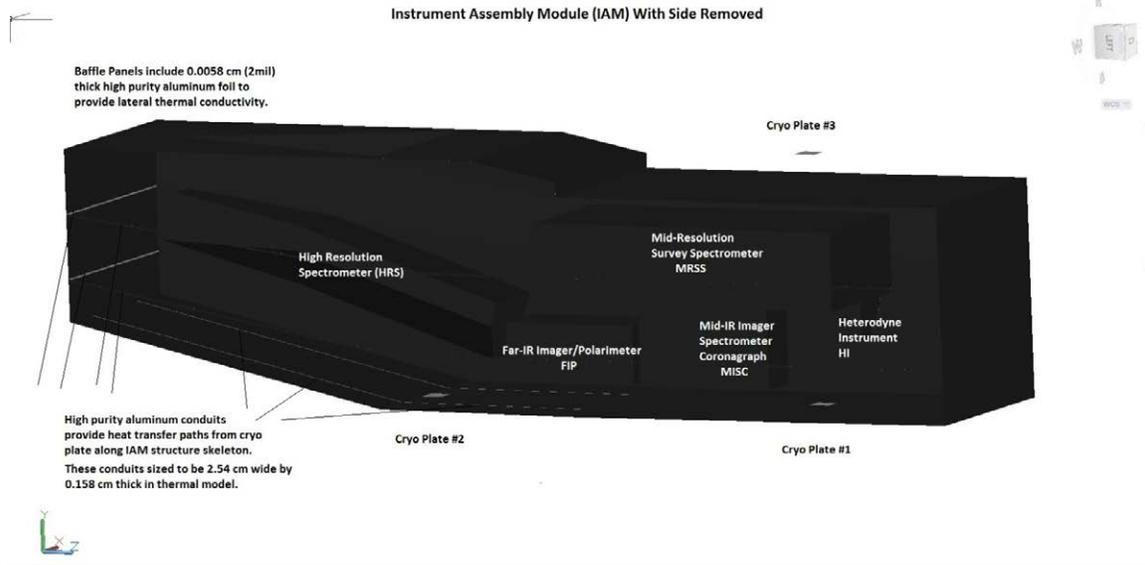

**Figure 3-57:** IAM with side removed exposes the internal geometry used in thermal modeling.

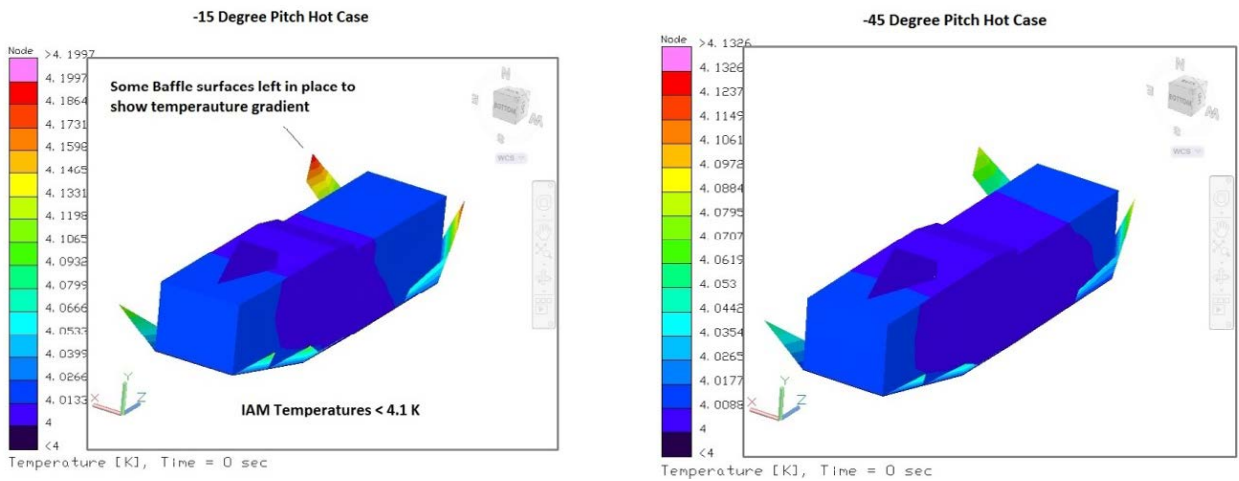

**Figure 3-58:** Predicted IAM temperature contours for two different observatory pitch angles show that the IAM temperature is maintained at <4.1 K.

### 3.3.2.2 Telescope

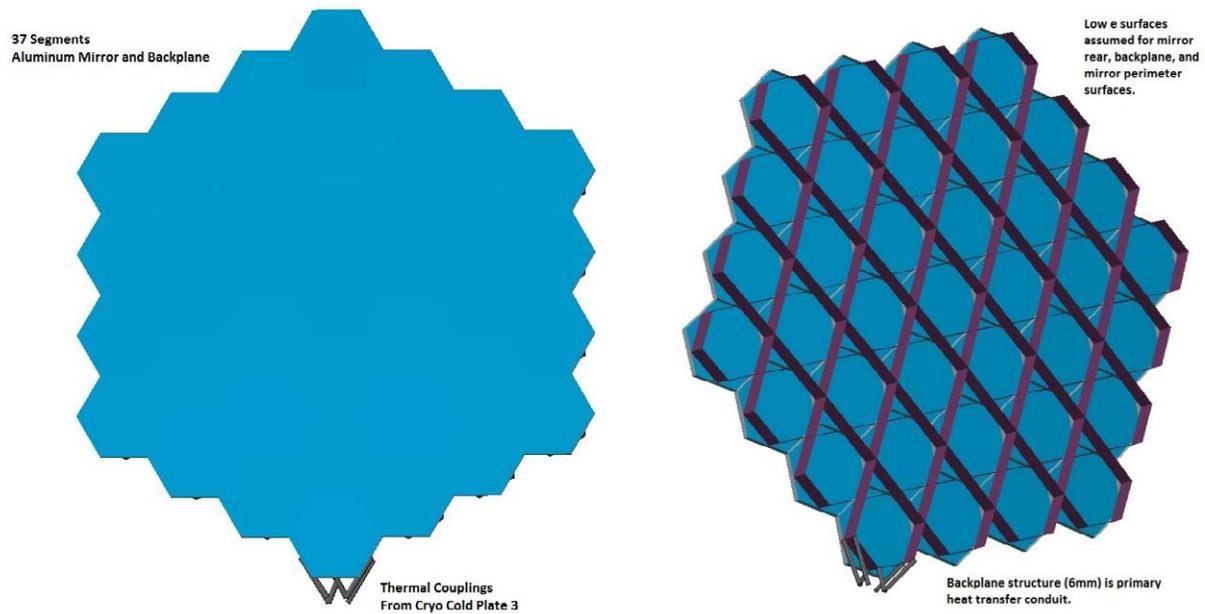

**Figure 3-59:** This illustration shows how the primary mirror is represented in the thermal model, with model nodes depicted.

### Thermal Control Strategy

**Figures 3-59** and **3-60** show the primary mirror thermal model geometry and assumptions. The telescope is actively cooled via Cryo Cold Plate #3 through heat straps that couple to the primary mirror segment located at the base. The backplane structure is assumed to be 6 mm thick aluminum forming continuous structure behind the telescope assembly. The gap between mirror segments is assumed 6mm. The backplane structure serves as the primary heat transfer conduit from the cryo cold sink to the telescope assembly. The thermal model assumes the back plane structure is continuous (i.e. no joints or gaps). Thermal couplings are assumed at the six back plane / mirror perimeter surface intersection points as shown in the figure below. Thermal straps bridge the gap between the back plane and individual primary mirror segments.

The thermal load to the telescope is minimized by the sun shade and baffle assemblies. The net telescope radiation coupling to its surrounding environment is minimized through use of low infrared emittance / highly specular thermal coatings.

**Figure 3-61** shows predicted primary mirror temperature contours for two observatory pitch angles, indicating that the temperature is very close to 4 K at all mirror locations, with a modest temperature gradient across the mirror.

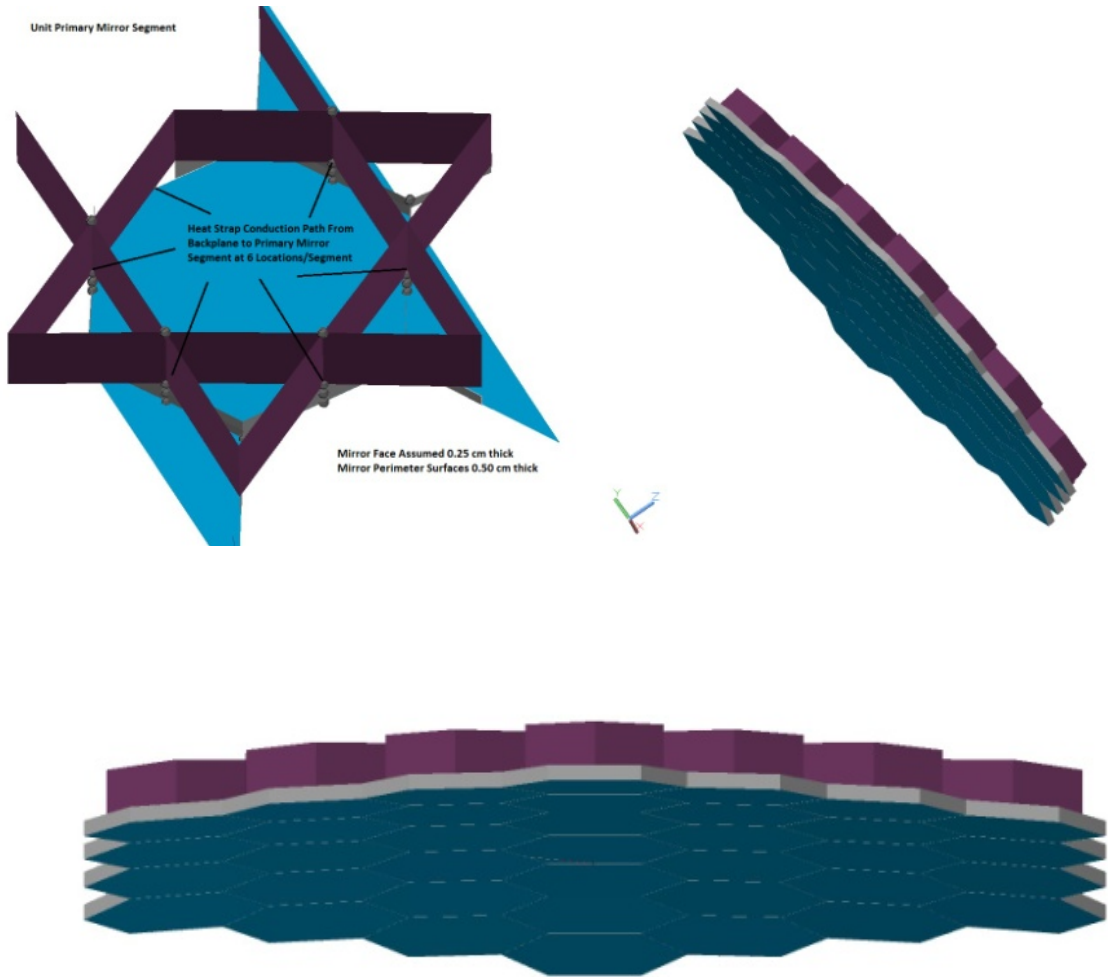

**Figure 3-60:** This illustration shows a section of the primary mirror backplane, indicating the locations of heat strap connections to a mirror segment.

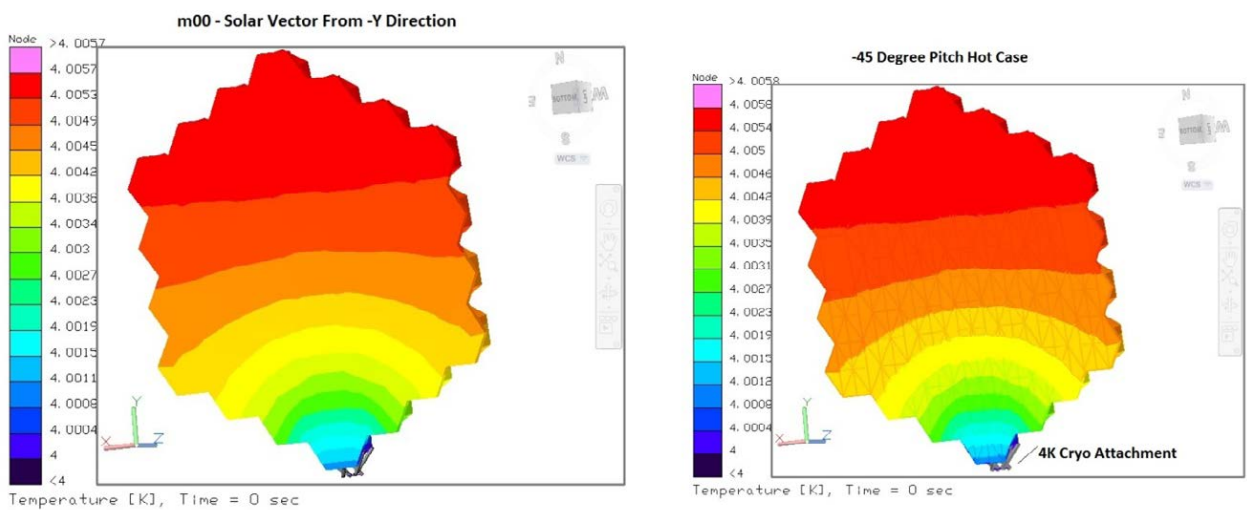

**Figure 3-61:** Predicted primary mirror temperature for two observatory pitch angles, 0 and -45 degrees.

### 3.3.2.3 Baffle

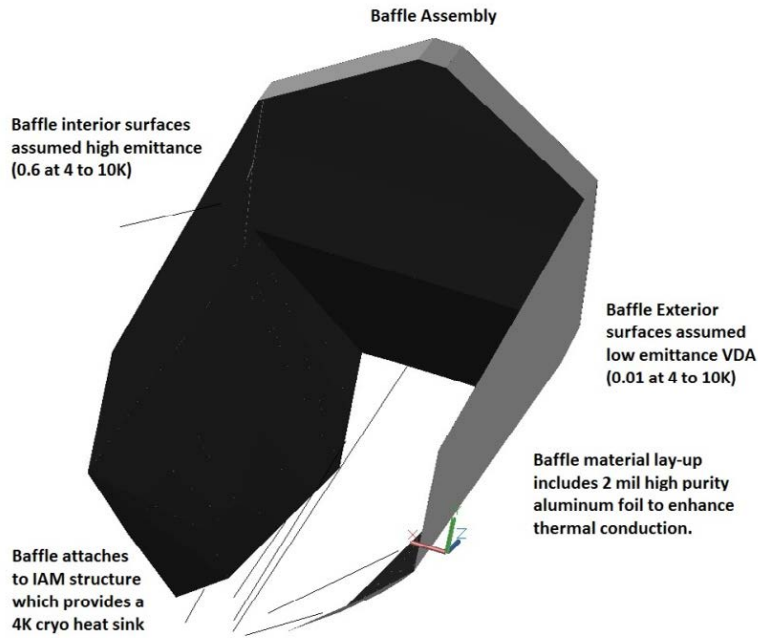

**Figure 3-62:** This illustration describes how the baffle is represented in the thermal model.

The baffle geometry is strategically devised to satisfy telescope and baffle view factor requirements. The baffle is thermal conductively coupled to the 4K IAM assembly structure. The baffle includes a 0.00508 cm (2 mil) thick aluminum foil layer which serves as the thermal conduit for baffle removal. It is assumed that the foil is thermally coupled along the entire interface with the IAM structure as shown in figure. Baffle surfaces viewing the telescope are assumed to be dark (high emittance) and surfaces viewing the sun shade are assumed to be low emittance (VDA).

**Figure 3-63** shows predicted baffle temperature contours for a +5 degree observatory pitch angle, indicating temperatures ranging from 4.0 to 4.7 K.

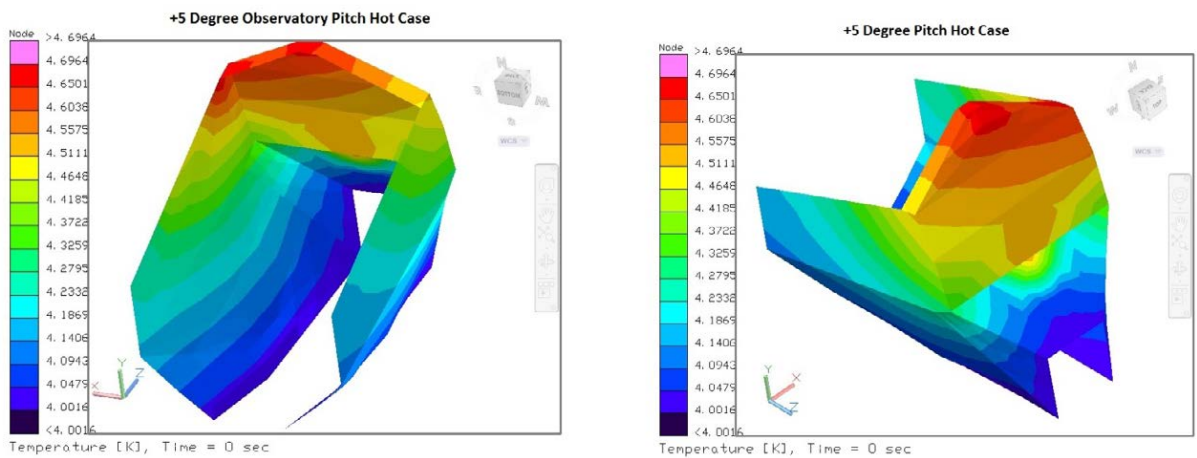

**Figure 3-63:** Thermal model predicted temperature contours for the +5 degree observatory pitch angle, shown from two perspectives, front and rear.

### 3.3.2.4 Deployed Tower Assembly (DTA)

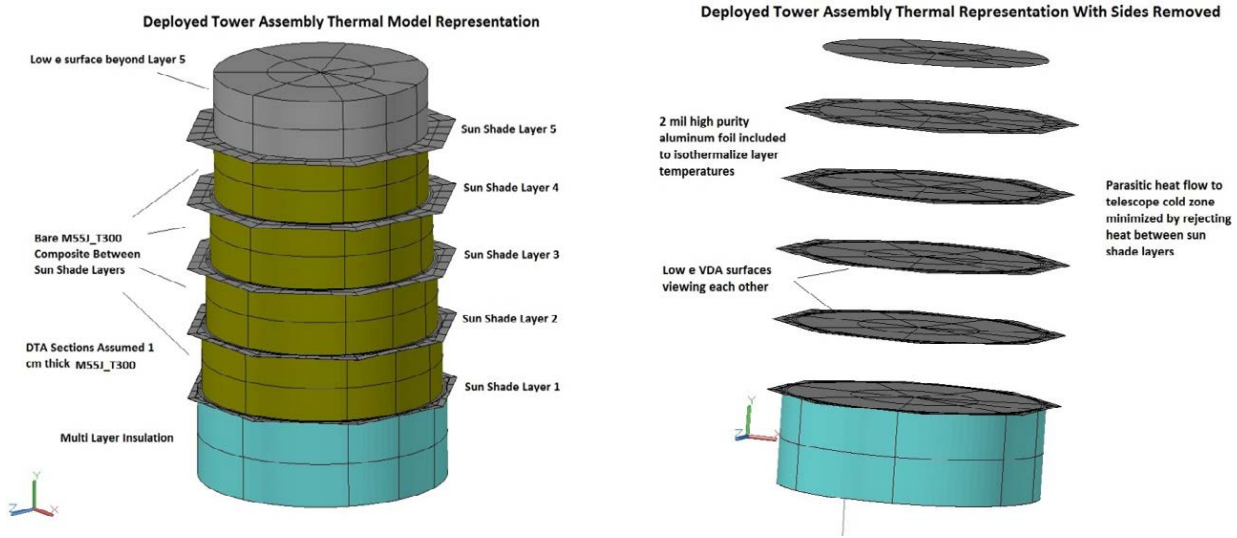

**Figure 3-64:** Thermal model predicted temperature contours for the +5 degree observatory pitch angle, shown from two perspectives, front and rear.

### Thermal Control Strategy

**Figure 3-64** states assumptions and illustrates how the DTA is represented in the thermal model. The overall system thermal objective is to minimize the parasitic heat load from the warm spacecraft to the cold telescope. Low conductivity composite materials minimize conductive heat transfer through the DTA to the telescope cold sections. The main structural cylinders are assumed to be 1 cm thick. Significant thermal resistance is assumed between these cylinders at the interfaces where the assembly deploys. MLI thermally decouples the DTA system from the environmental heat loads between Sun Shade Layer 1 and the spacecraft Bus. Thermal radiation to the cold sink between sun shade layers is promoted via bare composite (high emittance, 0.80 assumed) DTA surfaces to remove heat from the system. Thermal model results show this strategy to effectively reduce parasitic heat flow to the telescope Cold Zones. Internal to the DTA, low thermal emittance membranes separate each sun shade layer zone. These membranes include a 2 mil high purity aluminum foil for temperature equalization.

**Figure 3-65** shows predicted DTA temperatures, indicating a strong gradient between warm and cold zones.

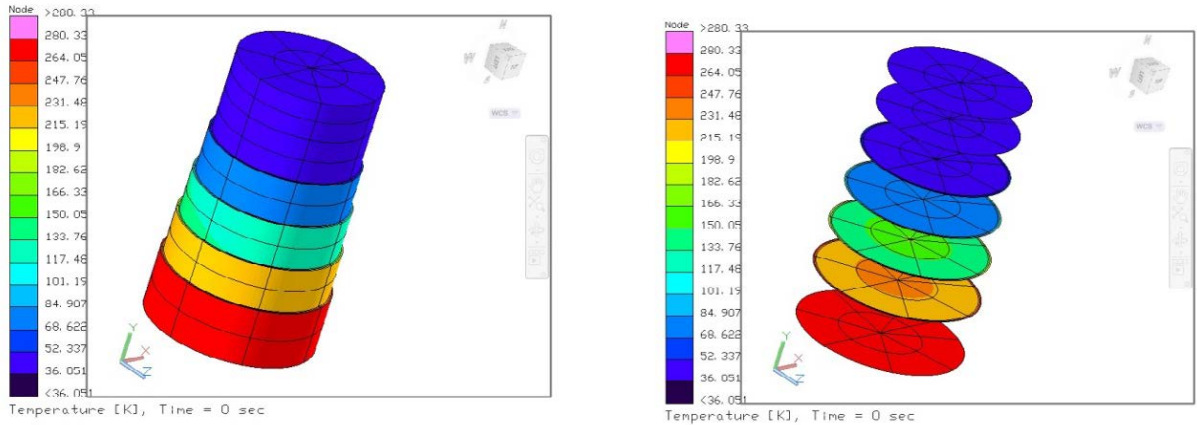

**Figure 3-65:** Thermal model predicted temperature contours for the +5 degree observatory pitch angle, shown from two perspectives, front and rear.

### 3.3.2.5 Spacecraft Bus

**Figure 3-66** shows the spacecraft thermal hardware components and their layout. Three science power dissipation modes seek to minimize the variation and magnitude of dissipated heat loads. Constant conductance aluminum ammonia heat pipes (CCHP's) are used to transport heat to Bus external panel radiators. The figure shows heat pipe routing and couplings. Dedicated CCHP's are routed directly to the MRSS base plate and to the cryo cooler mounts. Externally located CCHP's are also used to thermally couple some adjacent Bus panels. CCHP's are also assumed embedded within external panel radiators to increase radiator efficiency.

Spacecraft component distribution and a corresponding thermal block diagram are shown in **Figure 3-67**.

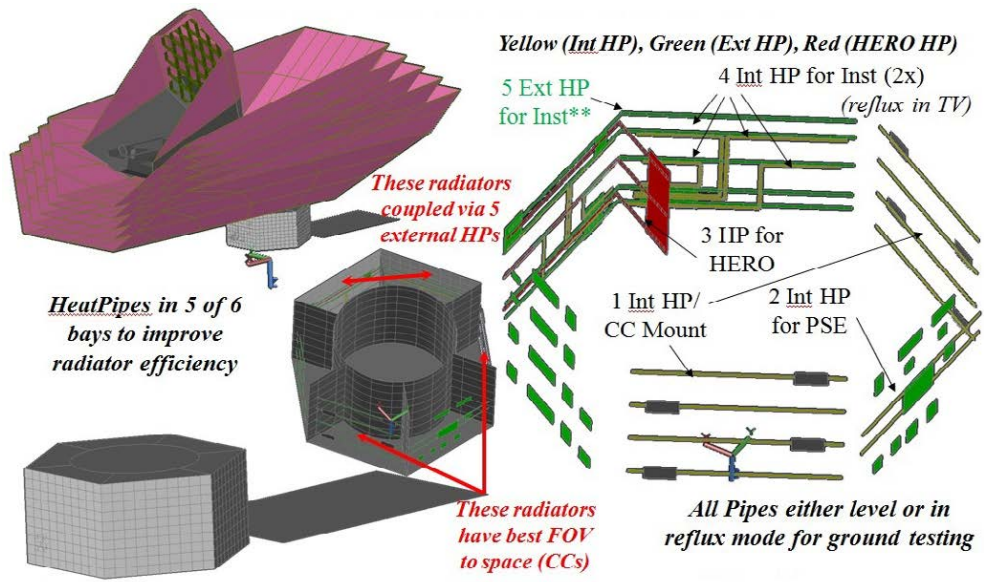

**Figure 3-66:** The spacecraft thermal hardware layout. The large spacecraft size provides sufficient radiating area to reject OST dissipated heat loads. Heat pipes, which transport heat efficiently, are employed to route heat from heat-dissipating components to the external radiators.

- -X,-Y Bay: 4 <u>cryocoolers</u> (4 each)
- -X Bay: PSE, CDH, SSR, Battery, ACE
- -X, +Y Bay: 4 <u>cryocoolers</u> (4 each)
- +X,+Y Bay: HRS, FIP
- +X,+Y Radial Panel: HERO
- +X Bay: MRSS, MISC
- +X,-Y Bay: <u>Comm</u>, ACS

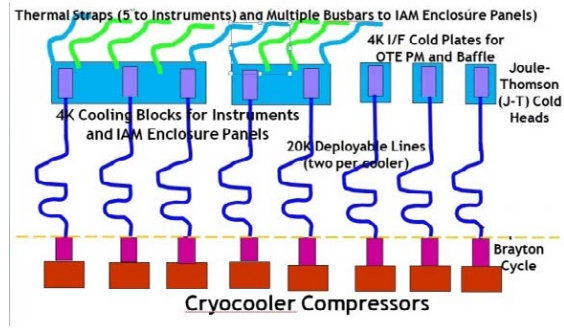

**Figure 3-67:** The spacecraft thermal block diagram shows how component dissipated heat loads are distributed to spacecraft external radiating panels.

The BUS structural central cylinder houses the propulsion tanks. MLI is located on the central cylinder to thermally isolate this zone.

**Figure 3-68** shows the spacecraft thermal model representation. **Table 3-10** lists assumed spacecraft component temperature requirements. **Table 3-11** shows thermal model results for the key design cases; temperatures are given for all components in the design hot and cold cases.

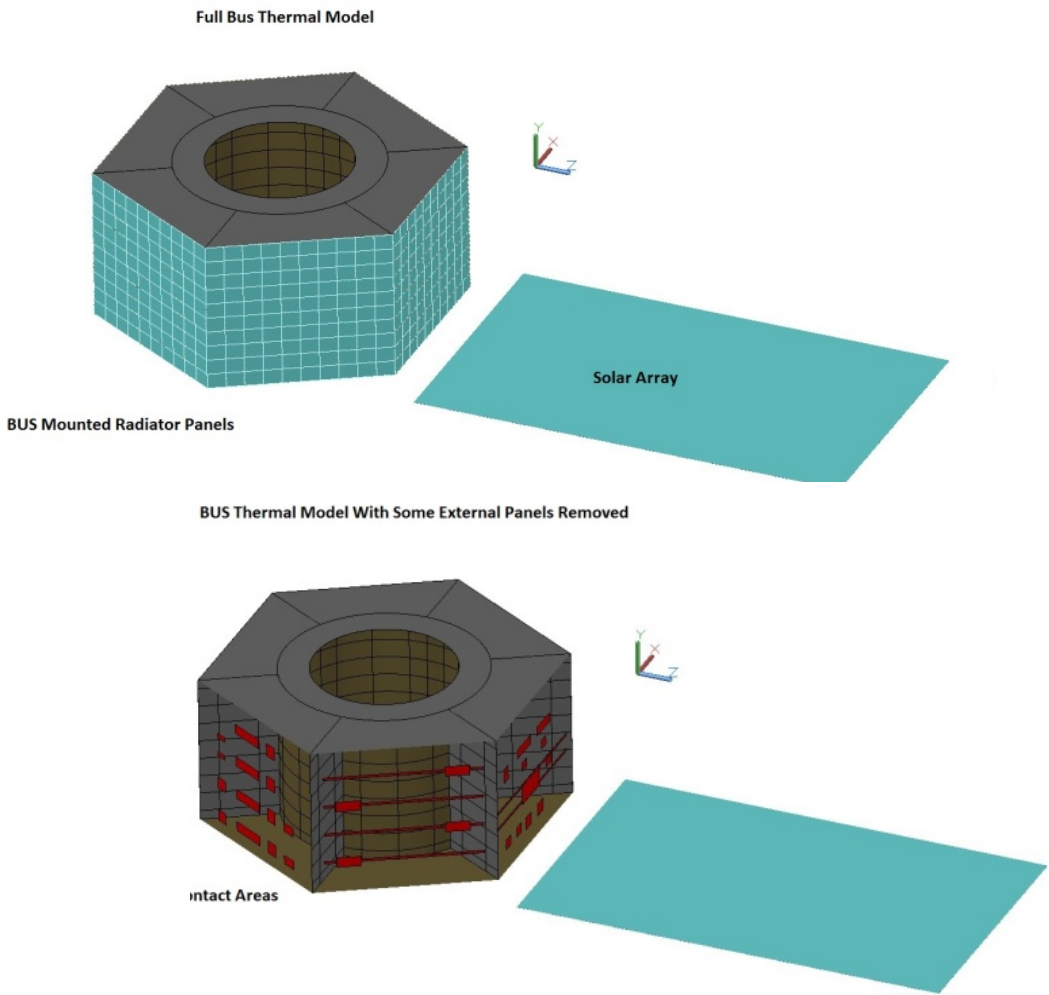

**Figure 3-68:** Spacecraft dissipated heat is transported to and rejected from externally located spacecraft radiating panels.

**Table 3-10:** Spacecraft

| Component Description | Survival Low | Operations Low | Operations High | Survival High |
|---|---|---|---|---|
| | C | C | C | C |
| **Instrument Components On Bus** | | | | |
| CryoCooler | -20 | -10 | 40 | 50 |
| FIP Electronics | -20 | -10 | 40 | 50 |
| HRS Electronics | -20 | -10 | 40 | 50 |
| HERO Electronics | -20 | -10 | 40 | 50 |
| MRSS Electronics | -20 | -10 | 40 | 50 |
| MISC Electronics | -20 | -10 | 40 | 50 |
| MISC Guider | -20 | -10 | 40 | 50 |
| **Bus Components** | | | | |
| Power Supply Electronics (PSE) | -20 | -10 | 40 | 50 |
| Battery | 0 | 10 | 30 | 40 |
| Command Data Handling 1(CDH1) | -20 | -10 | 40 | 50 |
| Command Data Handling 2(CDH2) | -20 | -10 | 40 | 50 |
| Attitude Control Electronics #1 (ACE1) | -20 | -10 | 40 | 50 |
| Attitude Control Electronics #2 (ACE2) | -20 | -10 | 40 | 50 |
| Solid State Recorder #1 (SSR1) | -20 | -10 | 40 | 50 |
| Solid State Recorder #2 (SSR2) | -20 | -10 | 40 | 50 |
| OTA Electronics | -20 | -10 | 40 | 50 |
| Solid State Inertial Reference Unit (SSIRU) | -20 | -10 | 40 | 50 |
| CMG-G | -20 | -10 | 40 | 50 |
| CMG-E | -20 | -10 | 40 | 50 |
| Reaction Wheels | -20 | -10 | 40 | 50 |
| Telescope ADU | -20 | -10 | 40 | 50 |
| SBand Transceiver | -20 | -10 | 40 | 50 |
| SBand Receiver #1 | -20 | -10 | 40 | 50 |
| SBand Receiver #2 | -20 | -10 | 40 | 50 |

**Table 3-11:** Component Thermal Model Predictions for Hot and Cold Design Cases

| Sheet: | All | Hot | Hot 5 Roll | Hot -5 Roll | Hot -5 Pitch | Hot 45 Pitch | All | Cold | Cold 5 Roll | Cold -5 Roll | Cold_ -5 Pitch | Cold 45 Pitch | Surv |
|---|---|---|---|---|---|---|---|---|---|---|---|---|---|
| Parameter: | **Max** | Max | Max | Max | Max | Max | **Min** | Min | Min | Min | Min | Min | Min |
| Node/Group | [°C] | [°C] | [°C] | [°C] | [°C] | [°C] | [°C] | [°C] | [°C] | [°C] | [°C] | [°C] | [°C] |
| **Instruments** | | | | | | | | | | | | | |
| FIP_ELEC | 33.8 | 28.5 | 31.5 | 28.0 | 33.8 | 6.3 | **-13.1** | 3.0 | 6.4 | 2.8 | 8.7 | -13.1 | -7.0 |
| HERO_ELEC | 32.4 | 27.4 | 30.4 | 27.1 | 32.4 | 6.0 | **-13.6** | 3.7 | 7.1 | 3.7 | 9.2 | -13.6 | -6.0 |
| HRS_ELEC | 31.4 | 26.0 | 29.0 | 25.5 | 31.4 | 1.3 | **-9.0** | 11.9 | 15.2 | 11.8 | 17.8 | -9.0 | -4.9 |
| MISC_ELEC | 33.1 | 27.8 | 31.0 | 27.3 | 33.1 | 3.8 | **-13.2** | 6.9 | 10.3 | 6.7 | 12.6 | -13.2 | -4.7 |
| MRSS_ELEC | 45.9 | 40.9 | 44.4 | 40.4 | 45.9 | 19.6 | **-16.4** | -0.7 | 3.2 | -0.9 | 4.9 | -16.4 | -8.5 |
| **Cryocooler** | | | | | | | | | | | | | |
| SC_CC_MOUNT | 49.3 | 35.1 | 41.9 | 38.3 | 35.3 | 49.3 | **-13.2** | -12.5 | -13.2 | -11.9 | -12.3 | 12.1 | -19.2 |
| **SC Avionics** | | | | | | | | | | | | | |
| SC_ACE | 17.7 | 15.3 | 16.3 | 15.5 | 15.0 | 17.7 | **2.8** | 2.8 | 3.5 | 3.7 | 2.9 | 6.0 | -5.1 |
| SC_ADU | 14.9 | 13.1 | 12.8 | 13.9 | 12.5 | 14.9 | **2.9** | 3.4 | 3.4 | 4.4 | 2.9 | 5.9 | -0.6 |
| SC_BAT | 25.5 | 18.7 | 19.5 | 19.1 | 18.3 | 25.5 | **9.1** | 9.3 | 9.9 | 9.9 | 9.1 | 17.0 | 1.2 |
| SC_CDH | 19.3 | 17.3 | 18.5 | 17.5 | 17.1 | 19.3 | **0.0** | 0.0 | 0.7 | 1.1 | 0.0 | 2.7 | -6.4 |
| SC_CMGE | 22.8 | 14.8 | 14.4 | 22.8 | 20.0 | -3.1 | **-17.8** | -9.0 | -9.2 | -0.7 | -4.6 | -17.8 | -23.3 |
| SC_CMGG | 28.4 | 20.6 | 20.1 | 28.4 | 25.6 | 2.0 | **-13.4** | -3.6 | -3.7 | 4.6 | 0.9 | -13.4 | -21.3 |
| SC_OTA | 35.2 | 28.2 | 28.5 | 35.2 | 32.7 | 14.6 | **-26.7** | -11.5 | -10.8 | -3.4 | -5.8 | -26.7 | -27.7 |
| SC_PSE | 31.1 | 23.9 | 24.6 | 24.4 | 23.5 | 31.1 | **14.4** | 14.6 | 15.2 | 15.2 | 14.4 | 22.6 | -1.7 |
| SC_RWE | 44.7 | 3.9 | 4.4 | 4.5 | 3.4 | 44.7 | **-11.4** | -11.0 | -11.0 | -10.0 | -11.4 | 31.5 | -18.0 |
| SC_SSIRU | -1.7 | -10.8 | -10.0 | -1.7 | -4.7 | -21.7 | **-33.2** | -22.1 | -23.5 | -13.6 | -16.9 | -33.2 | -27.6 |
| SC_SSR | 65.7 | 28.3 | 29.5 | 28.9 | 28.3 | 65.7 | **-3.8** | -3.3 | -3.0 | -2.9 | -3.8 | 36.9 | -14.3 |
| SC_TRANS | 25.6 | 18.6 | 18.1 | 25.6 | 23.7 | -1.3 | **-23.5** | -1.1 | -1.2 | 5.6 | 3.6 | -23.5 | -6.8 |
| **SC Panels** | | | | | | | | | | | | | |
| Panel1 | 46.8 | 32.6 | 39.3 | 31.0 | 32.8 | 46.8 | **-37.8** | -37.2 | -27.9 | -37.8 | -37.2 | -9.2 | -41.9 |
| Panel2 | 43.9 | 38.8 | 42.4 | 38.4 | 43.9 | 17.6 | **-31.6** | -17.5 | -9.9 | -17.6 | -12.1 | -31.6 | -23.3 |
| Panel3 | 34.7 | 29.5 | 32.6 | 29.1 | 34.7 | 7.7 | **-37.4** | -4.3 | -4.3 | -3.9 | 0.8 | -37.4 | -9.9 |
| Panel4 | 33.1 | 26.1 | 26.4 | 33.1 | 30.6 | 12.4 | **-38.6** | -27.7 | -28.7 | -18.7 | -22.2 | -38.6 | -34.3 |
| Panel5 | 43.6 | 28.6 | 27.3 | 35.4 | 28.8 | 43.6 | **-32.9** | -32.4 | -32.9 | -22.6 | -32.2 | -6.8 | -36.6 |
| Panel6 | 64.5 | 27.2 | 28.3 | 27.7 | 27.1 | 64.5 | **-26.9** | -26.6 | -25.6 | -26.2 | -26.9 | -15.5 | -31.7 |

### 3.3.2.6 Sun Shade

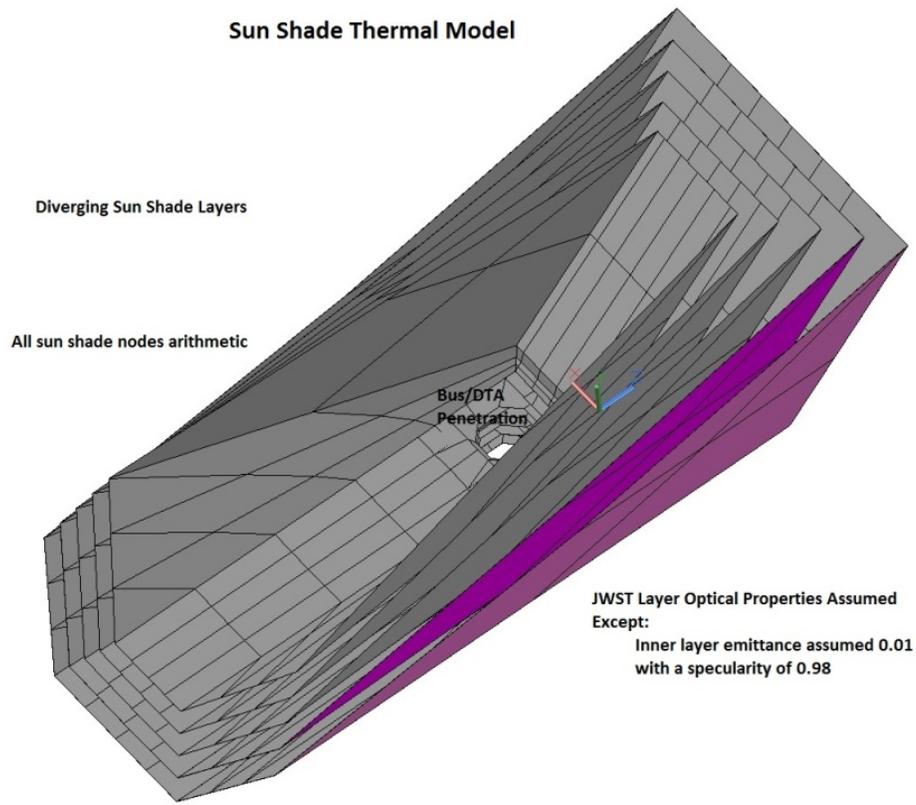

**Figure 3-69:** This sun shade thermal model illustration depicts the model geometry and shows key assumptions.

**Figure 3-69** shows the sun shade thermal model geometry. The optimal sun shade geometry minimizes the number of radiation bounces between sun shade layers while satisfying solar and lunar view angle exclusion requirements. The number of layers is an important parameter. We elected 5 sun shade layers. Improved thermal performance is realized as the spacing between layers is increased (75 cm). Diverging sun shade layers also improves thermal performance since the net radiation bounces to the space sink are reduced.

A computer program was written to generate sun shade and baffle geometry based on selected coordinates on the IAM/Baffle/Telescope assembly geometry provided by the mechanical team. A process was developed that enabled the sun shade geometry to be iterated for different input assumptions. The diagram shows the geometry selected for this study.

**Figure 3-70** shows the sun shade geometry parameters assumed for this study. Three divergent angle axes as shown on figure were selected. Angle A and Angle B were set to 1.5 degrees and Angle C was set to 1.0 degrees.

**Figure 3-71** shows temperature contours predicted by the sun shade thermal model for observatory pitch angles +5, 0, -15, and -45 degrees.

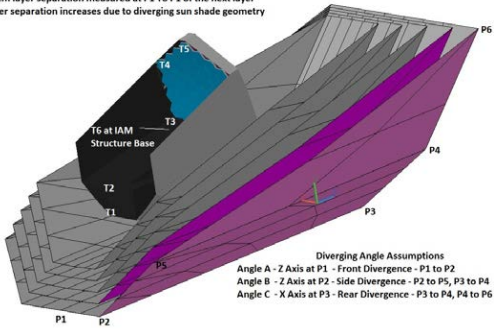

| | Input Data: | |
|---|---|---|
| | 35.0 | \|Front Solar Design Rotation Angle Measured From Z Plane |
| | 75.0 | \|Sun Shade Layer Spacing (cm) |
| | 1.1 | \|P2 X Axis Dist Multiplier |
| | 75.0 | \|Rear_Outer_Shade_Z_Coord |
| | 35.0 | \|Front Angle Viewed From Negative X  0-90 deg  where 90 is Perpendicular |
| | 44.0 | \|Rear Angle Viewed From Negative X  0-90 deg  where 90 is Perpendicular |
| | 45.0 | \|Angle (Degrees) Viewed From - Z  0-90 deg where 90 is Perpendicular |
| Sunshade Layer | | Divergent Angles: |
| | 1.50 | \|Divergance Angle For P1 Down center of bottom in Z direction |
| | 1.50 | \|Divergance Angle For P2 down -X side of shield bottom in Z direction |
| | 1.00 | \|Divergance Angle For P3 at the Angle Shade Layer in X Direction |
| | 0.00 | \|Divergance Angle for P6 at top on upper tip in Z direction |

**Figure 3-70:** The sun shade was optimized by iterating layer divergence angles and spacing.

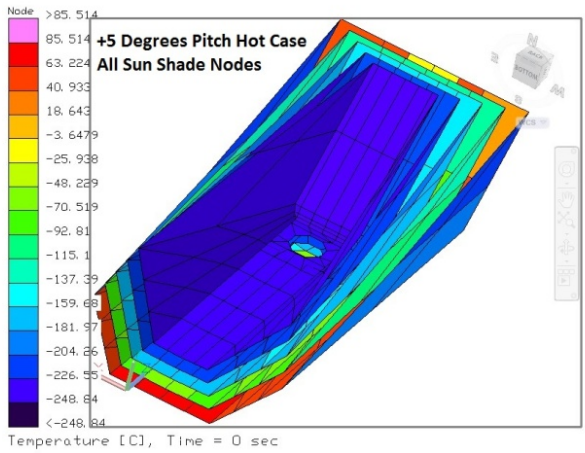

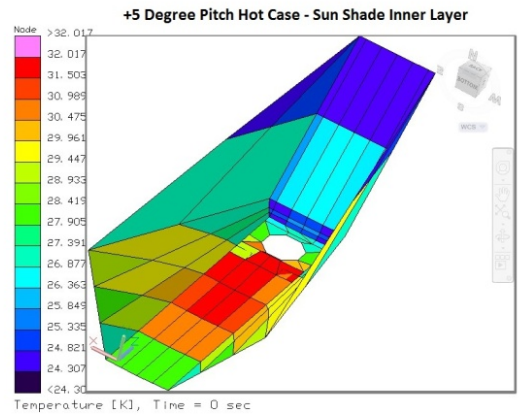

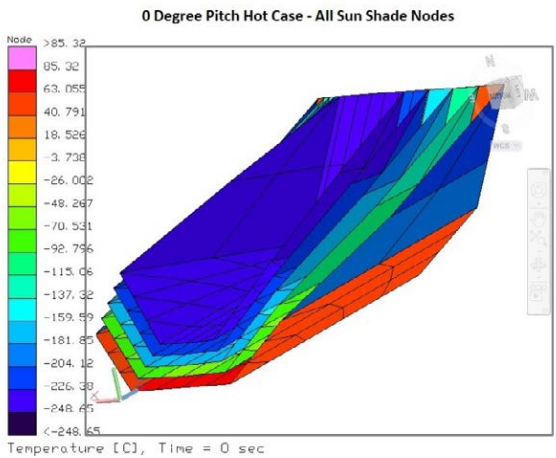

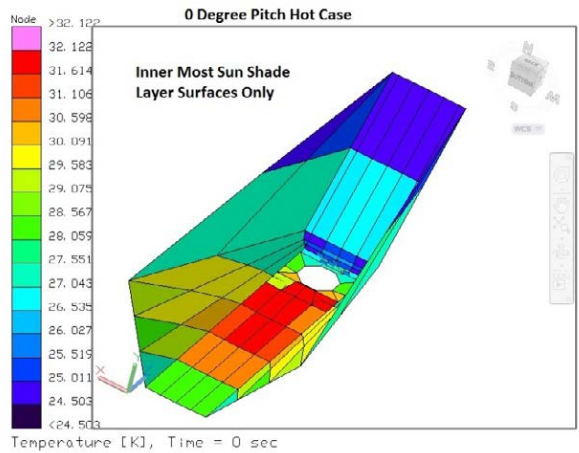

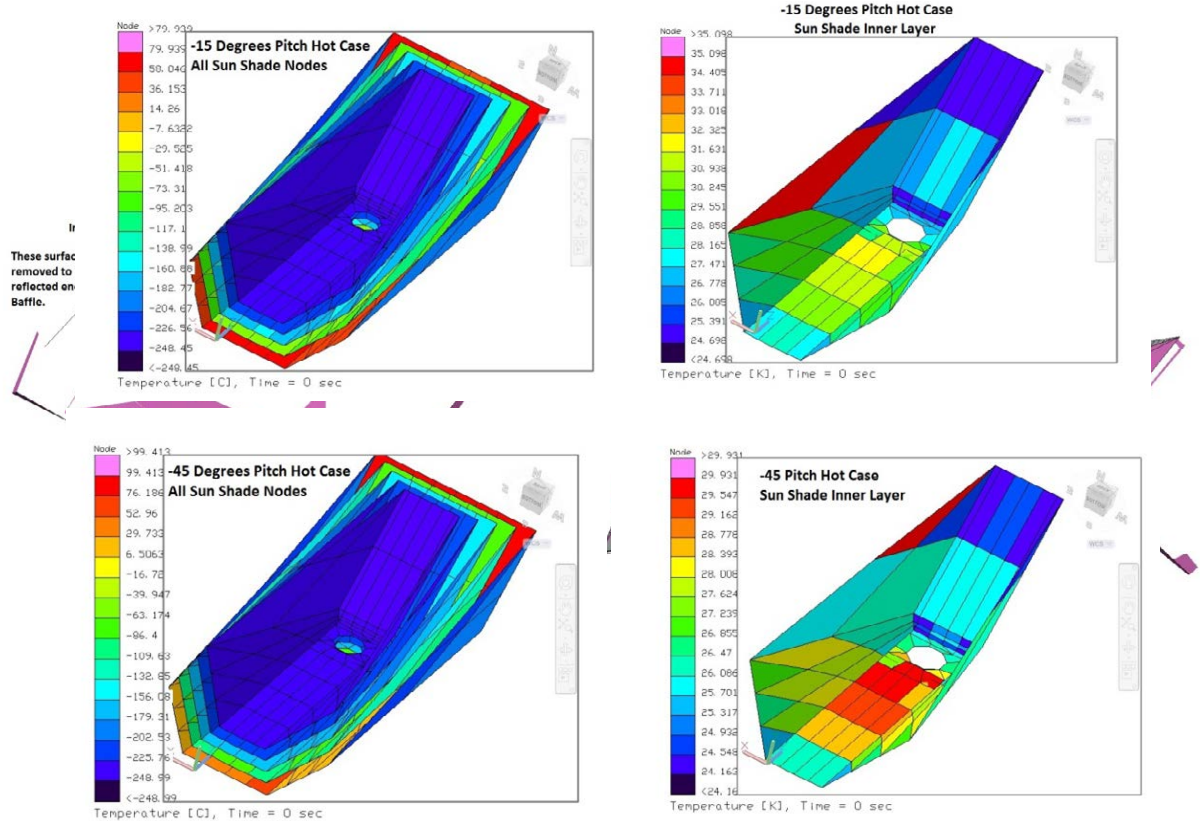

**Figure 3-71:** The thermal model predicts temperatures for the entire sun shade assembly. Right-hand panels show the temperatures of the shade layer closest to the cold zone on a stretched temperature scale.

### 3.3.2.7 Sun Shade Boom Deployment Mechanism

#### Thermal Control Strategy

**Figure 3-72** illustrates the sun shade boom deployment mechanism thermal model representation. The thermal model assumes that both sides of the Sun Shade Boom Deployment Structure have a low alpha high emittance silicon oxide thermal coating. The sun shade layer 1 thermal optical properties were assumed. It was also assumed to be 0.25 cm thick composite for purposes of calculating thermal conductances. A more thermal conductive material would reduce maximum predicted temperatures.

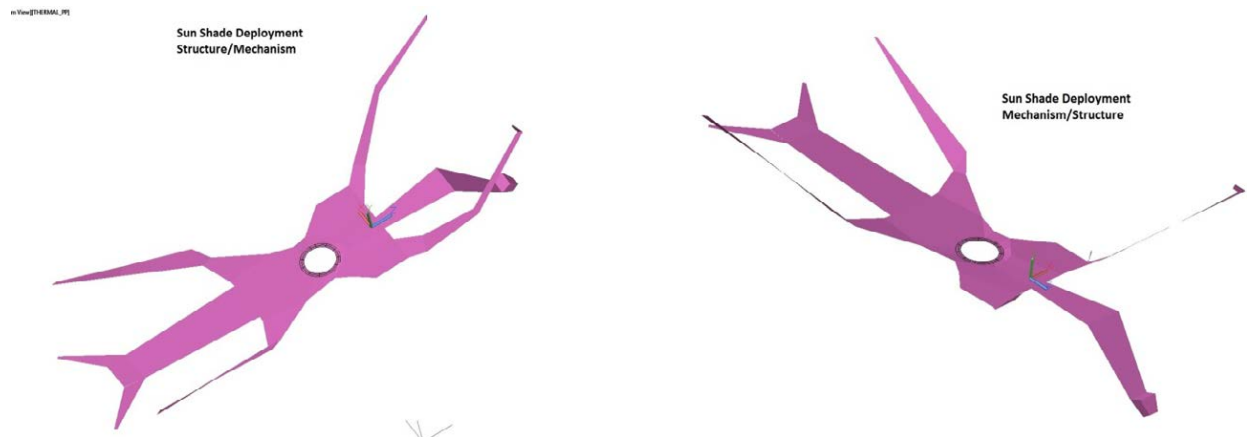

**Figure 3-72:** Thermal model representation of the sun shade deployment mechanism.

**Figure 3-73** shows predicted boom assembly temperatures for various modeled cases.

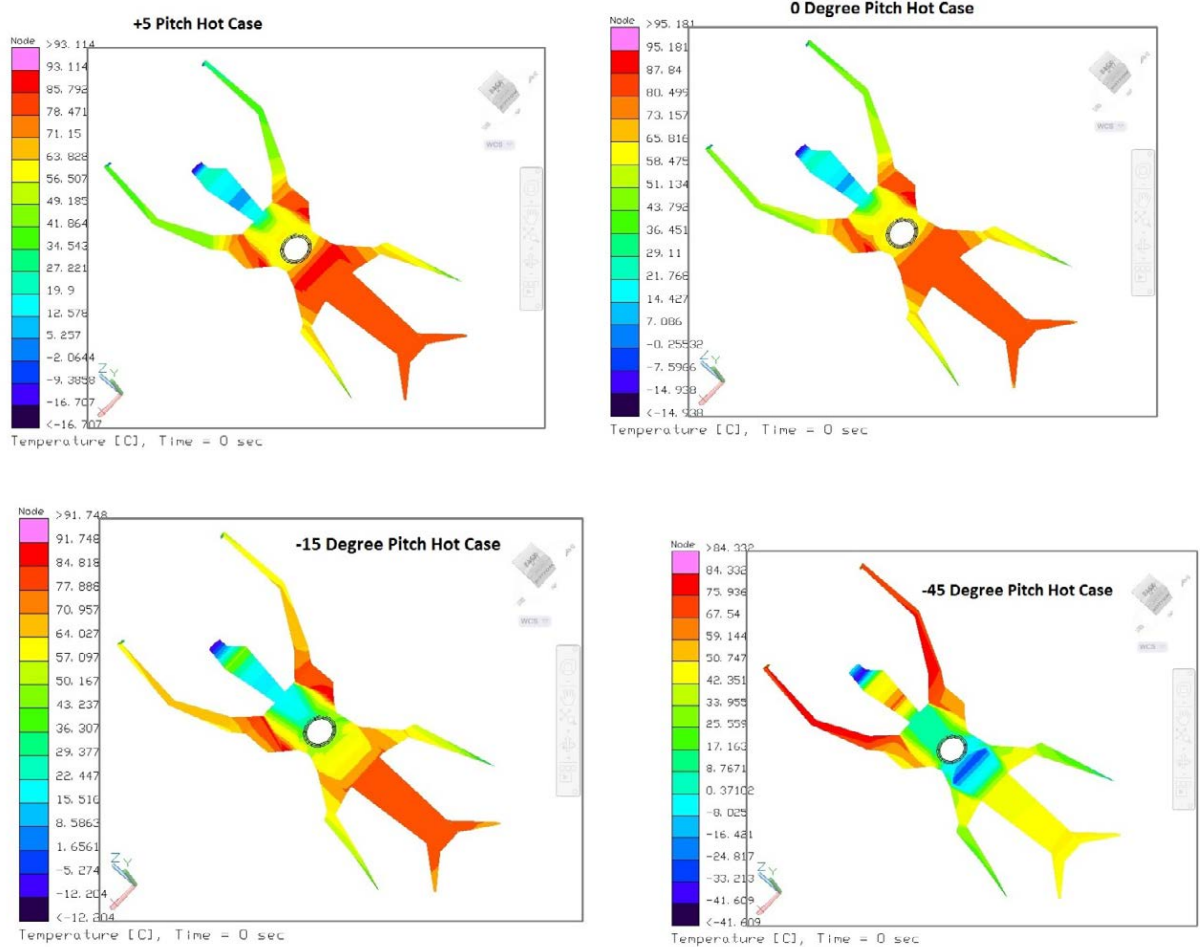

**Figure 3-73:** Sun shade deployment mechanism temperature predictions. Several different observatory pitch angles were modeled.

### 3.4 Observatory Mass and Power Summaries

Observatory mass and power summaries are provided in **Figure 3-24**. The observatory CBE wet mass is 29,926 kg, and the MEV wet mass is 36,634 kg. Observatory CBE power is 12,657 Watts and MEV power is 15,493 Watts. The nominal science mode operational power is ~9,226 Watts.

The observatory point design is preliminary at this early stage of development; hence, the design and materials selected are yet to be optimized. Optimization of the design and materials will reduce the mass and power. The conservative value of launch vehicle margin is assumed, which results in a 20% launch vehicle margin. The SLS performance is provided as a range at this juncture. If this point design were optimized; then it is expected that the launch vehicle margin of at least 30% could be achieved.

### 3.5 Observatory Integration and Test

The OST element- and observatory-level Integration and Testing (I&T) programs include the Instrument Accommodation Module (IAM), Optical Telescope Element (OTE), and Spacecraft Element (SCE). The observatory architecture consists of a cold zone (telescope and instruments operating at cryogenic temperatures) and a hot zone (spacecraft bus). The sun shield is located between the hot and cold zones.

### 3.5.1 I&T Overview

The major subassemblies of the fully-assembled OST observatory are the OTE and IAM (OTIAM) which consist of:

- Optical Telescope Element (OTE) with Primary Mirror (PM)
- Instrument Accommodation Module (IAM), which houses five instruments, the secondary mirror, tertiary mirror, and field steering mirror

Instruments:

- Medium Resolution Survey Spectrometer (MRRS) – JPL
- Hi Res (Far-IR) Spectrometer (HRS) – GSFC
- HERO – CNES
- FIR Imager/ Polarimeter (FIP) – GSFC
- MID-IR Imager Spectrometer/ Coronagraph (MISC) – JAXA

Spacecraft Element (SCE):

- Spacecraft bus, including the typical spacecraft bus subsystems (power, attitude control, communications, C&DH)
- Deployable Tower Assembly (DTA)
- Sun shield
- Thermal Radiator
- Solar Array
- Cryocoolers
- Instrument Warm Electronics

**Figure 3-74** shows the summary level I&T flow.

Verification of OST thermal performance by test with the observatory fully assembled would be far more challenging than for a typical spacecraft design due to the large size of the primary mirror and sunshield and the wide operational temperature gradient. As a result, the OST team will conduct separate thermal vacuum tests of the observatory's cold and hot zones. Protoflight level environmental tests will be performed at the element level and acceptance level dynamics tests will be performed at the observatory level. Simulation of missing interfaces during tests of the lower levels of assembly and testing of various non-flight models provide the data for final analysis and verification of overall observatory performance.

Commonality of test personnel and products across each level of assembly, including test sets, command and telemetry system/database, procedures, and automated test scripts will provide test consistency across levels of assembly, enable test product reuse, and reduce cost and schedule risk. Similarly, mechanical ground support equipment for flight hardware lifting, rotations, and deployments will be also used for mechanical operations and testing.

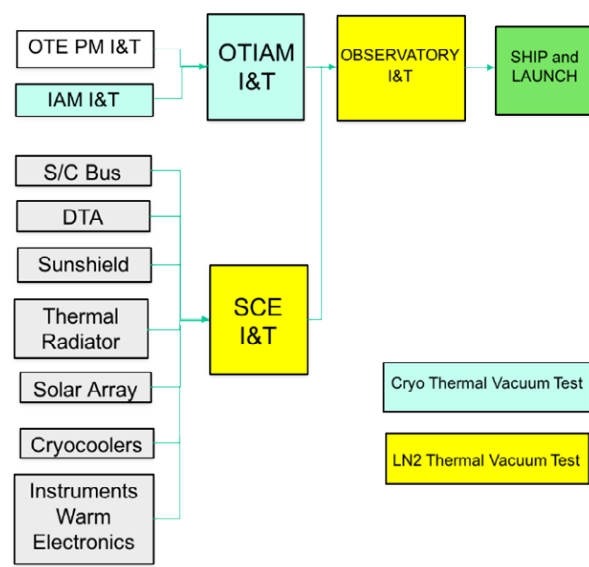

**Figure 3-74:** OST I&T flow is based on proven heritage processes from similar successful missions

Contamination control is another key process driver during OST I&T. External optic cleanliness can only be maintained through use of ISO Class 7 (Class 10 K) facilities and personnel protocols throughout all phases of I&T.

Testbeds, pathfinders, developmental test articles, mock-ups, and simulators reduce risk, provide a venue for tests that cannot be performed on the flight hardware, or simulate flight interfaces that are not available for a particular flight test. OST will utilize many of these proven options in addition to its flight I&T program.

The team will build a scale model of the sunshield, which will be cryo tested to validate thermal modeling and performance at operational temperatures. The reduced size will enable testing in an existing thermal vacuum chamber. Thermal modeling is especially important for OST since the flight sunshield will not be thermally tested in a deployed configuration.

An Engineering Model Test Bed (EMTB) will integrate engineering model electrical units together with flight-like harness and a duplication of the I&T electrical test environment. The EMTB is used to perform validation of flight software, as well as ensure the electrical compatibility of the observatory electronics. EMTB is a shared resource used by systems engineering to validate flight design objectives and to conduct test process demonstration during I&T.

The team will build a variety of spacecraft simulators and deliver them to lower levels of assembly. The simulators will provide a flight-like command and telemetry interface for testing "like-you-fly" before the instruments, IAM, or OTIAM are mated with the flight spacecraft. Spacecraft simulators will be used in conjunction with a common command and telemetry system and database at all levels of assembly to facilitate reuse of test products at each level of assembly.

### 3.5.2 Instruments Accommodation Module I&T

The IAM structure, which is comprised of an aft (AFT) and a forward (FWD) section (**Figures 3-75** and **3-76**), houses five instruments, the secondary mirror, tertiary mirror, and field steering mirror. The IAM instrument layout and installation approach is shown in **Figure 3-75**.

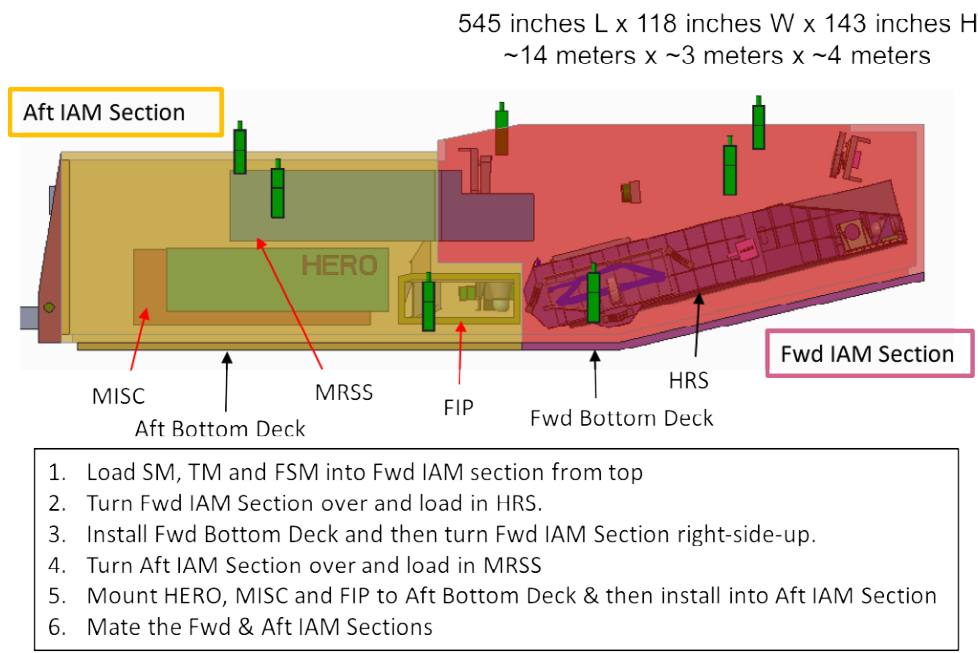

545 inches L x 118 inches W x 143 inches H
~14 meters x ~3 meters x ~4 meters

1. Load SM, TM and FSM into Fwd IAM section from top
2. Turn Fwd IAM Section over and load in HRS.
3. Install Fwd Bottom Deck and then turn Fwd IAM Section right-side-up.
4. Turn Aft IAM Section over and load in MRSS
5. Mount HERO, MISC and FIP to Aft Bottom Deck & then install into Aft IAM Section
6. Mate the Fwd & Aft IAM Sections

**Figure 3-75:** IAM instrument layout and installation approach is based on the team's experience with previous successful large telescope missions.

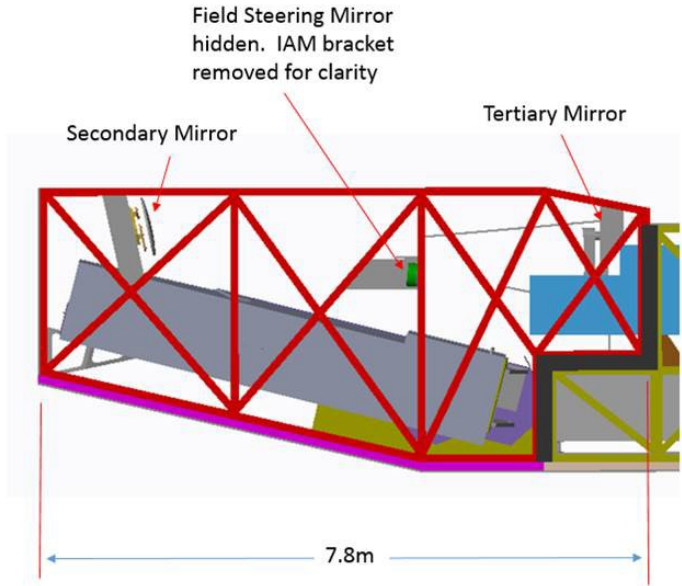

**Figure 3-76:** The IAM FWD section houses the optics (secondary, tertiary, and field steering mirrors).

IAM integration can be accommodated in existing facilities, including GSFC's 10/120 Whitehouse Cleanroom. The Whitehouse Cleanroom facility's cleanliness levels will be improved from 100 k to 10 K for OST. The IAM AFT and FWD sections are integrated in parallel in the same cleanroom. During integration, the team performs protoflight vibration and acoustics tests, in addition to Electromagnetic Interference (EMI)/Electromagnetic Compatibility (EMC) testing. Then cryo-vacuum testing is performed. Pre- and post-ambient functional testing is performed for these tests. OST also undergoes ambient metrology pre- and post-testing to ensure the hardware remains aligned. **Figure 3-77** shows the summary IAM I&T flow.

The environmental test campaign (except for the Cryo Thermal vacuum test) can be accommodated at GSFC using the large vibe table built for JWST, the acoustic test facility, and a customized EMItent. The cryo vacuum (CV) test can be accommodated at JSC's Chamber A. A specialized In Plant Transporter (IPT) would need to be designed and manufactured to move the OST flight hardware from the cleanroom to the vibration and acoustics facility. This IPT would require a portable clean tent to provide the necessary cleanliness during IAM transport.

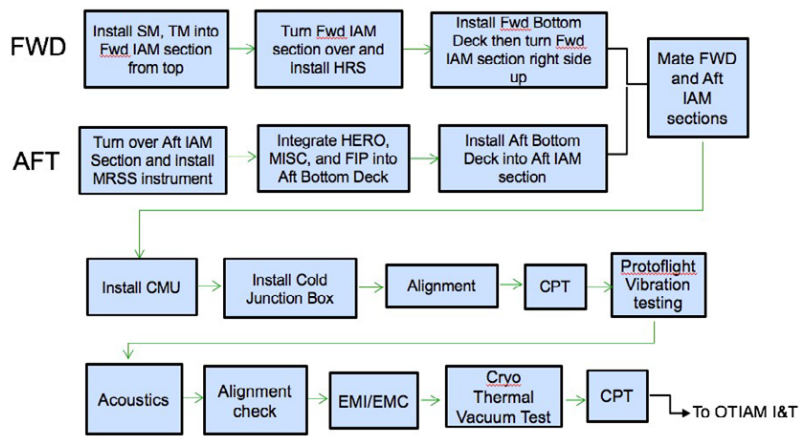

**Figure 3-77:** The IAM I&T summary flow is based on proven processes from missions with similar cleanliness requirements.

### 3.5.3 Optical Telescope Element with Primary Mirror

OTE I&T can begin with assembly of the structural subassemblies and deployment mechanisms at GSFC in the Spacecraft Systems Development and Integration Facility (SSDIF) (**Figure 3-78**). Flight harnesses and thermal components are installed. "Zero-g" offloaded deployment tests are performed on the primary mirror wings. A rollover fixture is used to orient the structure as required for access for assembly and optimal orientation for deployment tests. Then optics installation begins. All primary mirror assemblies are installed and aligned to the OTE structure. The completed OTE assembly is then ready for integration with the IAM at the OTIAM level of assembly. **Figure 3-79** shows the summary OTE I&T flow.

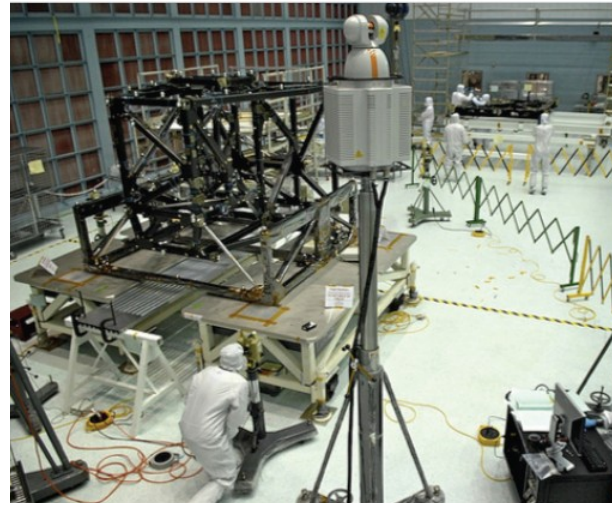

**Figure 3-78:** The SSDIF High Bay Cleanroom accommodates large structures like the OST OTE.

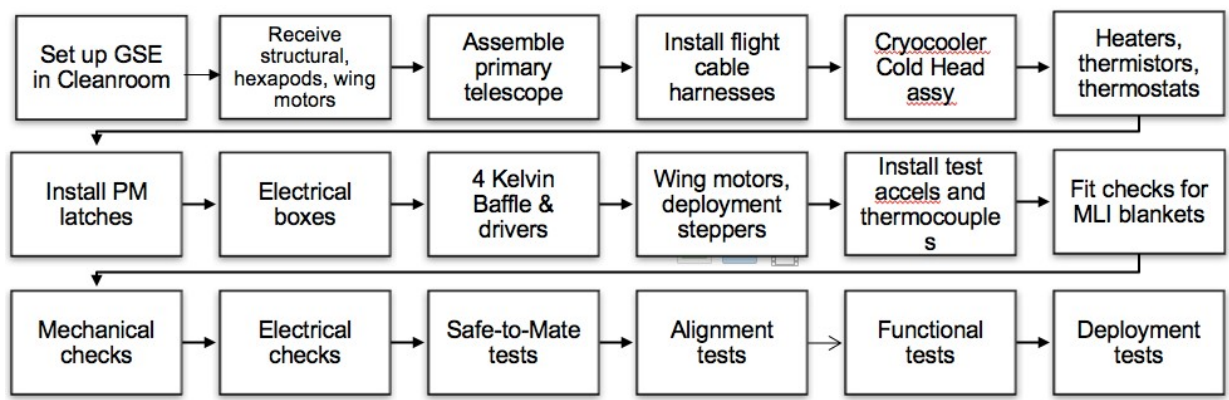

**Figure 3-79:** The summary OTE I&T flow is based on proven processes from JWST mission.

### 3.5.4 OTIAM I&T

The primary objective of the I&T program is to successfully deliver a fully tested and verified integrated Optical Telescope Element with Primary Mirror and Instrument Accommodation Module (OTIAM) that meets all functional requirements and performs as designed when placed on the SCE to create the full OST observatory. **Figure 3-80** shows the OTIAM I&T flow.

OTIAM integration can occur in GSFC's SSDIF (**Figure 3-78**). To accomplish the integration of all components will require developing specialized GSE. After ambient integration, the OTIAM is ready for mechanical testing that includes protoflight acoustics and sine vibe testing. Prior to mechanical testing, pre-environmental system functional testing is performed. A series of mirror and metrology measurements are made, and then the OTIAM is ready for acoustics and sine testing. A specialized In Plant Transporter (IPT) will need to be designed and manufactured to move the flight hardware from the cleanroom to the vibration and acoustics facility. This IPT will require a portable clean tent around OTIAM that provides the necessary cleanliness. After protoflight acoustics and sine testing, the hardware is returned to the cleanroom. The team makes post-environmental system functional tests, deployment tests, and metrology measurements. The hardware is then prepped for transportation and moved to JSC. Upon arrival at JSC, the flight OTIAM is removed from the transporter and prepped for installation into Chamber A (**Figure 3-81**). Preparations include deployments, sensor installation, metrology, and ambient functional testing. Once the hardware is installed into JSC Chamber A, OTIAM cryogenic testing can begin. The objective of cryogenic vacuum testing is to verify OTIAM-level requirements in the conditions of the expected flight environment, with emphasis on optical measurements that can be performed in this test configuration. The optical tests will verify OTIAM system optical workmanship, and provide optical test data to support integrated telescope modeling used to predict flight optical performance. After testing, OTIAM is removed from the chamber and prepped for transportation. The hardware is transported to GSFC, where it is mated to the SCE to create the OST observatory.

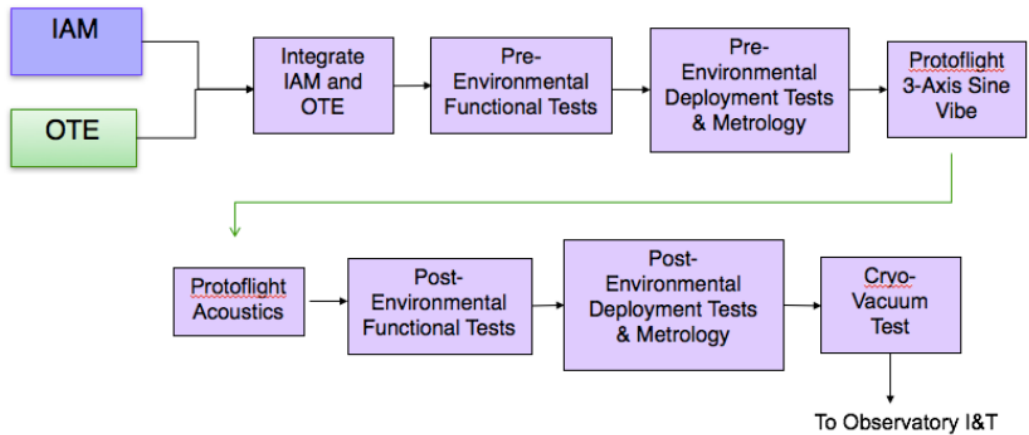

**Figure 3-80:** The Observatory OTIAM I&T flow is based on processes from previous successful large telescope missions, including JWST.

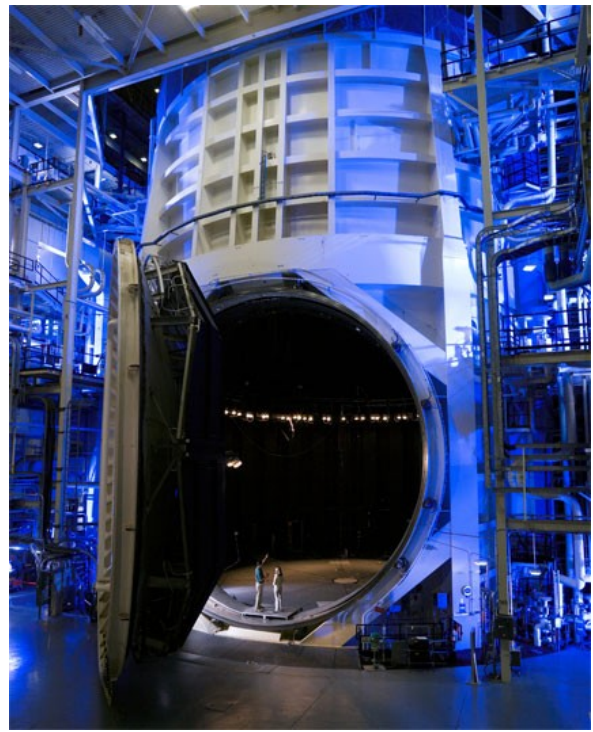

**Figure 3-81:** Chamber A at JSC can accommodate the OST OTIAM for cryogenic testing.

### 3.5.5 Spacecraft Element I&T

Spacecraft Element (SCE) I&T, which can be performed at GSFC's SSDIF, consists primarily of the spacecraft bus (including the typical spacecraft bus power, attitude control, communications, and C&DH subsystems), deployable tower assembly (DTA), sunshield, thermal radiator, solar array, cryocoolers, and instrument warm electronics. The objective of the SCE I&T is to deliver a fully verified and tested element that meets all functional requirements prior to integration with the OTIAM at the start of observatory I&T. **Figure 3-82** shows a summary SCE I&T flow.

SCE integration begins with propulsion system integration, followed by harness integration and thermal components and electronics assembly. A complete electrical integration of all bus- mounted electronics is performed in a "flat sat" configuration. Electrical equipment that will ultimately be flight-mounted to the bus structure is temporarily mounted to a ground support table to enable full flight electrical integration of the complement of bus-mounted equipment. Electrical system integration allows typical issues to be discovered during "flat sat" I&T, which is performed independently of the bus structure critical path. Major tests during SCE I&T include an initial ambient baseline electrical Comprehensive Performance Test (CPT–1), alignments, ambient deployments, RF compatibility, EMC/EMI, launch vehicle separation shock, acoustics, sine vibration, and thermal vacuum, followed by post-environmental functional and deployment testing. The TVAC test is performed without the solar array and sunshield. This testing can be accommodated at GSFC's Space Environment Simulator (SES) (**Figure 3-83**). IAM and OTE interfaces are electrically-simulated during functional and environmental testing. All environmental tests are performed at protoflight levels and a fully-tested and -verified SCE is ready for mate with the OTIAM in observatory I&T.

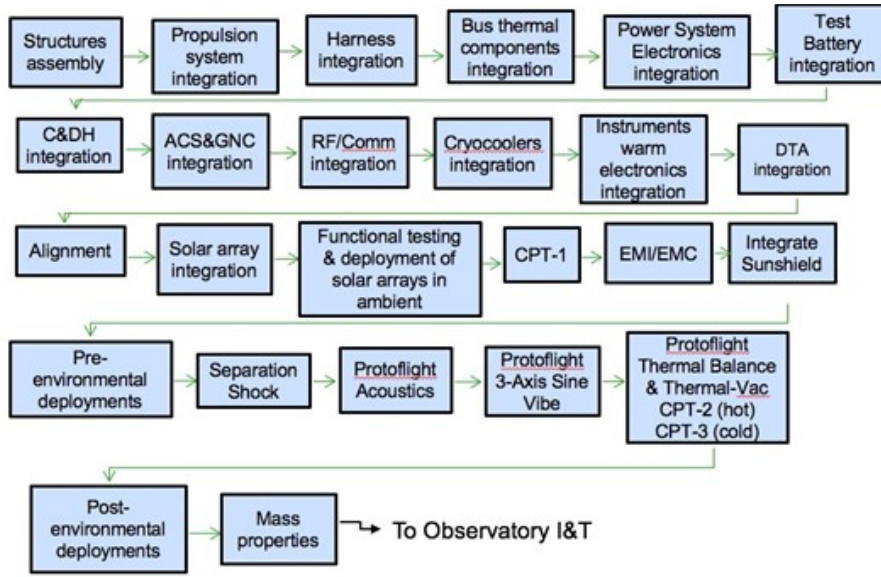

**Figure 3-82:** The summary SCE I&T flow has heritage from previous successful programs.

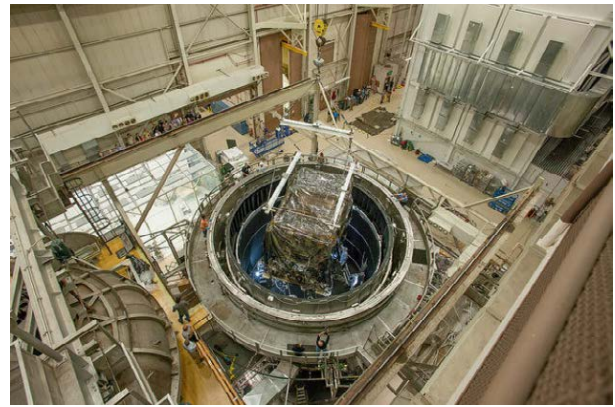

**Figure 3-83:** GSFC clean room facility, where hardware is mated to the SCE to create the OST observatory.

### 3.5.6 Observatory I&T and Summary

Observatory I&T integrates the fully-qualified OTIAM with the fully-qualified SCE. Full electrical and software compatibility across the interface has been previously validated on the EMTB. Previous testing at the OTIAM and SCE levels utilized high fidelity simulators to validate the interfaces through ambient and protoflight level environmental testing. Observatory I&T consists of ambient functional testing (CPT-4), alignments, and deployment testing to demonstrate the full end-to-end observatory functionality and compatibility, electrically and mechanically. EMI/EMC, Separation Shock, and acceptance-level acoustics and sine vibe tests are performed to verify the fully-assembled structure and post-environmental functional (CPT-5), deployments, and mass properties tests. Finally, the observatory is stowed in launch configuration and prepared for transport to the launch site. **Figure 3-84** shows the Observatory I&T flow.

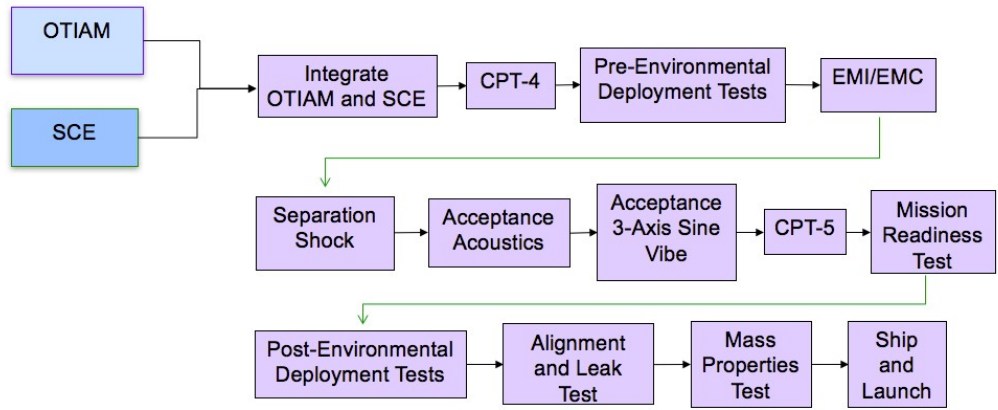

**Figure 3-84:** The summary Observatory I&T flow is based on processes from previous successful large telescope missions, including JWST.

OST I&T planning and flow benefit from incorporating proven JWST documentation, metrology, tools, and lessons learned. The large size of the deployed OTE and sunshield, combined with extreme temperature difference from the hot to cold zones, makes thermal testing of the full observatory impractical. Instead, the cold zone elements are processed through a separate protoflight test program with cryogenic thermal vacuum testing and the hot zone is processed through its own protoflight test program that is analogous to a typical spacecraft I&T flow. Testing thermal simulators, such as a scaled sunshield, is used to validate the thermal modeling and provides the ultimate verification of a deployed observatory and addresses aspects of the flight system not tested in full flight configuration. Based on heritage processes, OST implements many other testbeds and simulators to address cost/schedule risk and interface verification. Facilities to perform the complete OST I&T program already exist and are owned by NASA.

### 3.6 Launch Vehicle

OST Concept 1 baseline launch vehicle is the Space Launch System (SLS) with an 8.4-meter fairing (**Figure 3-85**). The SLS rocket is being developed by NASA for its maiden launch in November 2018. The 8.4-m fairing is expected to be available by 2022, or more than a decade before it is needed by OST. It is anticipated the SLS will be flight proven, with a number of launches occurring well before OST's September 2035 launch date.

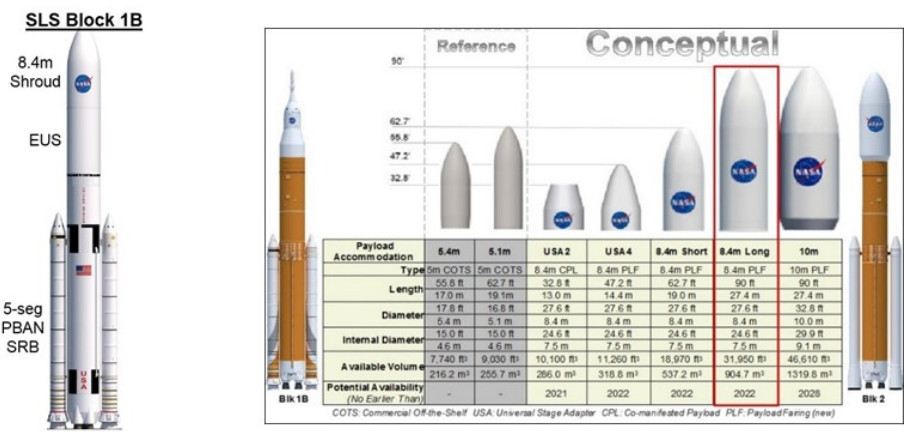

**Figure 3-85:** Space Launch System (SLS); an 8.4 m, five-segment system. The selected 8.4-m SLS fairing provides the necessary space to house the stowed observatory and the lift required to reach L2.

OST can be accommodated in the SLS 8.4 m x 27.4-m long fairing, and adequately within the fairing's static envelope as shown in **Figure 3-86**.

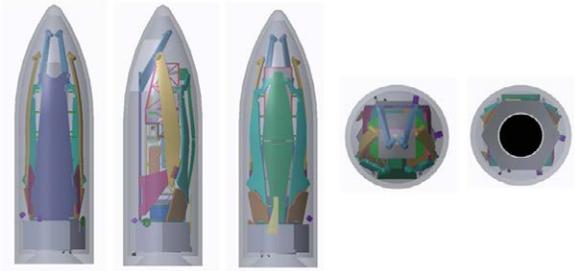

**Figure 3-86:** The OST observatory is accommodated in the SLS 8.4-m fairing.

## 3.7 Mission Operations and Ground System

OST ground system and mission operations drivers are:

- OST is a flagship mission and operations will focus not only on safe operations, but also on ensuring a scientifically-efficient mission, minimizing non-science time to the extent possible. This includes developing and testing procedures for potential anomalies to reduce recovery time and implementing a scheduling system that minimizes slew times.
- Large data volume. The daily science volume is ~39 Tbits per day (including margin), larger than any previous NASA science mission. This drives the space/ground link to use optical communications.
- L2 orbit. The orbit drives the sizing of the space/ground links.

### 3.7.1 Ground System

The ground system (**Figure 3-87**) is similar to those used by previous Astrophysics flagship missions, including HST and JWST.

### 3.7.2 Space/Ground Link

The science data downlink uses optical communications due to OST's large data volume. Optical ground stations exist at White Sands and Hawaii, and each terminal has a 0.6-meter telescope. These locations support the Laser Communications Relay Demonstration (LCRD) that launches in 2019. An additional terminal in South Africa would need to be built, either by OST or as a multi-user station. These three stations provide 20.5 hours of contact time per day. The downlink rate is 1 Gbps and 39 Tbits can be downlinked daily as long as the three stations are cloud-free at least 50% of the time.

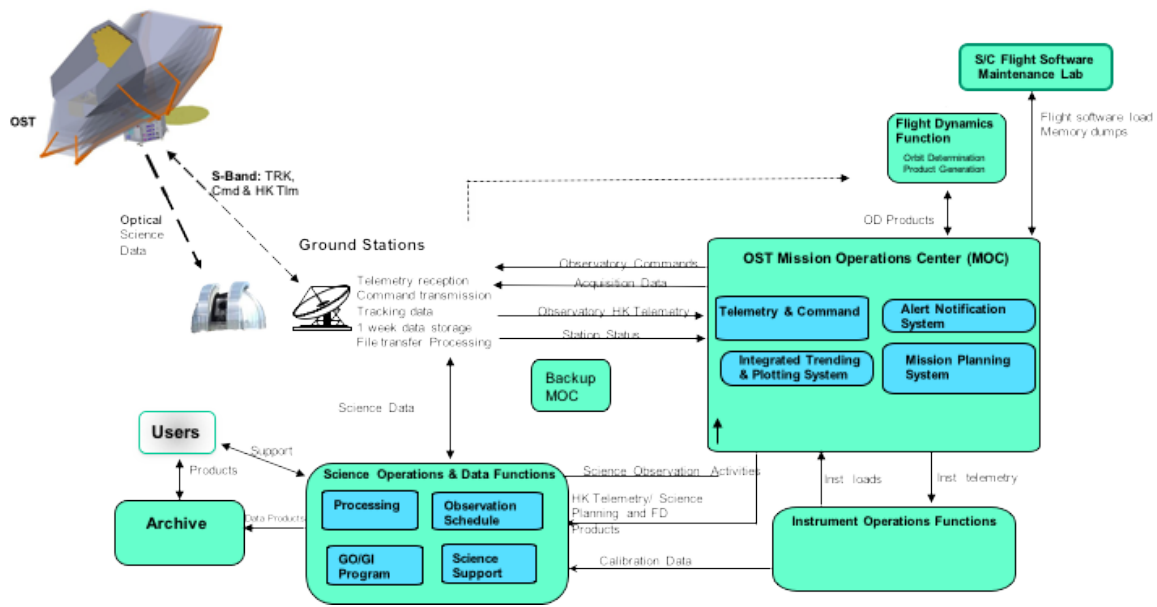

**Figure 3-87:** OST Ground System block diagram.

The three sites were selected for their dry climates, where clouds are less prevalent at night, when the stations are in view of L2. OST onboard storage is sized to store more than a day of data, reducing the risk of data loss when weather is cloudier than average.

OST uses S-band RF links for commands, housekeeping telemetry, and tracking. The baseline design uses three existing stations at White Sands, Hawaii, and Western Australia. The Western Australia station requires an upgrade to a 2 KW transmitter. These stations provide at least two 3-hour contacts per day for tracking data. OST also uses the uplink to send acknowledgments to close the data delivery protocol to ensure complete and error free science data reception.

Optical communications is a proven technology, but OST requires higher power and a longer lifetime than has currently been flown. NASA's Space Communications and Navigation organization is currently developing flight optical communications terminals for deep space missions and for optical relay terminals in geostationary orbit that would also meet OST's requirements. Psyche is slated to carry the deep space optical communications terminal in 2022 and the geostationary relay terminals are scheduled to be operational in 2025.

If optical communications technology does not mature as expected, Ka-band RF link is a backup option. To support 1 Gbps, the spacecraft would need a 3-meter High Gain Antenna and a 100 W transmitter. Packaging this large antenna in the fairing could be a challenge. There are also options to reduce the data volume by using compression or by limiting the MRSS instrument duty cycle.

### 3.7.3 Mission Operations Center

The MOC is responsible for the safe operation of OST and the health of the spacecraft. Its functions include:

- Scheduling Optical and RF Ground Stations. The daily data volume can vary by a factor of eight depending on which instruments are in use. MOC requests support as-needed based on data volume. The MOC also considers weather forecasts in the scheduling time when Hawaii and White Sands are both in view of OST.
- Commanding. The MOC is the sole source of commands for OST. The MOC integrates instrument and telescope commands from the Science Operations Center (SOC) with spacecraft commands and sends them to the ground stations for uplink.
- Monitoring. The MOC monitors housekeeping data and takes appropriate action in the event of a problem. The MOC maintains a comprehensive set of contingency procedures to reduce response time to anomalies.

The backup OST MOC can perform basic functions in the event the primary MOC is unavailable. Both MOCs will be developed and operated by an experienced organization with the necessary people, facilities, tools, and processes. MOC functions will benefit from significant reuse of software developed for similar missions.

### 3.7.4 Science Operations Center

Like the MOC, the OST SOC should be developed and operated by an experienced organization that can leverage past software, processes, and staff from similar missions. The SOC supports NASA HQ with evaluating proposal solicitations from Guest Observers (GO) and Guest Investigators (GI). The SOC supports these additional scientists with instrument user's guides, simulators, exposure time tools, and a help desk.

The SOC generates long- and short-term observing plans. The schedule must balance the need for time sensitive observations, such as coronagraph targets and targets near the ecliptic, which are only accessible for ~14 weeks/year, with the need for efficiency by minimizing the size of slews between targets. The SOC generates the instrument command load and provides it to the MOC.

The SOC processes science data and produces standard products. It sends data to an archive. The SOC also supports telescope alignment and instrument calibration.

OST generates a significant data volume: 1.8 PBytes per year of raw data. Data storage is not expected to be a significant challenge due to increases in data storage (including cloud storage), communications, and processing capabilities over time. The OST data volume is less than four times the data volume of WFIRST, which launches a decade earlier. Current astrophysics archives at IPAC and STScI hold over 1 and 2.5 PBytes respectively, and the STScI archive is expected to more than double over the next five years.

Existing scheduling processes should be sufficient for OST. The number of observations is less than HST and the L2 orbit has fewer constraints, reducing the complexity of scheduling algorithms.

### 3.7.5 Other Ground System Elements

Flight Dynamics is responsible for orbit determination and orbit maintenance maneuver planning. To support instrument commissioning and calibration, each instrument will have a function that monitors instrument performance. This function maintains instrument flight software, provides updates as required, and is the focal point for resolving instrument anomalies. Spacecraft flight software maintenance functions develop and test any flight software updates required after launch.

### 3.7.6 Operations

OST operates at L2 with a fixed telescope Gimbal angle. The sunshield is designed to allow a 50° pitch angle and ± 5° roll angle, and the observatory can yaw the full 360°. **Figure 3-88** shows the field of regard. A few times a year, the Moon impinges slightly on this field of regard. There are no eclipses in this orbit.

Instrument observations range from minutes to days for MISC coronagraph observations or MRSS mapping. Most observation durations are on the order of hours. **Figure 3-89** shows observations on notional day. MISC provides guiding for all observations. The baseline OST design has three instrument power modes: one each for MRSS and HERO, and a third for HRS, FIP, and

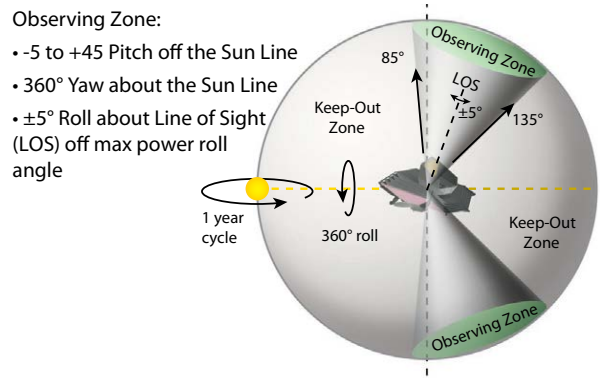

Observing Zone:
- -5 to +45 Pitch off the Sun Line
- 360° Yaw about the Sun Line
- ±5° Roll about Line of Sight (LOS) off max power roll angle

**Figure 3-88:** The OST Field of Regard views ~40% of the sky at a time in a band on the celestial sphere from 85° to 135° from the sun. All areas of the sky are visible from OST's position at L2, and can be viewed at least twice per year, each at least 50 days at a time.

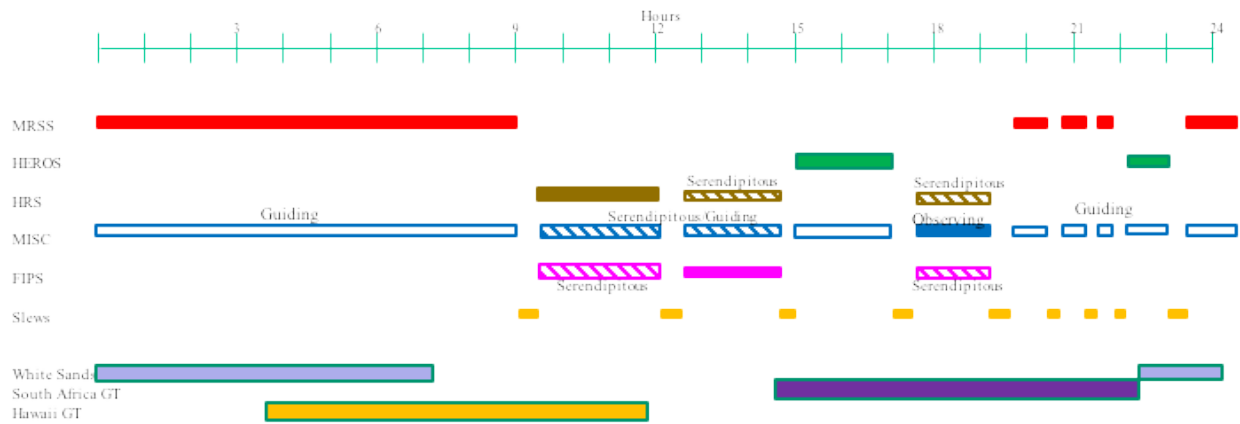

**Figure 3-89:** Notional day-in-the-life provides guidance for all observations.

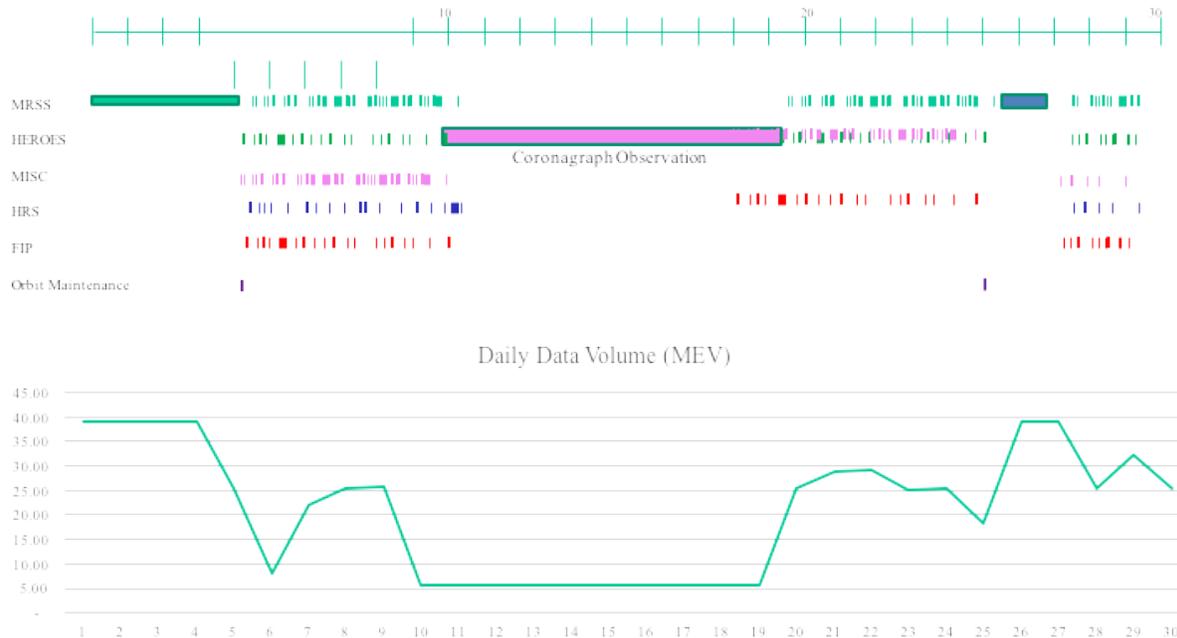

**Figure 3-90:** Notional 30 day observation schedule. Data volume varies based on duty cycles and mix of instruments.

MISC. In the third mode, multiple instruments may take data simultaneously. Pointing is driven by the primary observing instrument.

An average 30 days of observations provides opportunities for all instruments to reach their science goals (**Figure 3-90**). Data volume varies monthly based on instrument operations and observing schedules. When MRSS is on for a full day, data volume is 39 Tbits; during coronagraph observations, daily science data volume is only 5.6 Tbits.

In general, OST has no stringent requirements on data latency. The observatory can store 64 Tbits of science data. Data are downlinked up to 20.5 hours per day, but some data may be lost during initial transmission due to clouds. Retransmissions are requested during the two S-band contacts per day, but it may take up to two days for observation data to be downloaded, depending on cloud cover.

The SOC identifies Targets of Opportunity (TOO) when a short-term astronomical event occurs. The MOC and SOC work together to adjust the observatory's schedule and generate commands that allow OST to observe an event in less than 24 hours. TOOs are expected to occur about once per month.

OST is a large, complex observatory and its science efficiency will be about 65% – the other 35% of the time it is not taking science. Non-science activities include:

• Momentum dumps. Two 40-minute momentum dumps will occur per day on average.
• Slews. A 30° slew takes 21 minutes. A goal of the scheduling system is to minimize slew duration.
• Orbit maintenance. Orbit maintenance occurs every 3 weeks.
• Telescope realignment.
• Instrument calibrations.
• Weather. If it is cloudy at all three optical stations, the onboard storage will reach capacity. The efficiency number assumes 5% of time will be lost to onboard storage being at capacity.

The OST baseline mission is five years and the observatory has consumables to operate for 10 years. It is designed to be serviceable, so a longer lifetime is possible. Normal operation starts ~110 days after launch, shortly before L2 orbit insertion maneuver. This time is spent checking out the spacecraft, deploying the telescope and sunshield, waiting for the telescope and instruments to cool to the ~4 K operational temperature, aligning the telescope, and calibrating the instruments. **Table 3-12** provides a detailed list of activities during the post launch phase.

**Table 3-12:** Timeline from launch through L2 orbit.

| Time After Launch | Event |
|---|---|
| 30 minutes | Solar Array Deployment |
| Hours | Propulsion Checkout |
| 12 hours | Mid-course Correction |
| 1 day | HGA Deployment |
| Two weeks | Spacecraft Subsystem Checkout |
| Days 3 through 20 | Telescope and Sunshield Deployment:<br>Deploy forward and aft sunshade support structures<br>Deploy DTA and sunshade and 4 lower booms, forming sunshield basic shape. IAM is vertical.<br>Deploy IAM 90°<br>Release baffle attachments to IAM<br>Release mirror wing launch locks<br>Rotate mirror 75°<br>Deploy wings and baffle<br>Deploy sunshade 8 upper booms for 5 layer separation. Sunshade deployment complete. |
| Day 10 | MCC-1a Cleanup |
| Day 20 | MCC-2 |
| Day 21 | Deploy Momentum Trim Tab Final ACS Calibration |
| Day 21-35 | Initial Instrument Checkout (Warm) |
| Day 30 | Cryocoolers On |
| Day 60 | Cool down Complete MCC-3 |
| Day 60-72 | Telescope Alignment |
| Day 50-80 | Instrument checkout (Cold) and Calibration |
| Day 80-110 | Science Commissioning |
| Day 110+ | Normal Operations Commence |
| Day 113 | L2 Orbit Insertion |

### 3.8 On-Orbit Servicing

The OST mission design approach includes on-orbit servicing concept. While the current Concept 1 has yet to include on-orbit servicing hardware and scenario, the overall design concept allows implementation of servicing hardware without redesigning the entire system. The OST on-orbit servicing concept is based on system and subsystem modularity that can be accessed and replaced by a servicing robot or robot arm. On-orbit servicing will be limited to replacing non-functioning or outdated subsystem and/or instruments. Component replacement (*i.e.,* "card in a box") is not considered for OST on-orbit servicing. OST on-orbit servicing assumes replacement of an entire subsystem box or instrument, or replacing the entire Instrument Accommodation Module.

Prior to servicing instruments, the OST observatory cryo system is turned off and the cooler allowed to warm to room temperature. Prior to servicing spacecraft subsystem, power to the servicing subsystem is turned off. The OST observatory power system design allows partitioned powering for on-orbit servicing. The OST observatory includes appropriate grapple fixtures, hand holds, rails, etc., at correct locations for a servicing robot or robot arm. It assumes the servicing robot and robot arm is properly equipped with fixtures that can be mated with the OST observatory servicing hardware.

On-orbit servicing sequence:

1. Servicing Vehicle Rendezvous and OST capture: initiate master (servicing vehicle) and slave (OST observatory) form

2. Servicing Vehicle takes over OST thermal and pointing: perform thermal equalization of OST observatory for servicing
3. Servicing – remove and replace
4. Power on, checkout, and re-commission
5. De-mate servicing vehicle robot arm from OST observatory: null master and slave form
6. OST takes over pointing, thermal, and all other systems control

OST Concept 1 mission design requires additional design features to meet the on-orbit servicing requirements for instrument and subsystem replacements.

## 3.9 Materials

A number of materials were evaluated for the OST observatory optical and structural elements. These elements consist of the primary mirror, the backplane, the Instrument Accommodation Module (IAM), Bus, and the Deployable Tower Assembly (DTA).

### 3.9.1 Material Candidates: Optics

Mirror materials were evaluated first due to the inherent complexity of mirror manufacturing. Materials were selected for consideration based on their manufacturability for point to point hexagonal segments. Notable advantages and disadvantages are listed for each material with respect to performance and manufacturability (**Table 3-13**).

The first assessment was driven by the ~4 K temperature environment. Materials that were a concern for cryogenic temperatures were eliminated. This included the glasses and glass ceramics: ULE (titania-silicate glass), Zerodur (lithium-alumino-silicate glass-ceramic), and Borosilicate (glass with silica and boron trioxide). Though fused silica is a glass, it was not eliminated because of its prominent heritage as an optical substrate and potential to perform in cryogenic temperatures. Carbon fiber reinforced polymer (CFRP) was eliminated based on its anticipated poor optical quality and titanium because of its extremely high density. Five mirror material candidates remained at the time of the Concept 1 selection: beryllium, aluminum, AlBeMet (aluminum and beryllium metal matrix composite), silicon carbide, and fused silica.

**Table 3-13:** Five primary mirror material candidates remain after Concept 1 evaluation.

| Material | Advantages | Disadvantages |
|---|---|---|
| Beryllium | Superior stiffness, extremely lightweight, low CTE | Expensive, brittle, toxic, long machining time |
| Aluminum | Good structurally, good fabrication time, inexpensive | Heavy, reactive surface, low stiffness |
| AlBeMet | Good stiffness, low CTE, lightweight | Limited information (no heritage) |
| SiC | Excellent stiffness, excellent strength, low CTE | Heavy, expensive, long machining time |
| Fused Silica | Low CTE, lightweight, low | Low stiffness |
| Titanium | Excellent strength, good thermal properties | Extremely heavy, machining ability |
| ULE | Low CTE, lightweight | Poor cryo-survivability, low strength |
| Zerodur | Low CTE | Poor cryo-survivability, low strength |
| Borosilicate | Lightweight | Poor cryo-survivability, low strength |
| Composite/CFRP | Extremely lightweight, low CTE can be achieved | Surface quality (optical finish), creep |

### 3.9.2 Material Candidates: Structure

Materials were also assessed for the main structural elements of the OST observatory: the PM backplane, IAM, Bus, and DTA. An ideal structural material for spacecraft is strong, lightweight, and has a near-zero change in thermal expansion. For this reason, Carbon fiber reinforced polymer (CFRP) is used whenever possible. Aluminum, Invar, titanium and copper were also considered.

For OST, it is important to note that structures exist on the cold side and the warm side. The backplane and IAM are on the cold side. The Bus and DTA are on the warm side. A concern for CFRP is the potential for water off-gassing. While the observatory is deploying, the elevated temperature (due to sun exposure) may cause water vapor to off-gas from CFRP. On the cold side this vapor could condense onto the mirror and instruments and impact imaging quality. As a result, CFRP was avoided on the cold side.

Thermal conductivity is another important property for the backplane and primary mirror because it will be cooled conductively to 4 K. For this reason, a metal with good thermal conductivity at 4 K is an attractive material. The backplane and IAM would also ideally be made of the same material as the mirror segments to avoid CTE mismatch between structures. Thus, an athermal design on the cold side is favorable. On the other hand, the warm side needs to be thermally isolating. This makes CFRP advantageous for the Bus and DTA.

**Concept 1 Selection**

Materials selections for Concept 1 were made early on to accommodate the schedule of the Instrument Design Lab (IDL) study. After weighing the advantages of an athermal design on the cold side and the potential of CFRP water-off gassing, aluminum was selected for the primary mirror, backplane, IAM and other small structural components on the cold side. CFRP was chosen for the DTA and much of the Bus on the warm side.

Aluminum has flight mirror heritage on the Cosmic Background Explorer (COBE) mission. A clear disadvantage of aluminum is the high density leading to an overall large observatory mass. However, the SLS heavy launch vehicle was presumed for Concept 1, which made mass reduction less of a driver as far as material selection.

**On-going Materials Evaluation**

OST evaluations are on-going. The team has learned that specific stiffness and thermal stability are two derived parameters that are very important for mirror performance. Specific stiffness is the ratio of elastic modulus to density. Thermal stability is the ratio of thermal conductivity to the coefficient of thermal expansion. These parameters will be considered more heavily and help guide the materials evaluation as we look toward developing Concept 2. Finding a solution for CFRP on the cold side will also need to be explored. Concept 2 materials will need to be as lightweight as possible to help in keeping costs down.

Harris Space and Intelligence Systems Material are conducting further trades to evaluate the primary mirror materials. The five remaining materials are ranked based on performance, cost, schedule, and TRL (**Table 3-14**). Fused silica and silicon carbide are promising options.

**Table 3-14:** OST partners are conducting primary trades among potential mirror materials.

| Material | Performance | Schedule | Cost | TRL | Overall Comments |
|---|---|---|---|---|---|
| Fused Silica | 4 | 5 | 4 | 9 | Glass has max heritage as optic substrate |
| SiC (multiple) | 4 | 4 | 3 | 9 | Limited knowledge |
| Beryllium | 5 | 2 | 2 | 9 | JWST Heritage |
| Aluminum 6061 | 2 | 5 | 5 | 6 | Concept #1 found All-Al telescope is too massive |
| AlBeMet | 4 | 2 | 1 | 2 | No meter-class heritage |

## 3.10 Sun-Earth L2 Environment

### 3.10.1 Radiation

OST Concept 1 radiation analysis was based on the assumed launch date of September 1, 2035, the spacecraft's trajectory to the Sun-Earth L2 Halo Orbit, and the length of the mission while at $L_2$. The launch date determines the phase of the solar cycle when the mission begins, which determines the amount of exposure to solar energetic particle (SEP) radiation. The mission would begin during solar maximum when SEP radiation is expected, and extend into solar minimum when SEP radiation is negligible. The accumulated fluence of SEP was calculated from a statistical model known as ESP-PSY-CHIC, which is based on measurements during 21 solar maximum years [*M. Xapsos[561].4*]. The model provides fluences that will not be exceeded with a 95% confidence. Total Ionizing Dose (TID) of radiation on a silicon electronic part was calculated from the SEP fluences by using the SHIELDOSE-2 model for TID incurred by the specific material (Si) used to fabricate most electronic parts, assuming aluminum shielding of various thicknesses.

The trajectory included a brief time (~2 hours) when the spacecraft traversed Earth's trapped radiation belts of energetic electrons and protons, but the contribution to the total TID was negligible during this brief traversal. The accumulated TID was, for practical purposes, due to the time that the spacecraft would spend during the remainder of the mission while solar maximum was under way.

TID forecasts computed with the SHIELDOSE-2 model for each of the two mission length options (five years, ten years), for twenty-five different shielding thicknesses. If we imagine a silicon electronic component surrounded by a spherical shell of Aluminum shielding with a given thickness, then the entries under the green headings are the resulting TID forecasts. The thicknesses vary from 0.05 mm to 20 mm. A benchmark often used is the 2.5 mm thickness, which would result in about 19.8 krad (Si) for the five-year mission, or 27 krad (Si) for the ten-year mission. Doubling the mission length does not double the TID because of the end of the solar maximum, after which the SEP fluence typically drops to negligible levels.

Radiation hardness is usually assured by multiplying the TID by a 2X safety factor, and evaluating parts according to their reliability under those 2X doses. For example, 50 krad (Si) is a threshold that many parts fail to meet. For the five-year mission, the 2X TID behind 2.5 mm Al is 40 krad (Si), less than 50 Krad (Si); but for the ten-year mission the 2X TID is 54 krad (Si). To ensure reliability, the designers have the choice of purchasing parts with higher TID ratings, or to add more shielding than the 2.5 mm Al equivalent that typically is provided by box walls. A 3 mm thickness would bring the 2X TID below the 50 krad threshold, for example.

At a later stage of mission design, SEP ray tracings often are performed with a software tool such as the NOVICE system. A ray-tracing analysis would be more accurate than the SHIELDOSE model because it would account for effective shielding provided by other spacecraft components. TIDs from ray tracings typically are lower than the simple analysis herein provides.

The other radiation risk for OST is the impact of high-energy solar protons, heavy ions, and galactic cosmic ray flux, which are too energetic to mitigate with shielding. These particles cause Single Event Effects (SEE). Mitigation techniques for SEE are required for vulnerable components (*e.g.* memories, FPGAs, linear bipolar components, mixed signal devices, etc.). Major destructive concerns are single-event latchup, single-event gate rupture, and single-event burnout. Soft errors may be mitigated by redundant processors and memory designs that enable bits flipped by SEE to be corrected.

### 3.10.2 Reliability

The Optical Space Telescope (OST) mission is a five-year mission involving a deployable telescope and sunshield. To support OST science, a Spacecraft Bus was developed to support and minimize risk to the mission. A quantitative reliability model was developed to begin to characterize the risk posed by the spacecraft bus, as well as to compare the relative reliabilities of spacecraft subsystems and the reliability of different design configurations.

The OST architecture is fully redundant and consistent with a mission spacecraft bus designed according to NPR 8705.4 guidelines, with some critical items related to propulsion, mechanisms and cryogenics. Critical items with potential single-point failures (potentially non-credible/low probability, or waivable based on acceptable risk) include: fuel tanks and filters, S-band antenna gimbal, solar ar-ray gimbal, solar array deployment mechanisms, sunshield deployment mechanisms, and cryocoolers. Risks are being characterized and designs validated with appropriate Reliability Analyses, including Fault Tree Analysis (FTA), Failure Mode Effects and Criticality Analysis (FMECA), Parts Stress Analysis (PSA), Worst Case Analysis (WCA), and Probabilistic Risk Assessment (PRA). OST designs meet NASA and GSFC specifications including: EEE-INST-002, GSFC-STD-7000 General Environmental Verification Standard (GEVS), and GSFC-STD-1000.

### 3.10.3 Debris Analysis and Micrometeoroid Environment Orbital Debris Requirement

The NASA-STD-8719.14A, Process for Limiting Orbital Debris, contains the Agency requirements intended to: 1) limit the generation of orbital debris by NASA spacecraft and hardware, 2) protect those assets from existing space debris, and 3) protect human life from objects surviving atmospheric reentry. NASA missions are required to incorporate those requirements into the design early during the development process, to identify potential non-compliances and mitigate any existing risk.

Some requirements are applicable only to spacecraft operating in Earth or lunar orbit, but the Agency intends to protect other regions for the benefit of future missions. NASA-STD-8719.14A paragraph 4.2 states: "NASA should also limit debris generated at the Earth-Sun Lagrange points and in orbit around other celestial bodies." OST is passivated at End of Mission (EOM) to avoid future generation of debris by accidental explosion and performs a short propulsive maneuver at EOM to expedite the exit from L2.

Orbital debris requirements also apply to the launch vehicle upper stage. The OST mission prevents generation of debris in Earth orbit by the upper stage and ensures the upper stage proper disposal after spacecraft separation. The first assessment of the launch vehicle will be completed after launch vehicle selection. OST is compliant with applicable orbital debris limitation requirements.

The OST sunshield and baffle are made of thin Kapton membranes designed for passive thermal control and for preventing stray light from reaching the sensors. Due to their large area, the membranes are exposed to a significant number of micrometeoroid impacts. Hypervelocity impact analysis allows the team to estimate the number of micrometeoroid impacts capable of penetrating the layers; it is also useful to approximate the hole size to predict any detriment in the membrane's protection capability.

The amount of meteoroid impacts to the Primary Mirror is also of interest. In this case, the failure mode is not penetration (negligible probability), but surface damage, because of its potential to cause optical degradation.

The analysis estimates the micrometeoroid damage to the sunshield, baffle, and primary mirror for the nominal five year mission, and was extended to 10 years to observe long-term effects.

The OST mission is compliant with relevant orbital debris requirements, while most orbital debris requirements as defined in the NASA-STD-8719.14A do not apply to spacecraft operating in L2. OST is consistent with the NASA goal to protect areas of high scientific value for future missions. For fulfill this goal, the spacecraft does a short propulsive maneuver to expedite exit from the L2 orbit and minimize the possibility of drifting back to Earth. Additionally, the spacecraft performs passivation procedures at EOM, venting tanks after disposal maneuvers and depleting energy sources, to prevent accidental explosions.

Micrometeoroid impact analysis of the sunshield, baffle, and primary mirror demonstrates that the potential for particle impact damage that may prevent mission success is very low. Future analysis will cover additional surfaces and spacecraft components; however, the results from this preliminary analysis suggest a low probability of encountering issues related to exposure to the micrometeoroid environment.

# 4 - ENABLING TECHNOLOGIES

> Detector are the most important enabling technology for OST. With handoffs from ongoing detector and cryocooler technology maturation efforts and a roadmap for further maturation, all technology needed for OST will reach TRL by 2025.

## 4.1 Far-IR Direct Detectors, Multiplexing, and Readout

At far-infrared wavelengths, reaching the fundamental sensitivity limits set by astrophysical foregrounds requires use of a cold space telescope and sensitive detectors. The detector sensitivity needed to reach these limits, expressed as the Noise Equivalent Power (NEP) for imaging is around 3 x $10^{-19}$ W √Hz, whereas the NEP needed for moderate resolution spectroscopy is 3 x $10^{-20}$ W √Hz-.

Several direct detector technologies are currently in development. Transition Edge Sensors (TES) are based on the sharp change of resistance of a superconductor at its transition. Microwave Kinetic Inductance Detectors (MKIDs) use the superconductor's free electron population temperature dependence, and Quantum Capacitance Detectors (QCDs) use the Cooper-pair breaking that occurs with absorption of far IR photons. Each of these technologies have their own opportunities and challenges toward the OST goal of large scalable arrays (104 to 105) with sensitivities in Noise Equivalent Power (NEP) of 1x$10^{-19}$ to 1x$10^{-20}$ W/√Hz.

For Transition Edge Superconductor (TES) based pixels NEPs < 1x$10^{-19}$ W/√Hz have been demonstrated for tens of pixel channels (**Figure 4-1**). JPL, GSFC, and SRON are pushing the technology to enable >1000 pixels per channel with this resolution. Recently awarded SAT projects will greatly mature this technology for TESS and Microwave Kinetic Inductance Detectors (MKIDs).

For OST an enabling goal is to multiplex 1000 pixels per channel without cross-talk or loss of sensitivity. An enhancing goal is 3000 pixels per channel. This can be achieved with a combination of microwave SQUIDs, resonant circuits, parametric amplifiers, and/or High Electron Mobility Transitors (HEMTs). Microwave SQUID multiplexing is being developed by SRON in Europe and by NIST and Stanford University (Kent Irwin) in the US. The requirements for Far IR detectors and X-ray detectors for low temperature multiplexing have much overlap, and OST takes advantage of this synergy in the development.

Each HEMT or parametric amplifier is a channel. The HEMT or parametric amplifier is located at 4 K and provides the first level of amplification of the multiplexed signal that is sent to room temperature. For this reason, optimizing HEMTs for low power dissipation is important. The mul-tiplexed signal is currently sent to a low temperature HEMT for amplification and sent via low-loss coax cables to room temperature for processing. See **Figure 4-2** for low temperature HEMT package for 4 K.

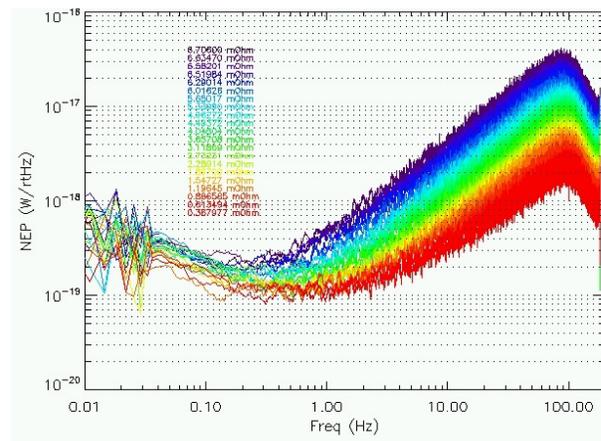

**Figure 4-1:** TES detector noise in multiplexed array for the far IR.

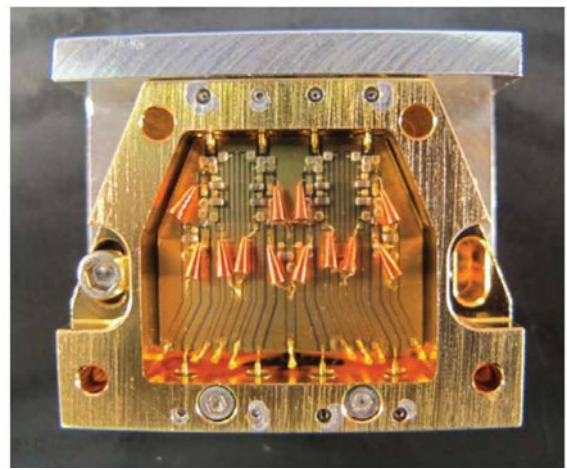

**Figure 4-2:** The low-temperature HEMT amplifier package for 4 K operation provides 5-10 dB of gain while using only 0.7 mW.

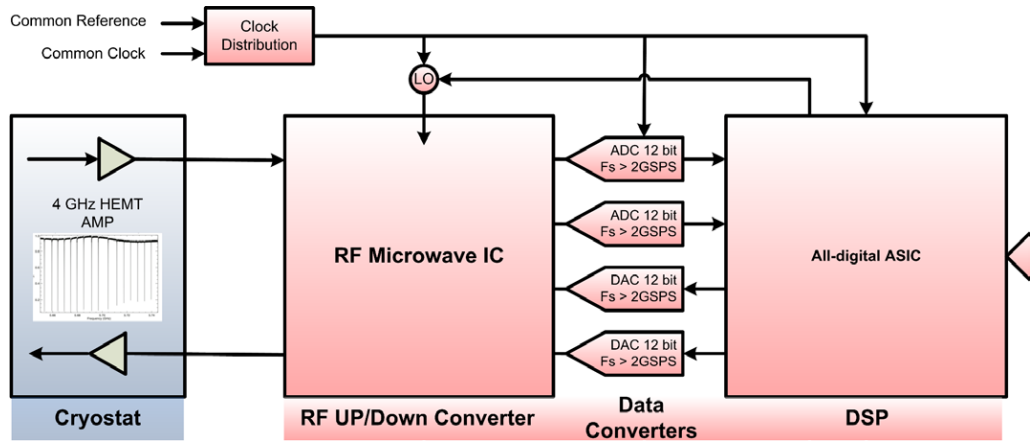

**Figure 4-3:** Schematic of ASIC-based room temperature readout. The design goal of this scheme is to readout a 1600 pixel channel using less than 30 W.

### 4.2 Low Power Room Temperature Readouts

The current state-of-the-art is to use FPGAs to read out multiple channels. The power required is approximately 120 W per channel. For OST the enabling goal is to dissipate less than 30 W per channel using ASICs (**Figure 4-3**).

### 4.3 Heterodyne Detectors

Current state-of-the-art, high-TRL heterodyne detectors have a noise equivalent temperature of <1000 K, exist in up to 16-pixel arrays, and have a frequency range of up to 2.1 THz. For Concept 1, OST needs to extend the range to 4.7 THz (63 μm wavelength) and increase the pixel count to 128 and decrease the power dissipated per pixel at low temperature from 1 mW to less than 0.5 mW (**Figure 4-4**).

### 4.4 Mid-IR Detectors

The mid-infrared instrument (MISC) would use arsenic-doped silicon impurity band conduction (Si:As IBC) detectors. These are being used in the MIRI instrument on JWST [*Love et al., 2010*] in a

1k x 1k format. To achieve the OST Concept 1 exoplanet detection goals in the mid-infrared, a factor of 10 improvement in stability is required over the current state-of-the-art as in the James Webb Space Telescope/Mid Infrared Instrument (JWST/MIRI). The team's JAXA partners are examining techniques to improve this stability at the instrument level for the MISC instrument; results of this study will be detailed in the OST final report.

### 4.5 Large Cryogenic Lightweight Optics

Concept 1 did not use low TRL technologies for the telescope, instead adopting an athermal, high thermal conductance design with backplane and mirrors fabricated from aluminum. The 37 segments for the Concept 1 primary will be actuated using JWST-like actuators (**Figure 4-5**). This represents a low cost/low risk strategy at the expense of a much heavier telescope. Better materials, such as silicon carbide (SiC) will be considered for Concept 2, as well as using mirror figure actuators, such as those under development at JPL to allow more aggressive light-weighting. Such figure control may be able to provide savings in mirror fabrication, allowing adjustability to remove figure errors that would otherwise need to be removed by cryo-null figuring.

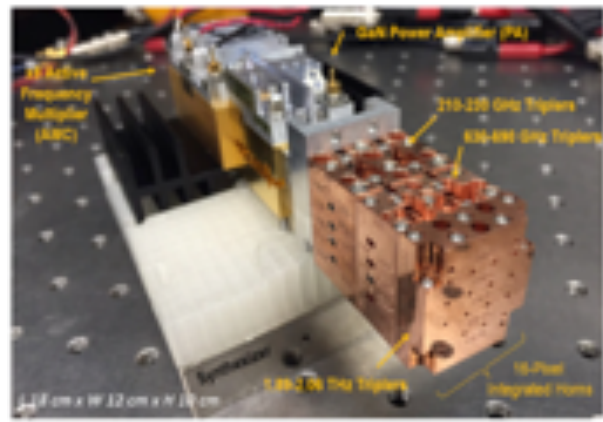

**Figure 4-4:** The 1.9 – 2.1 THz 16-pixel LO chain for SHASTA; individually adjustable power outputs were derived from a single 23-26 GHz source.

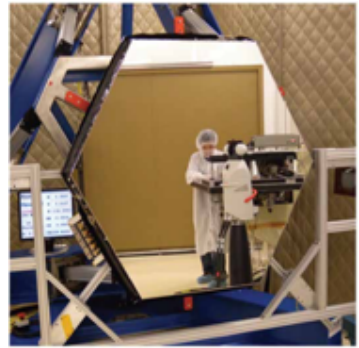

- Active mirror segments
- 37 to 414 actuators
- 0.5 to 1.35 m size demonstrated
- <14 nm rms SFE demonstrated
- 10-15 kg/m$^2$ substrate
- <25 kg/m$^2$ total
- Tested in 1G to 0G specs

**Figure 4-5:** The 37 segments for the Concept 1 primary will be actuated using JWST-like actuators

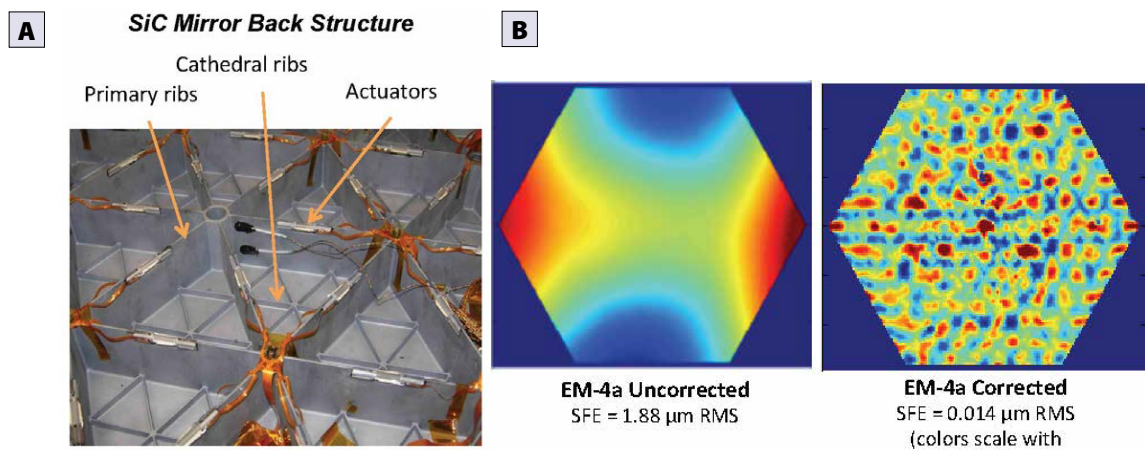

**Figure 4-6:** A) Actuated SiC mirror segments. B) Piezo-electric actuators are low dissipation and can be made to "set and hold."

## 4.6 Sub-Kelvin Cooling

For detector cooling to 50 mK, adiabatic demagnetization refrigerator (ADRs) are currently the only proven technology, although some work has been funded by ESA to develop a continuously recirculating dilution refrigerator (DR). A single shot DR was flown on Planck producing 0.1 μW of cooling at 100 mK for about 1.5 years, while a three-stage ADR was used on Hitomi producing 0.4 μW of cooling at 50 mK with an indefinite lifetime. In contrast, a TRL 4 continuous ADR has demonstrated 6μW of cooling at 50 mK with no life-limiting parts [*Shirron, 2002*]. This technology is currently being advanced toward TRL 6 by 2020 through SAT funding [*Tuttle, 2017*]. Demonstration of a 10 K upper stage for this machine, as is planned, would enable coupling to a higher temperature cryocooler, such as that of Creare, that has a near-zero vibration technology. The flight control electronics for this ADR is based on the flight-proven Hitomi ADR control, and has already achieved TRL 6 (**Figure 4-7**).

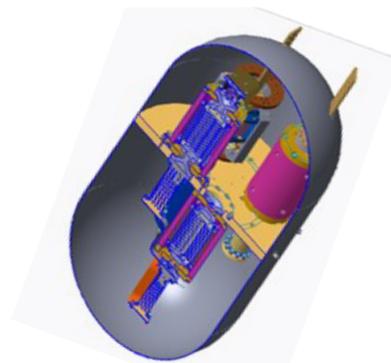

**Figure 4-7:** Continuous ADR (CADR) under development at NASA will provide 6 μW of cooling at 50 mK and has a pre-cooling stage that can be operated from 0.3 to 1.5 K. The team has a notional design for an enclosing magnetic shield for <1 μT fringing field.

**4.7 Cryocoolers for 4 K**

To achieve temperatures lower than about 30 K, mechanical cryocoolers are used. The state-of-the-art for these coolers are represented by Planck, JWST/MIRI, and Hitomi (ASTRO-H). Domestic coolers that could achieve 4.0 K are considered to be at TRL 4-5, having been demonstrated as a system in a laboratory environment [*Ross, 2003*], or as a variant of a cooler that has a high TRL (JWST/MIRI). Mechanical cryocoolers for higher temperatures have already demonstrated impressive on-orbit reliability. The moving components of a 4 K cooler are similar (expanders) or are exactly the same

## Recent Long-Life Space Cryocooler Flight Operating Experience as of May 2016

| Cooler / Mission | Hours/Unit | Comments |
|---|---|---|
| **Air Liquide Turbo Brayton (ISS MELFI 190K)** | 85,600 | Turn on 7/06, Ongoing, No degradation |
| **Ball Aerospace Stirling** | | |
| HIRDLS (60K 1-stage Stirling) | 83,800 | 8/04 thru 3/14, Instr. failed 2008; Data turned off 3/14 |
| TIRS cooler (35K two-stage Stirling) | 27,900 | Turn on 3/6/13, Ongoing, No degradation |
| **Creare Turbo Brayton (77K NICMOS)** | 57,000 | 3/02 thru 10/09, Off, Coupling to Load failed |
| **Fujitsu Stirling (ASTER 80K TIR system)** | 141,700 | Turn on 3/00, Ongoing, No degradation |
| **JPL Sorption (PLANCK 18K JT (Prime & Bkup))** | 27,500 | FM1 (8/10-10/13 EOM); FM2 failed at 10,500 h |
| **Mitsubishi Stirling (ASTER 77K SWIR system)** | 137,500 | Turn on 3/00, Ongoing, Load off at 71,000 h |
| **NGAS (TRW) Coolers** | | |
| CX (150K Mini PT (2 units)) | 161,600 | Turn on 2/98, Ongoing, No degradation |
| HTSSE-2 (80K mini Stirling) | 24,000 | 3/99 thru 3/02, Mission End, No degrad. |
| MTI (60K 6020 10cc PT) | 141,600 | Turn on 3/00, Ongoing, No degradation |
| Hyperion (110K Mini PT) | 133,600 | Turn on 12/00, Ongoing, No degradation |
| SABER on TIMED (75K Mini PT) | 129,600 | Turn on 1/02, Ongoing, No degradation |
| AIRS (55K 10cc PT (2 units)) | 121,600 | Turn on 6/02, Ongoing, No degradation |
| TES (60K 10cc PT (2 units)) | 102,600 | Turn on 8/04, Ongoing, No degradation |
| JAMI (65K HEC PT (2 units)) | 91,000 | 4/05 to 12/15, Mission End, No degrad. |
| GOSAT/IBUKI (60K HEC PT ) | 63,300 | Turn on 2/09, Ongoing, No degradation |
| STSS (Mini PT (4 units)) | 52,800 | Turn on 4/10, Ongoing, No degradation |
| OCO-2 (HEC PT) | 14,900 | Turn on 8/14, Ongoing, No degradation |
| Himawari-8 (65K HEC PT (2 units)) | 12,800 | Turn on 12/14, Ongoing, No degradation |
| **Oxford/BAe/MMS/Astrium/Airbus Stirling** | | |
| ISAMS (80 K Oxford/RAL) | 15,800 | 10/91 thru 7/92, Instrument failed |
| HTSSE-2 (80K BAe) | 24,000 | 3/99 thru 3/02, Mission End, No degrad. |
| MOPITT (50-80K BAe (2 units)) | 138,600 | Turn on 3/00, lost one disp. at 10,300 h |
| ODIN (50-80K Astrium (1 unit)) | 132,600 | Turn on 3/01, Ongoing, No degradation |
| AATSR on ERS-1 (50-80K Astrium (2 units)) | 88,200 | 3/02 to 4/12, No Degrad, Satellite failed |
| MIPAS on ERS-1 (50-80K Astrium (2 units)) | 88,200 | 3/02 to 4/12, No Degrad, Satellite failed |
| INTEGRAL (50-80K Astrium (4 units)) | 118,700 | Turn on 10/02, Ongoing, No degradation |
| Helios 2A (50-80K Astrium (2 units)) | 96,600 | Turn on 4/05, Ongoing, No degradation |
| Helios 2B (50-80K Astrium (2 units)) | 58,800 | Turn on 4/10, Ongoing, No degradation |
| SLSTR (50-80K Airbus (2 units)) | 1,400 | Turn on 3/16, Ongoing, No degradation |
| Planck (4K JT using 2 Astrium Comp.) | 38,500 | 5/09 thru 10/13, Mission End, No Degrad. |
| **Raytheon ISSC Stirling (STSS (2 units))** | 52,800 | Turn on 4/10, Ongoing, No degradation |
| **Rutherford Appleton Lab (RAL)** | | |
| ATSR 1 on ERS-1 (80K Integral Stirling) | 75,300 | 7/91 thru 3/00, Satellite failed |
| ATSR 2 on ERS-2 (80K Integral Stirling) | 112,000 | 4/95 thru 2/08, Instrument failed |
| **Sumitomo Stirling Coolers** | | |
| Suzaku (100K 1-stg) | 59,300 | 7/05 thru 4/12, Mission End, No degradation |
| Akari (20K 2-stg (2 units)) | 39,000 | 2/06 to 11/11 EOM, 1 Degr., 2nd failed at 13 kh |
| Kaguya GRS (70K 1-stg) | 14,600 | 10/07 thru 6/09, Mission End, No degradation |
| JEM/SMILES on ISS (4.5K JT) | 4,500 | Turn on 10/09, Could not restart at 4,500 h |
| **Sunpower Stirling** | | |
| RHESSI (75K Cryotel) | 124,600 | Turn on 2/02, Ongoing, Modest degradation |
| CHIRP (CryoTel CT-F) | 19,700 | 9/11 to 12/13, Mission End, No degradation |

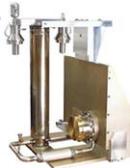
Creare NICMOS

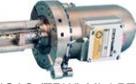
NGAS (TRW) Mini PT

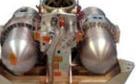
NGAS (TRW) AIRS PT

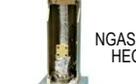
NGAS (TRW) HEC PT

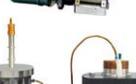
Astrium (BAe) 50-80K

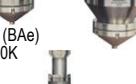
RAL ATSR

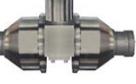
Sunpower RHESSI

Ross 05/18/16

**Figure 4-8:** Almost all cryocoolers have continued to operate normally on-orbit until turned off at the end of instrument life.

(compressors) as those that have flown. Further development of these coolers to maximize cooling power per input power for small cooling loads (<100 mW at 4 K) and lower mass is desired. There is also a need to minimize the exported vibration from the cooler system. The miniature reverse-brayton cryocoolers in development by Creare are examples of reliable coolers with negligible exported vibration. These coolers are at TRL 6 for 80 K and TRL 4 for 10 K operation (**Figure 4-8** and **Figure 4-9**).

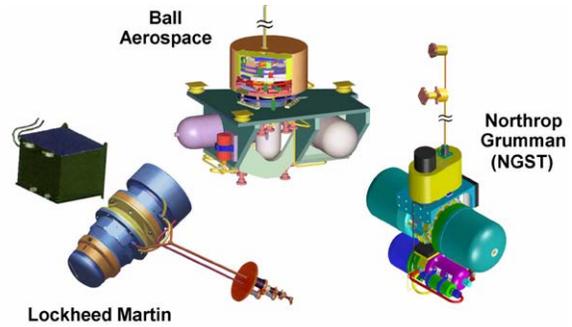

**Figure 4-9:** Three cryocoolers for 6 K cooling developed through the Advanced Cryocooler Technology Program (ACTDP).

## 5 - MANAGEMENT, SCHEDULE, AND COST

> Like the OST mission concept study, the OST mission will be developed by NASA in collaboration with international and industry partners, and responsive to community science priorities established in the 2020 Decadal Survey.

### 5.1 Concept 1 Study Team Organization

The OST Concept 1 study includes two intertwined components: the scientific case that motivates the observatory design, and an engineering study with enough detail to demonstsrate mission executability. The scientific case is driven by a community-based Science And Technology Definition Team (STDT) (**Appendix X**). The engineering study, centered at NASA's Goddard Space Flight Center (GSFC), includes input from study partners NASA's Jet Propulsion Laboratory (JPL), Infrared Processing and Analysis Center (IPAC), Japan Aerospace Exploration Agency (JAXA), Centre National d'Etudes Spatiales (CNES), Ball industry consortium (Ball), and Lockheed Martin (LM).

The OST team (**Figure 5-1**) is charged with completing the report for NASA Headquarters (HQ); the Chairs (Meixner, Cooray) are responsible for delivering the report and presenting it to the decadal committee. As the study center, GSFC (Carter, Leisawitz, Staguhn) leads the engineering and coordination effort. This diverse team was essential to a study driven by the exploration needs of the scientific community and designed with engineering rigor.

The OST management team (study office, community chairs, NASA HQ) and mission concept working group of engineers and instrument leads (led by Roellig), meet weekly via telecon/WebEx. To ensure connection with the astronomical community, OST has an external advisory group (**Appendix X.x**), advocacy and communication group (led by Battersby, Kataria, Narayan), and five science work-ing groups representing OST's major topics: galaxy and black hole formation and evolution (Pope, Vieira, Armus), formation of planetary systems (Pontoppidan, Su), exoplanets (Stevenson, Kaltenegger, Fortney, Kataria), Milkyway and nearby galaxies (Sandstrom, Battersby), and our Solar System (Milam, Bauer).

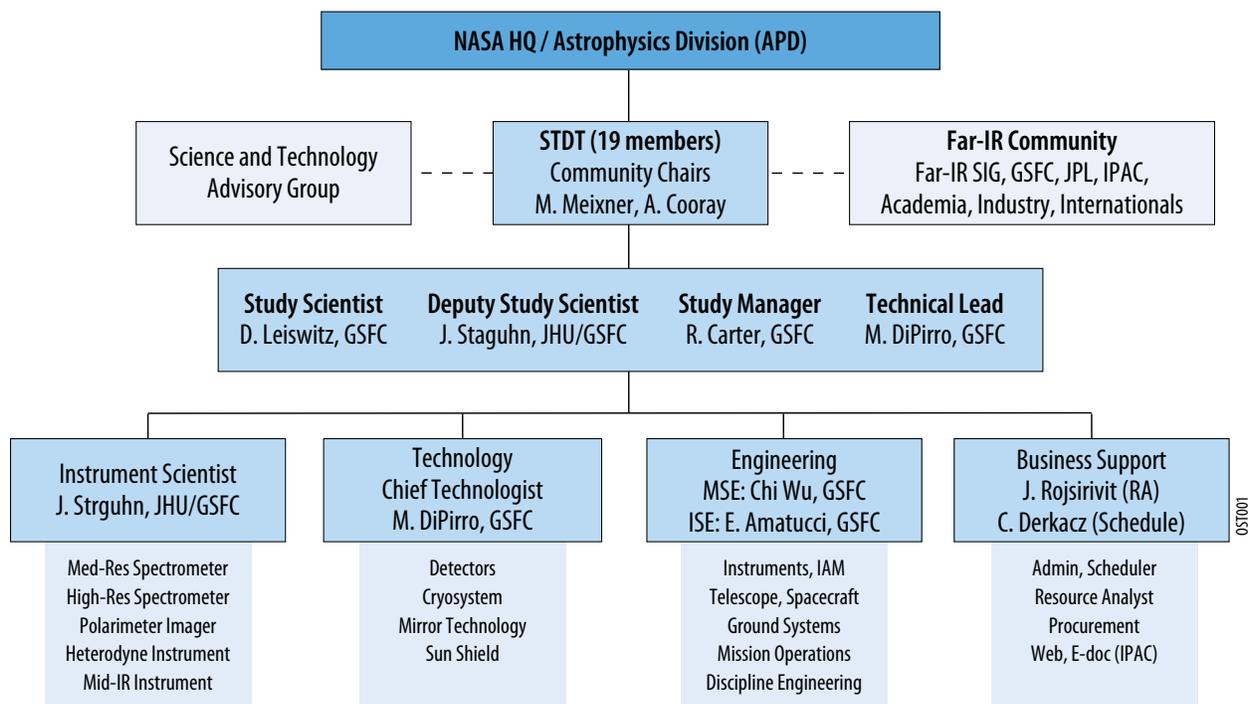

**Figure 5-1:** The OST Study Team brings together scientific and technology leaders to ensure OST represents the best interests of the astrophysics community.

Interested community members participate in regular telecons and public meetings held every 2 to 4 months. These meetings provided essential discussion, leading to meaningful mission decisions. The OST team also works with the Far-Infrared Science Interest Group (FIR-SIG).

## 5.2 Mission Implementation Organization

The Origins Space Telescope (OST) mission will be implemented in a partnership of the National Aeronautics and Space Administration (NASA), Japan Aerospace and Exploration Agency (JAXA), and Centre National d'Etudes Spatiales (CNES). These Agencies have extensive expertise and experience developing space science missions, spaceflight systems, and mid-to-far Infrared (IR) instrumentation. Successful relevant missions from this partnership include Herschel and Spitzer.

This report summarizes the organization and progress of the current OST study team. Mission roles and responsibilities for OST implementation will be finalized by NASA, JAXA, and CNES during Phase A based on the results of the technology review, mission studies, and programmatic considerations. In the scenario presented in this report, NASA leads the mission, including the spacecraft, cryocoolers, Deployment Module, mission system engineering, and mission integration and testing (I&T). In addition, NASA provides the Far-IR Imaging Polarimeter (FIP), High Resolution Spectrometer (HRS), and Medium Resolution Survey Spectrometer (MRSS) instruments. NASA also provides OST launch services. JAXA provides the Mid-IR Spectrometer and Coronagraph (MISC). CNES provides the Heterodyne instrument (HERO).

NASA's Goddard Space Flight Center (GSFC) and NASA's Jet Propulsion Laboratory (JPL) are partnering during the Study Phase to define and implement the initial steps for telescope primary mirror development, medium resolution spectrometer instrument, mission architecture, and systems engineering. The OST Study Team includes leading US aerospace companies Northrop Grumman, Ball Aerospace, Harris Engineering, and Lockheed Martin. These companies have contributed their expertise and experience in mission architecture, systems engineering, and mid/far-IR instrumentation. GSFC and JPL, along with these industry partners, have developed an OST mission architecture, Concept 1, that addresses the defined science cases and delivers the required mid- and far-IR science measurements.

OST uses design modularity to enable and facilitate development-sharing between the partners. GSFC or JPL have the capacity and expertise to lead the telescope development based on technical, cost, and schedule considerations at the time of the OST technology review in mid-2025. Instrument provider selection will occur through a coordinated multi-Agency Announcement of Opportunity (AO).

The lead Agency provides overall OST management and integration of the mission elements. The mission's lead agency also competitively selects an industry partner as prime contractor for spacecraft development and observatory integration. Formal communication interfaces are through the international partner organizations. All teams apply effective management and systems oversight lessons learned at all levels of the partnership and across all borders. To ensure clear communication, systems engineers from across the mission flight and ground segments are also members of the mission systems engineering team, and work together to define interfaces and perform system level analyses.

Engineers and managers across the mission, including all partner organizations, also participate in requirement, design, and other key reviews at the element and mission system level. Resources are applied early in the program to manage issues related to International Traffic in Arms Regulations (ITAR), Export Administration Regulations (EAR), and establishing Technical Assistance Agreements. The Program Management Council includes key stakeholders across the partner project teams and expedites issues across institutional boundaries.

## 5.3 Management Processes and Plans

The OST management process is focused on mission success and a strong commitment to continuously managing risk and staying within the allocated project resources. The OST team has formulated a mission design concept, technical approach, technology integration plan, risk management approach, budget, and a master schedule compatible with the Decadal Study guidelines.

The OST project is designed to be fully compliant with all requirements specified in NPR 7120.5, Program and Project Management Process and Requirements. The OST Project Plan, from Phase A through Phase E, is based on implemented plans of previous successful NASA-led missions. The Project Plan includes requirements, mission constraints, mission success criteria, descope options, integrated master schedule, mission technical and cost budgets, and a detailed implementation plan. Project planning addresses policies, guidelines, work breakdown structure (WBS), detailed schedules, receivables and deliverable agreements, document trees, decision-making priorities, configuration management, and quality assurance.

## 5.4 Top-Risks and Risk Mitigation

To successfully meet its mission criteria, the OST Concept 1 observatory program must deploy several key mission elements in-orbit, including the Telescope System, Deployment Tower Assembly (DTA), Instrument Accommodation Module (IAM), 4 Kelvin baffle, and five-layer sun shield. These five elements are the top mission risks due to their mission criticality. **Table 5-1** lists OST's top mission

**Table 5-9:** OST Concept 1 RIsks

| Risk # | Rank (LxC) | Risk Rating | System | Risk Statement | Mitigation |
|---|---|---|---|---|---|
| 1 | 1 (2x5) | Yellow | Observatory | If deployment tower fails to deploy, then the telescope and IAM will not reach mission operations configuration, resulting in complete loss of science | One time operation leverages heritage design. Extensive ground qualification program. |
| 2 | 1 (2x5) | Yellow | Telescope | If telescope wings fail to release from IAM, then the telescope will have no sight to science targets, resulting in complete loss of science | Leverages heritage design with redundant deployment mechanisms. Extensive ground qualification program. |
| 3 | 1 (2x5) | Yellow | Instrument Accommodation Module (IAM) | If IAM fails to deploy, then no instruments will be able to operate, resulting in complete loss of science | Leverages heritage design and incorporates redundant actuators. Extensive ground qualification program. |
| 4 | 1 (2x5) | Yellow | Observatory | If sun shields fail to deploy, then telescope and instrument temperature will be greater than 4 K, resulting in partial or complete loss of science | One-time operation incorporates redundant electronics for actuators. Extensive ground qualification program. |
| 5 | 2 (2x4) | Yellow | Observatory | If 4 K baffle fails to deploy, then the telescope will be exposed to higher IR background, resulting in diminished science | One-time operation. Extensive ground qualification program. |
| 6 | 2 (2x4) | Yellow | Far-IR Detectors | If detectors cannot achieve sky background-limited sensitivity, then discovery of weak signals will be limited, resulting in diminished science | Detector technology developments are included in the OST Technology Roadmap, and far-IR detector technology development is currently funded. Graceful degradation for diminished sensitivity. |
| 7 | 2 (2x4) | Yellow | Mid-IR Detectors | If 5 ppm stability cannot be achieved, then the number of discoveries of exoplanets with life signatures will be decreased, resulting in diminished science | Mid-IR detector technology development is included in the OST Technology Roadmap. |
| 8 | 3 (1x3) | Green | Cryocoolers | If cryocoolers fail to provide required 4 K environment for telescope and instruments, then far-IR science will be severely impacted, resulting in diminished science | Heritage, testing of existing hardware, and technology development. Cryocooler system is fully redundant. |

risks and mitigation strategies. **Figure 5-2** shows risk ranking in 5X5 Risk Matrix.

In addition to these primary systems, **Table 5-1** also includes developing the far-IR detectors to TRL 5 by mid-2020. The Microwave Kinetic Inductance Detector (MKID) and Transitions Edge Sensor (TES) bolometer detectors are currently at TRL 3 and 4, respectively, and are fully-funded to develop to TRL 5 in the next ~2 years. Either detector can provide the required far-IR measurements, and OST needs only one of the detectors to reach TRL 5 by 2025. This risk is further mitigated by Strategic Astrophysics Technology (SAT) funded technology developments for far-IR detectors.

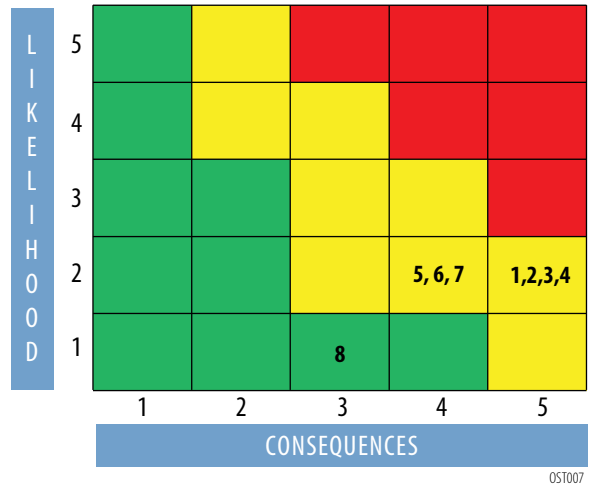

**Figure 5-2:** The OST Concept 1 Risk Matrix

OST risks also include cryocooler technology development. A SAT-funded development for the sub-Kelvin cooler is underway and is scheduled to reach at least TRL 5 by 2021. Analysis of existing cryo technology shows existing cryocoolers can meet OST's low-temperature requirements, further mitigating this risk.

OST mission risk is mitigated by the technology development efforts in place for all OST technologies below TRL 5, ensuring mission-enabling technologies will be available no later than 2025.

## 5.5 Mission Development Schedule

The overall OST mission development schedule is provided in **Figure 5-3**. Milestones and key decision points are consistent with NASA Procedural Requirements NPR-7120.5. The schedule supports a September 2025 launch, providing ~10 years for development, and schedule includes 10 months of funded reserve. The plan nominally includes 5 years of mission operations after launch, with an option to extend the science mission to 10 years.

## 5.6 Mission Cost

Mission development cost for OST Concept 1 is under development and not included in this interim report.

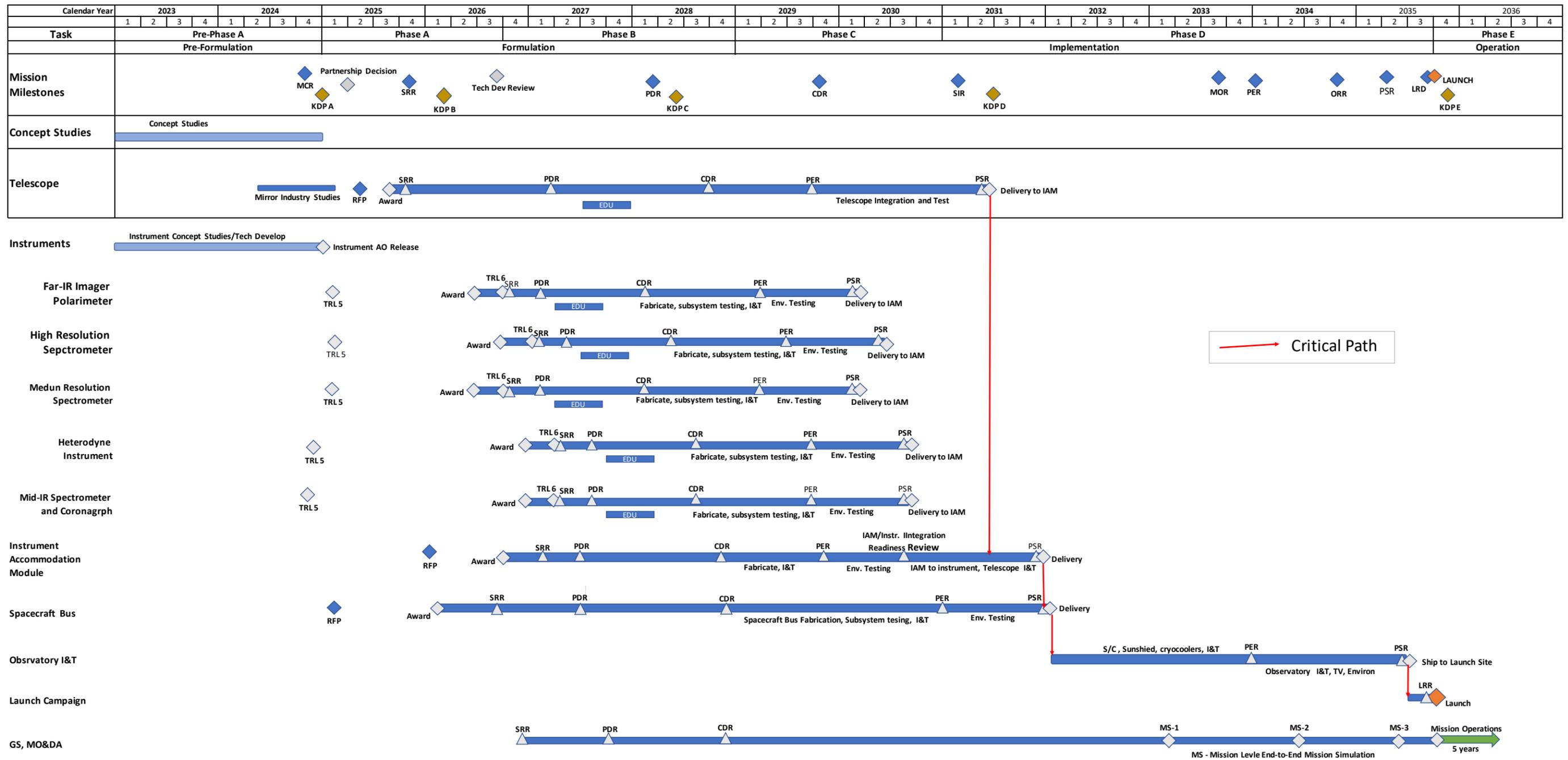

**Figure 5-3:** The OST Mission Development Schedule is derived from the successful schedules of similar scale missions managed by the partner Agencies.

# Appendix 1: References

## Appendix 2: Acronyms and Abbreviations

| Acronym | Definition |
|---------|------------|
| 2D | two dimensional |
| 3D | three dimensional |
| A/D | Analog/Digital |
| AB | asteroid belt |
| ACE | Attitude Control Electronics |
| ACS | Attitude Control System |
| ACT | Atacama Cosmology Telescope |
| ADC | analog-to-digital converter |
| ADR | Adiabatic Demagnetization Refrigerator |
| ADU | asynchronous data unit |
| AGB | asymptotic giant branch |
| AGN | Active Galactic Nuclei |
| AI&T | Assembly, Integration, and Test |
| Akari | aka ASTRO-F |
| Al2O$_3$ | aluminum oxide |
| AlBeMet | aluminum beryllium metal matrix |
| ALMA | Atacama Large Millimeter Array |
| AMES | NASA's Ames Research Center |
| amp | ampere |
| AO | Announcement of Opportunity |
| ARCADE | Absolute Radiometer for Cosmology, Astrophysics, and Diffuse Emission |
| As | arsenic |
| ASIC | application-specific integrated circuit |
| AST/RO | Antarctic Submillimeter Telescope and Remote Observatory |
| ASTRO-H | aka Hitomi |
| ATK | Alliant Techsystems |
| ATLAS | Atlas rocket |
| AU | Astronomical Unit |
| Ball | Ball Consortium |
| BEI | Baseline Execution Index |
| BHAR | black hole accretion rate |
| BHARD | black hole accretion rate density |
| BIB | Blocked Impurity Band |
| BOE | Basis of Estimate |
| BPM | binary pupil mask |
| BUG | Backshort Under Grid |
| C | carbon |
| C&DH | Command and Data Handling |
| C$_2$N$_2$ | cyanogen |
| C$_3$H$_4$ | propyne |
| C$_4$H$_2$ | diacetylene/butadiyne |
| CA | California |
| CaF$_2$ | calcium fluoride (fluorite) |
| CBE | Current Best Estimate |

| Acronym | Definition |
|---------|------------|
| CCB | Change Control Board |
| CCB | Configuration Control Board |
| CCHP | constant conductance heat pipe |
| CCR | Configuration Change Request |
| CDH | Command and Data Handling |
| CDR | Critical Design Review |
| CdTe | cadmium telluride |
| CEB | Cold Electronics Box |
| CEI | Current Execution Index |
| CEMA | Cost Estimating, Modeling, and Analysis |
| CFRP | carbon fiber reinforced polymer |
| cFS | Core Flight Executive Software System |
| cg | center of gravity |
| CH$_4$ | methane |
| CIRS | Composite Infrared Spectrometer |
| CLAES | Cryogenic Limb Array Etalon Spectrometer |
| CLASS | Cosmology Large Angular Scale Surveyor |
| CM | center of mass |
| cm | centimeter |
| CM | Configuration Management |
| CM&O | Center Management and Operations |
| CMB | cosmic microwave background |
| CMD | command |
| CMG | Control Moment Gyro |
| CMMI | Capability Maturity Model Integration |
| CMOS | complementary metal-oxide semiconductor |
| CNES | Centre National d'Etudes Spaciales |
| CO | carbon monoxide |
| CO | Contracting Officer |
| Co-I | Co-Investigator |
| CO$_2$ | carbon dioxide |
| COBE | Cosmic Background Explorer |
| CoC | Certificate of Compliance |
| Col | collimated |
| COPV | Composite Overwrapped Pressure Vessel |
| COR | Coronograph |
| Cornell | Cornell University |
| COSMOS | Cosmos Evolution Survey |
| COTS | Commercial-off-the-Shelf |
| CPS | Cycles Per Second |
| CPT | Comprehensive Performance Test |
| CPT | Current Procedural Terminology |
| cRIO | Compact Reconfigurable Input Output |
| CSA | Consumer Safety Agency |
| CSR | Concept Study Report |
| CSS | Coarse Sun Sensor |

| Acronym | Definition |
|---------|------------|
| Cu | copper |
| D | deuterium |
| D/H | deuterium to hydrogen ratio |
| DAC | digital-to-analog converter |
| DC | direct current |
| DCS | Data Cycle System |
| DDL | Detector Development Lab |
| deg | degree |
| DH | Data Handler |
| DIRBE | Diffuse Infrared Background Experiment |
| DITL | Day in the Life |
| DM | deformable mirror |
| DO | dissolved oxygen |
| DPI | Deputy Principal Investigator |
| DR | dilution refrigerator |
| Dr | Doctor |
| DTA | Deployable Tower Assembly |
| EAR | Export Administration Regulations |
| EGSE | Electrical Ground Support Equipment |
| EM | Engineering Model |
| EMC | Electromagnetic Compatibility |
| EMI | Electromagnetic Interference |
| EMTB | Engineering Model Test Bed |
| EOM | End of Mission |
| EOR | Epoch of Reionization |
| EPD | Emergency Power Disconnect |
| EPS | Electrical and Power System |
| EQFLUX | equivalent fluence model |
| ESA | European Space Agency |
| ETH | Swiss Federal Institute of Technology in Zurich |
| EXES | Echelon-Cross-Echelle Spectrograph |
| FIFI-LS | Far Infrared Field-Imaging Line Spectrometer |
| FIP | Far Infrared Imager/Polarimeter |
| FIR | far infrared |
| FIR-SIG | Far-Infrared Science Interest Group |
| FIRAS | Far Infrared Absolute Spectrometer |
| FIRST | Far-Infrared Spectroscopy of the Troposphere |
| FITS | Flexible Image Transport System |
| FM | Financial Manager |
| FMECA | Failure Mode Effects and Criticality Analysis |
| FOD | Foreign Object Debris |
| FoR | Field of Regard |
| FORCAST | Faint Object infraRed Camera for the SOFIA Telescope |
| FOV | Field of View |
| FP | Focal Plane |
| FPA | Focal Plane Assembly |
| FPI | Fabry-Perot Interferometer |
| FPOMA | Focal Plane Opto-Mechanical Assembly |

| Acronym | Definition |
|---------|------------|
| FSM | Field Steering Mirror |
| FSR | Free Spectral Range |
| FT | Functional Test |
| FTA | Fault Tree Analysis |
| FTE | Full Time Equivalent |
| FTS | Fourier Transform Spectrometer |
| FTS | Fourier Transform Spectroscopy |
| FY | Fiscal Year |
| GaAS | gallium arsenide |
| GEVS | Goddard Environmental Verification Specification |
| GFE | Government Furnished Equipment |
| GHE | gaseous helium |
| GHz | gigahertz |
| GI | Guest Investigator |
| GIDEP | Government-Industry Data Exchange Program |
| GO | Guest Observer |
| GOODS | Great Observatories Origins Deep Survey |
| GPM | Global Precipitation Measurement |
| GPR | Goddard Procedural Requirement |
| GREAT | German Receiver for Astronomy at Terahertz frequencies |
| GSE | Ground Support Equipment |
| GSFC | NASA's Goddard Space Flight Center |
| GUSTO | Galactic/Extragalactic Ultra-long Duration Balloon Spectroscopic Terahertz Observatory |
| H | hydrogen |
| H/W | Hardware |
| $H_2$ | hydrogen gas |
| $H_2O$ | water |
| HAWC+ | High-resolution Airborne Wideband Camera |
| HC | hydrocarbon |
| HCN | hydrogen cyanide |
| HD | hydrogen deuteride |
| HDO | high dissolved oxygen |
| He | helium |
| He4 | helium-4 |
| HEB | hot electron bolometer |
| HEMT | High Electron Mobility Transistors |
| HeNe | helium-neon |
| HERO | Heterodyne Receiver for OST |
| Herschel | Herschel Space Observatory |
| HFSW | high frequency surface wave |
| HGA | high gain antenna |
| HIFI | Heterodyne Instrument for the Far-Infrared |
| High-res | High Resolution |
| HIPO | High-speed Imaging Photometer for Occultation |
| HIRMES | High Resolution Mid-infrared Spectrometer |
| HK | housekeeping |
| HQ | Headquarters |

| Acronym | Definition |
|---------|------------|
| HRS | High Resolution Spectrometer |
| HRS-L | High Resolution Spectrometer long-wave channel |
| HRS-S | High Resolution Spectrometer short-wave channel |
| HST | Hubble Space Telescope |
| HUDF | Hubble Ultra Deep Field |
| HV | high voltage |
| HW | Hardware |
| HWP | half-wave plate |
| I/O | Input/Output |
| I&S | Imager and Spectrometer |
| I&T | Integration and Test |
| IAM | Instrument Accommodation Module |
| IC&DH | Instrument Control and Data Handling |
| ICD | Interface Control Document |
| ICDH | Instrument Control Data Handling |
| ICS | Instrument Concept Study |
| IDL | Instrument Design Laboratory |
| IF | intermediate frequency |
| IFU | Integral Field Unit |
| IMS | Integrated Master Schedule |
| IMU | inertial measurement unit |
| IPAC | Infrared Processing and Analysis Center |
| IR | infrared |
| IRAC | Infrared Array Camera |
| IRAS | Infrared Astronomical Satellite |
| IRC | Instrument Remote Control |
| IRD | Interface Requirements Document |
| IRMOS | Infrared Multiobject Spectrograph |
| IRS | Infrared Spectrograph |
| ISE | Instrument Systems Engineer |
| ISM | interstellar medium |
| ISO | Infrared Space Observatory |
| ISS | International Space Station |
| IT | Information Technology |
| ITAR | International Traffic in Arms Regulations |
| IV | current-voltage |
| IWA | inner working angle |
| JAXA | Japan Aerospace Exploration Agency |
| JCMT | James Clerk Maxwell Telescope |
| JHU | Johns Hopkins University |
| JPL | NASA's Jet Propulsion Laboratory |
| JSC | NASA's Johnson Space Center |
| JUICE | Jupiter Icy moons Explorer |
| JWST | James Webb Space Telescope |
| K | Kelvin |
| K | potassium |
| KAO | Kuiper Airborne Observatory |
| KB | Kuiper Belt |

| Acronym | Definition |
|---------|------------|
| KBO | Kuiper Belt Object |
| Kg | Kilograms |
| KID | kinetic inductance detector |
| km | kilometer |
| KPNO | Kitt Peak National Observatory |
| KWIC | Kuiper Widefield Infrared Camera |
| $L_2$ | Lagrange Point 2 |
| LADEE | Lunar Atmosphere Dust and Environment Explorer |
| LCRD | Laser Communications Relay Demonstration |
| LEON 3 | processor |
| LEV | $L_2$-Earth-Vehicle |
| LHe | liquid helium |
| LIRG | luminous infrared galaxy |
| LM | Lockheed Martin |
| $LN_2$ | liquid nitrogen |
| LNA | low-noise-amplifier |
| LO | local oscillator |
| LOS | line of sight |
| low-res | low-resolution |
| LPT | Limited Performance Test |
| LRO | Lunar Reconnaissance Orbiter |
| LRS | low-resolution spectrometer |
| LRU | Line Replaceable Units |
| LSST | Large Synoptic Survey Telescope |
| LTS | Long Term Support |
| LV | launch vehicle |
| LV | low voltage |
| m | meter |
| MAPTIS | Materials and Processes Technical Information System |
| MCCS | Mission Control and Communications System |
| MCE | Multi Channel Electronics |
| MCS | Mid-infrared Camera and Spectrometer |
| MEL | Master Equipment List |
| MEV | Maximum Expected Value |
| MGSE | Mechanical Ground Support Equipment |
| MICHI | Mid-Infrared Camera with High spectral resolution and IFU |
| Mid-IR | Mid-Infrared |
| Mid-res | Mid Resolution |
| MIL | military |
| MIL | multi-layer insulation |
| MIMIZUKU | Mid-infrared Multimode Imager for gaZing at the Un-Known Universe |
| min | minute |
| MIR | mid-infrared |
| MIRI | Mid-Infrared Instrument |
| MIRO | Microwave Instrument for Rosetta Orbiter |
| MISC | Mid-infrared Imager, Spectrometer, Coronagraph |

| Acronym | Definition |
|---------|------------|
| MKID | Microwave Kinetic Inductance Detector |
| mm | millimeter |
| MMS | Magnetospheric Multiscale |
| MMSM | minimum mass solar nebula |
| Mo/Au | molybdenum/gold |
| MOC | Mission Operations Center |
| MON-3 | Mixed Oxides of Nitrogen |
| MRSS | Medium Resolution Survey Spectrometer |
| MUSTANG | Modular Unified Space Technology Avionics for Next Generation |
| N | nitrogen |
| N/A | Not Applicable |
| $N_2H$ | diazenylium |
| $N_2O$ | nitrous oxide |
| Na | sodium |
| NASA | National Aeronautics and Space Administration |
| Nb | niobium |
| NEA | Non-Explosive Actuators |
| NEFD | noise equivalent flux density |
| NELF | Noise Equivalent Line Flux |
| NEP | Noise Equivalent Power |
| $NH_3$ | ammonia |
| NICER | Neutron star Interior Composition Explorer |
| NIRCam | Near Infrared Camera |
| NIST | National Institute of Standards and Technology |
| Nms | Newton meter second |
| NOVICE | Newly-Operating and Validated Instruments Comparison Experiement |
| NRA | NASA Research Announcement |
| O | oxygen |
| $O_3$ | ozone |
| OAN | Observatorio Astronomico Nacional |
| OAP | Off-Axis-Paraboloid |
| ODIN | Odin mission |
| OFSW | OST Flight Software |
| OIPB | Optical Image Performance Budget |
| OMB | Office of Business and Management |
| OPM | Office of Personnel Management |
| ORR | Operations Readiness Review |
| OST | Origins Space Telescope |
| OTE | Optical Telescope Element |
| OTIAM | Optical Telescope Element with Primary Mirror and Instrument Accommodation Module |
| OWA | outer working angle |
| PACS | Photoconductor Array Camera and Spectrometer |
| PAH | polycyclic aromatic hydrocarbons |
| pc | parsec |
| PCA | pressure control assembly |

| Acronym | Definition |
|---------|------------|
| PCS | Precision Cryogenic Systems |
| PDF | Probability Distribution Function |
| PDR | Preliminary Design Review |
| PDU | power data unit |
| PI | Principal Investigator |
| PIA | Propellant Isolation Assembly |
| PIPER | Primordial Inflation Polarization Explorer |
| PIR | Pre-Integration Review |
| PMD | Propellant Management Device |
| PMP | Project Management Plan |
| POD | Partnership Opportunity Document |
| POM | pick-off mirror |
| PR | Problem Report |
| PRA | Probabilistic Risk Analysis |
| PRS | Precision Rotation Stage |
| PS | Pressure System |
| PSA | Parts Stress Analysis |
| PSF | Point Spread Function |
| PSR | Pre-Ship Review |
| PT | Pulse Tube |
| PTC | Pulse-Tube Cooler |
| PV | Pressure Vessel |
| PVC | polyvinylchloride |
| PZT | Piezoelectric Transducer |
| Q | quality factor |
| QA | Quality Assurance |
| QCD | Quantum Capacitance Detector |
| QCL | quantum cascade laser |
| R&D | Research and Development |
| RAID | Redundant Array of Inexpensive Disks |
| RBSP | Radiation Belts Storm Probes (aka Van Allen Probes) |
| Res | resolution |
| RF | radio frequency |
| RFA | Request For Action |
| RJ | Rayleigh-Jeans |
| RMB | Risk Management Board |
| RMSWE | root mean square wavefront error |
| ROSES | Research Opportunities in Earth and Space Science |
| Rosetta | Rosetta mission |
| RP | Resolving Power |
| RWA | Reaction Wheel Assembly |
| RY | Real Year |
| s | second |
| S/W | Software |
| S&MA | Safety and Mission Assurance |
| SAFARI | imaging spectrometer instrument |
| $SAFIR_E$ | Submillimeter and Far-Infrared Experiment |
| SAO | Smithsonian Astronomical Observatory |

| Acronym | Definition |
| --- | --- |
| SAT | Strategic Astrophysics Technology |
| Sb | antimony |
| SC | spacecraft |
| SCE | Spacecraft Element |
| SCI | SPICA Coronagraphic Instrument |
| SCUBA | Sub-millimeter Common User Bolometer Array |
| SDL | Space Dynamics Laboratory |
| SDO | Scattered-Disk Object |
| SED | spectral energy distributions |
| $SEL_2$ | Sun-Earth-$L_2$ |
| SEMP | Systems Engineering Management Plan |
| SFR | star formation rate |
| SFRD | star formation rate density |
| SHARC | Submillimeter High Angular Resolution Camera |
| SHIELDOSE | space-shielding radiation dose |
| Si | silicon |
| SiC | silicon carbide |
| SiGe | silicon-germanium |
| SiN | silicon mononitride |
| SIS | solid-state imaging spectrometer |
| SLS | Space Launch System |
| SMA | Safety and Mission Assurance |
| SMART | Spectroscopic Modeling Analysis and Reduction Tool |
| SME | Subject Matter Expert |
| SMI | SPICA Mid-infrared Instrument |
| SMR | Sub-Millimetre Radiometer |
| SNR | Signal to Noise Ratio |
| SOC | Science Operations Center |
| SOFIA | Stratospheric Observatory for Infrared Astronomy |
| SPECULOOS | Search for habitable Planets Eclipsing Ultra-cOOl Stars |
| SPENVIS | SPace ENVironment Information System |
| SPICA | Space Infrared Telescope for Cosmology and Astrophysics |
| SPIFI | South Pole Imaging Fabry Perot Interferometer |
| $SPIR_E$ | Spectral and Photometric Imaging Receiver |
| Spitzer | Spitzer Space Telescope |
| SQUID | Superconducting Quantum Interference Device |
| SRON | Netherlands Institute for Space Research |
| SRR | Systems Requirements Review |
| SSIRU | Scalable Space Inertial Reference Unit |
| SSPC | Solid State Power Controllers |
| SSR | solid state recorder |
| ST | Star Tracker |
| STD | Standard |
| STDT | Science and Technology Definition Team |
| STM | Science Traceability Matrix |
| STO | Stratospheric TeraHertz Observatory |
| STScI | Space Telescope Science Institute |
| SVR | Servicing Vehicle Rendezvous |

| Acronym | Definition |
| --- | --- |
| SW | Software |
| SWAS | Submillimeter Wave Astronomy Satellite |
| SWI | Sub-millimeter Wave Instrument |
| SWS | Short Wavelength Spectrometer |
| TA | Telescope Assembly |
| TAAS | Telescope Assembly Alignment Simulator |
| TAO | Transient Astrophysics Observatory |
| TCMB | temperature cosmic microwave background |
| TDMS | Technical Data Management System |
| TES | Transition Edge Sensor |
| TESS | Transiting Exoplanet Survey |
| THz | terahertz |
| Ti | Titanium |
| TID | Total Ionizing Dose |
| TJ | triple junction |
| TLM | telemetry |
| TMA | three mirror anastigmat |
| TMT | Thirty Meter Telescope |
| TNO | trans-Neptunian object |
| TOO | Target of Opportunity |
| TPM | technical performance measure |
| TRAPPIST | TRAnsiting Planets And PlanetesImals Small Telescope |
| TRL | Technical Readiness Level |
| TT&C | Telemetry Tracking and Control |
| TTM | tip-tilt mirror |
| U | University |
| UARS | Upper Atmosphere Research Satellite |
| ULE | ultra-low expansion |
| UPS | Uninterruptible Power Supply |
| URLIG | ultraluminous infrared galaxy |
| UV | ultraviolet |
| V | volt |
| V&V | Verification and Validation |
| VDC | Volts Direct Current |
| VNIR | Visible and Near Infrared |
| VxWorks | operating system |
| W | Watts |
| WBS | Work Breakdown Structure |
| WCA | Worst Case Analysis |
| WFI | Wide-Field Imager |
| WFI-L | Wide-Field Imager-Long wavelength channel |
| WFI-S | Wide-Field Imager-Short wavelength channel |
| WFIRST | Wide-Field Infra-Red Survey Telescope |
| WISE | Wide Field Infrared Survey Explorer |
| WSC | White Sands Complex |
| $\Delta$ | delta (change) |
| $\mu m$ | micrometer |
| $\sigma$ | sigma (standard deviation) |

**Appendix 3: Supporting Documentation**

# OST Interim Report Authorship

A large team of scientists and engineers from NASA centers, academia, industry partners, and international institutions contributed to this report.

## *OST STDT*

| | | | |
|---|---|---|---|
| Cooray | Asantha | University of California, Irvine | Study Chair |
| Meixner | Margaret | STScI, Johns Hopkins University | Study Chair |
| Leisawitz | David, T. | NASA GSFC | Study Scientist |
| Armus | Lee | Caltech/IPAC | |
| Battersby | Cara | University of Connecticut | |
| Bauer | James | University of Maryland, College Park | |
| Bergin | Edwin | University of Michigan, Ann Arbor | |
| Bradford | Charles Matt | JPL/Caltech/NASA | |
| Ennico-Smith | Kimberly | NASA Ames | |
| Fortney | Jonathan J. | University of California, Santa Cruz | |
| Kataria | Tiffany | NASA JPL | |
| Melnick | Gary J. | Harvard, Smithsonian, CfA | |
| Milam | Stefanie N. | NASA GSFC | |
| Narayanan | Desika | University of Florida | |
| Padgett | Deborah | JPL/NASA | |
| Pontoppidan | Klaus | STScI | |
| Pope | Alexandra | University of Massachusetts, Amherst | |
| Roellig | Thomas | NASA Ames | |
| Sandstrom | Karin | University of California, San Diego | |
| Stevenson | Kevin B. | STScI | |
| Su | Kate Y. | University of Arizona | |
| Vieira | Joaquin | University of Illinois, Urbana-Champaign | |
| Wright | Edward | University of California, Los Angeles | |
| Zmuidzinas | Jonas | Caltech | |

### Non-voting STDT Members

| | | | |
|---|---|---|---|
| Sheth | Kartik | NASA Headquarters | Program Scientist |
| Benford | Dominic | NASA Headquarters | Deputy Program Scientist |
| Staguhn | Johannes G. | Johns Hopkins University/NASA GSFC | Deputy Study Scientist |
| Mamajek | Eric E. | NASA JPL | ExEP Deputy Program Scientist |
| Neff | Susan G. | NASA GSFC | COR Program |
| Carey | Sean | Caltech/IPAC | Communications Scientist |
| Burgarella | Denis | Laboratoire d'Astrophysique de Marseille | Collaborator |
| De Beck | Elvire | Chalmers Institute of Technology | SNSB Liaison |
| Gerin | Maryvonne | Sorbonne Université, Observatoire de Paris, Université PSL, CNRS, LERMA, F-75014, Paris, France | CNES Liaison |
| Helmich | Frank P. | SRON | Netherlands Institute for Space Research Liaison |
| Sakon | Itsuki | University of Tokyo | JAXA Liaison |
| Scott | Douglas | University of British Columbia | CAS Liaison |
| Wiedner | Martina C. | Sorbonne Université, Observatoire de Paris, Université PSL, CNRS, Laboratoire d'Etudes du Rayonnement et de la Matière en Astrophysique et Atmosphères (LERMA) F-75014, Paris, France | Instrument lead |
| Moseley | S. Harvey | NASA GSFC | HRS Instrument lead |

## *Engineering Study Team at NASA GSFC*

| | | |
|---|---|---|
| Amatucci | Ed | Instrument Systems Engineer (ISE) |
| Beltran | Porfi | Electrical and Data Systems |
| Bradley | Damon | Electrical and Data Systems |
| Carter | Ruth | Study Manager |
| Corsetti | James | Optics |
| Denis | Kevin | Detectors |
| DiPirro | Mike | System Architect and Chief Technologist |
| Fantano | Lou | Thermal |
| Folta | Dave | Flight Dynamics |
| Generie | Joe | Mechanical Systems |
| Hilliard | Larry | Instrument Electronics |
| Howard | Joe | Optics |
| Jamil | Anisa | Power System |
| Martins | Greg | Mechanical |
| Petro | Susanna | Integration and Test |
| Rampspacker | Dan | Propulsion |
| Sandin | Carly | Materials |
| Stoneking | Eric | ACS |
| Tompkins | Steve | Mission Ops and Ground System |
| Webster | Cassie | Flight Dynamics |
| Wu | Chi | Mission Systems Engineer (MSE) |

## Industry Partners

| | | | |
|---|---|---|---|
| Arenberg | Jon | Northrop Grumman | Systems Engineer |
| Pohner | John | Northrop Grumman | Thermal Engineer |
| Knollenberg | Perry | Northrop Grumman | Thermal Engineer |
| Harpole | George | Northrop Grumman | Thermal Engineer |
| Nguyen | Thanh | Northrop Grumman | Cryocooler Engineer |
| Chi | Danny | Northrop Grumman | Cryocooler Engineer |
| Petach | Michael | Northrop Grumman | Cryocooler Engineer |
| Lightsy | Paul | Ball Aerospace | Optical Engineer |
| Knight | Scott | Ball Aerospace | Systems Engineer |
| Lipscy | Sara | Ball Aerospace | Systems Scientist |
| East | Matthew | Harris | Optical Engineer |
| Allen | Lynn | Harris | Civil and Commercial Imaging Business Development Engineer |
| Mooney | Ted | Harris | Civil and Commercial Imaging Chief Technologist |
| Harvey | Keith | Harris | Thermal Systems Scientist |
| Feller | Greg | Lockheed Martin | Project Technical Lead |
| Stokowski | Larry | Lockheed Martin | Optical Engineer/Structures |
| Granger | Zac | Lockheed Martin | Optical Engineer |
| Steeves | John | NASA JPL | Optical Engineer |
| Nordt | Alison | Lockheed Martin | Project Manager |
| Dewell | Larry | Lockheed Martin | LM Fellow, Control Systems |
| Tajdaran | Kia | Lockheed Martin | Control Systems |
| Olson | Jeff | Lockheed Martin | Thermal, Cryocooler |
| Edgington | Samantha | Lockheed Martin | Instrument Scientist/Systems Engineer |
| Bell | Ray | Lockheed Martin | LM Senior Fellow |
| Jacoby | Mike | Lockheed Martin | Optical Engineer/Structures |

## Science and Technology Advisory Group

| Last | First | Institution |
|---|---|---|
| Arenberg | Jon | Northrop-Grumman |
| Carlstrom | John | U. Chicago |
| Ferguson | Harry | STScI |
| Greene | Tom | NASA Ames |
| Helou | George | IPAC |
| Kaltenegger | Lisa | Cornell |

| Lawrence | Charles | NASA JPL |
|----------|---------|----------|
| Lipscy | Sarah | Ball Aerospace |
| Mather | John | NASA GSFC |
| Moseley | Harvey | NASA GSFC |
| Rieke | George | U. Arizona |
| Rieke | Marcia | U. Arizona |
| Turner | Jean | UCLA |
| Urry | Meg | Yale |

# ORIGINS Space Telescope Master Schedule

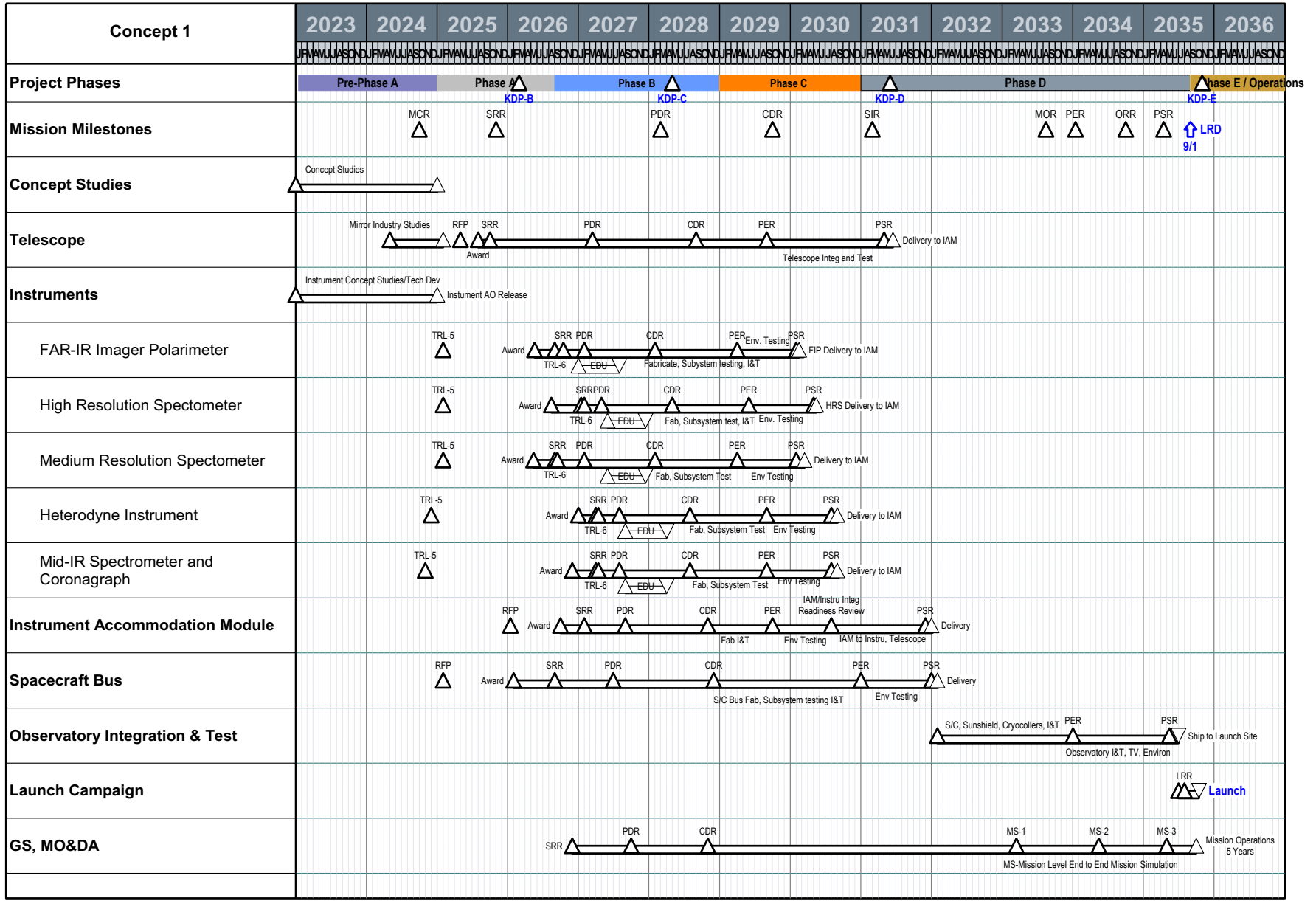

8/3/18

| Propsal Number and Title | Type | Vote Rank | Vote Mean | Vote Std Dev | Wavelength Range (µm) | Max Required Spatial Resolution (arc-sec) | Max Required Spatial Resolution (arc-sec: equ at 50) | Equivalent Single Dish Diameter (m) (1.22 lambda/D) | Interferometer Baseline (m) (0.87 lambda/b) | Spectral Resolving Power (lambda/d-lambda) | Min Sky Coverage (sq. deg) | Number of Targets | Continuum Sensitivity (µJy at X µm) | Line Sensitivity (W m-2) | Surface brightness sensitivity (MJy/str) | Comments | Normalized to Continuum sensitivity (µJy) sigma | Normalized to Line Sensitivity (W-m-2) sigma | Integration time needed for program - interferometer D=3.5m, b=20m (Hrs) | Integration time needed for program - interferometer D=6.5m, b=50m (Hrs) | Integration time needed for program - single dish D=5m (Hrs) | Integration time needed for program - single dish D=10m (Hrs) | Integration time needed for program - single dish D=15m (Hrs) | Notes |
|---|---|---|---|---|---|---|---|---|---|---|---|---|---|---|---|---|---|---|---|---|---|---|---|---|
| The Rise of Metals | G&H (Key for these codes?) | 1 | 8.78 | 0.94 | 10 - 400 | 5 @ 200 µm | 1.25 (why are these columns red?) | 10.00 | 7.5 | 5.00E+02 | 20 | 10000 | | 1.00E-21 | | Resolution quoted at 200 microns. | 1.67 @ 10 µm | 1E-21@200 | 100000 (why are these data in blue?) | 3.16E+04 | 1.90E+05 | 5.10E+04 | 2.11E+04 | |
| Water Content of Planet-Forming Disks | EP | 2 | 8.72 | 1.67 | 28 - 180 | 2 @ 30µm | 3.30 | 3.76 | 2.7 | 2.50E+04 | | 1000 | | 1.00E-21 | | A significant part of the science can be retained with 2.0" @30 micron, so this should be changed to be the requirement, with 0.5" @30 micron the desired value. | 2.3e2 @ 28 µm | 1.00E-21 | 6.70E+03 | #VALUE! | 6.70E+03 | 4.10E+02 | 8.40E+01 | Object positions known ahead of time |
| The First Dust | EU | 3 | 8.72 | 1.45 | 20 - 220 | 5 @ 100 µm | 2.50 | 4.96 | 3.5 | 5.00E+01 | | 100 | | 1.00E-20 | | | 3.33 @ 20 µm | 1.00E-20 | 1.00E+02 | 8.05E+00 | 1.00E+02 | 1.80E+00 | 1.23E+00 | Object positions known ahead of time, scaling from R=500 to R=100 with 5x in time, Discrepency |
| Direct Detection of Protoplanetary Disk Masses | EP | 4 | 8.17 | 1.89 | 112 | 3.0 @ 112 µm | 1.30 | 9.54 | 6.8 | 1.00E+04 | | 500 | | 1.00E-21 | | | 1.12E+02 | 1.00E-21 | | 2.25E+02 | 3.40E+03 | 2.05E+02 | 4.20E+01 | Object positions known ahead of time |
| Super Earth Biosignatures and Climates | EP | 5 | 7.89 | 1.88 | 9 - 40 | N/A | N/A | N/A | N/A | 5.00E+02 | | 30 | 0.2 with R=20, in one hour | | | Sensitivity=0.2 microjy in 1 hour at 15 micron with R=20 bins. This is about 10x that of JWST, and would allow the characterization of a 400 K super-Earth around an M star at 10 pc. The desired 15 ppm might enable a 300 K super-Earth at 5 pc. | 0.2 with R=20, in one hour | 3.30E-21 | 6.00E+02 | 6.00E+02 | 6.00E+02 | 6.00E+02 | 6.00E+02 | Assuming sum of 20 hours of transit time per target |
| Water Transport to Terrestrial Planetary Zone | MW | 6 | 7.83 | 1.76 | 270 - 600 | 30 @ 545µm | 2.80 | 4.43 | 2.3 | 5.00E+06 | | 120 | | 1.00E-19 | | | 4.5 e7 @ 30 µm | 1.00E-19 | | | 2.40E+03 | 6.00E+02 | 3.00E+01 | Adopting Tsys =2000 K * nu / 1.9 THz |
| Connection Between Black Hole Growth and Star Formation Over Cosmic Time | G&H | 7 | 7.72 | 1.71 | 30 - 300 | 5 @ 100 µm | 2.50 | 5.00 | 3.6 | 1.00E+02 | 10 | 20000 | | 1.00E-21 | | | 1.0 @ 30 µm | 1.00E-21 | | | 1.02E+05 | 2.56E+04 | 1.16E+04 | Time based on survey area at 200 microns, expect to find the number of objects listed in column N. Some "representative samples" of these tragets would be followed up with R=3,000 spectroscopy. This follow-up time is not included here. |
| Birth of Galaxies During Cosmic Dark Ages | EU | 8 | 7.67 | 2.35 | 32 - 275 | 5 @ 100 µm | 2.50 | 4.96 | 3.5 | 5.00E+02 | | 18 | | 4.00E-22 | | 5' Resolution at 100 microns | 2.13 @ 32 µm | 4.00E-22 | | | 3.00E+03 | 1.84E+02 | 3.80E+01 | Object positions known ahead of time, 4 pointings per cluster |
| Galaxy Feedback from SNe and AGN to z=3 | G&H | 9 | 7.61 | 1.5 | 20 - 300 | 5 @ 150 µm | 1.67 | 7.40 | 5.3 | 1.00E+03 | 3 | 2000 | 200 | 2.00E-21 | | Resolution quoted at 150 microns. The survey area is defined to provide sufficient sources in the highest luminosity bin, at z > 3, based on the Bethermin +12 models [Appendix X.x]. There are about 500-1000 ULIRGs per sq. degree at these redshifts. At least 10 redshift bins, 10 merger stage bins, and a range in AGN/SB fractional power, is desired. A sample that covers at least a few (5) degrees is needed. To defeat cosmic variance requires a few sq. degrees, at least. | 200 | 2.00E-21 | | | 7.68E+03 | 1.92E+03 | 8.11E+02 | Time based on survey area at 200 µm |
| Thermo-Chemical History of Comets and Water Delivery to Earth | SS | 10 | 7.61 | 1.85 | 500 - 1000 | 30 @ 545 µm | 2.80 | 4.43 | 3.2 | 1.00E+04 | | 50 | | 2.00E-21 | | | 3.33e3 @ 500 µm | 2.00E-21 | | | 1.36E+02 | 5.80E+00 | 1.70E+00 | Tracking is performed in linear segments at rates ranging from 0.1 milliarcsecond/sec to 1"/sec (from 0.36"/hr to 3600"/hr). |
| Star Formation and Multiphase ISM at Peak of Cosmic Star Formation | EU | 11 | 7.5 | 1.79 | 20 - 500 | 3 @ 150 µm | 1.00 | 13.80 | 9.6 | 3.00E+05 | 10 | 10000 | | 3.00E-21 | | Resolution quoted at lambda=100-150 microns; confusion not an issue as in a line survey, though good to resolve halos. | 6.0e1 @ 20 µm | 3.00E-21 | | | 1.07E+03 | 2.84E+03 | 1.20E+03 | Time based on survey area at 200 µm |
| Magnetic Fields and Turbulence - Role in Star Formation | MW | 12 | 7 | 2.17 | 200 - 500 | 2 @ 100 µm | 1.00 | 12.40 | 8.8 | 3.00E+03 | 10 | | 1000 @ 100 µm | 3.00E-19 | | | 1000 @ 100 µm | 3.00E-19 | | | 1.20E+03 | 1.48E+02 | 1.35E+02 | per 10 sq. deg, 1% pol on 1 mJy source |
| Galaxy Feedback Mechanisms at z<1 | MW | 13 | 6.94 | 1.92 | 20 - 350 | 5 @ 158 µm | 1.60 | 7.75 | 5.5 | 3.00E+03 | 0.1 | 200 | 507 @ 160µm | 1.00E-20 | | | 507 @ 160µm | 1.00E-20 | | | 2.84E+03 | 2.84E+03 | 2.84E+03 | Surface brightness example |
| Survey of Small Bodies in the Outer Solar System | SS | 14 | 6.89 | 2.42 | 100 | 3 @ 100 µm | 1.50 | 8.27 | 5.9 | 3.00E+00 | 1000 | 1000 | 50 @ 100 µm | | | | 50 @ 100 µm | 5.00E-19 | | | 4.80E+03 | 5.93E+02 | 7.30E+01 | These calculations are for 5 sigma 50 ujy at 100 microns. NOT 1 sigma, as the table specifies. So this goes much deeper than is required, but still tractable. Sensitivities are per beam. Here the 15 m column uses the 20 m speed. |
| Ice/Rock Ratio in Protoplanetary Disks | EP | 15 | 6.83 | 2.85 | 30 - 60 | 2.0 @ 43 µm | 2.33 | 5.32 | 3.8 | 2.50E+02 | | 1000 | | | | A significant part of the science can be retained with 2.0" @43 micron, so this should be changed to be the requirement, with 0.5" @43 micron the desired value to achieve the second science goal of spatially resolving SD disks. | 4 @ 60 µm | 6.00E-22 | | | | | | |
| Role of Environment in Galaxy Evolution | G&H | 16 | 6.83 | 2.43 | 80 - 500 | 5 @ 100 µm | 2.50 | 5.00 | 3.6 | 5.00E+02 | 10 | 1000 | | 1.00E-20 | | | 1.33e2 @ 80 µm | 1.00E-20 | | | | | | |
| Frequency of Kuiper Belt Analogues | EP | 17 | 6.67 | 2 | 60 | 1.7 @ 60 µm | 1.40 | 8.86 | 6.3 | 3.00E+00 | | 300 | | 4.25E-02 | | | 2.27 @ 60 µm | 3.78E-02 | | | | | | |
| Galaxies at Reionization | EU | 18 | 6.44 | 2.5 | 30 - 500 | 9.4 @ 400 µm | 1.18 | 0.95 | 0.64 | 5.00E+02 | | 1 | | 3.00E-22 | | | 7.5 @ 30 µm | 3.00E-22 | | | | | | |
| Formation and History of Low-Mass Ice Giant Planets | EP | 19 | 6.17 | 1.95 | 60 - 70 | 1 @ 65 µm | 0.77 | 16.10 | 11.5 | 5.00E+02 | | 150 | | | 0.85 | | 15.7 @ 65 µm | 7.25E-20 | | | | | | |
| Large-Scale Structure - Crucial FIR Link | EU | 20 | 6.11 | 2.25 | 80 - 500 | 5.0 @ 100 µm | 2.50 | 5.00 | 3.6 | 5.00E+02 | 3000 | | | 1.50E-19 | | | 2.0 e3 @ 80 µm | 1.50E-19 | | | | | | |

| Title | | | | | | | | | | | | | | | | Notes | | |
|---|---|---|---|---|---|---|---|---|---|---|---|---|---|---|---|---|---|---|
| Feedback on All Scales in the Cosmic Web | EU | 21 | 5.83 | 2.81 | 300 - 1000 | 60 @ 1000 μm | 10.00 | 4.10 | 2.9 | 3.00E+00 | 3000 | | | | 0.1 | Replaced header/title | 1.11e2 @ 300 μm | 1.11E-19 |
| Obscured AGN | MW | 22 | 5.78 | 1.83 | 20 - 500 | * | | | | 3.00E+03 | | 500 | 3000 | | | | 3000 @ 500 μm | 6.00E-21 |
| Regulating the Multiphase ISM | MW | 23 | 5.67 | 2.63 | 17 - 205 | 5.0 @ 158 μm | 1.60 | 7.75 | 5.5 | 1.00E+05 | 1.25 | 10 | | | 1.00E+04 | Surface brightness requirement is accurate for 5 km/s channel width. | 3.85e6 @ 158 μm | 9.24E-19 |
| Jupiter and Saturn Analogues | EP | 24 | 5.61 | 2.4 | 9 - 40 | 0.1 @ 10μm | 0.50 | 24.80 | 17.7 | 2.00E+02 | | | 1 | | | | 1.0 @ 10 μm | 1.50E-21 |
| Planetary Origins and Evolution of the Solar System | SS | 25 | 5.56 | 2.28 | 10 - 300 | 1 @ 50 μm | 1.00 | 12.40 | 8.84 | 1.00E+04 | | 20 | 100 | 1.00E-05 | | | 3.3e17 @ 10 μm | 1.00E-05 |
| Episodic Accretion in Protostellar Envelopes and Circumstellar Disks | EP | 26 | 5.33 | 2.3 | 50 - 300 | 5 @ 100 μm | 2.50 | 5.00 | 3.6 | 3.00E+00 | | | 2000 | | | Resolution quoted at 100 microns. Clarify survey area vs number of targets – Spitzer Orion Survey was ~9 sq.deg., ~3500 YSOs; this is one way to get the minimum. | 2e3 @ 100 μm | 2.00E-17 |
| Gas and Comets in ExoplanetarySystems | EP | 27 | 5.06 | 1.98 | 60 - 160 | 1 @ 63 μm | 0.79 | 15.60 | 11.1 | 1.00E+04 | | 100 | | 1.00E-21 | | The quoted 1" resolution is at 63 microns, and is required to achieve two of the science goals. | 2.0e2 @ 60 μm | 1.00E-21 |
| Find Planet IX | SS | 28 | 5.06 | 2.86 | 70 - 100 | 2 @ 100 | 1.00 | 12.40 | 8.8 | 3.00E+00 | 20000 | | 1000 | | | | 1000 @ 60 μm | 1.25E-12 |
| Stochastic vs. Secular Accretion in Forming Stars | MW | 29 | 4.78 | 2.6 | 50 - 350 | 3.0 @ 300 μm | 0.50 | 24.80 | 17.7 | 3 | | 1000 | 10000 @ 350 μm | | | Revised angular resolution requirement from discussions with the authors - 3" resolution will resolve 90% of the targets within 2 kpc. | 1e4 @ 350 μm | N/A |
| Cooling Power of Molecular Gas in Star-Forming Regions | EP | 30 | 4.67 | 1.85 | 50 - 500 | 1 @ 63 μm | 0.79 | 15.70 | 11.2 | 5.00E+03 | | 100 | | 1.00E-19 | | Number of targets (100) is for protostars. | 8.33e4 @ 50 μm | 1.00E-19 |
| Star Formation Efficiency Outside the Milky Way | MW | 31 | 4.5 | 1.58 | 10 - 500 | 1.0 @ 100 μm | 0.50 | 26.80 | 17.7 | 1.00E+05 | 5.50E-02 | 40 | | | 0.5 | | 0.92 @ 10 μm | 2.77E-24 |
| Comparative Climate and Thermal Evolution of Giant Planets | SS | 32 | 4.28 | 2.4 | 15 - 500 | 0.5 @ 100 μm | 0.25 | 49.50 | 35.4 | 3.00E+02 | | 4 | 20000 | | | Resolution quoted at lambda= 100 micron | 20000 @ 15 μm | 1.33E-13 |

# Far-IR Interferometer: Strawperson Mission Concepts

D. Leisawitz
13 August 2016

This white paper summarizes the measurement capabilities of two far-IR interferometry mission concepts. There are many design parameters, and different instrument options, but the two most fundamental parameters are light collecting telescope size and maximum baseline length. Light collecting area affects sensitivity, and maximum baseline affects angular resolution, as described in the white paper "Far-IR Interferometry: A Summary of Options and Trades," hereafter referred to as the "SOaT WP."

Here we will consider two configurations:

  A.  A system with two 6.5 m telescopes connected center-to-center by a 50 m long structure; and

  B.  A system with two 3.5 m telescopes connected center-to-center by a 20 m long structure.

In each case, the interferometer will be assumed to be equipped with a "double Fourier" beam combining (DFBC) instrument and at least one more instrument, which can be optimized for sensitivity, high-resolution spectroscopy, polarization measurements, etc. The SOaT WP gives the rationale for a two-telescope interferometer and Michelson (as opposed to Fizeau) beam combination. In this case, the beams are combined in a pupil plane, and the combined light is focused onto two focal planes, one each in the two Michelson output ports; no light is wasted. At each focal plane, there will be multiple detector arrays, one for each wavelength octave.

There is no fundamental limit to the wavelength range, but technical challenges (e.g., mirror quality, alignment tolerances, number of detector pixels) generally increase as wavelength decreases. Tentatively, we will assume a minimum wavelength of 10 μm and a maximum wavelength of 640 μm (6 octaves). The DFBC will contain a single scanning optical delay line that adds and subtracts optical path length between the two arms of the interferometer, one from each telescope. The delay line will provide a scan range long enough to equalize the external optical path length at all field angles in the field of view (FoV), and to provide R ~ 3000 Michelson (FTS) spectroscopy at all wavelengths from 10 to 640 μm in all field positions. As noted in the SOaT WP, a single scanning mechanism provides the optical delay in all wavelength bands.

In both configurations, the interferometer will be astrophysical background photon noise limited. This implies that the telescopes and optical components will be cryocooled to a temperature ~4.5 K, and the detectors will be sufficiently sensitive so as not to limit sensitivity. The temperature requirement is the same as that for a single-aperture telescope. In the DFBC, if broadband light (one octave in wavelength for each detector array) reaches the detectors, the NEP requirement is relaxed relative to a system (or instrument) in which filters or spectral dispersion limit the bandwidth. The shortest wavelength detector array will have the largest number of pixels, and that number will be chosen to provide the desired FoV, which can be several arcminutes in diameter. Here we assume that a 1024 x 1024 pixel array is used for the shortest wavelength channel (10 – 20 μm). Then each successive longer wavelength band will be measured with a detector array that is smaller by a factor of 2, down to a 32 x 32 pixel array in the 320 – 640 μm band, and the field of view (Nyquist sampled Airy disk at the geometric mean wavelength) will be 9.3 arcmin in diameter in system A, and 17.4 arcmin in system B. The pixel count can be adjusted to increase or reduce the FoV, as described in section III of the SOaT WP.

We will assume that in both A and B concepts the telescopes can be moved (e.g., via repositioning along the structure and rotation) to measure any baseline angle in a range of baseline lengths ranging from $b_{min}$ to $b_{max}$, and that a zero spacing FTS spectrum can be measured by splitting the light from a single telescope and scanning the delay line. Since the telescope diameter in both configurations A and B is greater than the size of a centrally-located instrument module, $b_{min}$ is 0. In other words, there will be no central hole in the *u-v* plane. The maximum baseline length is the length of the structure plus one mirror diameter. Table 1 summarizes the baselines accessible in each of the two strawperson configurations.

Table 1. *u-v* plane coverage, image resolution, and field of view in two configurations

| Configuration | $b_{min}$ (m) | $b_{max}$ (m) | Position angles | Image resolution at 50 μm (") | FoV diameter (arcmin) |
|---|---|---|---|---|---|
| **System A** | 0 | 56.5 | All angles accessible | 0.16 | 9.3 |
| **System B** | 0 | 23.5 | All angles accessible | 0.38 | 17.4 |

In operation, the interferometer with its DFBC will have flexibility to provide *u-v* plane coverage ranging from one baseline (a single point in the *u-v* plane) to complete and uniform *u-v* plane coverage, in the latter case (**"full imaging mode"**) by sampling baselines every half mirror diameter (*i.e.*, Nyquist sampling in two dimensions). The image resolution at 50 μm in the case of complete and uniform *u-v* plane coverage out to $b_{max}$ is given in arcseconds in Table 1, and the resolution scales linearly with wavelength (see section II of the SOaT WP), as shown in Figure 1.

**"Snapshot mode"** interferometric observations, described in the SOaT WP, can be useful to survey an area much larger than the instantaneous FoV and obtain astrometric and spectroscopic reconnaissance information on point sources, perhaps inspiring more detailed followup observations in selected fields of interest. In this mode, we will assume that a pair of orthogonal baselines will be observed, in which case positional information will be available to ~λ/2$b_{max}$, or 0.1" or 0.2" for systems A or B, respectively. The DFBC can provide a spectrum at the full spectral resolution, ~3000, and across the entire wavelength range, 10 – 640 μm and the entire field of view.

The DFBC can also be operated in a **"broadband spectral imaging mode"** in which the delay line is scanned over a relatively narrow range to equalize optical path length over a restricted part of the FoV and measure a low-resolution spectrum (R ~ 300), thus concentrating the integration time on source and providing correspondingly better sensitivity. This is especially useful when the science target subtends a small angle, and when low-resolution spectroscopy is sufficient (e.g., to map PAH features in galaxies or water ice broadband spectral features in protoplanetary disks).

Yet another option is one in which a filter or a dispersing element is used to limit the light reaching a detector pixel to that in a narrow spectral band surrounding a line of interest (*e.g.*, the HD 112 μm line or the [O I] line at 63 μm). The sensitivity can be greatly improved in this **"spectral line mapping"** mode, which yields an image of the FoV in a selected spectral line.

The final operating mode we consider here is one in which the delay line is deactivated and the DFBC is used as a broadband imaging instrument rather than an interferometric beam combiner. Regardless of the operating mode, total power images are *always* available at the two DFBC Michelson output ports at resolution $1.22\lambda/D_{tel}$, where $D_{tel}$ is either 3.5 or 6.5 m, but in **"broadband, low-resolution imaging mode"** there are no interferograms and the entire exposure time is dedicated to capturing images in each of the detector arrays. Images at the complementary Michelson output ports will be coadded to take full advantage of the light from both telescopes.

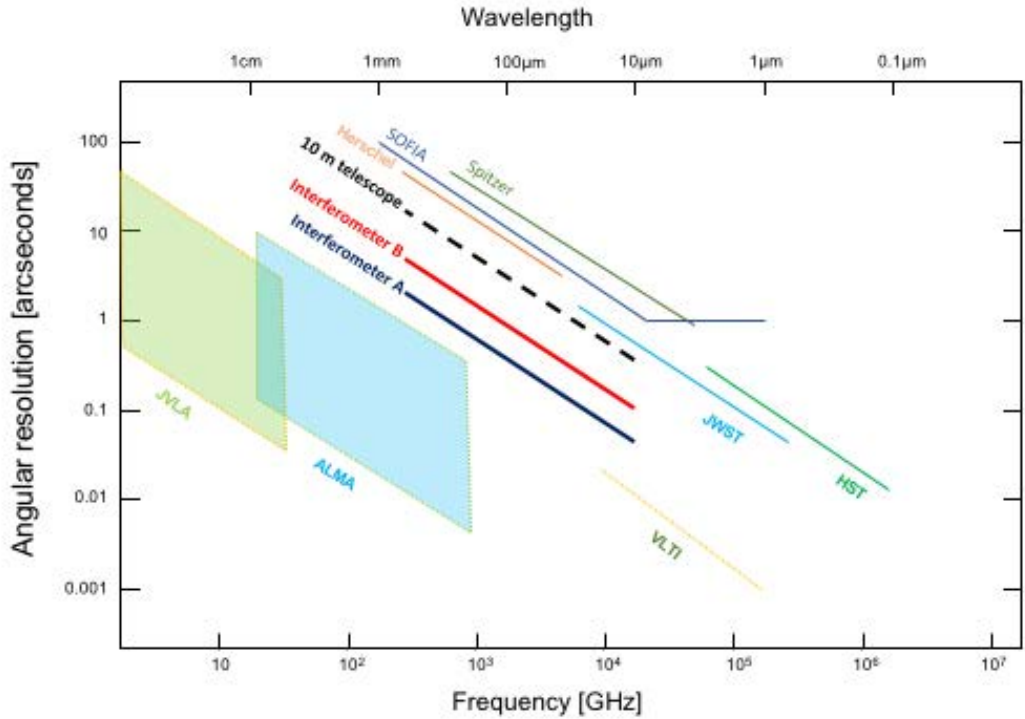

Figure 1. Interferometer concept B provides an order of magnitude better angular resolution than *Herschel*, the largest space telescope flown to date. Concept A provides even better resolution, comparable to that of JWST, but at 10x longer wavelengths. The angular resolution of a 10 m single-aperture telescope is shown for comparison.

In summary, myriad options exist wherein the baseline coverage and the delay line scan range can be chosen based on prior information about the target of interest. For this reason, it is not practical to document the interferometer's performance for every possible permutation of observing parameters. The observing modes mentioned above are simple limiting cases, and we summarize the interferometer's sensitivity in each of these cases in Tables 2 and 3 for concepts A and B, respectively. An important variable is the fraction of time $f_t$ spent collecting data that are useful to the measurement of interest (*e.g.*, time spent on fringes from a single object or resolution element). In these tables MDFD refers to the minimum detectable flux density and MDLF refers to the minimum detectable line flux, and both are quoted at the $5\sigma$ level. The sensitivity performance is traceable viable MDFD and MDLF values quoted by Juanola-Parramon in her white paper, and in turn traceable to the sensitivity performance of a single aperture telescope as derived by Bradford et al.

Table 2. System A, alternative DFBC operating modes

| DFBC Mode | Efficiency $f_i$ by octave | Number of baselines | Minimum observation time assuming 1 minute per delay scan (hrs) | MDFD at 5σ in 1 hour (μJy) | MDLF at 5σ in minimum observation time ($10^{-20}$ W m$^{-2}$) |
|---|---|---|---|---|---|
| Total power images | 1 in all bands | N/A | N/A | 0.74, 1.8, 6.0, 20 in 1 hour at 35, 71, 141, 283 μm | N/A |
| Full spatio-spectral imaging of FoV (Nyquist) | 0.28 0.44 0.61 0.76 0.86 0.93 | 151 | 2.5 | | 3.8, 4.0, 5.8, 9.3 in min obs time at 35, 71, 141, 283 μm |
| Full spatio-spectral imaging of FoV (half Nyquist) | 0.28 0.44 0.61 0.76 0.86 0.93 | 38 | 0.63 | | 7.6, 7.9, 12, 19 in min obs time at 35, 71, 141, 283 μm |
| Broadband spectral imaging (Nyquist), narrow FoV | 1 in all bands | 151 | 2.5 | | 2.5, 3.1, 5.1, 8.7 in min obs time at 35, 71, 141, 283 μm |
| Spectral line mapping of the FoV (e.g., 112 μm) | 0.76 | 151 | 2.5 | | 0.3 at 112 μm |
| Snapshot survey, full FoV | 0.28 0.44 0.61 0.76 0.86 0.93 | 2 | 0.033 | | 33, 34, 51, 81 in 2 minutes at 35, 71, 141, 283 μm |

Table 3. System B, alternative DFBC operating modes

| DFBC Mode | Efficiency $f_i$ by octave | Number of baselines | Minimum observation time assuming 1 minute per delay scan (hrs) | MDFD at 5σ in1 hour (µJy) | MDLF at 5σ in minimum observation time ($10^{-20}$ W m$^{-2}$) |
|---|---|---|---|---|---|
| Total power images | 1 in all bands | N/A | N/A | 2.5, 6.1, 20, 68 in 1 hour at 35, 71, 141, 283 µm | N/A |
| Full spatio-spectral imaging of FoV (Nyquist) | 0.34<br>0.50<br>0.67<br>0.80<br>0.89<br>0.94 | 90 | 1.5 | | 15, 17, 25, 40 in min obs time at 35, 71, 141, 283 µm |
| Full spatio-spectral imaging of FoV (half Nyquist) | 0.34<br>0.50<br>0.67<br>0.80<br>0.89<br>0.94 | 23 | 0.38 | | 31, 33, 49, 79 in min obs time at 35, 71, 141, 283 µm |
| Broadband spectral imaging (Nyquist), narrow FoV | 1 in all bands | 90 | 1.5 | | 11, 14, 22, 38 in min obs time at 35, 71, 141, 283 µm |
| Spectral line mapping of the FoV (e.g., 112 µm) | 0.67 | 90 | 1.5 | | 1.3 at 112 µm |
| Snapshot survey, full FoV | 0.34<br>0.50<br>0.67<br>0.80<br>0.89<br>0.94 | 2 | 0.033 | | 100, 110, 170, 270 in 2 minutes at 35, 71, 141, 283 µm |

The sensitivity values quoted in tables 2 and 3 are for point sources. The sensitivity to a spatially resolved source scales in proportion to the number of resolution elements into which the source is resolved, as described in the SOaT WP. For example, if a source is resolved into ten spatial resolution elements, then the detectable flux per beam (at 5σ) is ten times greater than the sensitivity values quoted above, and one can do better only by considering an interferometer with more collecting area or by spending more observing time on the source.

Next we consider the mapping speed. The time per FoV in snapshot survey mode is limited either by a requirement to reach a certain sensitivity level, or in the extreme, when sensitivity is not a driving requirement, to approximately 2 minutes per FoV (2 optical delay line scans, 1 each at orthogonal angles). The entire field would be surveyed at one baseline angle, then the interferometer would be rotated 90 degrees, and the field re-observed. In total power image mode, either a requirement to reach a certain sensitivity level or the sky confusion noise floor would determine the mapping speed. For comparison with the Bradford et al. calculations for a single aperture telescope, Table 4 shows the time to survey 1 square degree to $5\sigma = 10^{-19}$ W m$^{-2}$ in snapshot survey mode, which yields astrometric as well as spectroscopic information. The time to survey 1 square degree to $5\sigma = 10$ μJy in total power image mode is also tabulated. Although an interferometer would not be well suited to observing a large fraction of the sky, multiple square degree fields can be surveyed in a reasonable amount of time.

Table 4. Mapping speed

| Configuration | DFBC mode of observation | Time to survey 1 deg$^2$ to $5\sigma = 10^{-19}$ W m$^{-2}$ (hours) | Time to survey 1 deg$^2$ to $5\sigma = 10$ μJy (hours) |
|---|---|---|---|
| **System A** | Snapshot survey | 15, 16, 35, 89 at 35, 71, 141, 283 μm | N/A |
| **System A** | Total power imaging | N/A | 0.2, 1.3, 15, 164 at 35, 71, 141, 283 μm |
| **System B** | Snapshot survey | 39, 45, 100, 263 at 35, 71, 141, 283 μm | N/A |
| **System B** | Total power imaging | N/A | 0.7, 4.1, 44, 509 at 35, 71, 141, 283 μm |

In addition to the Double Fourier Beam Combiner (DFBC), the interferometer can be equipped with one or more additional instruments to provide, for example, higher spectral resolution observations. These instruments can use the light collected by one or both of the telescopes that comprise the interferometer, and the sensitivities will scale linearly with collecting area. Table 5 shows the size of a single aperture telescope that provides total collecting area equivalent to that of the pair of telescopes in System A and System B. For these instruments, the sensitivity calculations by Bradford et al. are applicable when they are scaled by the collecting area.

Table 5. Size of a single-aperture telescopes providing equal light collecting area

| Configuration | Interferometer light-collecting telescope size (m) | Single-aperture telescope with the same collecting area, $D_{tel}$ (m) |
|---|---|---|
| **System A** | 3.5 (like Herschel) | 4.9 |
| **System B** | 6.5 (like JWST) | 9.2 |

# Far-IR Surveyor Single-Aperture Concepts and Capabilities

*updated 3 August 2016*

C.M. Bradford + Caitlin Casey, Paul Goldsmith...

August 3, 2016

## Abstract

For FIRS STDT in considering scientific capabilities, architecture selection.

## 1    Concept

We outline broadly the capabilities of an actively-cooled cryogenic telescope . Basic parameters are:

- Telescope size from 5 to 20 meters. Actively cooled to 4.5 K. Temperature impact on sensitivity considered.

- Consider wavelengths from $10\,\mu$m to 1 mm.

- 3 types of instruments considered: broad-band imagers, direct-detection spectrometers, and high-resolution (heterodyne) receivers.

  - Estimates of broadband source confusion are included (from Caitlin Casey).

  - Direct-detection spectrometer backend has R∼500, but a high-resolution mode is envisioned for some of the field

  - Spectrometer spatial formats expected of at least 100 beams, more may be possible.

  - Large (∼100 pixel) heterodyne arrays also under consideration, and sensitivities are provided.

- Key aspect is the large format sensitive direct detectors that are on the near horizon – envision total pixel counts among all instrument approaching 1 million, most in spectroscopy.

- Combination of per-pixel sensitivity and format yields massive new capability for 3-D surveys, which will not be limited by confusion.

## 2 Sensitivity and Speed

### 2.1 Background model

The ultimate sensitivity limit on a cold space telescope with ideal detectors is set by the background: zodiacal (solar system) dust, potentially Galactic dust (cirrus), and the microwave background. In practice the the telescope emission contributes some if it is not cooled below the 2.73 K (the background temperature), but for 4-K and small emissivity, this is only a small perturbation.

The background levels depends in detail on the position and time of observation in the earth's orbit. The ecliptic poles offer the lowest zodi backgrounds, so we consider this as the sensitivity for deep field observation. For a fiducial for the FIR Surveyor, we use intensities provided by Bidushi Bhattacharya and William Reach in 2006, based on COBE DIRBE measurements ([2, 6, 7, 5, 4, 3]) as part of preparation for Herschel; they refer to a typical line of sight toward the North Ecliptic Pole from an Earth-Sun L2 orbit. (This estimate has also been used for the BLISS-SPICA sensitivity estimates [1], as well as the CALISTO white paper in 2015.) Figure 1 plots the intensities used this estimate. A glance at the DIRBE annual average maps indicates that these estimates are reasonable; slightly lower intensities ($\sim$2–3$\times$) may be found at $\lambda \sim 100 \mu m$, but this is only a factor of 1.4–1.7 in NEP. Sky averages are, of course, much brighter than this model.

Conversion of these intensities into photon occupation number and the resulting photon noise is straightforward and is provided in Appendix **??**.

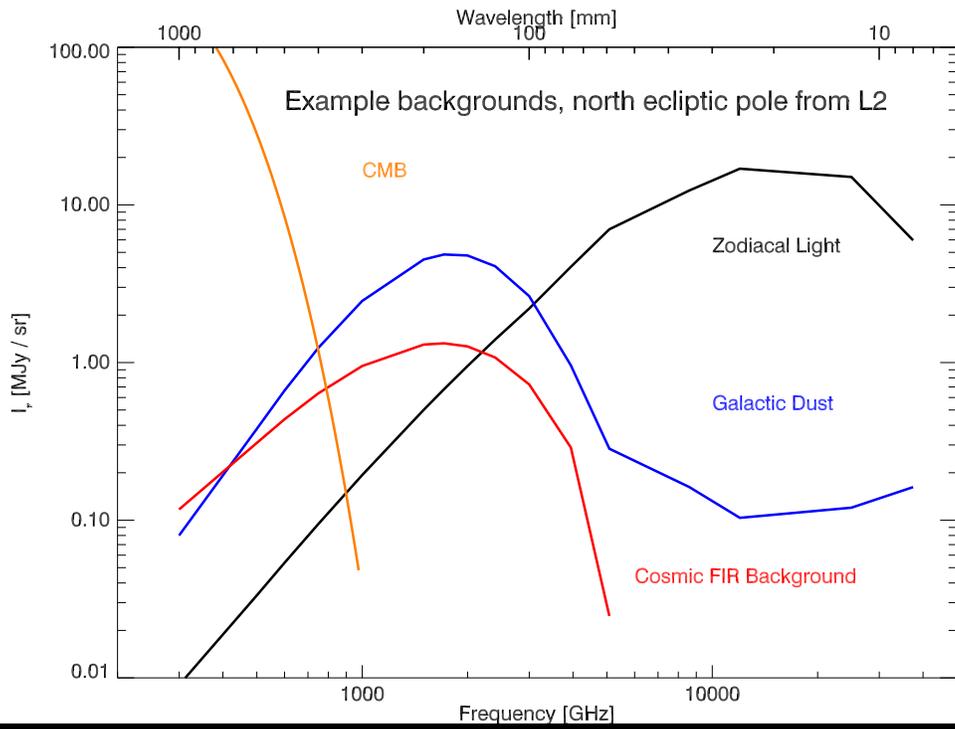

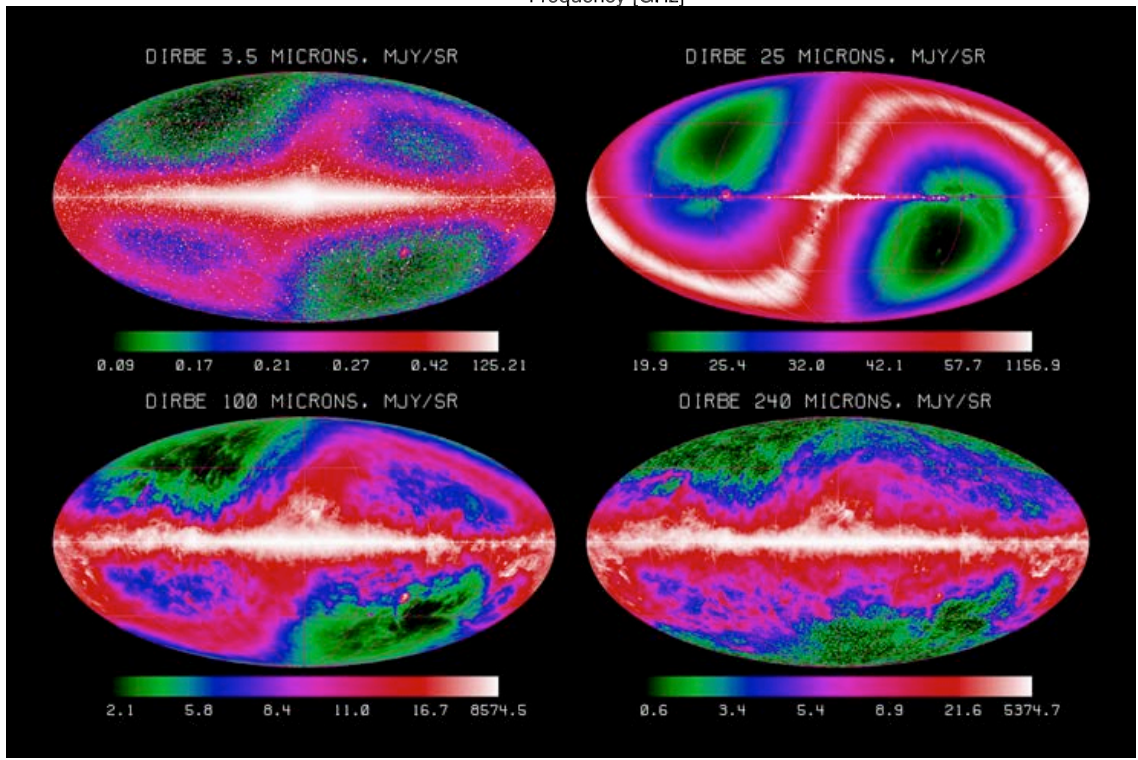

Figure 1: TOP: background model for sensitivity calculations. From Bhattacharya and Reach (personal communication 2006), for viewing the north ecliptic pole from earth-sun L2. Bottom, DIRBE annual average sky maps for comparison.

## 2.2 Direct-detection instrument parameters

Sensitivities for all instruments are based on a diffraction-limited, high-Strehl telescope with good point source coupling. Heterodyne spectrometers will be treated separately. For the direct-detection imagers and spectrometers, we assume:

1. Background from a single mode (diffraction-limited beam) of radiation, including a 4.5 K telescope with 3% emissivity.
2. An aperture efficiency of 75% to the point source.
3. An instrument transmission of 25% for the spectrometers and 40% for the imagers.
4. Spectrometer channel bandwidth of 1 part in 500, imager bandwidth of 10%.
5. Detector NEP of $2 \times 10^{-20}\,\mathrm{W\,Hz^{-1/2}}$ for the spectrometers (and perfect detectors for the imagers).
6. Single polarization (appropriate for spectrometers, conservative for imagers unless configured as polarimeters).
7. A factor of 2 penalty for extraction of the point-source signal to account for chopping (conservative given multi-beam systems, particularly for the imaging).
8. A further factor of 1.8 to account for additional non-idealities: in particular, extraction of the signal using multiple detectors.

For the moment, we will not consider instrument stability or low-level systematics, so that throughout, one may scale per-beam sensitivity figures with $1./\sqrt{\mathrm{time}}$, and mapping speeds as $\mathrm{Sky\ Area}[\mathrm{sr\ or\ deg^2}]/\sqrt{\mathrm{time}}$.

## 2.3 Broad-band imaging and confusion

Imaging with the cold telescope is very fast, so large surveys are possible. Source confusion is a sensitivity limitation in the 2-D dataset, and we include confusion limits corresponding to 15 beams per source from Caitlin Casey, estimated with her source-count models.

Figure 2 shows the raw sensitivity [in mJy units] in 1 second, along with the confusion limit. Figure 3 shows fiducial surveys that could be carried out in 1000 hours: one shallow covering 1000 square degrees, and one deep covering 10 square degrees. Here we assume a camera focal plane with a fixed size on the sky of 10,000 beams at 100 $\mu$m (e.g. 100 × 100 beams). This corresponding to approximately 14 square arc minutes on the 10-meter telescope, and of course the number of beam scales as $1/\lambda^2$. It is envisioned that rapid, large-field imaging programs could be carried out in a drift-scan mode in which the observatory rotates and the arrays are read out continuously, fast enough to (as with CMB experiments such as Planck, as well as Herschel SPIRE). At wavelengths below 30 microns, the doped-silicon photoconductors could be used – these have large formats, but might require a different scan approach or perhaps a scan mirror, as they require typically stare.

The details of camera pixel sampling are not essential to the basic confusion limit and speed figures, at least for large maps. For completeness, we note that we are assuming a single spatial mode of radiation per detector. The actual size of the array could be larger than the 14 square arcminutes if, for example, the detectors are coupled with feedhorns, but this doesn't change the speed figure.

Obviously, scaling the mapping speeds to areas smaller than the ∼10 square arcminute field size is not appropriate.

Finally, in Table 1 we cast the confusion limits into luminosity depths ($\nu L_\nu = \nu F_\nu \times 4\pi d_{\text{luminosity}}^2$). (An improvement would be to tabulate the depths based on actual SEDs).

*ToDo: Investigate detector time constants required for the rapid surveys, and number of turnarounds assuming square fields of the 1 and 100 square degree options.*

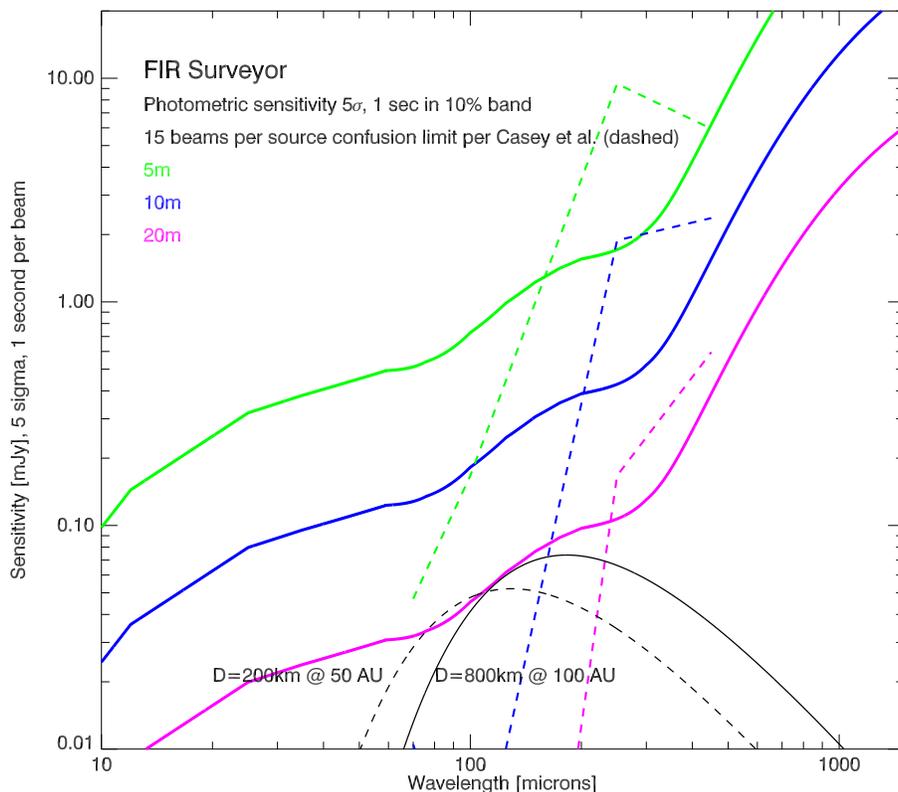

Figure 2: Broad-band imaging sensitivity for the Far-IR Surveyor, and confusion limits (15 beams per source) from Caitlin Casey. Also shown are blackbody fluxes for fiducial solar system objects, assuming T=278 K $/\sqrt{d[\text{AU}]}$ (this may be too warm....)

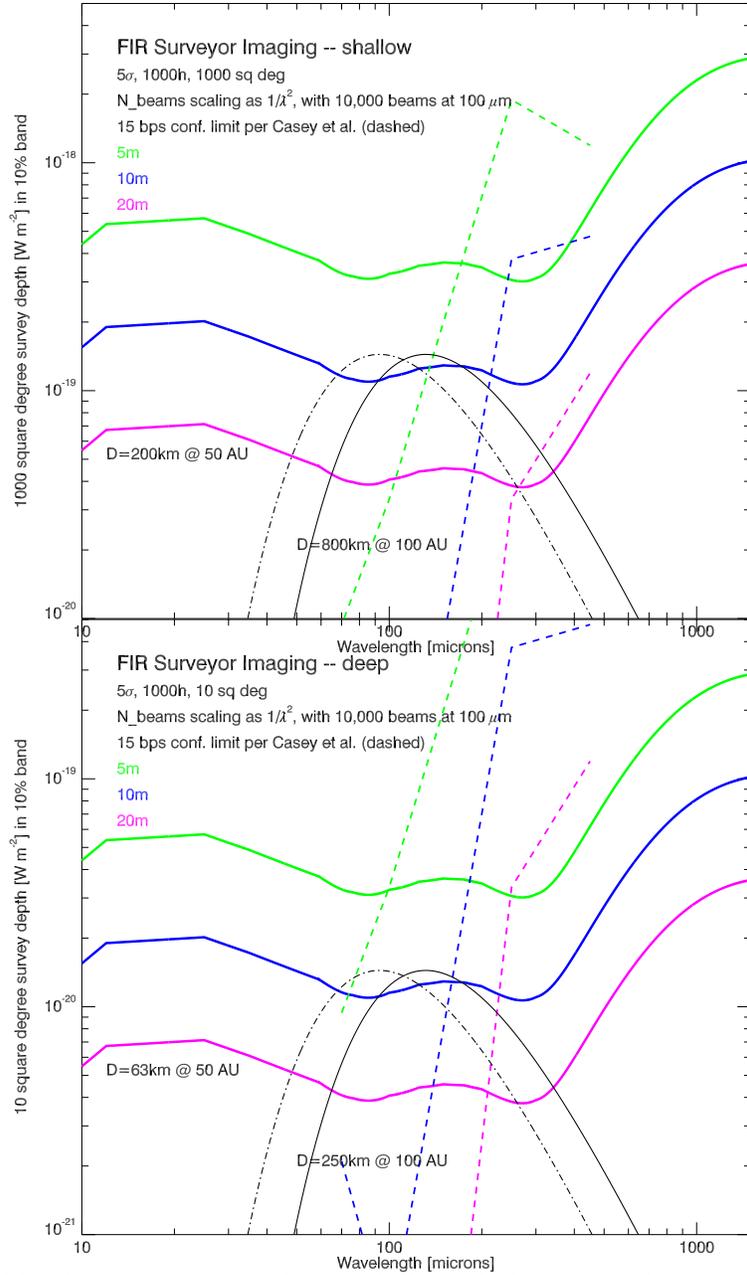

Figure 3: Fiducial broad-band imaging surveys. Above is shallow, which assumes 1 hour per square degree. Below is deep, which assumes 100 hours per square degree. We assume a camera focal plane with a fixed size on the sky, it is 10,000 beams at 100 $\mu$m (100 × 100 beams) corresponding to approximately 14 square arc minutes on the 10-meter telescope, scaling as $1/D^2$

### 2.3.1 Polarimetry

The above sensitivities assume single-polarization, in anticipation that the camera pixels would be compatible with a polarimeter. Crude estimates of polarization mapping capability to a given depth can be obtained by simply multiplying the unpolarized time estimates corresponding to that depth by 2 and by the inverse square of the fractional polarization. For example, for a $5\sigma$ detection on a 1% polarization signal, the time required would be 20,000 × that required to detect the full continuum signal at the same significance. The 2 is because the measurement requires a difference between two measurements (e.g. Stokes Q, U). (*should check this with a polarimetry expert*)

Table 1: FIR Surveyor: Broad-Band Confusion Limits and Luminosity Depths

| λ | | 5m | | | | 10m | | | | 20m | | |
|---|---|---|---|---|---|---|---|---|---|---|---|---|
| μm | depth [mJy] | fr sky | $\nu L_\nu$ z=2 | $\nu L_\nu$ z=6 | dth[mJy] | fr sky | $\nu L_\nu$ z=2 | $\nu L_\nu$ z=6 | dth[mJy] | fr sky | $\nu L_\nu$ z=2 | $\nu L_\nu$ z=6 |
| 70 | 4.7E-02 | 9.2E-03 | 3.3E+07 | 4.6E+08 | 1.0E-02 | 3.7E-03 | 7.3E+06 | 1.0E+08 | 2.6E-03 | 1.8E-03 | 1.8E+06 | 2.5E+07 |
| 100 | 1.7E-01 | 5.7E-02 | 1.2E+08 | 1.6E+09 | 1.9E-03 | 5.8E-05 | 1.3E+06 | 1.8E+07 | 4.2E-06 | 2.3E-09 | 2.9E+03 | 4.1E+04 |
| 250 | 9.4E+00 | 8.0E+01 | 6.6E+09 | 9.2E+10 | 1.9E+00 | 2.5E+01 | 1.3E+09 | 1.8E+10 | 1.7E-01 | 1.6E+00 | 1.2E+08 | 1.6E+09 |
| 450 | 5.9E+00 | 1.0E+00 | 4.2E+09 | 5.8E+10 | 2.4E+00 | 1.3E+00 | 1.7E+09 | 2.3E+10 | 6.0E-01 | 6.4E-01 | 4.2E+08 | 5.8E+09 |

Notes: Confusion depths in mJy from Caitlin Casey (see Figure 4). Fraction of sky is the fraction of sky that can be covered to this depth ($5\,\sigma$) in 1000 hours with the camera described above (14 square arcmin on the 10-m, scales as $1/D^2$). Approximately luminosity limits are just $\nu F_\nu$ converted into luminosity with the luminosity distance – does not take into account SED details, in particular the benefit of the submillimeter negative K-correction which would be important for the larger telescopes at the long wavelengths.

```
# diameter[m]  conflim_5bps[mJy]  conflim_10bpc[mJy]  conflim_15bpc[mJy]
#100um
       5.00000     0.00530885         0.0595662          0.167881
       6.00000     0.00133352         0.0188365          0.0668344
      10.0000      1.49624e-05        0.000334966        0.00188365
      12.0000      2.98538e-06        6.68344e-05        0.000421697
      15.0000      1.05925e-06        9.44061e-06        5.95662e-05
      20.0000      1.05925e-06        1.05925e-06        4.21696e-06
#250um
       5.00000     2.66073            6.68344            9.44061
       6.00000     1.67881            4.21696            6.68344
      10.0000      0.298538           0.944062           1.88365
      12.0000      0.149624           0.530885           1.05925
      15.0000      0.0595662          0.237137           0.473151
      20.0000      0.0211349          0.0749894          0.167881
#450um
       5.00000     2.98538            4.73151            5.95662
       6.00000     2.11349            3.75838            4.73151
      10.0000      0.841395           1.49624            2.37137
      12.0000      0.595662           1.18850            1.67881
      15.0000      0.375838           0.749894           1.05925
      20.0000      0.211349           0.421696           0.595662
#850um
       5.00000     2.98538            4.21696            5.30885
       6.00000     2.66073            3.75838            4.21696
      10.0000      1.18850            2.11349            2.66073
      12.0000      0.841395           1.49624            2.11349
      15.0000      0.595662           1.05925            1.49624
      20.0000      0.334965           0.668344           0.944062
#1100um
       5.00000     1.88365            2.66073            3.34965
       6.00000     1.33352            2.11349            2.66073
      10.0000      0.530885           1.05925            1.49624
      12.0000      0.334965           0.749894           1.05925
      15.0000      0.211349           0.473151           0.668344
      20.0000      0.105925           0.237137           0.375838
```

Figure 4: Confusion limit table from Caitlin Casey.

## Table 2: Far-IR Surveyor Spectrometer Backends: R=500 Strawman (on 10m)

| Parameter | 30 $\mu$m | 50 $\mu$m | 100 $\mu$m | 200 $\mu$m | 400 $\mu$m | Scaling w/ $D_{\rm eff}$ |
|---|---|---|---|---|---|---|
| Photon-noise limited NEP [W Hz$^{-1/2}$] | 4.6e-20 | 3.6e-20 | 2.8e-20 | 3.0e-20 | 3.7e-20 | ... |
| Beam size | 0.70$''$ | 1.16$''$ | 2.3$''$ | 4.7$''$ | 9.3$''$ | $\propto D^{-1}$ |
| Instantaneous FOV [sq deg] | 1.1e-6 | 3.0e-6 | 1.19e-5 | 4.7e-5 | 1.9e-4 | $\propto D^{-2}$ |
| Line sensitivity [W m$^{-2}$], 5$\sigma$, 1h | 9.7e-22 | 7.7e-22 | 6.4e-22 | 6.8e-22 | 8.1e-22 | $\propto D^{-2}$ |
| Pt. sce. map. speed [deg$^2$/($10^{-19}$W m$^{-2}$)$^2$/sec] | 1.5e-5 | 6.7e-5 | 3.8e-4 | 1.4e-3 | 3.8e-3 | $\propto D^2$ |
| Surface bright. sens. per pix [W m$^{-2}$ sr$^{-1}$], 5$\sigma$, 1h | 1.7e-10 | 4.9e-11 | 1.0e-11 | 2.7e-12 | 8.0e-13 | $\propto D^0$ |
| Surface bright. sens. per pix [MJy/sr $\sqrt{\rm sec}$], 1$\sigma$ | 5.1 | 2.4 | 1.01 | 0.53 | 0.32 | $\propto D^0$ |

Numbers are for a 10-meter telescope, can be scaled per the last column. Notes: Sensitivities assume single-polarization instruments with a product of cold transmission and detector efficiency of 0.25, and an aperture efficiency of 0.75. FOV estimate 100 beams. (Each beam has $\theta_{\rm FWHM}$ of 1.13$\lambda$/D, appropriate for an assumed 10dB edge taper, and a solid angle of $\frac{\pi}{4\ln 2}\,\theta_{\rm FWHM}^2$.) Sensitivities assume background is subtracted with no penalty.

## 2.4 Direct-Detection Spectroscopy

Broad-band moderate-resolution spectroscopy throughout the far-IR will be an especially powerful feature of the Far-IR Surveyor, as confusion is negligible for spectroscopy (see Figure 6). The concept is a suite of 5–8 grating spectrometer modules (or equivalent), each covering a large fraction of an octave of bandwidth instantaneously at a modest resolving power (e.g. R=500). The modules abut one another spectrally so that they combine to cover the full 1–2 decades of the Far-IR Surveyor band (e.g. 10–500 $\mu$m). Each spectrometer is optimized for both point source spectroscopy, but also has at least 100 spatial beams. A goal is to align the fields of the various spectrometer bands using dichroic and polarization splitters — this would enable the fastest possible full-band spectroscopy of a given source. Figure 5 shows the spectral-line sensitivity and spectral-line mapping speed of such a system.

**3-D survey speed, example depths.** The 3-D survey speed is tremendous, particularly at the long wavelengths, and multiple-depth 'wedding cake' surveys are envisioned. Table 3 indicates the depths possible in 3000 hours dedicated to various survey areas. We also attempt to estimate the number of galaxy detections, based on the line counts. The wide surveys will detect millions of galaxies blindly in the fine-structure lines, while the deep surveys allow access to the early-Universe and faint galaxy populations. In addition to the detected sources, the 3-D maps will be ideal for tomography (3-D intensity mapping) techniques which can measure clustering and thus total luminosity density of the undetected sources.

**Sensitivity to broad features such as PAHs.** The brightest of the PAH features carry the most luminosity of any spectral feature (up to 4% of the total IR), and so will be the easiest to detect. The PAH features have a natural width of 3–10%, corresponding to R=30 to 10, so whether observed via binning a R=500 system, or with a dedicated low-res system, a PAH feature will be subject to the background noise in this larger bandwidth. Thus for detection of PAHs, the tabulated W m$^{-2}$ sensitivities must be multiplied by $\sim \sqrt{500/30} \sim 4$ (considering the narrow single features which would be matched to R=30).

**High-resolution mode.** While the native grating resolution is only a few hundred, it is possible to provide higher resolution by adding a spectral element in front of the grating. In particular, a high-efficiency, folded-path Fourier-transform spectrometer (FTS) (e.g. a Martin-Puplett which offers near unity transmission in a single polarization) can offer at least 10× increase in resolving power relative to the native grating system, without a large volume, and without a substantial time (sensitivity) penalty in W m$^{-2}$. Thus resolving powers in the few thousand range should be possible with the FTS + grating mode. The field of view through this system is a subject for study. Fields of several ($\sim$10) beams about a central field position in each band are certainly possible, but it will not likely support a full 100-beam-long slit for an echelle grating spectrometer without an image-slicing element.

Table 3: Far-IR Surveyor Example 3000-hour Spectroscopic Surveys (on 10m)

| 30 $\mu$m | 50 $\mu$m | 100 $\mu$m | 200 $\mu$m | 400 $\mu$m | approx. galaxy yield |
|---|---|---|---|---|---|
| LSS survey: 5000 square degrees – 5$\sigma$ depths [W m$^{-2}$] | | | | | |
| 6.1e-19 | 2.9e-19 | 1.21e-19 | 6.4e-20 | 3.8e-20 | ~1 billion |
| wide: 100 square degrees – 5 $\sigma$ depths [W m$^{-2}$] | | | | | |
| 8.6e-20 | 4.1e-20 | 1.7e-20 | 9.0e-21 | 5.4e-21 | ~60 million |
| deep: 3 square degrees – 5 $\sigma$ depths [W m$^{-2}$] | | | | | |
| 1.5e-20 | 7.1e-21 | 3.0e-21 | 1.6e-21 | 9.3e-22 | ~4 million |
| ultra-deep: 110 square arcminutes – 5 $\sigma$ depths [W m$^{-2}$] | | | | | |
| 1.5e-21 | 7.1e-22 | 3.0e-22 | 1.6e-22 | 9.3e-23 | ? |

Numbers are for a 10-meter telescope, using R=500 sensitivities. Assumes that background is subtracted with no penalty. Depth scales as 1/D$_{\rm tel}^2$. Instrument is assumed to be 100 beams in all bands. Survey yield is based on line count estimates at 200 microns, integrating over 1 octave of bandwidth (e.g. 140–280 $\mu$m) based on line counts from Murphy et al.

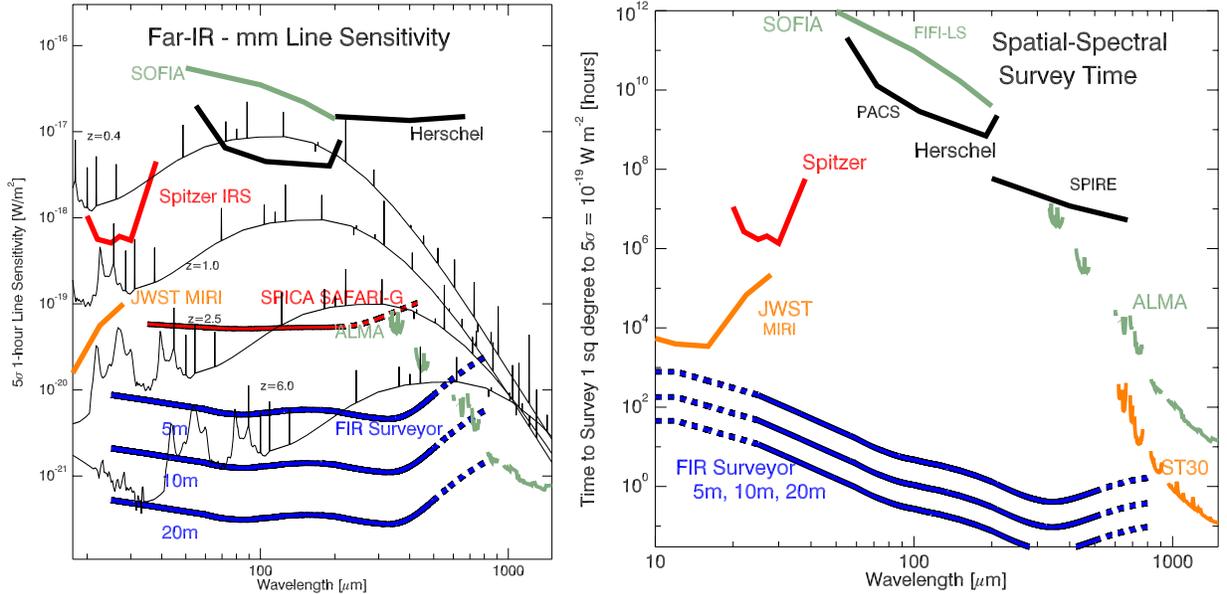

Figure 5: Spectroscopic sensitivities in the far-IR and submillimeter. Left shows the sensitivity in W m$^{-2}$ for a single pointed observation. Galaxy spectra assuming L = $10^{12}$ L$_\odot$ at various redshifts are overplotted using light curves, with continuum smoothed to R=500. The right panel shows the speed for a blind spatial-spectral survey reaching a depth of $10^{-19}$ W m$^{-2}$ over a square degree, including the number of spatial beams and the instantaneous bandwidth. We assuming R=500 grating spectrometers with 100 beams (a conservative figure) and 1:1.5 instantaneous bandwidth. Detectors are assumed to operate with NEP = $2\times10^{-20}$ WHz$^{-1/2}$, a figure which has been demonstrated in the lab. The SPICA / SAFARI-G curve refers to the new configuration: a 2.5-meter telescope with a suite of R=300 grating spectrometer modules with 4 spatial beams, and detectors with NEP=$2\times10^{-19}$ WHz$^{-1/2}$. ST30 represents a 30-meter class wide-field submillimeter telescope in the Atacama, such as CCAT, equipped with 100 spectrometer beams, each with 1:1.5 bandwidth. ALMA band averaged sensitivity, and survey speed based on 16 GHz in the primary beam.

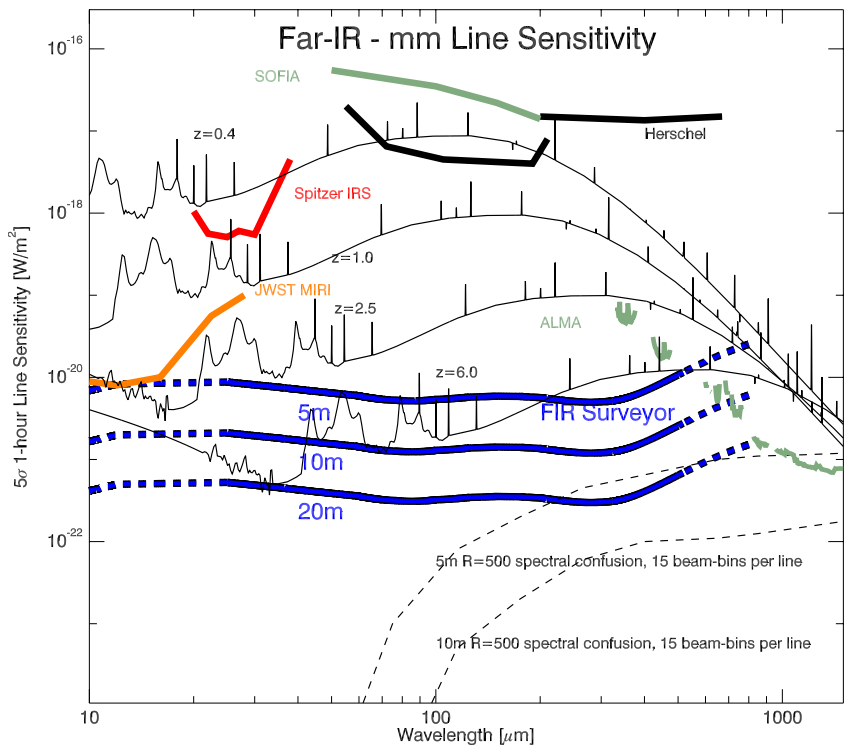

Figure 6: Source confusion for moderate-resolution spectroscopy, based on line-count models—luminosity functions with line templates incorporated. When each spatial pixel is divided into many R~ few × 100 bins, the density of line emitters is much less than unity, even for very interesting depths.

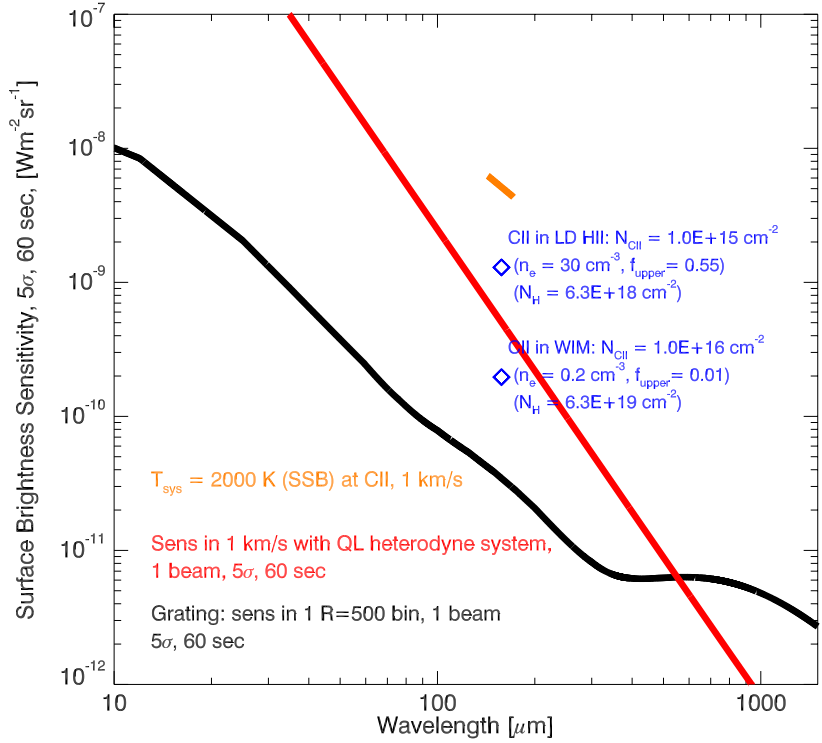

Figure 7: Surface brightness sensitivity for diffuse line emission, for both direct-detection and heterodyne instrumentation. Heterodyne systems assume 1 km/s linewidth, and their sensitivity in $\mathrm{W\,m^{-2}\,sr^{-1}}$ will scale as $\sqrt{\delta v}$. The [CII] column density point assumes a temperature of 1000 K, and an electron density of 1 $\mathrm{cm^{-3}}$, so that the occupation in the upper level is only 8%. Thermalized [CII] with levels populated according to the statistical weights would be $\sim 8\times$ brighter.

## 2.5 Heterodyne Spectroscopy

Heterodyne instrumentation is definitely a possibility for a single-dish telescope, and is the only way to achieve sub-km/s resolution that is desired for some applications. While the sensitivities for heterodyne systems do not benefit from the cold platform, it should be possible to include heterodyne instruments in such a way that they do not impact the overall mission. As an example, at 158 $\mu$m for [CII], Paul Goldsmith has suggested that a $\sim$100-pixel system with per-pixel system temperatures of 2000 K should be obtainable (single-sideband). The line sensitivity ($1\sigma$) of such a system for a narrow-band measurement would be

$$\delta\mathrm{F}_{\mathrm{RMS,het}}\left[\mathrm{Wm}^{-2}\right] = 2\frac{\mathrm{T}_{\mathrm{sys}}}{\sqrt{\delta\nu\,\mathrm{t}_{\mathrm{int}}}} \times \frac{\mathrm{k_B}\delta\nu}{\mathrm{A}_{\mathrm{eff}}} = 4.2\times10^{-20}\,\mathrm{Wm}^{-2}\sqrt{\frac{\delta\mathrm{v}}{1\,\mathrm{km/s}}}\sqrt{\frac{1\mathrm{hour}}{\mathrm{t}_{\mathrm{int}}}}\left(\frac{10\mathrm{m}}{\mathrm{D}_{\mathrm{tel}}}\right)^2 \qquad (1)$$

where the second equality holds for CII assuming the 2000 K $\mathrm{T}_{\mathrm{sys}}$. The factor of 2 accounts for the fact that the receiver operates single-polarization, but the flux refers to both polarizations.

This sensitivity is based on Herschel HIFI receiver sensitivities. However, the ultimate limit for a heterodyne system is the quantum noise, which could yield $T_{\mathrm{rec}} = h\nu/k$ (e.g. Kerr, Feldman, and Pan, 1997), which would provide a system temperature which is perhaps $2\times\mathrm{T}_{\mathrm{rec}}$ to account for differencing or non-idealities, beam coupling, etc. The flux sensitivity of this best-case system assuming $\mathrm{T}_{\mathrm{sys}} = 2h\nu/k$ is then

$$\delta\mathrm{F}_{\mathrm{RMS,het,QL}}\left[\mathrm{Wm}^{-2}\right] = 2\frac{\mathrm{T}_{\mathrm{sys}}}{\sqrt{\delta\nu\,\mathrm{t}_{\mathrm{int}}}} \times \frac{\mathrm{k_B}\delta\nu}{\mathrm{A}_{\mathrm{eff}}} = 1.5\times10^{-21}\,\mathrm{Wm}^{-2}\left(\frac{\nu}{1\,\mathrm{THz}}\right)^{3/2}\sqrt{\frac{\delta\mathrm{v}}{1\,\mathrm{km/s}}}\sqrt{\frac{1\mathrm{hour}}{\mathrm{t}_{\mathrm{int}}}}\left(\frac{10\mathrm{m}}{\mathrm{D}_{\mathrm{tel}}}\right)^2 \qquad (2)$$

In Figures 8 and 7 we plot these $\mathrm{T}_{\mathrm{sys}} = 2h\nu/k$ sensitivities in $\mathrm{W\,m}^{-2}$ units, assuming the 1 km/s velocity resolution channel.

Brightness temperature units are perhaps a more natural unit for high-resolution spectroscopy. For reference, we note that the temperature sensitivity of the fiducial quantum limited system is just:

$$\delta\mathrm{T}_{\mathrm{RMS,het,QL}}\left[\mathrm{mK}\right] = \frac{\mathrm{T}_{\mathrm{sys}}}{\sqrt{\delta\nu\,\mathrm{t}_{\mathrm{int}}}} = 0.88\mathrm{mK}\sqrt{\frac{\nu}{1\,\mathrm{THz}}}\sqrt{\frac{1\,\mathrm{km/s}}{\delta\mathrm{v}}}\sqrt{\frac{1\mathrm{hour}}{\mathrm{t}_{\mathrm{int}}}} \qquad (3)$$

and is, of course, independent of aperture.

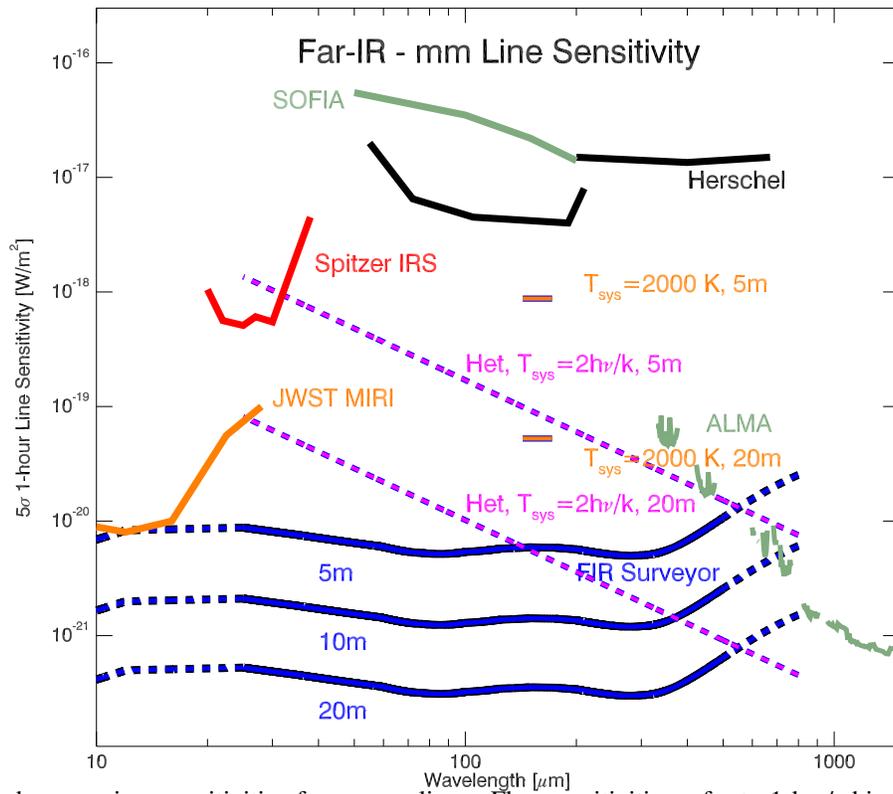

Figure 8: Heterodyne receiver sensitivities for narrow lines. Flux sensitivities refer to 1 km/s bins, and sensitivity in these units scales as $1/\sqrt{\delta v}$. 5-meter and 20-meter sensitivities are shown. Gold shows $T_{sys} = 2000$K, as is currently achievable for [CII]. Magenta shows a quantum limited system with $T_{sys} = 2\,h\nu/k$. See Figure 5 caption for explanation of the blue grating spectrometer curves.

## 3 Telescope Concepts

### 3.1 Impact of Telescope Temperature

Actively cooling the telescope to near the microwave background temperature is vital for obtaining the full sensitivity available. Figure 9 shows this effect as colored curves diverging from the 5-meter telescope sensitivity. These calculations assume an emissivity of 4% for the telescope, a figure which is challenging but doable for an on-axis telescope. An off-axis system could be better.

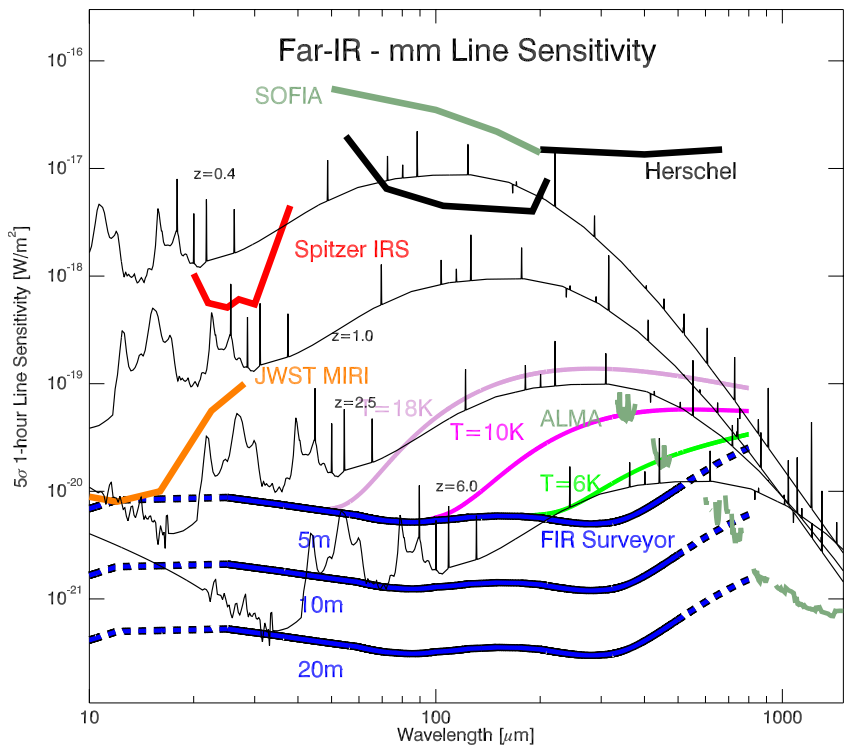

Figure 9: Impact of telescope temperature on direct-detection sensitivity. Curves refer to the 5-meter telescope, and assume 4% emissivity. Even 10 K results in nearly an order of magnitude sensitivity penalty in the 200–400 $\mu$m band ( 2 orders of magnitude in speed).